\documentclass[amsmath, amsfonts]{article}
\usepackage{graphicx}
\usepackage{epsfig}
\usepackage{bm}
\usepackage{euscript}
\usepackage{dcolumn}
\usepackage{color}
\usepackage{amsmath}
\usepackage{amssymb}
\usepackage[varg]{txfonts}

\def\Tr{\mathrm{Tr}}
\def\tr{\mathrm{tr}}
\def\R{\mathrm{Re}}
\def\I{\mathrm{Im}}
\def\d{\mathrm{d}}
\def\D{\mathbf{D}}
\def\rt{(\mathbf{r},t)}
\def\rtp{(\mathbf{r},t')}
\def\qo{(\mathbf{q},\omega)}
\def\qom{(-\mathbf{q},-\omega)}
\def\Qk{\hat{Q}_{\EuScript{K}}}
\def\tauel{\tau_{\mathrm{el}}}

\begin{document}

\title{{\bf Keldysh technique and non--linear $\sigma$--model:\\
basic principles and applications}}

\author{Alex Kamenev and Alex Levchenko\\
\hskip0cm {\small {\it Department of Physics, University of
Minnesota, Minneapolis, MN 55455, USA}}}

\date{April 27, 2009}

\maketitle

\begin{abstract}

The purpose of this review  is to provide  a comprehensive
pedagogical introduction into Keldysh technique for interacting
out--of--equilibrium fermionic and bosonic systems. The emphasis is
placed on a functional integral representation of underlying
microscopic models. A large part of the review is devoted to
derivation and applications of the non--linear $\sigma$--model for
disordered metals and superconductors. We discuss such topics as
transport properties, mesoscopic effects,  counting statistics,
interaction corrections, kinetic equation, {\em etc.} The chapter
devoted to disordered superconductors includes Usadel equation,
fluctuation corrections, time--dependent Ginzburg--Landau theory,
proximity and Josephson effects, {\em etc.} (This review is a
substantial extension of arXiv:cond-mat/0412296.)

\bigskip

\end{abstract}

\newpage

\tableofcontents

\newpage

\section{Introduction}\label{sec_intro}
\subsection{Motivation and outline}\label{sec_motivation}

This review  is devoted to the Keldysh formalism for the treatment
of out--of--equilibrium  interacting many--body systems. The
technique takes its name from the 1964 paper of
L.~V.~Keldysh~\cite{Keldysh-1964}. Among earlier approaches that are
closely related to the Keldysh technique, one should mention
Konstantinov and Perel~\cite{KonstantinovPerel},
Schwinger~\cite{Schwinger}, Kadanoff and Baym~\cite{KadanoffBaym},
and Feynman and Vernon~\cite{Feynman}. Classical counterparts of the
Keldysh technique are extremely useful and interesting on their own
right. These include the Wild diagrammatic technique~\cite{Wyld} and
Matrin--Siggia--Rose (MSR) method~\cite{MSR} for  stochastic systems
(see also related work of DeDominicis~\cite{DeDominicis}).

There is a number of  presentations of the method in the existing
literature~\cite{Lifshitz,RammerSmith,SmithJensen-book,Mahan,Rammer-book,SchwabRaimondi,Spicka}.
The emphasis of this review, which is a substantially extended
version of Les Houches Session \textit{LXXXI}
lectures~\cite{Kamenev}, is on the functional integration approach.
It makes the structure and the internal logic of the theory
substantially more clear and  transparent. We focus on  various
applications of the method, exposing connections to other techniques
such as the equilibrium Matsubara method~\cite{Matsubara,AGD} and
the classical Langevin and Fokker--Planck (FP)
equations~\cite{Gardiner,Risken}. The major part of the review is
devoted to a detailed derivation of the non--linear $\sigma$--model
(NLSM)~\cite{HorbachSchon,KamenevAndreev,ChamonLudwigNayak,
Feigel'manLarkinSkvortsov}, which is probably the most powerful
calculation technique in the theory of disordered metals and
superconductors. This part may be considered as complimentary to
earlier presentations of the replica
\cite{Wegner,EfetovLarkinKhmelnitskii,Finkel'stein,BelitzKirkpatrick,Lerner}
and the supersymmetric \cite{Efetov,Efetov-book,Mirlin} versions of
the $\sigma$--model.

Our aim is to expose the following applications and advantages of
the Keldysh formulation of the many--body theory.
\begin{itemize}
\item Treatment of systems away from  thermal equilibrium,
either due to the presence of external  fields, or in a transient
regime.

\item An alternative to replica and
supersymmetry methods in the theory of systems with quenched disorder.

\item Calculation of the full counting statistics of a
quantum  observable, as opposed to its average value or
correlators.

\item Treatment of equilibrium problems, where
Matsubara analytical continuation may prove to be cumbersome.
\end{itemize}
Our intent is not to cover all applications of the technique that
have appeared previously in the literature. We rather aim at a
systematic and self--contained exposition, helpful for  beginners.
The choice of cited literature is therefore very partial and
subjective. It is mainly intended to provide more in--depth details
about the chosen examples, rather than a comprehensive literature
guide.

The outline of the present review is as follows. We first introduce
the essential elements of the Keldysh method: concept of the closed
contour Sec.~\ref{sec_contour}, Green's functions, {\em ext}.,
starting from a simple example of non--interacting system of bosons,
Sec.~\ref{sec_bosons}, and fermions, Sec.~\ref{sec_fermion}. Boson
interactions, the diagrammatic technique and  quantum kinetic
equation are discussed in  Sec.~\ref{sec_int_bos}. Section
\ref{sec_environment} is devoted to a  particle in contact with a
dissipative environment (bath). This example is used to establish
connections with the classical methods (Langevin, Fokker--Planck,
Martin--Siggia--Rose) as well as with the equilibrium Matsubara
technique. Non--interacting fermions in  presence of quenched
disorder are treated in Sec.~\ref{sec_NLSM} with the help of the
Keldysh non--linear $\sigma$--model. It is generalized to include
Coulomb interactions in Sec.~\ref{sec_int_ferm} and superconducting
correlations in Sec.~\ref{sec_SuperCond}. All technicalities are
accompanied by examples of applications, intended to illustrate
various aspects of the method. We cover spectral statistics  in
mesoscopic samples, universal conductance fluctuations (UCFs),  shot
noise and full counting statistics of electron transport,
interaction corrections to the transport coefficients in disordered
metals and superconductors, Coulomb drag, {\em etc}. We also devote
much attention to derivations of effective phenomenological models,
such as Caldeira--Leggett, time dependent Ginzburg--Landau (TDGL),
Usadel, {\em etc}.,  from the microscopic Keldysh formalism.

\subsection{Closed time contour}\label{sec_contour}

Consider a quantum many--body system  governed by a (possibly
time--dependent) Hamiltonian $\hat{H}(t)$. Let us assume that in the
distant past $t=-\infty$ the system was in a state, specified by a
many--body density matrix $\hat{\rho}(-\infty)$. The precise form of
the latter is of no importance. It may be, e.g., the equilibrium
density matrix associated with the Hamiltonian $\hat{H}(-\infty)$.
The density matrix evolves according to the Von Neumann equation
$\partial_{t}\hat{\rho}(t)=-i\big[\hat{H}(t),\hat{\rho}(t)\big]$,
where we set $\hbar=1$. It is formally solved by
$\hat{\rho}(t)=\hat{\mathcal{U}}_{t,-\infty}\hat{\rho}(-\infty)
\big[\hat{\mathcal{U}}_{t,-\infty}\big]^{\dag}=
\hat{\mathcal{U}}_{t,-\infty}\hat{\rho}(-\infty)
\hat{\mathcal{U}}_{-\infty,t}$, where the evolution operator is
given by the time--ordered exponent:
\begin{eqnarray}
\hat{\mathcal{U}}_{t,t'}=\mathbb{T}\exp\left(-i\int^{\,t}_{t\,'}
\hat{H}(\tau)\mathrm{d}\tau\right)=\lim_{N\rightarrow\infty}
e^{-i\hat{H}(t)\delta_{t}}e^{-i\hat{H}(t-\delta_{t})\delta_{t}}\ldots
e^{-i\hat{H}(t\,'+\delta_{t})\delta_{t}},
\end{eqnarray}
where an infinitesimal time-step is $\delta_{t}=(t-t')/N$.

One is usually interested in  calculations of  expectation
value for some observable $\hat{\mathcal{O}}$ (say density or
current) at a time $t$, defined as
\begin{equation}\label{contour-O-average}
\big\langle\hat{\mathcal{O}}(t)\big\rangle\equiv
\frac{\Tr\{\hat{\mathcal{O}}\hat{\rho}(t)\}}
{\Tr\{\hat{\rho}(t)\}}=\frac{1}{\Tr\{\hat{\rho}(t)\}}
\Tr\big\{\hat{\mathcal{U}}_{-\infty,t}\hat{\mathcal{O}}
\hat{\mathcal{U}}_{t,-\infty}\hat{\rho}(-\infty)\big\}\,,
\end{equation}
where the traces are performed over the many--body Hilbert space.
The expression under the last trace describes (read from right to
left) evolution from $t=-\infty$, where the initial density matrix
is specified, forward to $t$, where the observable is calculated,
and then backward to $t=-\infty$. Such forward--backward evolution
is avoided in the equilibrium by a specially designed trick.

Let us recall how it works, for example, in the zero--temperature
quantum field theory~\cite{AGD}. The latter deals with the
expectation values of the type
$\langle\mathrm{GS}|\hat{\mathcal{O}}|\mathrm{GS}\rangle=
\langle0|\hat{\mathcal{U}}_{-\infty,t}\hat{\mathcal{O}}
\hat{\mathcal{U}}_{t,-\infty}|0\rangle$, where
$|\mathrm{GS}\rangle=\hat{\mathcal{U}}_{t,-\infty}|0\rangle$ is a
ground state of  full interacting system. The only time dependence
allowed for the Hamiltonian is an adiabatic switching of
interactions on and off in the distant past and future,
respectively. The evolution operator therefore describes the
evolution of a simple non--interacting ground state $|0\rangle$
toward $|\mathrm{GS}\rangle$ upon adiabatic switching of the
interactions. Now comes the trick: one inserts the operator
$\hat{\mathcal{U}}_{+\infty,-\infty}$ in the left--most position to
accomplish the evolution along the entire time axis. It is then
argued that
$\langle0|\hat{\mathcal{U}}_{+\infty,-\infty}=\langle0|e^{iL}$. This
argument is based on the assumption that the system adiabatically
follows its ground state upon slow switching of the interactions
"on" and "off" in the distant past and future, respectively.
Therefore, the only result of evolving the non--interacting
ground--state along the entire time axis is acquiring a phase factor
$e^{iL}$. One can then compensate for the added evolution segment by
dividing this factor out. As the result:
$\langle\mathrm{GS}|\hat{\mathcal{O}}|\mathrm{GS}\rangle=
\langle0|\hat{\mathcal{U}}_{+\infty,t}\hat{\mathcal{O}}
\hat{\mathcal{U}}_{t,-\infty}|0\rangle/e^{iL}$ and one faces
description of the evolution along the forward time axis without the
backward segment. However, it comes with the price: one has to take
care of the denominator (which amounts to subtracting of the
so--called disconnected diagrams).

Such a trick does not work in a non--equilibrium situation with a
truly time--dependent Hamiltonian. If the system was driven out of
equilibrium, then the final state of its evolution does not have to
coincide with the initial one. In general, such a final state
depends on the peculiarities of the switching procedure as well as
on the entire history of the system. Thus, one can not get rid of
the backward portion of the evolution history contained
in~\eqref{contour-O-average}. Schwinger~\cite{Schwinger} was the
first to realize that this is not an unsurmountable obstacle. One
has to accept that the evolution in the non--equilibrium quantum
field theory takes place along the closed time contour. Along with
the conventional forward path, the latter contains the backward
path. In this way one avoids the need to know the state of the
system at $t=+\infty$.

\begin{figure}
\begin{center}\includegraphics[width=10cm]{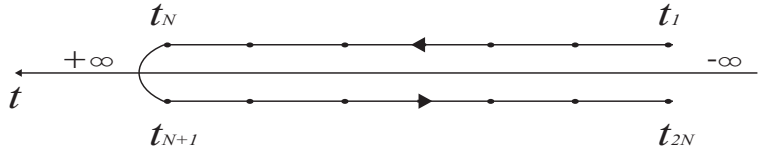}\end{center}
\caption{The closed time contour $\mathcal{C}$. Dots on the forward
and the backward branches of the contour denote discrete time
points. \label{Fig-Contour}}
\end{figure}

It is still convenient to extend the evolution
in~\eqref{contour-O-average} to $t=+\infty$ and back to $t$. It is
important to mention that this operation is identical and does not
require any additional assumptions. Inserting
$\hat{\mathcal{U}}_{t,+\infty}\hat{\mathcal{U}}_{+\infty,t}=\hat{1}$
to the left of $\hat{\mathcal{O}}$ in~\eqref{contour-O-average}, one
obtains
\begin{equation}
\big\langle\hat{\mathcal{O}}(t)\big\rangle=\frac{1}{\Tr\{\hat{\rho}(-\infty)\}}
\Tr\big\{\hat{\mathcal{U}}_{-\infty,+\infty}\hat{\mathcal{U}}_{+\infty,t}
\hat{\mathcal{O}}\hat{\mathcal{U}}_{t,-\infty}\hat{\rho}(-\infty)\big\}\,.
\end{equation}
Here we also used that, according to the Von Neumann equation, the
trace of the density matrix is unchanged under the unitary
evolution. As a result, we have obtained the evolution along the
closed time contour $\mathcal{C}$ depicted in
Figure~\ref{Fig-Contour}.

The observable $\hat{\mathcal{O}}$ is inserted at time $t$,
somewhere along the forward branch of the contour. Note that
inserting the unit operator
$\hat{\mathcal{U}}_{t,+\infty}\hat{\mathcal{U}}_{+\infty,t}=\hat{1}$
to the right of $\hat{\mathcal{O}}$, we could equally well arrange
to have an observable on the backward branch of the contour. As we
show below, the most convenient choice is to take a half--sum of
these two equivalent representations. The observable may be also
generated by adding to the Hamiltonian a source term
$\hat{H}_{\mathcal{O}}(t)\equiv\hat{H}(t)\pm\hat{\mathcal{O}}\eta(t)/2$,
where the plus (minus) signs refer to the forward (backward) parts
of the contour. One needs to calculate then the generating
functional $Z[\eta]$ defined as the trace of the evolution operator
along the contour $\mathcal{C}$ with the Hamiltonian
$\hat{H}_{\mathcal{O}}(t)$. Since the latter is non--symmetric on
the two branches, such a closed contour evolution operator is not
identical to unity. The expectation value of the observable may then
be generated as the result of functional differentiation
$\big\langle\hat{\mathcal{O}}(t)\big\rangle=\delta
Z[\eta]/\delta\eta(t)|_{\eta=0}$. We first omit the source term and
develop a convenient representation for the partition function
\begin{equation}\label{contour-Z-0}
Z[0]\equiv\frac{\Tr\{\hat{\mathcal{U}}_{\mathcal{C}}\hat{\rho}(-\infty)\}}
{\Tr\{\hat{\rho}(-\infty)\}}=1\,,
\end{equation}
where
$\hat{\mathcal{U}}_{\mathcal{C}}=\hat{\mathcal{U}}_{-\infty,+\infty}
\hat{\mathcal{U}}_{+\infty,-\infty}=\hat{1}$. The source term,
breaking the forward--backward symmetry, will be discussed at a
later stage. Note that since $Z[0]=1$, the observable may be equally
well written in the form, which is more familiar from the
equilibrium context:
$\big\langle\hat{\mathcal{O}}(t)\big\rangle=\delta\ln
Z[\eta]/\delta\eta(t)|_{\eta=0}$. The  logarithm is {\em optional}
in the theory with the closed time contour.

The need to carry  the evolution along the two--branch contour
complicates the non--equilibrium theory in comparison with the
equilibrium theory. The difficulties may be substantially reduced by
a proper choice of variables based  on the forward--backward
symmetry of the theory. There are also good news: there is no
denominator $e^{iL}$, unavoidably present in the single--branch
contour theory. (One should not worry about
$\Tr\{\hat\rho(-\infty)\} $ in~(\ref{contour-Z-0}). Indeed, this
quantity refers entirely to $t=-\infty$, before the interactions
were adiabatically switched "on". As a result, it is trivially
calculated and never represents a problem.) The absence of the
denominator dramatically simplifies description of systems with the
quenched disorder. It is the denominator, $e^{iL}$, which is the
main obstacle in performing the disorder averaging of the
expectation values of observables. To overcome this obstacle the
replica~\cite{Edwards,Wegner,EfetovLarkinKhmelnitskii} and the
supersymmetry~\cite{Efetov,Efetov-book} tricks were invented. In the
closed time contour theory  the denominator is absent and thus there
is no need in any of these tricks.

\section{Bosons}\label{sec_bosons}
\subsection{Partition function}\label{sec_bosons-1}

Let us consider the simplest  many--body system: bosonic particles
occupying a single quantum state with  energy $\omega_0$. Its
secondary quantized Hamiltonian has the form
\begin{equation}\label{boson-H}
\hat{H} = \omega_0\, \hat{b}^{\dagger} \hat{b}\, ,
\end{equation}
where $\hat{b}^{\dagger}$ and $\hat{b}$ are bosonic creation and
annihilation operators with the commutation relation
$[\hat{b},\hat{b}^{\dagger}]=1$. Let us define the partition
function as
\begin{equation}\label{boson-Z}
Z=\frac{\Tr\big\{\hat{\mathcal{U}}_{\mathcal{C}}\hat{\rho}\big\}}{\Tr\{\hat{\rho}\}}\,.
\end{equation}
If one assumes that all external fields are exactly the same on the
forward and backward branches of the contour, then
$\hat{\mathcal{U}}_{\ \mathcal{C}}=1$ and, therefore, $Z=1$. The
initial density matrix $\hat{\rho}=\hat{\rho}(\hat{H})$ is some
operator--valued function of the Hamiltonian. To simplify the
derivations one may choose it to be the equilibrium density matrix,
$\hat{\rho}_0 = \exp\{-\beta(\hat{H}-\mu\hat
{N})\}=\exp\{-\beta(\omega_0-\mu)\hat{b}^\dagger\hat{b}\}$. Since
arbitrary external perturbations may be switched on (and off) at a
later time, the choice of the equilibrium  initial density matrix
does not prevent one from treating non--equilibrium dynamics. For
the equilibrium initial density matrix one finds
\begin{equation}\label{boson-density-trace}
\Tr\{\hat{\rho}_0\}=\sum\limits_{n=0}^\infty
e^{-\beta(\omega_0-\mu)n}= [1-\rho(\omega_0)]^{-1}\, ,
\end{equation}
where $\rho(\omega_0)=e^{-\beta(\omega_0-\mu)}$. An important point
is that, in general, $\Tr\{\hat{\rho}\}$ is an interaction- and
disorder--independent constant. Indeed, both interactions and
disorder are supposed to be switched on (and off) on the forward
(backward) parts of the contour sometime after (before) $t=-\infty$.
This constant is, therefore, frequently omitted without causing a
confusion.

The next step is to divide the $\mathcal{C}$ contour into $(2N-2)$
time--steps of length $\delta_t$, such that $t_1=t_{2N}=-\infty$ and
$t_{N}=t_{N+1}=+\infty$ as shown in Figure~\ref{Fig-Contour}. One
then inserts the resolution of unity in the over--complete coherent
state basis\footnote{The Bosonic coherent state $|\phi\rangle$
($\langle\phi|\,$), parameterized by a complex number $\phi$,   is
defined as a right (left) eigenstate of the annihilation (creation)
operator: $\hat{b}|\phi\rangle =\phi|\phi\rangle$
($\langle\phi|\hat{b}^\dagger= \langle\phi|\bar\phi\, $). Matrix
elements of a {\em normally ordered} operator, such as Hamiltonian,
take the form $\langle\phi|\hat
H(\hat{b}^\dagger,\hat{b})|\phi'\rangle = H(\bar\phi,\phi')
\langle\phi|\phi'\rangle$. The overlap between two coherent states
is $\langle\phi|\phi'\rangle=\exp\{\bar\phi\phi'\}$. Since the
coherent state basis is over--complete, the trace of an operator,
$\hat A$, is calculated with the weight: $\Tr\{\hat A\}=\pi^{-1}
\int\!\!\!\int d(\mbox{Re} \phi)\, d(\mbox{Im} \phi)\,
e^{-|\phi|^2}\, \langle\phi|\hat A|\phi\rangle
$.}~\cite{NegeleOrland}
\begin{equation}\label{boson-unity}
\hat{1}=\iint\frac{\mathrm{d}(\mathrm{Re}\phi_{j})\mathrm{d}(\mathrm{Im}\phi_{j})}
{\pi}\ e^{-|\phi_{j}|^{2}}|\phi_{j}\rangle\langle\phi_{j}|
\end{equation}
at each point $j=1,2,\ldots, 2N$ along the contour. For example, for
$N=3$ one obtains the following sequence in the expression for
$\Tr\{\hat{\mathcal{U}}_{\ \mathcal{C}}\hat{\rho}_0\}$ (read from
right to left):
\begin{equation}
  \langle\phi_6|\hat{\mathcal{U}}_{-\delta_t}|\phi_5\rangle
  \langle\phi_5|\hat{\mathcal{U}}_{-\delta_t}|\phi_4\rangle
  \langle\phi_4|\hat 1|\phi_3\rangle
  \langle\phi_3|\hat{\mathcal{U}}_{+\delta_t}|\phi_2\rangle
  \langle\phi_2|\hat{\mathcal{U}}_{+\delta_t}|\phi_1\rangle
  \langle\phi_1|\hat{\rho}_0|\phi_6\rangle\, ,
\end{equation}
where $\hat{\mathcal{U}}_{\pm \delta_t}$ is the evolution operator
during the time interval $\delta_t$ in the positive (negative) time
direction. Its matrix elements are given by:
\begin{eqnarray}\label{boson-matrix-element}
\left\langle\phi_{j+1}\left|\hat{\mathcal{U}}_{\pm\delta_t}\right|\phi_j\right\rangle\equiv
\left\langle \phi_{j+1}\left| e^{\mp i\hat H(\hat{b}^\dagger,
\hat{b})\delta_t}\right|\phi_{j}\right\rangle\approx \left\langle
\phi_{j+1}\left|\big( 1\mp i\hat H(\hat{b}^\dagger,
\hat{b}\big)\delta_t\right |\phi_{j}\right\rangle\nonumber\\=
\big\langle \phi_{j+1}|\phi_{j}\big\rangle\big(1\mp  i
H(\bar\phi_{j+1},\phi_j)\delta_t\big) \approx \big\langle
\phi_{j+1}| \phi_{j}\big\rangle\, e^{\mp i
H(\bar\phi_{j+1},\phi_j)\delta_t}\, ,
\end{eqnarray}
where the  approximate equalities  are valid up to the linear order
in $\delta_t$. Obviously this result is not restricted to the toy
example~\eqref{boson-H}, but holds for any \textit{normally ordered}
Hamiltonian. Note that there is no evolution operator inserted
between $t_N$ and $t_{N+1}$. Indeed, these two points are physically
indistinguishable and thus the system does not evolve during this
time interval. Employing the following properties of coherent
states: $\langle\phi|\phi'\rangle=\exp\{\bar\phi\phi'\}$ along with
$\langle\phi| e^{-\beta(\omega_0-\mu) \hat{b}^\dagger
\hat{b}}|\phi'\rangle = \exp\left\{\bar\phi\phi' \rho(\omega_0)
\right\}$, and collecting all of the matrix elements along the
contour, one finds for the partition function~\eqref{boson-Z},
\begin{equation}\label{boson-Z-discrete}
Z=\frac{1}{\Tr\{\hat{\rho}_{0}\}}\iint\prod\limits^{2N}_{j=1}
\left[\frac{\mathrm{d}(\R\phi_{j})\mathrm{d}(\I\phi_{j})}{\pi}\right]
\exp\left(i\sum\limits^{2N}_{j,j\,'=1}\bar{\phi}_{j}\,
G^{-1}_{jj\,'}\,\phi_{j\,'}\right)\, ,
\end{equation}
where the $2N \times 2N$ matrix $iG^{-1}_{jj\,'}$ stands for
\begin{equation}\label{boson-G-matrix}
  iG^{-1}_{jj\,'}\equiv\
\left[\begin{array}{ccc|ccc}
 \,\,-1\,\,   &\,\,\,\,      &\,\,\,\,    &\,\,\,\,   &\,\,\,\,   &\,\,\,\,   \rho(\omega_0) \\
 \,\, 1\!-\!h\,\, & -1   &    &   &   &                 \\
      &  \,\,1\!-\!h \,\,& \,\,\,-1\,\,\, &   &   &                 \\ \hline
     &     &  1 & -1 &    &                 \\
     &     &    &\,\,1\!+\!h\,\, & -1 &                 \\
     &     &    &    & \,\,1\!+\!h\,\,&  -1
 \end{array} \right]\, ,
\end{equation}
and $h\equiv i\omega_0\delta_t$. It is straightforward to evaluate
the determinant of such a
matrix
\begin{equation}\label{boson-determinant}
\mathrm{Det}\big[i\hat{G}^{-1}\big]=(-1)^{2N} - \rho(\omega_0)(1-h^2)^{N-1}\approx
1-\rho(\omega_0)\, e^{(\omega_0\delta_t)^2(N-1)}\to 1-
\rho(\omega_0) \, ,
\end{equation}
where one  used that $\delta_t^2N\to 0$ if $N\to \infty$ (indeed,
the assumption was $\delta_tN \to \mathrm{const}$). Employing the
fact that the Gaussian integral in~\eqref{boson-Z-discrete} is equal
to the inverse determinant of  $i\hat{G}^{-1}$ matrix, see Appendix
\ref{app_Gaussian}, along with~\eqref{boson-density-trace}, one
finds
\begin{equation}\label{boson-Z-norm}
Z=\frac{\mathrm{Det}^{-1}\big[i\hat{G}^{-1}\big]}{\Tr\{\hat{\rho}_0\}}
= 1\, ,
\end{equation}
as it should be, of course. Note that keeping the upper--right
element of the discrete matrix~\eqref{boson-G-matrix} is crucial to
maintaining this normalization identity.

One may  now take the limit $N\to \infty$ and formally write the
partition function in the continuum  notations, $\phi_j\to \phi(t)$,
as
\begin{equation}\label{boson-Z-1}
Z=\int\D[\bar{\phi}\phi]\,
\exp\left(iS[\bar{\phi},\phi]\right)=\int\D[\bar{\phi}\phi]
\exp\left(i\int_{\mathcal{C}}\d t \,
\big[\bar{\phi}(t)\,\hat{G}^{-1}\phi(t)\big] \right)\,,
\end{equation}
where according to~\eqref{boson-Z-discrete}-- \eqref{boson-G-matrix}
the action is given by
\begin{equation}\label{boson-S}
S[\bar{\phi},\phi] =
\sum\limits_{j=2}^{2N}\left[i\bar\phi_j\,\frac{\phi_j-\phi_{j-1}}{\delta
t_j} -\omega_0\bar\phi_j\phi_{j-1}\right]\delta
t_j+i\,\bar\phi_1\Big[\phi_1-\rho(\omega_0)\phi_{2N}\Big] \, ,
\end{equation}
with $\delta t_j\equiv t_j-t_{j-1}=\pm \delta_t$. Thus, the continuum
form of the operator $\hat G^{-1}$ is
\begin{equation}\label{boson-G-continious}
\hat G^{-1}= i\partial_t - \omega_0 \,.
\end{equation}
It is important to remember that this continuum  notation is only an
abbreviation that represents  the large discrete
matrix~\eqref{boson-G-matrix}. In particular, the upper--right
element of the matrix (the last term in~\eqref{boson-S}), which
contains the information about the distribution function, is
seemingly absent in the continuum
notation~\eqref{boson-G-continious}.

To avoid integration along the closed time contour, it is convenient
to split the bosonic field $\phi(t)$ into the two components
$\phi_{+}(t)$ and $\phi_{-}(t)$ that reside on the forward and the
backward parts of the time contour, respectively. The continuum
action may be then rewritten as
\begin{equation} \label{boson-S-plus-minus}
S[\bar{\phi},\phi]=\int_{-\infty}^{+\infty} \d t\,
\big[\bar\phi_{+}(t)(i\partial_t - \omega_0) \phi_{+}(t)-
\bar\phi_{-}(t)(i\partial_t - \omega_0) \phi_{-}(t)\big]\, ,
\end{equation}
where the relative minus sign comes from the reversed direction of
the time integration on the backward part of the contour. Once
again, the continuum  notations are somewhat misleading. Indeed,
they create an undue impression that  $\phi_{+}(t)$ and
$\phi_{-}(t)$ fields are completely uncorrelated. In fact, they are
correlated owing to the presence of the non--zero off--diagonal
blocks in the discrete matrix~\eqref{boson-G-matrix}. It is
therefore desirable to develop a continuum  representation that
automatically takes into account the proper regularization. We shall
achieve it in the following sections. First, the Green's functions
should be discussed.

\subsection{Green's functions}\label{sec_bosons-2}

According to the basic properties of the Gaussian integrals, see Appendix \ref{app_Gaussian},  the
correlator of the two bosonic fields is given by
\begin{equation}\label{boson-G-fun}
\big\langle\phi_{j}\,\bar\phi_{j\,'}\big\rangle \equiv \int
\D[\bar{\phi}\phi] \,\,\phi_{j}\,\bar\phi_{j\,'} \exp\left(i
\sum\limits_{j,j\,'=1}^{2N}
\bar\phi_{j}\,G^{-1}_{jj\,'}\,\phi_{j\,'}\right)= iG_{jj\,'} \, .
\end{equation}
Note the absence of the factor $Z^{-1}$ in comparison with the
analogous definition in the equilibrium theory~\cite{NegeleOrland}.
Indeed, in the present construction $Z=1$. This seemingly minor
difference turns out to be the major issue in the theory of
disordered systems (see further discussion in
Section~\ref{sec_NLSM}, devoted to fermions with the quenched
disorder). Inverting the discrete matrix in~\eqref{boson-G-matrix},
one finds
\begin{equation}\label{boson-G-matrix-inverted}
  iG_{jj\,'}=\frac{1}{1-\rho}
 \left[\begin{array}{lcc|ccc}
 1   &  \,\,\rho\,e^{h}\,\,    & \,\, \rho\,e^{2h}\,\,  &\,\, \rho\,e^{2h}\,\,  &\,\, \rho\,e^{h}\,\,  &\,\, \rho\, \\
 e^{-h}\,\, & 1  &  \rho\,e^{h}   & \rho\,e^{h}   & \rho  &  \,\,   \rho\,e^{-h}  \,           \\
 e^{-2h}\,\,    &  \,e^{-h} & 1 & \rho  & \rho\,e^{-h}   &   \rho\,e^{-2h}               \\ \hline
  e^{-2h}   &   \,e^{-h}   &  1 & 1 &\,\, \rho\,e^{-h}\,\,    &      \rho\,e^{-2h}            \\
   e^{-h}   &   1  &  \,e^{h}   &\,e^{h} & 1 &  \rho\,e^{-h}                \\
   1  &  \,e^{h}    &  \,e^{2h}   & \,e^{2h}    & \,e^{h}&  1
 \end{array} \right]\,,
\end{equation}
where $\rho\equiv \rho(\omega_0)$, and following the discussion
after~\eqref{boson-determinant}, we have put $(1\pm h)^j\, \approx\,
e^{\pm jh}\,$ and $(1-h^2)^j\approx 1$. In terms of the fields
$\phi_{j\pm}$ (hereafter $j=1,\ldots, N$ and therefore the $2N\times
2N$ matrix above is labeled as $1,\ldots, N-1,N,N,N-1,\ldots, 1$)
the corresponding correlators read as
\begin{subequations}\label{boson-corr-fun}
\begin{equation}
\hskip-3cm \langle \phi_{j+}\bar{\phi}_{j\,'-}\rangle\equiv
iG^{<}_{jj\,'}= n_B\, \exp\{-(j-j\,')h\}\,,
\end{equation}
\begin{equation}
\hskip-2.2cm \langle \phi_{j-}\bar\phi_{j\,'+}\rangle\equiv
iG^{>}_{jj\,'}=(n_B+1)\, \exp\{-(j-j\,')h\}\,,
\end{equation}
\begin{equation}
\langle\phi_{j+}\bar\phi_{j\,'+}\rangle\equiv
iG^{\mathbb{T}}_{jj\,'}= {1\over 2}\,
\delta_{jj\,'}+\theta(j-j\,')iG^{>}_{jj\,'}+
\theta(j\,'-j)iG^{<}_{jj\,'}\,,
\end{equation}
\begin{equation}
\langle\phi_{j-}\bar\phi_{j\,'-}\rangle \equiv
iG^{\widetilde{\mathbb{T}}}_{jj\,'}={1\over 2}\,\delta_{jj\,'}+
\theta(j\,'-j)iG^{>}_{jj\,'} +\theta(j-j\,')iG^{<}_{jj\,'}\,,
\end{equation}
\end{subequations}
where the  bosonic occupation number $n_B$ stands for $n_B(\omega_0)
\equiv \rho/(1-\rho)$ and symbols $\mathbb{T}$ (respectively
$\widetilde{\mathbb{T}})$ denote time--ordering (respectively
anti--time--ordering). The step--function $\theta(j)$ is defined
such that $\theta(0)=1/2$, so $\theta(j)+\theta(-j)\equiv 1$.

Obviously not all four Green's functions defined above are
independent. Indeed, a direct inspection shows that
\begin{subequations}
\begin{equation}\label{boson-relation}
G^{\mathbb{T}}
+G^{\widetilde{\mathbb{T}}}-G^{>}-G^{<}=-i\delta_{jj\,'}\,,\qquad\,\,
\end{equation}
\begin{equation}\label{boson-relation-1}
G^{\mathbb{T}}-G^{\widetilde{\mathbb{T}}}=\mbox{sign}(j-j\,')\left(G^{>}-
G^{<}\right)\,,
\end{equation}
\end{subequations}
where $\mbox{sign}(j)=\theta(j)-\theta(-j)$. We would like  to
perform a linear transformation of the fields to benefit explicitly
from these relations. This is achieved by the Keldysh rotation
\begin{equation}\label{boson-rotation}
\phi^{cl}_{j}={1\over \sqrt{2}}\big(\phi_{j+} +\phi_{j-}\big) \, ,
\,\,\,\,\,\,\,\,\, \phi^{q}_{j}={1\over\sqrt{2}}\big(\phi_{j+}-
\phi_{j-}\big)\,,
\end{equation}
with the analogous transformation for the conjugated fields. The
superscripts {\em ``cl''} and {\em ``q''} denote the {\em classical}
and the {\em quantum} components of the fields, respectively. The
rationale for this notation will become clear shortly. First, a
simple algebraic manipulation
with~(\ref{boson-corr-fun}a)--(\ref{boson-corr-fun}d) shows that
\begin{equation}\label{boson-G-fun-1}
-i\big\langle\phi^{\alpha}_{j}\,\bar\phi^{\,\beta}_{j\,'}\big\rangle
= \left(\begin{array}{cc} G^K_{jj\,'} & G^{R}_{jj\,'} \\ \frac{}{} &
\frac{}{}
\\G^{A}_{jj\,'} & -{i\over 2}\,\delta_{jj\,'}
\end{array}\right)\, ,
\end{equation}
where hereafter $\alpha, \beta = (cl,q)$. The explicit form  of the
$(q,q)$ element of this matrix is a manifestation of
identity~\eqref{boson-relation}. Superscripts $R,A$ and $K$ denote
the {\em retarded, advanced} and {\em Keldysh} components of the
Green's function, respectively. These three Green's functions are
the fundamental objects of the Keldysh technique. They are defined
as
\begin{subequations}\label{boson-G-fun-RAK}
\begin{equation}
\hskip-1.2cm
G^{R}_{jj\,'}=-i\big\langle\phi^{cl}_{j}\,\bar\phi^{q}_{j\,'}\big\rangle
=\theta(j-j\,')\left(G^{>}_{jj'}-G^{<}_{jj'}\right)
=-i\theta(j-j\,')\, e^{-(j-j\,')h}\, ,
\end{equation}
\begin{equation}
\hskip-1.5cm
G^{A}_{jj\,'}=-i\big\langle\phi^{q}_{j}\,\bar\phi^{cl}_{j\,'}\big\rangle
= \theta(j\,'-j)\left(G^{<}_{jj'}-G^{>}_{jj'}\right)=i\theta(j\,'-j)
e^{-(j-j\,')h}\,,
\end{equation}
\begin{equation}
G^{K}_{jj\,'}=-i\big\langle\phi^{cl}_{j}\,\bar\phi^{cl}_{j\,'}\big\rangle =
-{i\over 2}\,\delta_{jj\,'}+G^{>}_{jj'}+G^{<}_{jj'}=-{i\over
2}\,\delta_{jj\,'}-i\left(2n_B+1\right) e^{-(j-j\,')h}\,.
\end{equation}
\end{subequations}
Since by definition $\left[G^<\right]^\dagger =-G^>$
[cf.~\eqref{boson-corr-fun}], one may notice that
\begin{equation}\label{boson-G-conjugation}
G^{A} =\big[G^R\big]^{\dagger}\,, \hskip 2cm
G^K=-\big[G^K\big]^\dagger\,.
\end{equation}
The retarded (advanced) Green's function  is lower (upper)
triangular matrix in the time domain. Since a  product of any number
of triangular matrices is again a triangular matrix, one obtains the
simple rule:
\begin{subequations}\label{bosons-G-traces}
\begin{equation}
G_1^{R}\circ G_2^{R}\circ \ldots \circ G_l^{R}  = G^R \,,
\end{equation}
\begin{equation}
G_1^{A}\circ G_2^{A}\circ \ldots \circ G_l^{A}  = G^A \,,
\end{equation}
\end{subequations}
where the circular multiplication sign is understood as a
convolution in the time domain (i.e. it implies integration over an
intermediate time).

\begin{figure}
\begin{center}\includegraphics[width=10cm]{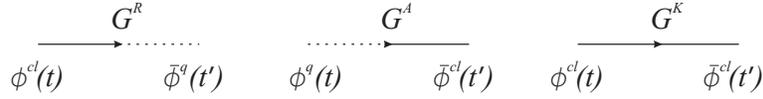}\end{center}
\caption{Graphic representation of $G^R$, $G^A$, and $G^K$. The full
line represents the classical field component $\phi^{cl}$, while the
dashed line the quantum component $\phi^{q}$. \label{Fig-boson-G}}
\end{figure}

One can now take the continuum limit ($N\to \infty$, while
$N\delta_t\to \mbox{const}$) of the Green's functions. To this end,
one defines $t_j=j\delta_t$ and notices that $\exp\{-(j-j')h\}\to
\exp\{-i\omega_0(t-t')\}$. A less trivial observation is that the
factors $\delta_{jj\,'}$, see~\eqref{boson-G-fun-1} and
\eqref{boson-G-fun-RAK}, may be omitted in the continuum limit. The
reason for this is twofold: (i) all observables are given by the
{\em off--diagonal} elements of the Green's functions, e.g. the mean
occupation number at the moment $t_j$ is given by $\langle
n_B(t_j)\rangle =iG^{\mathbb{T}}_{jj+1}=iG^<_{jj+1}$; (ii) the
intermediate expressions contain multiple sums (integrals) of the
form $\delta_t^2\sum_{j,j\,'} \delta_{jj\,'}G_{j\,'j}\to
\delta_t^2N\to 0$. As a result the proper continuum limit of the
relations derived above is
\begin{equation}\label{boson-G-fun-2}
-i\big\langle\phi^{\alpha}(t)\,\bar\phi^{\,\beta}(t')\big\rangle =
G^{\alpha\beta}(t,t') =\left(\begin{array}{cc}
G^K(t,t') & G^{R}(t,t') \\
G^{A}(t,t') & 0
\end{array}\right)\, ,
\end{equation}
where
\begin{subequations}\label{boson-G-fun-RAK-cont}
\begin{equation}
\hskip-2.5cm G^{R}=-i\theta(t-t')\,e^{-i\omega_0(t-t')}\to
(\epsilon-\omega_0+i0)^{-1}\,,
\end{equation}
\begin{equation}
\hskip-2.5cm G^{A}=\phantom{-}i\theta(t'-t)\,e^{-i\omega_0(t-t')}\to
(\epsilon-\omega_0- i0)^{-1}\,,
\end{equation}
\begin{equation}
G^{K}=-i\left[2n_B(\omega_0)+1\right] e^{-i\omega_0(t-t')}\to-2\pi i
[2n_B(\epsilon)+1]\delta(\epsilon-\omega_0)\,.
\end{equation}
\end{subequations}
The Fourier transforms with respect to $t-t'$ are given for each of
the three Green's functions. An important property of these Green's
functions is [cf.~\eqref{boson-G-fun-RAK}]
\begin{equation}\label{boson-GR-plus-GA}
G^R(t,t)+G^A(t,t) = 0\, .
\end{equation}
It is useful to introduce  graphic representations for the three
Green's functions. To this end, let us denote the classical
component of the field by a full line and the quantum component by a
dashed line. Then the retarded Green's function is represented by a
full arrow--dashed line, the advanced by a dashed arrow--full line
and the Keldysh by full arrow--full line, see
Figure~\ref{Fig-boson-G}. Note that the dashed arrow--dashed line,
which would represent the $\langle \phi^q \bar\phi^q\rangle$ Green's
function, is absent in the continuum limit. The arrow shows the
direction from $\phi^{\alpha}$ towards $\bar\phi^{\,\beta}$.

Notice that the retarded and advanced components only contain
information about the spectrum and are independent of the occupation
number, whereas the Keldysh component does depend on it. In thermal
equilibrium $\rho=e^{-\beta\epsilon}$, while
$n_B=(e^{\beta\epsilon}-1)^{-1}$ and therefore
\begin{equation}\label{boson-G-FDT}
G^{K}(\epsilon)=\left[G^{R}(\epsilon)-G^{A}(\epsilon)\right]
\mathrm{coth}\frac{\epsilon}{2T}\, .
\end{equation}
The last equation constitutes the statement of the {\em
fluctuation--dissipation theorem} (FDT). The FDT is, of course, a
general property of  thermal equilibrium that is not restricted to
the toy example considered here. It implies the rigid relation
between the response and correlation functions in equilibrium.

In general, it is convenient to parameterize the anti--Hermitian,
see~\eqref{boson-G-conjugation}, Keldysh Green's function by a
Hermitian matrix $F=F^\dagger$, as follows
\begin{equation}\label{boson-G-FDT-general}
G^{K}=G^{R}\circ F-F\circ G^{A}  \, ,
\end{equation}
where $F=F(t,t')$, and the circular multiplication sign implies
convolution. The Wigner transform (see below),
$\mathbf{f}(\tau,\epsilon)$, of the matrix $F$ is referred to as the
\textit{distribution function}. In thermal equilibrium
$\mathbf{f}(\epsilon) = \mbox{coth}(\epsilon/2T)$ (
see~\eqref{boson-G-FDT}).

\subsection{Keldysh action and causality}\label{sec_bosons-3}

One would like to have a continuum  action, written in terms of
$\phi^{cl},\phi^{q}$, that properly reproduces the
correlators~\eqref{boson-G-fun-2} and \eqref{boson-G-fun-RAK-cont}.
To this end, one formally inverts the correlator
matrix~\eqref{boson-G-fun-2}, and uses it in the Gaussian action
\begin{equation}\label{boson-S-1}
S[\phi^{cl},\phi^{q}]=\iint_{-\infty}^{+\infty}\d t\, \d t'\,
\left(\bar{\phi}^{cl}_t,\bar{\phi}^{q}_t\right)
\left(\begin{array}{cc}
0   & \big[G^{-1}_{t,t\,'}\big]^{A}  \\
  \big[G^{-1}_{t,t\,'}\big]^{R}  & \big[G^{-1}_{t,t\,'}\big]^K
\end{array}\right)
\left(\begin{array}{c} \phi^{cl}_{t\,'} \\ \phi^{q}_{t\,'}
\end{array}\right) ,
\end{equation}
where
\begin{subequations}\label{boson-G-fun-2-inverted}
\begin{equation}
\big[G^{-1}\big]^{R(A)} = \big[G^{R(A)}\big]^{-1} = \epsilon-\omega_0\pm i0\to
\delta_{t,t'}\left(i\partial_t-\omega_0 \pm i0\right)\,,
\end{equation}
\begin{equation}
\hskip-3.3cm \big[G^{-1}\big]^K = \big[G^R\big]^{-1}\circ F -F \circ
\big[G^A\big]^{-1}\, ,
\end{equation}
\end{subequations}
where we used that the Fourier transform of $\epsilon$ is $
\delta_{t,t'}i\partial_t$ and
parametrization~\eqref{boson-G-FDT-general} was employed in the last
line. It is important to mention that the actual discrete matrix
action~\eqref{boson-Z-discrete}--\eqref{boson-G-matrix}, being
transformed to $\phi^{cl},\phi^{q}$ according
to~\eqref{boson-rotation}, does {\em not} have the structure
of~\eqref{boson-S-1}. The action \eqref{boson-S-1} should be viewed
as a formal construction  devised to reproduce the continuum limit
of the correlators according to the rules of the Gaussian
integration. It is, however, fully self--consistent in the following
sense:  (i) it does not need to appeal to the discrete
representation for a regularization; (ii) its general structure  is
intact in every order of the perturbative renormalization.

Here we summarize the  main features of the action
\eqref{boson-S-1}, which, for lack of a better terminology, we call
the {\em causality structure}.
\begin{itemize}
\item The $cl-cl$ component is zero.
It  reflects the fact that for a pure classical field configuration
($\phi^q=0$) the action is zero. Indeed, in this case
$\phi_+=\phi_-$ and the action on the forward part of the contour is
canceled  by that on the backward part (safe for the boundary terms,
that may be omitted in the continuum limit). The very general
statement is, therefore, that
\begin{equation}\label{boson-causality}
S\big[\phi^{cl},0\big] = 0\, .
\end{equation}
Obviously this statement should not be restricted to the Gaussian
action of the form given by Eq.~\eqref{boson-S-1}.

\item The $cl-q$ and $q-cl$ components are mutually Hermitian conjugated
upper and lower (advanced and retarded) triangular matrices in the
time representation. This property is responsible for the causality
of the response functions as well as for protecting the $cl-cl$
component from a perturbative renormalization (see below). Relation
\eqref{boson-GR-plus-GA} is necessary for the consistency of the
theory.

\item The $q-q$ component is an anti--Hermitian matrix
[cf.~\eqref{boson-G-conjugation}]. In our example
$\big[G^K\big]^{-1}=i0F$, where $F$ is a Hermitian matrix, with a
positive--definite spectrum. It is responsible for the convergence
of the functional integral. It also keeps the information about the
distribution function.
\end{itemize}

\subsection{Free bosonic fields}\label{sec_bosons-4}

It is a straightforward matter to generalize the entire construction
to bosonic systems with more than one degree of freedom. Suppose the
states are labeled by an index $\mathbf{k}$, that may be, e.g., a
momentum vector. Their energies are given by a function
$\omega_\mathbf{k}$, for example
$\omega_{\mathbf{k}}=\mathbf{k}^2/(2m)$, where $m$ is the mass of
bosonic atoms. One introduces next a doublet of complex fields
(classical and quantum) for every state $\mathbf{k}\,$,
$(\phi^{cl}(\mathbf{k},t), \phi^{q}(\mathbf{k},t))$, and writes down
the action in the form of~\eqref{boson-S-1} including a summation
over the index $\bf{k}$. Away from  equilibrium, the Keldysh
component may be non--diagonal in the index $\mathbf{k}$:
$F=F(\mathbf{k},\mathbf{k}';t,t')$. The retarded (advanced)
component, on the other hand, has the simple form
$[G^{R(A)}]^{-1}=i\partial_t -\omega_\mathbf{k}$.

If $\mathbf{k}$ is momentum, it is instructive to perform the
Fourier transform to the real space and to deal with
$(\phi^{cl}(\mathbf{r},t), \phi^q(\mathbf{r},t))$. Introducing a
combined time--space index $x=(\mathbf{r},t)$, one may write down
for the action of the free complex bosonic field (atoms)
\begin{equation}\label{boson-S-2}
S_{0}[\phi^{cl},\phi^{q}] = \iint\d x\,\d
x'\,\big(\bar\phi^{cl}_x,\bar\phi^{q}_x\big) \left(\begin{array}{cc}
0   & \big[G^{A}_{x,x'}\big]^{-1}  \\
  \big[G^{R}_{x,x'}\big]^{-1}  & \big[G^{-1}_{x,x'}\big]^K
\end{array}\right)
\left(\begin{array}{c} \phi^{cl}_{x\,'} \\ \phi^{q}_{x\,'}
\end{array}\right),
\end{equation}
where in the continuum  notations
\begin{equation}\label{boson-G-Real-Space}
\big[G^{R(A)}\big]^{-1}(x,x\,')=\delta(x-x\,')\left( i\partial_{t}
+{1\over 2m}
\partial_{\mathbf{r}}^2 +\mu\right)\, ,
\end{equation}
while in the discrete form it is a lower (upper) triangular matrix
in time (not in space). The $\big[G^{-1}\big]^K$ component for the
free field is only the regularization factor, originating from the
(time) boundary terms. It is, in general, non--local in $x$ and
$x'$, however, being a pure boundary term it is frequently omitted.
It is kept here as a reminder that the inversion, $\hat{G}$, of the
correlator matrix must posses the causality
structure~\eqref{boson-G-fun-2}. We have introduced the chemical
potential $\mu$ into~\eqref{boson-G-Real-Space}, understanding that
one may want to consider an effective Hamiltonian
$\hat{H}-\mu\hat{N}$, where $\hat{N}$ is the total particle number
operator. The new term may be considered as a mean to enforce a
ceratin particle number with the help of the Lagrange  multiplier
$\mu$. For discussion of real bosonic fields see Appendix
\ref{app_SPQM}.

\section{Collisions and kinetic equation for
bosons}\label{sec_int_bos}
\subsection{Interactions}\label{sec_int_bos-1}

\begin{figure}
\begin{center}\includegraphics[width=10cm]{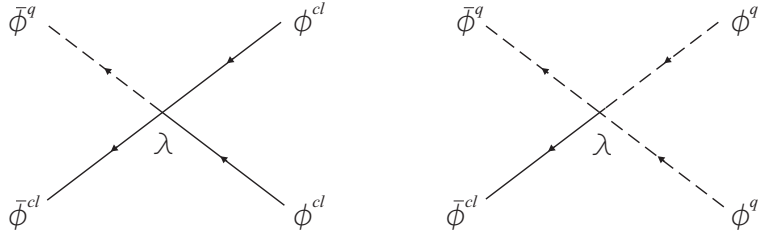}\end{center}
\caption{Graphic representation of the two interaction vertices
 of the $|\phi|^4$ theory. There are also two complex conjugated vertices
 with a reversed direction of all arrows. \label{Fig-Phi4}}
\end{figure}

The short range two--body collisions of bosonic atoms are described
by the local \textit{four--boson} Hamiltonian
$\hat{H}_{\mathrm{int}}=\lambda\sum_{\mathbf{r}}
\hat{b}^{\dagger}_{\mathbf{r}} \hat{b}^{\dagger}_{\mathbf{r}}
\hat{b}_{\mathbf{r}} \hat{b}_{\mathbf{r}}$, where index $\mathbf{r}$
``enumerates'' spatial locations. The interaction constant,
$\lambda$, is related to a commonly used $s$--wave scattering
length, $a_s$, as $\lambda=4\pi a_s/m$ (see~\cite{Leggett}). The
corresponding term in the continuum  Keldysh action takes the form
\begin{equation}\label{int-bos-S}
S_{\mathrm{int}}[\phi_{+},\phi_{-}]=-\lambda\int\d\mathbf{r}\int_{\cal
C}\d t\,(\bar\phi
\phi)^2=-\lambda\int\d\mathbf{r}\int_{-\infty}^{+\infty}\d t\,
\big[(\bar{\phi}_{+}\phi_{+})^2-(\bar{\phi}_{-}\phi_{-})^2\big]\,.
\end{equation}
It is important to remember that there are no interactions in the
distant past, $t=-\infty$ (while they are present in the future,
$t=+\infty$). The interactions are supposed to be adiabatically
switched on and off on the forward and backward branches,
respectively. This guarantees that the off--diagonal blocks of the
matrix~\eqref{boson-G-matrix} remain intact.  Interactions modify
only those matrix elements of the evolution
operator~\eqref{boson-matrix-element} that are away from
$t=-\infty$. It is also worth remembering that in the discrete time
form the $\bar\phi$ fields are taken one time step $\delta_t$ {\em
after} the $\phi$ fields along the Keldysh contour ${\cal C}$.
Performing the Keldysh rotation~\eqref{boson-rotation}, one finds
\begin{equation}\label{int-bos-S-1}
S_{\mathrm{int}}[\phi^{cl},\phi^{q}]=-\lambda\int\d\mathbf{r}
\int^{+\infty}_{-\infty}\d t\,\Big[ \bar{\phi}^{cl}\bar{\phi^{q}}
\big[\big(\phi^{cl}\big)^2+\big(\phi^{q}\big)^{2}\big]+c.c.\Big] \,,
\end{equation}
where $c.c.$ denotes the complex conjugate of the first term. The
collision action~\eqref{int-bos-S-1} obviously satisfies the
causality condition~\eqref{boson-causality}. Diagrammatically the
action \eqref{int-bos-S-1} generates two types of vertices depicted
in Figures~\ref{Fig-Phi4} (as well as two complex conjugated
vertices, obtained by reversing the direction of the arrows): one
with three classical fields (full lines) and one quantum field
(dashed line) and the other with one classical field and three
quantum fields.

Let us demonstrate that an addition of the collision term to the
action does not violate the fundamental normalization, $Z=1$. To
this end, one may expand $\exp(iS_{\mathrm{int}})$ in powers of
$\lambda$ and then average term by term with the Gaussian
action~\eqref{boson-S-2}. To show that the normalization, $Z=1$, is
not altered by the collisions, one needs to show that $\langle
S_{\mathrm{int}}\rangle = \langle S_{\mathrm{int}}^{\,2}\rangle =
\ldots=0$. Applying the Wick theorem, one finds for the terms that
are linear order in $\lambda$: $\big\langle\bar\phi^q
\bar\phi^{cl}\big(\phi^{cl}\big)^2 +c.c.\big\rangle \sim \big[
G^R(t,t)+G^A(t,t)\big] G^K(t,t)=0$, and $\big\langle \bar\phi^{q}
\bar\phi^{cl} \big(\phi^{q}\big)^{2}+c.c\big\rangle=0$. The first
term vanishes owing to the identity \eqref{boson-GR-plus-GA}, while
the second vanishes because $\big\langle
\phi^{q}\bar\phi^{q}\big\rangle=0$ (even if one appeals to the
discrete version~\eqref{boson-G-fun-1}, where
$\big\langle\phi^{q}_{j}\bar\phi^{q}_{j\,'}\big\rangle=-i\delta_{jj\,'}/2\neq
0$, this term is still identically zero, since it is given by
$\sum_{jj\,'}\delta_{jj\,'}(G^A_{j\,'j}+G^R_{j\,'j})=0$,
cf.~\eqref{boson-GR-plus-GA}). There are two families of terms that
are second order in $\lambda$. The first one is
$\big\langle\bar\phi^{q}_{1}
\bar\phi^{cl}_{1}\big(\phi^{cl}_{1}\big)^2
\phi^{q}_{2}\phi^{cl}_{2}\big(\bar\phi^{cl}_{2}\big)^2\big\rangle\sim G^R(t_2,t_1)\\
G^A(t_2,t_1)[G^K(t_1,t_2)]^2$, while the  second is
$\big\langle\bar\phi^{q}_{1}
\bar\phi^{cl}_{1}\big(\phi^{cl}_{1}\big)^2
\phi^{q}_{2}\phi^{cl}_{2}\big(\bar\phi^{q}_{2}\big)^2\big\rangle
\sim [G^R(t_{1},t_{2})]^2G^R(t_{2},t_{1})G^A(t_{2},t_{1})$, where
$\phi^{\alpha}_{1,2}\equiv\phi^\alpha_{j_{1,2}}$. Both of these
terms are zero, because $G^R(t_2,t_1)\sim\theta(t_2-t_1)$, while
$G^A(t_2,t_1)\sim G^R(t_1,t_2)^*\sim\theta(t_1-t_2)$ and thus their
product has no support~\footnote{Strictly speaking,
$G^R(t_{2},t_{1})$ and $G^A(t_{2},t_{1})$ are both
 simultaneously non--zero at the diagonal: $t_{1}=t_{2}$. The
 contribution of the diagonal to the integrals, however, is $\sim
 \delta_t^2N\to 0$, when $N\to \infty$.}.
It is easy to see that, for exactly the same reasons, all
higher--order terms vanish and thus the normalization is unmodified
(at least in the perturbative expansion).

\begin{figure}
\begin{center}\includegraphics[width=10cm]{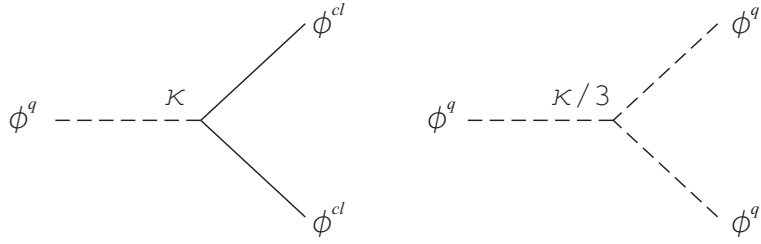}\end{center}
\caption{Graphic
representation of the two interaction vertices
 of the $\phi^3$ theory. Note the relative factor of one--third
 between them. \label{Fig-Phi3}}
\end{figure}

As another example, consider a real boson field, see Appendix
\ref{app_SPQM}, with the cubic non--linearity
\begin{equation}\label{int-bos-S-Phi-3}
S_{\mathrm{int}}=
\frac{\kappa}{6}\int\d\mathbf{r}\int_{\mathcal{C}}\d t\,
\phi^{3}=\frac{\kappa}{6}\int\d\mathbf{r} \int^{+\infty}_{-\infty}\d
t\, \big[\phi^{3}_{+}-\phi^{3}_{-}\big]=
\kappa\int\d\mathbf{r}\int^{+\infty}_{-\infty}\d t\,
\left[\big(\phi^{cl}\big)^{2}\phi^{q}+\frac{1}{3}
\big(\phi^{q}\big)^{3}\right].
\end{equation}
The causality condition~\eqref{boson-causality} is again satisfied.
Diagrammatically the cubic non--linearity generates two types of
vertices, Figure~\ref{Fig-Phi3}: one with two classical fields (full
lines) and one quantum field (dashed line), and the other with three
quantum fields. The former vertex carries the factor $\kappa$, while
the latter has  weight  $\kappa/3$. Note that for the real field the
direction of lines is not specified by arrows.

\subsection{Saddle point equations}\label{sec_int_bos-2}

Before developing  the perturbation theory further, one has to
discuss the saddle points of the action. According
to~\eqref{boson-causality}, there are no terms in the action that
have zero power of both $\bar\phi^{q}$ and $\phi^{q}$. The same is
obviously true regarding $\delta S/ \delta \bar\phi^{cl}$ and
therefore one of the saddle point equations
\begin{equation}
{\delta S\over \delta \bar\phi^{cl} } = 0\,
\end{equation}
may always be solved by
\begin{equation}\label{int-bos-Phi-q-0}
\Phi^{q}=0 \, ,
\end{equation}
irrespectively of what the  classical component, $\Phi^{cl}$, is. By
capital letter $\Phi^{cl(q)}$ we denote solutions of the saddle
point equations. One may check that this is indeed the case for the
action given by e.g.~\eqref{boson-S-2} plus \eqref{int-bos-S-1}.
Under condition~\eqref{int-bos-Phi-q-0} the second saddle point
equation takes the form
\begin{equation}\label{int-bos-GP}
 {\delta S\over \delta \bar\phi^{q} }=
\left(\big[G^R\big]^{-1} - \lambda\,|\Phi^{cl}|^2\right)
\Phi^{cl}=\left(i\partial_t +
\frac{1}{2m}\partial^{2}_{\mathbf{r}}+\mu - \lambda\,
|\Phi^{cl}|^2\right)\Phi^{cl}=0 \, .
\end{equation}
This is  the non--linear time--dependent Gross--Pitaevskii equation,
which determines the  classical field configuration, provided some
initial and boundary conditions are specified.

The message is that among the possible solutions of the saddle point
equations for the Keldysh action, there is always one with  zero
quantum component and with  classical component that obeys the
classical (non--linear) equation of motion. We shall call such a
saddle point {\em ``classical''}. Thanks to~\eqref{boson-causality}
and \eqref{int-bos-Phi-q-0}, the action on the classical saddle
point field configurations is identically zero. As was argued above,
the perturbative expansion in small fluctuations around the
classical saddle point leads to a properly normalized partition
function, $Z=1$. This seemingly excludes the possibility of having
any other saddle points. However, this conclusion is premature. The
system may posses ``non--classical'' saddle points, such that
$\Phi^{q}\neq 0$. Such saddle points do not contribute to the
partition function (and thus do not alter the fundamental
normalization, $Z=1$), however, they may contribute to observables
and correlation functions. In general, the action on a {\em
non--classical} saddle point is non--zero. Its contribution is thus
associated with exponentially small (or oscillatory) terms. Examples
include tunneling, thermal activation (considered in the next
section), oscillatory  contributions to the level statistics, {\em
etc}.

Let us develop now a systematic perturbative expansion in deviations
from the {\em classical} saddle point:
$\phi^{cl}=\Phi^{cl}+\delta\phi^{cl}$ and $\phi^q=0+\delta\phi^q$.
As was discussed above, it does not bring any new information about
the partition function. It does, however, provide information about
the Green's functions (and thus various observables). Most notably,
it generates the kinetic equation for the distribution function. To
simplify the further consideration, we restrict ourselves to
situations where no Bose condensate is present, i.e. $\Phi^{cl}=0$
is the proper solution of the classical saddle point equation
\eqref{int-bos-GP}. In this case $\phi^\alpha =\delta\phi^\alpha$
and thus the $\delta$--symbol may be omitted.

\subsection{Dyson equation}\label{sec_int_bos-3}

The next goal is to calculate the {\em dressed} Green's function,
defined as
\begin{equation}\label{int-bos-G-dressed}
\mathbf{G}^{\alpha\beta}(t,t')=-i\int\D[\bar{\phi}\phi]
\,\phi^\alpha(t)\,\bar{\phi}^{\,\beta}(t')\,
\exp\big(iS_0+iS_{\mathrm{int}}\big)\, ,
\end{equation}
here $\alpha,\beta=(cl,q)$ and the action is given
by~\eqref{boson-S-2} and \eqref{int-bos-S-1}. To this end, one may
expand the exponent in powers of $S_{\mathrm{int}}$. The functional
integration with the remaining Gaussian action is then performed
using the Wick theorem, see Appendix \ref{app_Gaussian}. This leads
to the standard diagrammatic series. Combining all one--particle
irreducible diagrams into the self--energy matrix $\hat{\Sigma}$,
one obtains
\begin{equation}\label{int-bos-G-expansion}
\hat{{\bf G}}=\hat{G}+\hat{G}\circ\hat{\Sigma}\circ
\hat{G}+\hat{G}\circ\hat{\Sigma}\circ \hat{G}\circ\hat{\Sigma}\circ
\hat{G} +\ldots= \hat{G}\circ\left(\hat 1+\hat{\Sigma}\circ\hat{{\bf
G}}\right) \, ,
\end{equation}
where $\hat G$ is given by~\eqref{boson-G-fun-2} and the circular
multiplication sign implies convolution in times and space domain as
well as a $2\times 2$ matrix multiplication. The only difference
compared with the textbook diagrammatic
expansion~\cite{Mahan,AGD,NegeleOrland} is the presence of the
$2\times 2$ Keldysh matrix structure. The fact that the series is
arranged as a sequence of matrix products is of no surprise. Indeed,
the Keldysh index, $\alpha=(cl,q)$, is just one more index in
addition to time, space, spin, etc. Therefore, as with any other
index, there is a summation
 over all of its intermediate values, hence the matrix
multiplication. The concrete form of the self--energy matrix,
$\hat{\Sigma}$, is specific to the Keldysh technique and is
discussed below in some details.

Multiplying both sides of~\eqref{int-bos-G-expansion} by
$\hat{G}^{-1}$ from the left, one obtains the Dyson equation for the
exact dressed Green's function, $\hat{{\bf G}}$, in the form
\begin{equation}\label{int-bos-Dyson}
\left(\hat{G}^{-1} -\hat{\Sigma}\right)\circ \hat{{\bf G}}=\hat 1\, ,
\end{equation}
where $\hat 1$ is a unit matrix. The very non--trivial feature of
the Keldysh technique is that the self--energy matrix,
$\hat{\Sigma}$, possesses the same causality structure as
$\hat{G}^{-1}$ (see Eq.~\eqref{boson-S-1}), namely
\begin{equation}\label{int-bos-Sigma}
\hat{\Sigma}=\left(\begin{array}{cc}
0   & \Sigma^{A}  \\
  \Sigma^{R}  & \Sigma^K
\end{array}\right)\, ,
\end{equation}
where $\Sigma^{R(A)}$ are lower (upper) triangular matrices in the
time domain, while $\Sigma^K$ is an anti--Hermitian matrix. This
fact is demonstrated below. Since both $\hat{G}^{-1}$ and
$\hat{\Sigma}$ have the same structure, one concludes that the
dressed Green's function, $\hat{{\bf G}}$, also possesses the
causality structure, like~\eqref{boson-G-fun-2}. As a result, the
Dyson equation acquires the form
\begin{equation}\label{int-bos-Dyson-matrix}
\left(\begin{array}{cc}
0   & \big[G^{A}\big]^{-1}- \Sigma^{A}  \\
  \big[G^{R}\big]^{-1}- \Sigma^{R}  & -\Sigma^K
\end{array}\right)\circ
\left(\begin{array}{cc}
{\bf G}^K   & {\bf G}^R  \\
  {\bf G}^A  & 0
\end{array}\right) =\hat 1 ,
\end{equation}
where we have taken into account that $\big[G^{-1}\big]^K$ is a pure
regularization ($\sim i0F$) and thus may be omitted in the presence
of a non--zero $\Sigma^K$. Employing the specific form of
$\big[G^{R(A)}\big]^{-1}$ (see~\eqref{boson-G-Real-Space}), one
obtains for the retarded (advanced) components
\begin{equation}\label{int-bos-Dyson-RA}
\left( i\partial_t + {1\over 2m}\partial_\mathbf{r}^2 +\mu -
\Sigma^{R(A)}\right)\circ {\bf G}^{R(A)} =
\delta(t-t')\delta(\mathbf{r}-\mathbf{r}') \, .
\end{equation}
Provided the self--energy component $\Sigma^{R(A)}$ is known (in
some approximation), equation~\eqref{int-bos-Dyson-RA} constitutes a
closed equation for the retarded (advanced) component of the dressed
Green's function. The latter carries the information about the
spectrum of the interacting system.

To write down equation for the Keldysh component we parameterize it
as ${\bf G}^K={\bf G}^R\circ {\bf F}- {\bf F}\circ {\bf G}^A$,
cf.~\eqref{boson-G-FDT-general}, where ${\bf F}$ is a Hermitian
matrix in the time domain. The equation for the Keldysh component
then takes the form $\big(\big[G^R\big]^{-1} -\Sigma^R\big)\circ
\big({\bf G}^R\circ {\bf F}-{\bf F} \circ {\bf
G}^A\big)=\Sigma^K\circ{\bf G}^A$. Multiplying it from the right by
$\big(\big[G^A\big]^{-1}-\Sigma^A\big)$ and
employing~\eqref{int-bos-Dyson-RA}, one finally finds
\begin{equation}\label{int-bos-Dyson-F}
\left[{\bf F}\,,\, \left( i\partial_t + {1\over
2m}\partial^{2}_\mathbf{r} \right) \right]=\,
\Sigma^K-\left(\Sigma^{R}\circ {\bf F} - {\bf F}
\circ\Sigma^A\right)\, ,
\end{equation}
where $[\,\,,\,]\,$ denotes for the commutator. This equation is the
quantum kinetic equation for the distribution matrix ${\bf F}$. Its
left--hand side is called the {\em kinetic} term, while the
right--hand side is the {\em collision integral} (up to a factor).
As is shown below, $\Sigma^K$ has the meaning of an ``incoming''
term, while $\Sigma^{R}\circ {\bf F} - {\bf F} \circ\Sigma^A$ is an
``outgoing'' term. In equilibrium these two channels cancel each
other (the kinetic term vanishes) and the self--energy has the same
structure as the Green's function: $\Sigma^K=\Sigma^{R}\circ {\bf F}
- {\bf F} \circ\Sigma^A$. This is not the case, however, away from
the equilibrium.

\begin{figure}
\begin{center}\includegraphics[width=10cm]{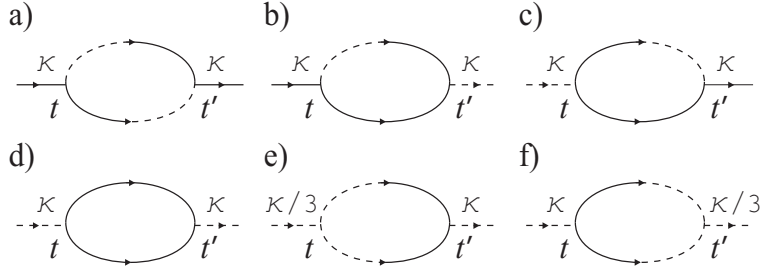}\end{center}
\caption{Self--energy diagrams for the $\phi^3$
theory.\label{Fig-Sigma}}
\end{figure}

\subsection{Self--energy}\label{sec_int_bos-4}

Let us demonstrate  that the self--energy matrix, $\hat{\Sigma}$,
indeed possesses the causality structure~\eqref{int-bos-Sigma}. To
this end, we consider the real boson field with the $\kappa\phi^3$
non--linearity~\eqref{int-bos-S-Phi-3}, and perform calculations up
to the second order in the parameter $\kappa$. Employing the two
vertices of figure~\ref{Fig-Phi3} one finds the following.

(i) The \textit{$cl-cl$} component is given by the single diagram,
depicted in Figure~\ref{Fig-Sigma}a. The corresponding analytic
expression is $\Sigma^{cl-cl}\!(t,t')=4i \kappa^2
G^R\!(t,t')G^A\!(t,t')\!=\!0$. Indeed, the product
$G^R(t,t')G^A(t,t')$ has no support (see  footnote in section
\ref{sec_int_bos-1}).

(ii) The \textit{$cl-q$} (advanced) component is given by the single
diagram, Figure~\ref{Fig-Sigma}b. The corresponding expression is
\begin{equation}\label{int-bos-Sigma-A}
\Sigma^A(t,t') = 4i \kappa^2 G^A(t,t')G^K(t,t')\,.
\end{equation}
Since $\Sigma^A(t,t')\sim G^A(t,t')\sim\theta(t'-t)$, it is, indeed,
an advanced (upper triangular) matrix. There is a combinatoric
factor of four, associated with the diagram (four ways of choosing
external legs $\times$ two internal permutations $\times$ 1/(2!) for
having two identical vertices).

(iii) The \textit{$q-cl$} (retarded) component is given by the
diagram of Figure~\ref{Fig-Sigma}c
\begin{equation}\label{int-bos-Sigma-R}
\Sigma^R(t,t') = 4i\kappa^2 G^R(t,t')G^K(t,t')\,,
\end{equation}
that could be obtained, of course, by the Hermitian conjugation
of~\eqref{int-bos-Sigma-A} with the help
of~\eqref{boson-G-conjugation}:
$\Sigma^R=\big[\Sigma^A\big]^\dagger$. Since $\Sigma^R(t,t')\sim
G^R(t,t')\sim\theta(t-t')$, it is indeed a retarded (lower
triangular) matrix.

(iv) The \textit{$q-q$} (Keldysh) component is given by the three
diagrams, Figure~\ref{Fig-Sigma}d--\ref{Fig-Sigma}f. The
corresponding expression (sum of these diagrams) is
\begin{eqnarray}\label{int-bos-Sigma-K}
\Sigma^K(t,t') = 2i\kappa^2 \big[G^K(t,t')\big]^2 +
6i\left({\kappa\over 3}\right)\kappa \big[G^A(t,t')\big]^2
+ 6i\left({\kappa\over 3}\right)\kappa \big[G^R(t,t')\big]^2\nonumber
\\
= 2i\kappa^2\left(\big[G^K(t,t')\big]^2+\big[G^R(t,t') -
G^A(t,t')\big]^2\right)\,.
\end{eqnarray}
The combinatoric factors are two for diagram
Figure~\ref{Fig-Sigma}d, and six for Figure~\ref{Fig-Sigma}e and f.
In the last equality the fact that $G^R(t,t')G^A(t,t')=0$, owing to
the absence of support in the time domain, has been used again.
Employing~\eqref{boson-G-conjugation}, one finds
$\Sigma^K=-\big[\Sigma^K\big]^\dagger$.

This demonstrates that the self--energy $\hat{\Sigma}$ possesses the
same structure as $\hat{G}^{-1}$. One may check that the statement
holds in  higher orders as well.
In~\eqref{int-bos-Sigma-A}--\eqref{int-bos-Sigma-K} one has omitted
the spatial coordinates, which may be restored in an obvious way.

\subsection{Kinetic equation}\label{sec_int_bos-5}

To make further progress in the discussion of the kinetic equation
it is  convenient to  perform the Wigner transformation (WT)
\footnote{ The Wigner transform of a matrix
$A(\mathbf{r},\mathbf{r}')$ is defined as $
a(\mathbf{R},\mathbf{k})\equiv\int\d \mathbf{r}_1  \,
A\left(\mathbf{R}+{\mathbf{r}_1\over 2},
\mathbf{R}-{\mathbf{r}_1\over 2}\right)\,\exp\{i\mathbf{kr}_1\}.$
One may show  that the Wigner transform of the matrix $C=A\circ B$,
which means $C(\mathbf{r},\mathbf{r}')=\int\d\mathbf{r}''
A(\mathbf{r},\mathbf{r}'')B(\mathbf{r}'',\mathbf{r}')$, is equal to
$$
c(\mathbf{R},\mathbf{k})=\iint\d\mathbf{r}_1\d\mathbf{r}_2 \iint{\d
\mathbf{k}_1\d\mathbf{k}_2\over (2\pi)^{2d}}\,\,
a\left(\mathbf{R}+{\mathbf{r}_1\over
2},\mathbf{k}+\mathbf{k}_1\right)
b\left(\mathbf{R}+{\mathbf{r}_2\over
2},\mathbf{k}+\mathbf{k}_2\right)
\exp\{i(\mathbf{k}_1\mathbf{r}_2-\mathbf{k}_2\mathbf{r}_1)\}\,.
$$
Expanding the functions under the integrals in $\mathbf{k}_i$ and
$\mathbf{r}_i$, one finds: $c(\mathbf{R},\mathbf{k})=
a(\mathbf{R},\mathbf{k})\,b(\mathbf{R},\mathbf{k})+(2i)^{-1}\big(\nabla_{\bf{R}}
a\nabla_\mathbf{k} b - \nabla_\mathbf{k} a\nabla_{\bf{R}} b\big)
+\ldots.$ }. The WT of a distribution function matrix, $\mathbf
{F}(\mathbf{r},\mathbf{r}';t,t')$, is a function
$\mathbf{f}(\mathbf{R},\mathbf{k};\tau,\epsilon)$, where $\tau$ and
$\bf{R}$ are central  time and coordinate, respectively. According
to the definition~\eqref{boson-G-FDT-general}, the ${\bf f}$
function appears in a product with $G^R-G^A$. The latter is a
sharply peaked function at $\epsilon=\omega_\mathbf{k}$ for free
particles, while for the interacting systems this is still the case
as long as quasi--particles are well--defined. One therefore
frequently writes ${\bf f}(\bf{R},\mathbf{k},\tau)$, understanding
that $\epsilon=\omega_\mathbf{k}$.

To rewrite the kinetic term (the left--hand side
of~\eqref{int-bos-Dyson-F}) in the Wigner representation, one
notices that the WT of $i\partial_t$ is $\epsilon$, while the WT of
$\partial^2_\mathbf{r}$ is $-\mathbf{k}^2$. Then, e.g., $[{\bf
F},\partial_\mathbf{r}^2]_-\to [\mathbf{k}^2,\mathbf{f}]_-
+i\nabla_\mathbf{k} \mathbf{k}^2\nabla_{\mathbf{R}}
\mathbf{f}=2i\mathbf{k}\nabla_{\mathbf{R}}\mathbf{f}$, where the
commutator vanishes, since WTs commute.  In a similar way: $[{\bf
F},i\partial_t]_-\to -i\partial_\tau {\bf f}$. If there is a scalar
potential
$V(\mathbf{r})\hat{b}^\dagger_\mathbf{r}\hat{b}_\mathbf{r}$ in the
Hamiltonian, it translates into the term
$-V(\bar\phi^{cl}\phi^{q}+\bar\phi^{q}\phi^{cl})$ in the action and
thus $-V(\mathbf{r})$ is added to $\big[G^{R(A)}\big]^{-1}$. This,
in turn, brings the term $-[{\bf F},V]_-$ to the left--hand side of
the Dyson equation \eqref{int-bos-Dyson-F} or, after the WT,
$i\mathbf{E}\nabla_\mathbf{k}{\mathbf{f}}$, where
$\mathbf{E}\equiv-\nabla_{\mathbf{R}} V$ is the electric field. As a
result, the WT of the Dyson equation \eqref{int-bos-Dyson-F} takes
the form
\begin{equation}\label{int-bos-kinetic-eq}
\Big(\partial_\tau-v_\mathbf{k}\nabla_{\mathbf{R}} -
\mathbf{E}\nabla_\mathbf{k}\Big)\,\mathbf{f}(\mathbf{R},\mathbf{k},\tau)
= I_{\mathrm{coll}}[\mathbf{f}]\, ,
\end{equation}
where $v_\mathbf{k}\equiv \mathbf{k}/m$ and $I_{\mathrm{coll}}[{\bf
f}]$ is the WT of the right--hand side of~\eqref{int-bos-Dyson-F}
(times $i$). This is the kinetic equation for the distribution
function.

For  real bosons with the dispersion relation
$\epsilon=\omega_\mathbf{k}$, see Appendix \ref{app_SPQM},
the kinetic term takes the form $[\epsilon^2 -
\omega_\mathbf{k}^2,{\bf F}]_-\to2i\big(\epsilon\,
\partial_\tau - \omega_\mathbf{k}(\nabla_\mathbf{k}\omega_\mathbf{k})
\nabla_{\mathbf{R}}\big)\,{\bf f}= 2i\epsilon\big(\partial_\tau -
v_\mathbf{k}\nabla_{\mathbf{R}}\big)\,{\bf f}$, where
$v_\mathbf{k}\equiv \nabla_\mathbf{k}\omega_\mathbf{k}$ is the group
velocity. As a result, the kinetic equation takes the form: $\big(
\partial_\tau - v_\mathbf{k}\,\nabla_{\mathbf{R}}\big)\,
\mathbf{f}(\mathbf{R},\mathbf{k},\tau)= I_{\mathrm{coll}}[{\bf f}]$,
where the collision integral $I_{\mathrm{coll}}[{\bf f}]$ is the WT
of the right--hand side of~\eqref{int-bos-Dyson-F}, divided by
$-2i\epsilon$.

Let us discuss the collision integral now, using the $\phi^{3}$
theory calculations of Section~\ref{sec_int_bos-4} as an example. To
shorten the algebra, let us consider a system that is spatially
uniform and isotropic in the momentum space. One thus focuses on the
energy relaxation only. In this case the distribution function is
$\mathbf{f}(\mathbf{R},\mathbf{k},\tau)=\mathbf{f}(\tau,\omega_\mathbf{k})
=\mathbf{f}(\tau,\epsilon)$, where the dependence on the modulus of
the momentum is substituted by the $\omega_\mathbf{k}=\epsilon$
argument.
Employing~\eqref{int-bos-Sigma-A}--\eqref{int-bos-Sigma-K}, one
finds for the WT of the right--hand side of\footnote{Only products
of WTs are retained, while all the gradient terms are neglected, in
particular ${\bf G}^K\to {\bf f}\,({\bf g}^R-{\bf g}^A)$. The
energy--momentum representation is used, instead of the time--space
representation as
in~\eqref{int-bos-Sigma-A}--\eqref{int-bos-Sigma-K}, and in the
equation for $\Sigma^{R}\circ {\bf F} -{\bf F} \circ\Sigma^A$ one
performs a symmetrization between the $\omega$ and $\epsilon-\omega$
arguments.}~\eqref{int-bos-Dyson-F}:
\begin{subequations}\label{int-bos-Sigma-WT}
\begin{equation}
\Sigma^{R}\circ {\bf F} -{\bf F} \circ\Sigma^A \to-2i\,{\bf
f}(\tau,\epsilon)\int\d\omega\, M(\tau,\epsilon,\omega)\, \Big[{\bf
f}(\tau,\epsilon-\omega)+{\bf f}(\tau,\omega)\Big]\,,
\end{equation}
\begin{equation}
\hskip-2.5cm \Sigma^{K}\to  -2i\int\d\omega\,
M(\tau,\epsilon,\omega)\, \Big[{\bf f}(\tau,\epsilon-\omega){\bf
f}(\tau,\omega)+1\Big]\,,
\end{equation}
\end{subequations}
where the transition rate  is given by
\begin{equation}\label{int-bos-M}
M(\tau,\epsilon,\omega) = 2\pi\kappa^2\sum_{q}\, {\bf
\Delta_{g}}(\tau,\epsilon-\omega;\mathbf{k}-\mathbf{q})\,{\bf
\Delta_{g}}(\tau,\omega;\mathbf{q})\, .
\end{equation}
Here ${\bf \Delta_{g} }\equiv i({\bf g}^R - {\bf g}^A)/(2\pi)$ and
${\bf g}^{R(A)}(\tau,\epsilon;\mathbf{k})$ are the WT of the
retarded (advanced) Green functions $\mathbf{G}^{R(A)}$. One has
substituted the dressed Green's functions
into~\eqref{int-bos-Sigma-A}--\eqref{int-bos-Sigma-K} instead of the
bare ones to perform a partial resummation of the diagrammatic
series. (This trick is sometimes called the {\em self--consistent
Born approximation}. It still neglects the vertex corrections.)
Assuming the existence of well--defined quasi--particles at all
times, one may regard ${\bf \Delta_{g}}(\tau,\epsilon,\mathbf{k})$
as a sharply peaked function at $\epsilon=\omega_\mathbf{k}$. In
this case~\eqref{int-bos-M} simply reflects the fact that an initial
particle with $\epsilon=\omega_\mathbf{k}$ decays into two real (on
mass--shell) particles with energies $\omega=\omega_\mathbf{q}$ and
$\epsilon-\omega=\omega_{\mathbf{k}-\mathbf{q}}$. As a result, one
finally obtains for the kinetic equation
\begin{equation}\label{int-bos-kinetic-eq-fin}
{\partial {\bf f}(\epsilon)\over \partial\tau}= \int\!\! {\d\omega}\,\,
{M(\epsilon,\omega)\over \epsilon}\,\, \Big\{ {\bf
f}(\epsilon-\omega){\bf f}(\omega)+1 -{\bf f}(\epsilon)\big[{\bf
f}(\epsilon-\omega)+{\bf f}(\omega)\big]\Big\},
\end{equation}
where the time arguments are suppressed for brevity. Due to the
identity: $\coth(a-b)\,\coth( b)+1=\coth(a)\big(\coth
(a-b)+\coth(b)\big)$, the collision integral is identically
nullified by ${\bf f}(\epsilon)=\coth(\epsilon/2T)$ where $T$ is a
temperature. This is the thermal equilibrium distribution function.
According to the kinetic equation \eqref{int-bos-kinetic-eq-fin}, it
is stable for any temperature (the latter is determined either by an
external reservoir, or, for a closed system, from the conservation
of total energy). Since the equilibrium distribution obviously
nullifies the kinetic term, according to~\eqref{int-bos-Dyson-F} the
{\em exact} self--energy satisfies
$\Sigma^K=\coth(\epsilon/2T)\big[\Sigma^R-\Sigma^A\big]$. Since also
the bare Green's functions obey the same
relation~\eqref{boson-G-FDT}, one concludes that in  thermal
equilibrium the {\em exact} dressed Green's function satisfies
\begin{equation}\label{int-bos-FDT}
{\bf G}^K=\big({\bf G}^R - {\bf G}^A\big)\,\coth\, {\epsilon \over
2\,T}\, .
\end{equation}
This is the statement of the {\em fluctuation--dissipation theorem}
(FDT). Its consequence is that in  equilibrium the Keldysh component
does not contain any additional information with respect to the
retarded component. Therefore, the Keldysh technique may be, in
principle, substituted by a more compact construction: the Matsubara
method. The latter does not work, of course, away from equilibrium.

Returning to the kinetic equation \eqref{int-bos-kinetic-eq-fin},
one may identify ``in'' and ``out'' terms in the collision integral.
It may be done by writing the collision integral in terms of the
occupation numbers $\mathbf{n}_{\epsilon}$, defined as ${\bf
f}_{\epsilon}=1+2\,{\bf n}_{\epsilon}$. The expression in the curly
brackets on the right--hand side of~\eqref{int-bos-kinetic-eq-fin}
takes the form $4\left[{\bf n}_{\epsilon-\omega}{\bf n}_\omega- {\bf
n}_\epsilon({\bf n}_{\epsilon-\omega}+{\bf n}_\omega+1)\right]$. The
first term ${\bf n}_{\epsilon-\omega}{\bf n}_\omega$ gives a
probability that a particle with energy $\epsilon-\omega$ absorbs a
particle with energy $\omega$ to populate a state with energy
$\epsilon$: this is the ``in'' term of the collision integral. It
may be traced back to the $\Sigma^K$ part of the self--energy. The
second term $ -{\bf n}_\epsilon({\bf n}_{\epsilon-\omega}+{\bf
n}_\omega+1)$ says that a state with energy $\epsilon$ may be
depopulated either by stimulated emission of particles with energies
$\epsilon-\omega$ and $\omega$, or by spontaneous emission. This is
the ``out'' term, that may be traced back to the $\Sigma^{R(A)}$
contributions.

Finally, let us discuss the approximations involved in the Wigner
transformations. Although~\eqref{int-bos-Dyson-F} is formally exact,
it is very difficult to extract any useful information from it.
Therefore, passing to an approximate, but much more tractable, form
such as~\eqref{int-bos-kinetic-eq} or \eqref{int-bos-kinetic-eq-fin}
is highly desirable. In doing it, one has to employ the approximate
form of the WT. Indeed, a formally infinite series in
$\nabla_\mathbf{k}\nabla_{\mathbf{R}}$ operators is truncated,
usually by the first non--vanishing term. This is a justified
procedure as long as $\delta \mathbf{k}\,\delta \mathbf{R}\gg 1$,
where $\delta \mathbf{k}$ is a characteristic microscopic scale of
the momentum dependence of ${\bf f}$, while $\delta\mathbf{R}$ is a
characteristic scale of its spatial variations. One may ask if there
is a similar requirement in the time domain: $\delta\epsilon\,
\delta \tau\gg 1$, with $\delta\epsilon$ and $\delta \tau$ being the
characteristic energy and the time scales of ${\bf f}$,
correspondingly. Such a requirement is very demanding, since
typically $\delta\epsilon \approx T$ and at low temperature it would
only allow very slow processes to be treated: with $\delta\tau \gg
1/T$. Fortunately, this is not the case. Because of the peaked
structure of ${\bf \Delta_g}(\epsilon,\mathbf{k})$, the energy
argument $\epsilon$ is locked to $\omega_\mathbf{k}$ and does not
have its own dynamics as long as the peak is sharp. The actual
criterion is therefore that $\delta\epsilon$ is much larger than the
width of the peak in ${\bf \Delta_g}(\epsilon,\mathbf{k})$. The
latter is, by definition, the quasi--particle life--time,
$\tau_{qp}\,$, and therefore the condition is $\tau_{qp}\gg 1/T$.
This condition is indeed satisfied by many systems where the
interactions  are not too strong.

\section{Particle in contact with an environment}\label{sec_environment}
\subsection{Quantum dissipative action}\label{sec_environment-1}

Consider a particle with the coordinate $\Phi(t)$, placed in a
potential $U(\Phi)$ and attached to a harmonic string
$\varphi(x,t)$. The particle may represent a collective degree of
freedom, such as the phase of a Josephson junction or the charge on
a quantum dot. On the other hand, the string serves to model a
dissipative environment. The advantage of the one--dimensional
string is that it is the simplest continuum system, having a
constant density of states at small energies. Owing to this property
it mimics, for example, interactions with a Fermi sea. A continuum
reservoir with a constant density of states at small energies is
sometimes called an ``Ohmic''  environment (or bath). The
environment is supposed to be in  thermal equilibrium.

The Keldysh action of such a system is given by the three terms
$S=S_{\mathrm{p}}+S_{\mathrm{str}}+S_{\mathrm{int}}$, where (see
Appendix \ref{app_SPQM})
\begin{subequations}\label{environment-S-Str-Particle}
\begin{equation}
S_{\mathrm{p}}[\Phi]=\int_{-\infty}^{+\infty}\d
t\left[-2\,\Phi^{q}{\d^{\,2} \Phi^{cl}\over \d t^2}
-U\left(\Phi^{cl}+\Phi^{q}\right)+ U(\Phi^{cl}-\Phi^{q}) \right]\,,
\end{equation}
\begin{equation}
\hskip-2.9cm S_{\mathrm{str}}[\varphi]=\int_{-\infty}^{+\infty} \d
t\int\d x\ \vec{\varphi}^{T}(x,t) \hat{D}^{-1}\vec{\varphi}(x,t)\,,
\end{equation}
\begin{equation}
\hskip-2cm S_{\mathrm{int}}[\Phi,\varphi]=
2\sqrt{\gamma}\int_{-\infty}^{+\infty}\d t
\left.\vec{\Phi}^{T}(t)\,\hat{\sigma}_x\,\partial_x\vec{\varphi}(x,t)\right|_{x=0}\,.
\end{equation}
\end{subequations}
Here we have introduced vectors of classical and quantum components,
e.g., $\vec{\Phi}^T\equiv (\Phi^{cl},\Phi^{q})$ and the string
correlator, $\hat{D}^{-1}$, that has typical bosonic
form~\eqref{boson-S-2},  with
$\big[D^{R(A)}\big]^{-1}=-\partial^{2}_{t}+v^{2}_{s}\partial^{2}_{x}$,
which follows from~\eqref{app-SPQM-S-Phonons-1}. The
$S_{\mathrm{p}}$ represents a particle (see corresponding discussion
in~\eqref{app-SPQM-S-3} in Appendix~\ref{app_SPQM}). The
$S_{\mathrm{str}}$ is the action of the
string~\eqref{app-SPQM-S-Phonons-1}. The interaction term between
the particle and the string is taken to be the local product of the
particle coordinate and the string stress at $x=0$ (so the force
acting on the particle is proportional to the local stress of the
string). In the time domain the interaction is instantaneous,
$\Phi(t)\partial_x\varphi(x,t)|_{x=0}\to
\Phi_+\partial_{x}\varphi_+-\Phi_-\partial_{x}\varphi_-$ on the
Keldysh contour. Transforming to the classical--quantum notation
leads to
$2(\Phi^{cl}\partial_{x}\varphi^{q}+\Phi^{q}\partial_{x}\varphi^{cl})$,
that satisfies the causality condition~\eqref{boson-causality}. In
the matrix notation it takes the form
of~(\ref{environment-S-Str-Particle}c). The interaction constant is
denoted $\sqrt{\gamma}$.

One may now integrate out the degrees of freedom of the harmonic
string to reduce the problem to the particle coordinate only.
According to the standard rules of  Gaussian integration (see
Appendix~\ref{app_Gaussian}), this leads to the so--called
dissipative action for the particle
\begin{subequations}\label{environment-S-diss}
\begin{equation}
\hskip-.7cm S_{\mathrm{diss}}=\gamma\iint_{-\infty}^{+\infty}\d
t\,\d t' \, \vec{\Phi}^T(t) \hat{\mathfrak{D}}^{-1}(t-t')
\hat\Phi(t') \, ,
\end{equation}
\begin{equation}
\hat{\mathfrak{D}}^{-1}(t-t\,')=-\left.\hat{\sigma}_{x}\,\partial_x\,\partial_{x\,'}
\hat{D}(x-x\,';t-t\,')\right|_{x=x'=0}\, \hat{\sigma}_{x}\,.
\end{equation}
\end{subequations}
The straightforward matrix multiplication shows that the dissipative
correlator $\hat{\mathfrak{D}}^{-1}$ possesses the standard
causality structure. Fourier transforming its retarded (advanced)
components, one finds
\begin{equation}\label{environment-L-RA}
\big[\mathfrak{D}^{R(A)}(\epsilon)\big]^{-1}=
-\sum\limits_k\frac{k^2}{(\epsilon\pm i0)^2-k^2} = \pm\, {i\over
2}\,\epsilon + \mbox{const}\, ,
\end{equation}
where we put $v_s=1$ for brevity. The $\epsilon$--independent
constant (same for $R$ and $A$ components) may be absorbed into the
redefinition of the harmonic part of the potential
$U(\Phi)=\mbox{const}\, \Phi^2+\ldots$ and, thus, may be omitted. In
equilibrium the Keldysh component of the correlator is set by the
FDT
\begin{equation}\label{environment-L-K}
\big[\mathfrak{D}^{-1}\big]^{K}(\epsilon)=\left(
\big[\mathfrak{D}^{R}\big]^{-1}-\big[\mathfrak{D}^{A}\big]^{-1}
\right)\coth{\epsilon\over 2\,T} = i\epsilon\,\coth{\epsilon\over
2\,T}\, .
\end{equation}
It is an anti--Hermitian operator with a positive--definite
imaginary part, rendering  convergence of the functional integral
over $\Phi$.

In the time representation the retarded (advanced) component of the
correlator takes a  time--local form
$\big[\mathfrak{D}^{R(A)}\big]^{-1}= \mp {1\over
2}\,\delta(t-t')\,\partial_{t}$. On the other hand, at low
temperatures the Keldysh component is a non--local function, that
may be  found by the inverse Fourier transform
of~\eqref{environment-L-K}:
\begin{equation}\label{eqenvironment-L-K-time}
\big[\mathfrak{D}^{-1}\big]^{K}(t-t') =i\left[(2T+C)\delta(t-t')-
\frac{\pi T^2}{\sinh^2[\pi T(t-t')]}\right]\,,
\end{equation}
where the infinite constant $C$ serves to satisfy the condition
$\int\d t\big[\mathfrak{D}^{-1}(t)\big]^{K}=
\big[\mathfrak{D}^{-1}(\epsilon=0)\big]^{K}=2iT$. Finally, for the
Keldysh action of the particle connected to a string, one obtains
\begin{eqnarray}\label{environment-S-diss-fin}
S[\Phi]=\int_{-\infty}^{+\infty}\d t\left[-2\,\Phi^q\left({\d^{\,2}
\Phi^{cl}\over \d t^2} +{\gamma\over 2}{\d \Phi_{cl}\over \d
t}\right)- U\left(\Phi^{cl}+\Phi^{q}\right)+ U(\Phi^{cl}-\Phi^{q})
\right]\nonumber
\\
 +2i\gamma\int^{+\infty}_{-\infty}\d t\left[T\big(\Phi^{q}(t)\big)^2+
  \frac{\pi T^2}{2} \int_{-\infty}^{+\infty}\d t'\,
\frac{\big(\Phi^q(t)-\Phi^q(t')\big)^2}{\sinh^2[\pi
T(t-t')]}\right]\,.
\end{eqnarray}
This action satisfies all the causality criterions  listed in
Sec.~\ref{sec_bosons-3}. Note that, in the present case, the Keldysh
($q-q$) component is not just a regularization factor, but rather a
quantum fluctuations damping term, originating from the coupling to
the string. The other manifestation of the string is the presence of
the friction term, $\sim \gamma\partial_t$ in the $R$ and the $A$
components. In equilibrium the friction coefficient and fluctuations
amplitude are connected rigidly by the FDT. The quantum dissipative
action~\eqref{environment-S-diss-fin} is a convenient playground to
demonstrate various approximations and connections to other
approaches.

\subsection{Classical limit}\label{sec_environment-2}

The {\em classical} saddle point equation (the one that takes
$\Phi^{q}(t)=0$) has the form:
\begin{equation}\label{environment-Newton-eq}
\left. -{1\over 2}\frac{\delta S[\Phi]}{\delta
\Phi^{q}}\right|_{\Phi^{q}=0} = {\d^{\,2} \Phi^{cl}\over \d t^2}
+{\gamma\over 2}{\d \Phi^{cl}\over \d t} + {\partial
U(\Phi^{cl})\over
\partial \Phi^{cl} } =0 \, .
\end{equation}
This is the deterministic classical equation of motion. In the
present case it happens to be  Newton equation with the viscous
force $-(\gamma/2) \dot \Phi^{cl}$. This approximation neglects both
{\em quantum} and {\em thermal} fluctuations.

One may keep the thermal fluctuations, while completely neglecting
the quantum ones. To this end, it is convenient to restore the
Planck constant in the action \eqref{environment-S-diss-fin} and
then take the limit $\hbar\to 0$. For dimensional reasons, the
factor $\hbar^{-1}$ should stand in front of the action. To keep the
part of the action responsible  for the classical equation of motion
\eqref{environment-Newton-eq} free from the Planck constant  it is
convenient to  rescale the variables as $\Phi^{q} \to \hbar
\Phi^{q}$. Finally, to keep proper  units, one needs to substitute
$T\to T/\hbar$ in the last term of~\eqref{environment-S-diss-fin}.
The limit $\hbar\to 0$ is now straightforward: (i) one has to expand
$U(\Phi^{cl}\pm \hbar\Phi^{q})$ to the first order in
$\hbar\Phi^{q}$ and neglect all higher order terms; (ii) in the
$\hbar\to0$ limit, which is equivalent to the $T\to\infty$, the
non--local part of the action \eqref{environment-S-diss-fin} drops
out, while the local term $\propto \big(\Phi^{q}(t)\big)^2$
survives. As a result, the classical limit of the dissipative action
is
\begin{equation}\label{environment-S-diss-class}
S[\Phi]=2\int_{-\infty}^{+\infty}\d t\left[ -\Phi^{q}\left(
 {\d^{\,2} \Phi^{cl}\over \d t^2}+{\gamma\over 2}{\d
\Phi^{cl}\over \d t}+{\partial U(\Phi^{cl})\over
\partial \Phi^{cl} } \right)+i\gamma\, T\, \big(\Phi^{q}\big)^2 \right].
\end{equation}
Physically the limit $\hbar\to 0$ means that $\hbar\Omega\ll T$,
where $\Omega$ is a characteristic classical frequency of the
particle. This condition is necessary for the last term
of~\eqref{environment-S-diss-fin} to take the time--local form. The
condition for neglecting the higher--order derivatives of $U$ is
$\hbar\ll \gamma  \big(\tilde{\Phi}^{cl}\big)^2$, where
$\tilde\Phi^{cl}$ is a characteristic classical amplitude of the
particle motion.

\subsection{Langevin equation}\label{sec_environment-3}

One way to proceed with the classical action
\eqref{environment-S-diss-class} is to note that the exponent of its
last term (times $i$) may be identically rewritten in the following
way
\begin{equation}
\exp\left(-2\gamma T\int^{+\infty}_{-\infty}\d
t\,\big[\Phi^{q}(t)\big]^{2}\right)=\int\D[\xi]\
\exp\left(-\int^{+\infty}_{-\infty}\d t\left[{\xi^2(t)\over 2\gamma
T} - 2i \xi(t)\Phi^{q}(t)\right]\right)\, .
\end{equation}
This identity is  called the Hubbard--Stratonovich transformation,
while $\xi(t)$ is an auxiliary Hubbard--Stratonovich field. The
identity is proved by completing the square in the exponent on the
right--hand side and performing the Gaussian integration at every
instance of time. There is a constant multiplicative factor hidden
in the integration measure, $\D[\xi]$.

Exchanging the order of the functional integration over $\xi$ and
$\Phi$, one finds for the partition function:
\begin{eqnarray}\label{environment-Langevin-Z}
 Z\!&=&\!\int\D[\xi]\,
 \exp\left(-{1\over 2\gamma T}\int^{+\infty}_{-\infty}\d t\, \xi^2(t)\right)\nonumber\\
 &\times&\! \int\D\big[\Phi^{cl}\big]\int\D\big[\Phi^{q}\big]\,
 \exp\left(-2i\int^{+\infty}_{-\infty}\d t\, \Phi^{q}(t)
 \left[ {\d^{\,2}\Phi^{cl}\over \d t^2} +{\gamma\over 2}{\d
\Phi^{cl}\over \d t}+ {\partial U(\Phi^{cl})\over
\partial \Phi^{cl} }-\xi(t) \right]\right).
\end{eqnarray}
Since the  exponent depends  linearly on $\Phi^{q}(t)$, the
integration over $\D\big[\Phi^{q}\big]$ results in the
$\delta$--function of the expression in the square brackets. This
functional $\delta$--function enforces its argument to be zero at
every instant  of time. Therefore, among all  possible trajectories
$\Phi^{cl}(t)$, only those that satisfy the following equation
contribute to the partition function:
\begin{equation}\label{environment-Langevin-eq}
 {\d^{\,2} \Phi^{cl}\over \d t^2} +{\gamma\over 2}{\d \Phi^{cl}\over
\d t} + {\partial U(\Phi^{cl})\over \partial \Phi^{cl} } = \xi(t) \,.
\end{equation}
This is  Newton equation with a time--dependent external force
$\xi(t)$. Since, the same arguments are applicable to any
correlation function of the classical fields, e.g. $\big\langle
\Phi^{cl}(t)\Phi^{cl}(t')\big\rangle$, a solution strategy is as
follows: (i) choose some realization of $\xi(t)$; (ii)
solve~\eqref{environment-Langevin-eq} (e.g. numerically); (iii)
having its solution, $\Phi^{cl}(t)$, calculate the correlation
function; (iv) average the result over an ensemble of realizations
of the force $\xi(t)$. The statistics of the latter is dictated by
the weight factor in the $\D[\xi]$ functional integral
in~\eqref{environment-Langevin-Z}. It states that $\xi(t)$ is a
Gaussian short--range (white) noise with the correlators
\begin{equation}\label{environment-Langevin-noise}
\langle \xi(t) \rangle = 0\,, \hskip 1cm \langle \xi(t)
\xi(t')\rangle = \gamma T \delta(t-t')\,.
\end{equation}
Equation \eqref{environment-Langevin-eq} with the white noise on the
right--hand side is called the Langevin equation. It describes
classical Newtonian dynamics in presence of stochastic thermal
fluctuations. The fact that the noise amplitude is proportional to
the friction coefficient, $\gamma$, and  the temperature is a
manifestation of the FDT. The latter holds as long as the
environment (string) is at thermal equilibrium.

\subsection{Martin--Siggia--Rose method}\label{sec_environment-4}

In the previous section we derived the Langevin equation for a
classical coordinate, $\Phi^{cl}$, from the action written in terms
of $\Phi^{cl}$ and another field, $\Phi^q$. An inverse procedure of
deriving the effective action from the Langevin equation is known as
Martin--Siggia--Rose (MSR) technique~\cite{MSR}. It is sketched here
in the form suggested by DeDominicis~\cite{DeDominicis}.

Consider  a Langevin equation
\begin{equation}
\hat {\cal O}\big[\Phi^{cl}\big] = \xi(t)\, ,
\end{equation}
where $\hat {\cal O}\big[\Phi^{cl}\big]$ is a non--linear
differential operator acting on the coordinate $\Phi^{cl}(t)$, and
$\xi(t)$ is a white noise force, specified
by~\eqref{environment-Langevin-noise}. Define the ``partition
function'' as
\begin{equation}\label{environment-Z-MSR}
Z[\xi]=\int\D[\Phi^{cl}]\, {\cal J}[\hat {\cal O }
]\,\delta\big(\hat {\cal O}\big[\Phi^{cl}\big] - \xi(t)\big) \equiv
1\, .
\end{equation}
It is identically equal to unity by virtue of  the integration of
the $\delta$--function, provided ${\cal J}[\hat {\cal O}]$ is the
Jacobian of the operator $\hat {\cal O}\big[\Phi^{cl}\big]$. The way
to interpret~\eqref{environment-Z-MSR} is to discretize the time
axis, introducing $N$--dimensional vectors
$\Phi^{cl}_j=\Phi^{cl}(t_j)$ and $\xi_j=\xi(t_j)$. The operator
takes the form: ${\cal O}_i=\mathcal{O}_{ij}\Phi^{cl}_j+{1\over
2}\Gamma_{ijk}\Phi^{cl}_j\Phi^{cl}_k +\ldots$, where  summations are
taken  over repeated indexes. The Jacobian, ${\cal J}$, is given by
the absolute value of the determinant of the following $N\times N$
matrix: $J_{ij}\equiv
\partial {\cal O}_i/ \partial\Phi^{cl}_j=
\mathcal{O}_{ij}+\Gamma_{ijk}\Phi^{cl}_k+\ldots$. It is  possible to
choose a proper (retarded) regularization  where the  $J_{ij}$
matrix is a lower triangular matrix with a unit main diagonal
(coming entirely from the $\mathcal{O}_{ii}=1$ term). One then finds
that, in this case, ${\cal J}=1$. Indeed, consider, for example,
$\hat {\cal O}\big[\Phi^{cl}\big]=\partial_t\Phi^{cl}-U(\Phi^{cl})$.
The retarded regularized version of the Langevin equation is
$\Phi^{cl}_{i}
=\Phi^{cl}_{i-1}+\delta_t(U(\Phi^{cl}_{i-1})+\xi_{i-1})$. Clearly in
this case $J_{ii}=1$ and $J_{i,i-1}=-1-\delta_t
U'(\Phi^{cl}_{i-1})$, while all other components are zero; as a
result ${\cal J}=1$.

Although the partition function \eqref{environment-Z-MSR} is
trivial, it is clear that all  meaningful observables and the
correlation functions may be obtained by inserting a set of factors:
$\Phi^{cl}(t)\Phi^{cl}(t')\ldots$ into the functional
integral~\eqref{environment-Z-MSR}. Having this in mind, let us
proceed with the partition function. Employing the integral
representation of the $\delta$--function with the help of an
auxiliary field $\Phi^{q}(t)$, one obtains
\begin{equation}
Z[\xi]=\int\D[\Phi^{cl},\Phi^{q}]\,\exp\left(-2i\int\d
t\,\Phi^{q}(t) \big[\hat {\cal O}^R[\Phi^{cl}(t)] -
\xi(t)\big]\right)\, ,
\end{equation}
where $\hat {\cal O}^R$ stands for the retarded regularization of
the $\hat {\cal O}$ operator and thus one takes ${\cal J}=1$. One
may average now over the white
noise~\eqref{environment-Langevin-noise}, by performing the Gaussian
integration over $\xi$
\begin{eqnarray}\label{environment-Z-MSR-average}
Z&=&\int\D[\xi] \exp\left(-{1\over 2\gamma T}\int\d
t\,\xi^2(t)\right)Z[\xi]\nonumber
\\
&=&\int\D[\Phi^{cl},\Phi^{q}]\,
 \exp\left(-\int\d t \,\left[ 2i\,\Phi^{q}(t)\hat{\cal O}^R\big[\Phi^{cl}(t)\big]+2\gamma
T\big[\Phi^q(t)\big]^{2}\right] \right)\, .
\end{eqnarray}
The exponent in~\eqref{environment-Z-MSR-average} is exactly the
classical limit of the Keldysh action,
cf.~\eqref{environment-S-diss-class}, including the retarded
regularization of the differential operator. The message is that MSR
action is nothing else but the classical (high--temperature) limit
of the Keldysh action. The MSR technique provides a simple way to
transform from a classical stochastic problem to its proper
functional representation. The latter is useful for analytical
calculations. One example is given below.

\subsection{Thermal activation}\label{sec_environment-5}

\begin{figure}
\begin{center}\includegraphics[width=10cm]{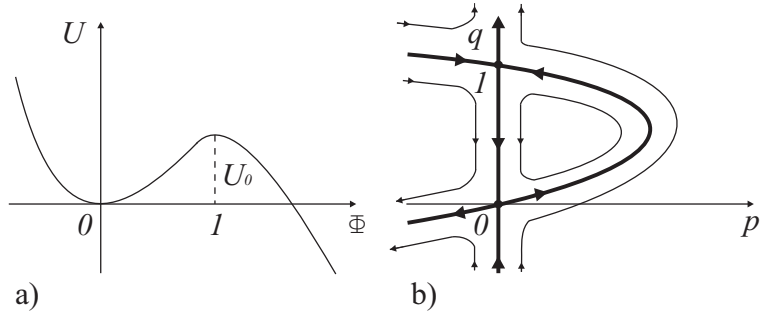}\end{center}
\caption{ a) A potential with a meta--stable minimum. b) The phase
portrait of the Hamiltonian system~\eqref{environment-activation-H}.
Thick lines correspond to zero energy, arrows indicate evolution
direction.\label{Fig-Activation}}
\end{figure}

Consider a particle in a meta--stable potential well, plotted in
Figure~\ref{Fig-Activation}a. The potential has a meta--stable
minimum at $\Phi=0$, and a maximum at $\Phi=1$ with the relative
hight $U_0$. Let us also assume that the particle's motion is
over--damped, i.e. $\gamma\gg \sqrt{U''}$. In this case one may
disregard the inertia term, leaving only viscous relaxation
dynamics. The classical dissipative action
\eqref{environment-S-diss-class} takes the form
\begin{equation}\label{environment-S-diss-viscous}
S[\Phi] = 2\int_{-\infty}^{+\infty}\d t\left[ -\Phi^{q}(t)
\left({\gamma\over 2}{\d \Phi_{cl}\over \d t}+ {\partial
U(\Phi^{cl})\over
\partial \Phi^{cl} } \right) + i\gamma\, T\, \big[\Phi^{q}(t)\big]^2\, \right]\, .
\end{equation}
The corresponding saddle point equations are
\begin{subequations}\label{environment-escape-eq}
\begin{equation}
{\gamma\over 2}\,\dot\Phi^{cl}= - {\partial U(\Phi^{cl})\over
\partial \Phi^{cl} } + 2i\gamma T\, \Phi^q\,,
\end{equation}
\begin{equation}
\hskip-1.2cm
{\gamma\over 2}\, \dot\Phi^{q}=  \Phi^q \,{\partial^2
U(\Phi^{cl})\over
\partial (\Phi^{cl})^2 } \,.
\end{equation}
\end{subequations}
These equations possess the {\em classical} solution:
$\Phi^q(t)\equiv 0$ whereas $\Phi^{cl}(t)$ satisfies the classical
equation of motion: ${\gamma\over 2}\,\dot\Phi^{cl}= -\partial
U(\Phi^{cl})/\partial \Phi^{cl}$. For the  initial condition
$\Phi^{cl}(0)<1$ the latter equation predicts the viscous relaxation
towards the minimum at $\Phi^{cl}=0$. According to this equation,
there is no possibility to escape from this minimum. Therefore, the
classical solution of~\eqref{environment-escape-eq} does {\em not}
describe thermal activation. Thus, one has to look  for another
possible solution of~\eqref{environment-escape-eq}, the one with
$\Phi^q\neq 0$.

To this end, let us perform a linear change of variables:
$\Phi^{cl}(t)=q(t)$ and $\Phi^q(t)=p(t)/(i\gamma)$. Then the
dissipative action \eqref{environment-S-diss-viscous} acquires the
form of a Hamiltonian action $iS=-\int\d t\,\big(p\,\dot{q} - H(p,q)
\big)$ where the effective  Hamiltonian
\begin{equation}\label{environment-activation-H}
H(p,q)\equiv {2\over \gamma}\left[ -p\, {\partial U(q)\over
\partial q} + Tp^{\,2} \right] ,
\end{equation}
is introduced.  It is straightforward to see that in terms of the
new variables the equations of motion \eqref{environment-escape-eq}
take the form of the Hamilton equations: $\dot q=\partial H/\partial
p$ and $\dot p=-\partial H/\partial q$. One needs, thus, to
investigate the Hamiltonian system with the
Hamiltonian~\eqref{environment-activation-H}. To visualize it, one
may plot its phase portrait, consisting of lines of constant energy
$E=H(p(t),q(t))$ on the $(p,q)$ plane, see
Figure~\ref{Fig-Activation}b. The topology is determined by the two
lines of zero energy, $p=0$ and $Tp=\partial U(q)/\partial q$, that
intersect at the two stationary points of the potential, $q=0$ and
$q=1$. The $p=0$ line corresponds to the classical (without Langevin
noise) dynamics (note that the action is identically zero for motion
along this line) and thus $q=0$ is the stable point, while $q=1$ is
the unstable one. Owing to Liouville theorem, every fixed point must
have one stable and one unstable direction. Therefore, along the
``non--classical'' line $p=T^{-1}\partial U(q)/\partial q$, the
situation is reversed: $q=0$ is unstable, while $q=1$ is stable. It
is clear now that to escape from the bottom of the potential well,
$q=0$, the system must evolve along the non--classical line of zero
energy until it reaches the top of the barrier, $q=1$, and then
continue to move according to the classical equation of motion (i.e.
moving along the classical line $p=0$). There is a non--zero action
associated with the motion along the non--classical line $$ iS=
-\int\d t\, p\dot q=-\int_0^1 p(q)\d q=-{1\over T}\int_0^1 {\partial
U(q)\over \partial q}\d q = -\,{U_0\over T}\,,$$ where one has used
that $H=0$ along the trajectory. As a result, the thermal escape
probability is proportional to $e^{iS}=e^{-U_0/T}$, which is nothing
but the thermal activation exponent.

Amazingly, this trick of rewriting viscous (or diffusive) dynamics
in a Hamiltonian form works in a wide class of problems, see
e.g.~\cite{Elgart}. The price one has to pay is the doubling of the
number of degrees of freedom: $q$ and $p$ in the Hamiltonian
language, or ``classical'' and ``quantum'' components in the Keldysh
language.

\subsection{Fokker--Planck equation}\label{sec_environment-6}

Another way to approach the action
\eqref{environment-S-diss-viscous} is to notice that it is quadratic
in $\Phi^q$ and therefore the $\D[\Phi^q]$ integration may be
performed explicitly. To shorten notation and emphasize the relation
to the classical coordinate, we follow the previous section and use
the notation $\Phi^{cl}(t)\equiv q(t)$. Performing the Gaussian
integration over $\Phi^q$ of $\exp\big(iS[\Phi]\big)$, with
$S\big[\Phi^{cl},\Phi^q\big]$ given
by~\eqref{environment-S-diss-viscous}, one finds the action,
depending on $\Phi^{cl}\equiv q$ only
\begin{equation}\label{environment-S-FP}
iS[q] = -{1\over 2\gamma T}\int_{-\infty}^{+\infty}\d t\,
\left({\gamma\over 2}\,\dot q+U'_q\right)^2 \, .
\end{equation}
One may now employ  the same trick, which allows to pass from the
Feynman path integral to the Schr\"odinger
equation~\cite{FeynmanHibbs}. Namely, let us introduce the ``wave
function'', $\mathcal{P}(q,t)$, that is a result of the functional
integration of $\exp(iS[q])$ over all trajectories that at time
$t+\delta_t$ pass through the point $q_N\equiv q$. Considering
explicitly the  last time--step, $\delta_t$, integration, one may
write ${\cal P}(q_N,t+\delta_t)$ as an integral of ${\cal
P}(q_{N-1},t)={\cal P}(q-\delta_q,t)$ over $\delta_q\equiv
q-q_{N-1}$:
\begin{eqnarray}
\hskip-.4cm {\cal
P}(q,t+\delta_t)\!\!\!\!\!&=&\!\!\!\!\!C\int_{-\infty}^{\infty}
\d[\delta_q] \exp\left(-{\delta_t\over 2\gamma T}\left[{\gamma\over
2}\,{\delta_q\over\delta_t}+ U_q'(q-\delta_q)\right]^2\right){\cal
P}(q-\delta_q,t)\nonumber
\\
&=&\!\!\!\!\!C\int_{-\infty}^{\infty} \d[\delta_q]
\exp\left(-{\gamma\over 8 T}
{\delta_q^2\over\delta_t}\right)\left[\exp\left(\,-{\delta_q\over
2T}\, U'_q(q-\delta_q) - {\delta_t\over 2\gamma T} \left( U'_q
\right)^2\right){\cal P}(q-\delta_q,t)\right]\,,
\end{eqnarray}
where the integration measure $C$ is determined by the condition:
$C\int\d[\delta_q]\,
\exp\big(-\gamma\delta_q^2/(8T\delta_t)\big)=1$. Expanding the
expression in the square brackets on the right--hand side of the
last equation to the second order in $\delta_q$ and the first order
in $\delta_t$, one finds
\begin{eqnarray}
{\cal P}(t+\delta_t)
&=&
\left( 1+{\langle \delta_q^2\rangle\over
2T}U''_{qq} + {1\over 2} {\langle \delta_q^2\rangle\over 4T^2}
\left( U'_q \right)^2 - {\delta_t\over 2\gamma T} \left( U'_q
\right)^2\right){\cal P}+{\langle \delta_q^2\rangle\over
2T}U'_q{\cal P}'_q \nonumber
+{\langle \delta_q^2\rangle\over 2}{\cal P}''_{qq}  \\
&=&
{\cal P}(t)+\frac{2\delta_t}{\gamma}\left(U''_{qq}\,{\cal P}+U'_q {\cal
P}'_q+ T{\cal P}''_{qq}\right)\,,
\end{eqnarray}
where $$\langle \delta_q^2\rangle\equiv C\int_{-\infty}^{\infty}
\d[\delta_q] \, \delta^2_q \,
\exp\left(-\frac{\gamma\delta_q^2}{8T\delta_t}\right) =
4T\delta_t/\gamma\,.$$ Finally, rewriting the last expression in the
differential form, one obtains
\begin{equation}
{\partial {\cal P}\over\partial t}={2\over \gamma}\left[
 {\partial \over \partial
q}{\partial U\over \partial q}  + T{\partial^2\over \partial
q^2}\right] {\cal P}
 ={2\over \gamma}\,
 {\partial \over \partial
q}\left[ {\partial U\over \partial q}\,{\cal P}  + T\,{\partial
{\cal P}  \over
\partial q}\right]\,.
\end{equation}
This is the Fokker--Planck (FP) equation for  the evolution of the
probability distribution function, ${\cal P}(q,t)$. The latter
describes the probability  to find the particle at a point
$q=\Phi^{cl}$ at time $t$. If one starts from an initially sharp
(deterministic) distribution, ${\cal P}(q,0)= \delta(q-q(0))$, then
the first term on the right--hand side of the FP equation describes
the viscous drift of the particle in the potential $U(q)$. Indeed,
in the absence of the second term ($T=0$), the equation is solved by
${\cal P}(q,t)= \delta(q-q(t))$, where $q(t)$ satisfies the
deterministic equation of motion\footnote{To check this statement
one may substitute ${\cal P}(q,t)= \delta(q-q(t))$ into the $T=0$ FP
equation: $\delta'_q(q-q(t))(-\dot q(t)) =(2/\gamma)\left[U''_{\!
qq}\delta(q-q(t)) + U'_q\delta'_q(q-q(t))\right]$. Then multiplying
both parts of this equation by $q$ and integrating over $\d q$ (by
performing integration by parts), one finds $\dot q(t) =
-(2/\gamma)U'_q(q(t))$.} $(\gamma/2)\dot q(t)=-\partial
U(q(t))/\partial q$ . The second term on the right--hand side
describes the diffusion spreading of the probability distribution
owing to the thermal stochastic noise $\xi(t)$. For a confining
potential $U(q)$ (such that $U(\pm\infty)\to \infty$) the stationary
solution of the FP equation is the equilibrium Boltzmann
distribution: ${\cal P}(q)\sim \exp\{-U(q)/T\}$.

The FP equation  may be considered as the (imaginary time)
Schr\"odinger equation: $\dot {\cal P}=\hat H {\cal P}$, where the
Hamiltonian, $\hat H$, is nothing but the ``quantized'' version of
the classical Hamiltonian~\eqref{environment-activation-H},
introduced in the previous section. The ``quantization'' rule is
$p\to \hat p\equiv -\partial/\partial q$, so the canonical
commutation relation $[q,\hat p]=1$, holds. Notice that before
applying this quantization rule, the corresponding classical
Hamiltonian must be {\em normally ordered}. Namely, the momentum
$\hat p$ should be to the left of the coordinate $q$,
cf.~\eqref{environment-activation-H}. Using the commutation
relation, one may rewrite the quantized Hamiltonian as $$\hat H
=T\hat p^2-\hat pU'_q= T\left(\hat p - U'_q/(2T)\right)\left(\hat p
- U'_q/(2T)\right) - (U'_q)^2/(4T) + U''_{qq}/2$$ (we put
$\gamma/2=1$) and perform the canonical transformation $Q=q$ and
$\hat P =\hat p -U'_q/(2T)$. In terms of these new variables the
Hamiltonian takes the familiar form $\hat H = T\hat P^2 +V(Q)$,
where $V(Q)= - (U'_Q)^2/(4T) + U''_{QQ}/2$, while the ``wave
function'' transforms as $\tilde {\cal P}(Q,t) = e^{U(Q)/(2T)} {\cal
P}$.

\subsection{From Matsubara to Keldysh}\label{sec_environment-7}

In some applications it may be convenient to derive an action in the
equilibrium  {\em Matsubara} technique~\cite{Matsubara,AGD} and
change to the Keldysh representation at a later stage to tackle
out--of--equilibrium problems. This section intends to illustrate
how such a transformation may be carried out. To this end, consider
the following bosonic Matsubara action:
\begin{equation}\label{environment-S-Matsubara}
S[\Phi_m]=\gamma\, T\!\! \sum\limits_{m=-\infty}^\infty\!{1\over
2}\,  |\epsilon_m||\Phi_m|^2\, ,
\end{equation}
where $\epsilon_m=2\pi T m$ and $\Phi_m=\Phi_{-m}^*=\int_0^{\beta}
d\tau \Phi(\tau) e^{i\epsilon_m \tau}$ are the Matsubara components
of a real bosonic field, $\Phi(\tau)$, with the periodic boundary
conditions $\Phi(0)=\Phi(\beta)$. Note that, owing to the absolute
value sign, $|\epsilon_m|\neq i\partial_\tau$. In fact, in the
imaginary time representation the kernel
$\mathbb{K}_{m}=|\epsilon_{m}|$ of the
action~\eqref{environment-S-Matsubara} acquires the form
$\mathbb{K}(\tau)=\sum_{m}|\epsilon_m|e^{-i\epsilon_{m}\tau}=C\delta(\tau)-\pi
T^2/\sin^{2}(\pi T\tau)$, where the infinite constant $C$ is chosen
to satisfy $\int^{\beta}_{0}\d\tau\mathbb{K}(\tau)=\mathbb{K}_0=0$.
As a result, in the imaginary time representation the
action~\eqref{environment-S-Matsubara} has the following non--local
form
\begin{eqnarray}\label{environment-S-CL}
S[\Phi]&=&\frac{\gamma
T}{2}\iint^{\beta=1/T}_{0}\d\tau\d\tau'\,\Phi(\tau)\mathbb{K}(\tau-\tau')\Phi(\tau')
\nonumber\\
&=&{\gamma\over 4\pi}\iint_{0}^{\beta} \d\tau\, \d\tau' \,
\frac{\pi^2 T^2}{\sin^2[\pi T(\tau-\tau')]}
\,\Big(\Phi(\tau)-\Phi(\tau')\Big)^2\, .
\end{eqnarray}
This action is frequently named after Caldeira and
Leggett~\cite{CaldeiraLeggett}, who used it to investigate the
influence of dissipation on quantum tunneling.

To transform to the Keldysh representation one needs to double the
number of degrees of freedom: $\Phi\to
\vec{\Phi}=(\Phi^{cl},\Phi^q)^T$. Then according to the causality
structure, Section~\ref{sec_bosons-4}, the general form of the time
translationally invariant Keldysh action is
\begin{equation}
S\big[\Phi^{cl},\Phi^{q}\big] = \gamma\int{\d\epsilon\over 2\pi}\,
\big(\Phi^{cl}_{\epsilon},\Phi^q_{\epsilon}\big)
\left(\begin{array}{cc}
0   & \big[\mathfrak{D}^{A}(\epsilon)\big]^{-1}  \\
 \big[\mathfrak{D}^{R}(\epsilon)\big]^{-1}  & \big[\mathfrak{D}^{-1}(\epsilon)\big]^{K}
\end{array}\right)
\left(\begin{array}{c} \Phi^{cl}_{\epsilon} \\ \Phi^q_{\epsilon}
\end{array}\right) \, ,
\end{equation}
where $[\mathfrak{D}^{R(A)}(\epsilon)]^{-1} $ is the analytic
continuation of the Matsubara correlator $|\epsilon_m|/2$ from the
{\em upper (lower)} half--plane of the complex variable $\epsilon_m$
to the real axis: $-i\epsilon_m\to \epsilon$, see~\cite{AGD}. As a
result, $\big[\mathfrak{D}^{R(A)}(\epsilon)\big]^{-1}=\pm
i\epsilon/2$. In equilibrium the Keldysh component follows from the
FDT: $\big[\mathfrak{D}^{-1}(\epsilon)\big]^{K}=
([\mathfrak{D}^{R}]^{-1}-[\mathfrak{D}^{A}]^{-1}) \coth\,
(\epsilon/2\,T)=i\epsilon\,\coth\, (\epsilon/2\,T)$,
cf.~\eqref{environment-L-RA} and \eqref{environment-L-K}. Therefore,
the Keldysh counterpart of the Matsubara
action~\eqref{environment-S-Matsubara} or \eqref{environment-S-CL},
is the already familiar dissipative
action~\eqref{environment-S-diss-fin}, (without the potential and
inertial terms, of course). One may now include external fields and
allow the system to deviate from the equilibrium.

\subsection{Dissipative chains and membranes}\label{sec_environment-8}

Instead of dealing with a single particle connected to a bath, let
us now consider a chain or a lattice of coupled particles, with each
one connected to a bath. To this end, we (i) supply a spatial index,
${\bf r}$, to the field $\Phi(t)\to \Phi(\mathbf{r},t)$, and (ii)
adds the harmonic interaction potential between neighboring
particles: $\sim D(\Phi(\mathbf{r},t)-\Phi(\mathbf{r}
+\mathbf{1},t))^2\to D(\partial_\mathbf{r}\Phi)^2$ in the continuum
limit, where $D$ is the rigidity of the chain or membrane. By
changing to the classical--quantum components and performing the
spatial integration by parts [cf.~\eqref{app-SPQM-S-Phonons-1}], the
gradient term translates to:
$D\left(\Phi^q\partial^2_\mathbf{r}\Phi^{cl} +
\Phi^{cl}\partial^2_\mathbf{r}\Phi^{q}\right)$. Thus, it modifies
the retarded and advanced components of the correlator, but it does
{\em not} affect the $(q-q)$ Keldysh component:
\begin{equation}
\big[\mathfrak{D}^{R(A)}\big]^{-1}={1\over 2}\,
\delta(t-t')\,\delta(\mathbf{r}-\mathbf{r}')\big( \mp
\partial_{t} + D\,\partial^2_{\mathbf{r}} \big)\, .
\end{equation}
In the Fourier representation
$\big[\mathfrak{D}^{R(A)}(\mathbf{k},\epsilon)\big]^{-1}={1\over 2}
\big(\pm i\epsilon -D\mathbf{k}^2\big)$. In equilibrium the Keldysh
component is not affected by the gradient terms, and is given
by~\eqref{environment-L-K} (in the real space representation it
acquires the factor $\delta(\mathbf{r}-\mathbf{r}')$). In
particular, its classical limit is
$\big[\mathfrak{D}^{-1}\big]^K=i2T\delta(t-t')\delta(\mathbf{r}-\mathbf{r}')$,
cf.~\eqref{eqenvironment-L-K-time}. As a result, the action of a
classical elastic membrane  in contact with a bath is
\begin{equation}\label{environment-S-chain}
S[\Phi^{cl},\Phi^{q}]=2\iint\d\mathbf{r}\d t\left[-\Phi^q
\left(\partial_t \Phi^{cl}-D\partial^2_\mathbf{r}\Phi^{cl}+{\partial
U(\Phi^{cl})\over
\partial \Phi^{cl} }\right)+i2T\big[\Phi^{q}\big]^2\right]\,,
\end{equation}
where the inertia terms have been neglected and we put $\gamma/2=1$
for brevity. One may introduce now an auxiliary
Hubbard--Stratonovich field $\xi(\mathbf{r},t)$ and write the
Langevin equation according to Section~\ref{sec_environment-4}:
\begin{equation}
\partial_{t}\Phi^{cl} -D\partial^2_\mathbf{r}\Phi^{cl}+{\partial U(\Phi^{cl})\over
\partial \Phi^{cl} } =\xi(\mathbf{r},t)\, ,
\end{equation}
where $\xi$ is a  Gaussian noise  with short--range correlations
$\langle
\xi(\mathbf{r},t)\xi(\mathbf{r}',t')\rangle=2T\delta(t-t')
\delta(\mathbf{r}-\mathbf{r}')$.

Let us consider an elastic chain  placed in the bottom of  the
($\mathbf{r}$--independent) meta--stable potential well, depicted in
Figure~\ref{Fig-Activation}a. If a sufficiently large piece of the
chain thermally escapes from the well, it may find it favorable to
slide down the potential, pulling the entire chain out of the well.
To find the shape of such an optimally large critical domain and its
action, let us change to the Hamiltonian variables of
Section~\ref{sec_environment-7}: $q(\mathbf{r},t)\equiv
\Phi^{cl}(\mathbf{r},t) $ and $p(\mathbf{r},t)\equiv
2i\Phi^q(\mathbf{r},t)$. The action \eqref{environment-S-chain}
takes the Hamiltonian form $iS=-\iint\d\mathbf{r}\d t\,\big(p\,\dot
q - H(p,q) \big)$ with
\begin{equation}\label{environment-chain-H}
H\equiv p\, D\partial_{\mathbf{r}}^2 q - p\, {\partial U(q)\over
\partial q} +  Tp^{\,2}  ,
\end{equation}
and the corresponding equations of motion are
\begin{subequations}
\begin{equation}
\dot q = \frac{\delta H}{\delta p}\, =\, D\partial^2_\mathbf{r} q -
U'_q(q)
+2Tp\,,
\end{equation}
\begin{equation}
\hskip-.4cm \dot p =-\frac{\delta H}{\delta q} =
-D\partial^2_\mathbf{r} p + p\,U''_{\!qq}(q)\,.
\end{equation}
\end{subequations}
These are complicated partial differential equations, that cannot be
solved in general. Fortunately, the shape of the optimal critical
domain can be found. As was discussed in
Section~\ref{sec_environment-7}, the minimal action trajectory
corresponds to a motion with zero energy, $H=0$. According to
Eq.~\eqref{environment-chain-H}, this is the case  if either $p=0$
(classical zero--action trajectory), or $Tp= U'_q(q)-D
\partial_\mathbf{r}^2 q $ (finite--action escape trajectory). In the
latter case the equation of motion for $q(\mathbf{r},t)$ takes the
form of the classical equation in the {\em reversed time}: $\dot q =-
D\partial^2_\mathbf{r} q+U'_q(q)=Tp\,$. Thanks to the last equality
the equation of motion for $p(\mathbf{r},t)$ is automatically
satisfied \footnote{Indeed, $T\dot p =\partial_t \dot
q=-D\partial^2_\mathbf{r} \dot q+\dot q
U''_{\!qq}=T(-D\partial^2_\mathbf{r} p + pU''_{\!qq} )$. This
non--trivial fact reflects the existence of an  accidental
conservation law: $H\big(p\rt,q\rt\big)=0$ -- {\em locally}! While
from the general principles only the total global energy has to be
conserved.}.
In the reversed time dynamics the $q(\mathbf{r},t)=0$
configuration is unstable and therefore the chain develops a
``tongue'' that grows until it reaches the stationary shape:
\begin{equation}\label{environment-domain-eq}
- D\partial^2_\mathbf{r} q + U'_q(q)=0\, .
\end{equation}
The solution of this equation with the boundary conditions
$q(\pm\infty)=0$ gives the shape of the critical domain. Once it is
formed, it  may grow further according to the classical equation
$\dot q = D\partial^2_\mathbf{r} q - U'_q(q)$ and $p=0$ with zero
action. The action along the non--classical escape trajectory, paid
to form the ``tongue'' is ($H(p,q)=0$):
\begin{equation}\label{environment-S-chain-static}
iS=-\iint\d\mathbf{r}\d t\,p\, \dot q
=-{1\over T} \iint\d\mathbf{r}\d t\,
\left(-D \partial^2_\mathbf{r} q + U'_q(q)\right) \dot q
=-{1\over T}\int\d\mathbf{r}\,\Big({D\over 2} (\partial_\mathbf{r}
q)^2 + U(q)\Big)\, ,
\end{equation}
where in the last equality an explicit integration over time was
performed. The escape action is given therefore by the static
activation expression that includes both the elastic and the
potential  energies. The optimal
domain~\eqref{environment-domain-eq}, is found by the minimization
of this static action \eqref{environment-S-chain-static}. One
arrives, thus, at a thermodynamic Landau--type description of the
first--order phase transitions. Note that the effective
thermodynamic description appears owing to the assumption that
$H(p,q)=0$, when  all the processes take  infinitely long time.

\section{Fermions}\label{sec_fermion}
\subsection{Partition function}\label{sec_fermion-1}

Consider a single  quantum state with  energy $\epsilon_0$. This
state is populated by spin--less fermions (particles obeying the
Pauli exclusion principle). In fact, one may have either zero, or
one particle in this state. The secondary quantized Hamiltonian of
such a system has the form
\begin{equation}\label{fermion-H}
\hat{H}=\epsilon_0\,\hat{c}^\dagger \hat{c}\, ,
\end{equation}
where $\hat{c}^\dagger$ and $\hat{c}$ are creation and
annihilation operators of fermions on the state $\epsilon_0$. They obey standard
anti--commutation relations: $\{\hat{c}\,,\hat{c}^\dagger\}=1$ and
$\{\hat{c}\,,\hat{c}\}=\{\hat{c}^\dagger\,,\hat{c}^\dagger\}=0$,
where $\{\,\, ,\,\}$ stands for the anti--commutator.

One can now consider  the evolution operator along the Keldysh
contour, $\mathcal{C}$ and the corresponding partition function,
$Z=1$, defined in exactly the same way as for bosonic
systems~\eqref{boson-Z}. The trace of the equilibrium density matrix
is $\Tr\{\hat{\rho}_0\}=1+\rho(\epsilon_0)$, where the two terms
stand for the empty and the singly  occupied states. One divides the
Keldysh contour onto $(2N-2)$ time intervals of  length
$\delta_t\sim 1/N\to 0$ and introduces resolutions of unity in $2N$
points along the Keldysh contour, $\mathcal{C}$; see
Figure~\ref{Fig-Contour}. The only difference from the bosonic case
of Section~\ref{sec_bosons-1} is that now one uses the resolution of
unity in the fermionic coherent state basis\footnote{The fermionic
coherent state $|\psi\rangle\equiv (1-\psi
\hat{c}^\dagger)|0\rangle$, parameterized by a Grassmann number
$\psi$ (such that $\{\psi,\psi'\}=\{\psi,\hat{c}\}=0$), is an
eigenstate of the annihilation operator:
$\hat{c}|\psi\rangle=\psi|\psi\rangle$. Similarly,
$\langle\psi|\hat{c}^\dagger=\langle\psi|\bar{\psi}$, where
$\bar{\psi}$ is another Grassmann number, {\em unrelated} to $\psi$.
The matrix elements of a {\em normally ordered} operator, such as
e.g. the Hamiltonian, take the form
$\langle\psi|\hat{H}(\hat{c}^\dagger,\hat{c})|\psi'\rangle =
H(\bar{\psi},\psi') \langle\psi|\psi'\rangle$. The overlap between
any two coherent states is
$\langle\psi|\psi'\rangle=1+\bar\psi\psi'=\exp\{\bar\psi\psi'\}$.
The trace of an operator, $\hat{\mathcal{O}}$, is calculated as
$\Tr\big\{\hat{\mathcal{O}}\big\}= \langle 0| \hat{\mathcal{O}}|0
\rangle + \langle 1| \hat{\mathcal{O}}|1 \rangle = \langle 0|
\hat{\mathcal{O}}|0 \rangle + \langle 0|\hat{c}\,
\hat{\mathcal{O}}\, \hat{c}^\dagger |0 \rangle=  \iint\d\bar{\psi}\,
\d\psi\,
e^{-\bar{\psi}\psi}\langle-\psi|\hat{\mathcal{O}}|\psi\rangle $,
where the Grassmann integrals are {\em defined} as $\int\! \d\psi\,
1=0$ and $\int\! \d\psi\, \psi =1$.}
\begin{equation}\label{fermion-unity}
\hat{1}=\iint\d\bar{\psi}_j\, \d\psi_j\,\,
e^{-\bar{\psi}_{j}\,\psi_{j}}\, |\psi_j\rangle\langle\psi_j|\, ,
\end{equation}
where $\bar{\psi}_j$ and $\psi_j$ are {\em mutually independent}
Grassmann variables. The rest of the algebra goes through exactly as
in the bosonic case, see Section~\ref{sec_bosons-1}. As a result,
one arrives at
\begin{equation}\label{fermion-Z}
Z={1\over \Tr\{\hat{\rho}_{0}\}}\iint
\prod\limits_{j=1}^{2N}\left[\d\bar{\psi}_j\, \d\psi_j \right] \,\,
\exp\left(\,i \sum\limits_{j,j\,'=1}^{2N} \bar{\psi}_{j}\,
G^{-1}_{jj\,'}\,\psi_{j\,'}\right)\, ,
\end{equation}
where the $2N \times 2N$ matrix $G^{-1}_{jj\,'}$ is
\begin{equation}\label{fermion-G-matrix}
  iG^{-1}_{jj\,'}\equiv
\left[\begin{array}{rrr|rrr}
 -1   &      &    &   &   &   -\rho(\epsilon_0) \\
  1\!-\!h & -1   &    &   &   &                 \\
      &  1\!-\!h & -1 &   &   &                 \\ \hline
     &     &  1 & -1 &    &                 \\
     &     &    &1\!+\!h & -1 &                 \\
     &     &    &    & 1\!+\!h&  -1
 \end{array} \right]\, ,
\end{equation}
and $h\equiv i\epsilon_0\delta_t$. The only difference from the
bosonic case is the negative sign in front of  $\rho(\epsilon_0)$
matrix element, originating from the minus sign in the
$\langle-\psi_{2N}|$ coherent state in the expression for the
fermionic trace. To check the normalization, let us evaluate the
determinant of such a matrix
\begin{equation}\label{fermion-determinant}
\mathrm{Det}\big[i\hat{G}^{-1}\big]=1+\rho(\epsilon_0)(1-h^2)^{N-1}
\approx1+ \rho(\epsilon_0)\,e^{(\epsilon_0\delta_t)^2(N-1)}\to
1+\rho(\epsilon_0)\,.
\end{equation}
Employing the fact that the fermionic Gaussian integral is given by
the determinant (unlike the inverse determinant for bosons) of the
correlation matrix, (see Appendix~\ref{app_Gaussian} for details),
one finds
\begin{equation}\label{fermion-Z-unity}
Z=\frac{\mathrm{Det}\big[i\hat{G}^{-1}\big]}{\Tr\{\hat{\rho}_0 \} }
= 1\, ,
\end{equation}
as it should be. Once again,  the upper--right element of the
discrete matrix~\eqref{fermion-G-matrix} is crucial to maintain the
correct normalization. Taking the limit $N\to \infty$ and
introducing the continuum  notation, $\psi_j\to \psi(t)$, one
obtains
\begin{equation}\label{fermion-Z-1}
Z=\int\D[\bar{\psi} \psi]\, \exp\left(iS[\bar{\psi},\psi]\right) =
\int\D[\bar{\psi} \psi] \, \exp\left(i \int_{\mathcal{C}}\d t\,\big[
\bar\psi(t)\,\hat{G}^{-1}\psi(t)\big] \right),
\end{equation}
where according to~\eqref{fermion-Z} and \eqref{fermion-G-matrix}
the action is given by
\begin{equation}\label{fermion-S}
S[\bar{\psi},\psi]=
\sum\limits_{j=2}^{2N}\left[i\bar{\psi}_j\,\frac{\psi_j-\psi_{j-1}}{\delta
t_j} -\epsilon_0\bar{\psi}_{j}\,\psi_{j-1}\right]\delta
t_j\,+i\,\bar{\psi}_1\Big[\psi_1+\rho(\epsilon_0)\psi_{2N}\Big]\, ,
\end{equation}
with $\delta t_j\equiv t_j-t_{j-1}=\pm \delta_t$. Thus the continuum
form of the operator $\hat G^{-1}$ is the same as for
bosons~\eqref{boson-G-continious}: $\hat G^{-1}=
i\partial_t-\epsilon_0$. Again the upper--right element of the
discrete matrix (the last term in~\eqref{fermion-S}), which contains
information about the distribution function, is seemingly absent in
the continuum notation.

Splitting the Grassmann field $\psi(t)$ into the two components
$\psi_{+}(t)$ and $\psi_{-}(t)$ that reside on the forward and the
backward parts of the time contour, respectively, one may rewrite
the action as
\begin{equation}\label{fermion-S-1}
S[\bar{\psi},\psi]=\int_{-\infty}^{+\infty}\d t\,\left[
\bar{\psi}_{+}(t)(i\partial_t-\epsilon_0)
\psi_{+}(t)-\bar{\psi}_{-}(t)(i\partial_t-\epsilon_0)
\psi_{-}(t)\right] ,
\end{equation}
where the dynamics of $\psi_{+}$ and $\psi_{-}$ are actually {\em
not} independent from each other, owing to the presence of non--zero
off--diagonal blocks in the discrete
matrix~\eqref{fermion-G-matrix}.

\subsection{Green's functions and Keldysh rotation}\label{sec_fermion-2}

The four fermionic Green's functions:
$G^{\mathbb{T}(\widetilde{\mathbb{T}})}$ and $G^{<(>)}$ are defined
in the same way as their bosonic counterparts,
see~\eqref{boson-corr-fun},
\begin{subequations}\label{fermion-corr-fun}
\begin{equation}
\hskip-2cm \langle\psi_{+}(t)\bar{\psi}_{-}(t\,')\rangle\equiv
iG^{<}(t,t')=-n_{F}\exp\{-i\epsilon_{0}(t-t')\}\,,
\end{equation}
\begin{equation}
\hskip-1.7cm \langle\psi_{-}(t)\bar{\psi}_{+}(t\,')\rangle\!\equiv
iG^{>}(t,t')=(1\!-\!n_{F})\exp\{-i\epsilon_{0}(t-t')\}\,,
\end{equation}
\begin{equation}
\langle\psi_{+}(t)\bar{\psi}_{+}(t\,')\rangle\equiv
iG^{\mathbb{T}}(t,t')
=\theta(t-t')iG^{>}(t,t')+\theta(t'-t)iG^{<}(t,t')\,,
\end{equation}
\begin{equation}
\langle\psi_{-}(t)\bar{\psi}_{-}(t\,')\rangle\equiv
iG^{\widetilde{\mathbb{T}}}(t,t')=
\theta(t'-t)iG^{>}(t,t')+\theta(t-t')iG^{<}(t,t')\,.
\end{equation}
\end{subequations}
The difference, however, is in the minus sign in the expression for
$G^<$, due to the anti--commutation relations, and Bose occupation
number is exchanged for the Fermi occupation number: $n_B\to
n_F\equiv \rho(\epsilon_0)/(1+\rho(\epsilon_0))$. Equations
\eqref{boson-relation} and \eqref{boson-relation-1} hold for the
fermionic Green's functions as well.

It is customary to perform the  Keldysh rotation in the fermionic
case in a different manner from the bosonic one. Define the new
fields as
\begin{equation}\label{fermion-rotation}
\psi_{1}(t)={1\over \sqrt{2}}\big(\psi_{+}(t)+\psi_{-}(t)\big)\,,
\qquad
\psi_{2}(t)={1\over\sqrt{2}}\big(\psi_{+}(t)-\psi_{-}(t)\big)\,.
\end{equation}
Following Larkin and Ovchinnikov~\cite{LarkinOvchinnikov}, it is
agreed that the \textit{bar}--fields transform in a different way:
\begin{equation}\label{fermion-rotation-1}
\bar{\psi}_{1}(t)={1\over\sqrt{2}}\big(\bar{\psi}_{+}(t)-
\bar{\psi}_{-}(t)\big)\,,\qquad \bar{\psi}_{2}(t)=
{1\over\sqrt{2}}\big(\bar{\psi}_{+}(t)+\bar\psi_{-}(t)\big)\,.
\end{equation}
The point is that the Grassmann fields $\bar\psi$ are {\em not}
conjugated to $\psi$, but rather are completely independent fields,
that may be  transformed  in an arbitrary manner (as long as the
transformation matrix has a non--zero determinant). Note that there
is no issue regarding the convergence of the integrals, since the
Grassmann integrals are always convergent. We also avoid the
subscripts $cl$ and $q$, because the Grassmann variables never have
a classical meaning. Indeed, one can never write a saddle point or
any other equation in terms of $\bar\psi,\psi$, rather they must
always be integrated out in some stage of the calculations.

Employing~\eqref{fermion-rotation}, \eqref{fermion-rotation-1} along
with Eq.~\eqref{fermion-corr-fun}, one finds
\begin{equation}\label{fermion-G-fun}
-i\big\langle\psi_{a}(t)\bar{\psi}_{b}(t\,')\big\rangle=
G_{ab}(t,t') = \left(\begin{array}{cc}
G^R(t,t') & G^{K}(t,t') \\
0                & G^{A}(t,t')
\end{array}\right)\, ,
\end{equation}
where hereafter $a,b  = (1,2)$. The fact that the $(2,1)$ element of
this matrix is zero is a manifestation of
identity~\eqref{boson-relation}. The {\em retarded, advanced} and
{\em Keldysh} components of the Green's function
\eqref{fermion-G-fun} are expressed in terms of
$G^{\mathbb{T}(\widetilde{\mathbb{T}})}$ and $G^{<(>)}$ in exactly
the same way as their bosonic analogs~\eqref{boson-G-fun-RAK}, and
therefore posses the same symmetry properties:
\eqref{boson-G-conjugation}--\eqref{boson-GR-plus-GA}. An important
consequence of~\eqref{bosons-G-traces} and \eqref{boson-GR-plus-GA}
is
\begin{equation}\label{fermion-traces}
\Tr\left\{G_{ab}^{(1)}\circ G_{bc}^{(2)}\circ\ldots\circ G_{za}^{(l)}
\right\}(t,t) =0\, ,
\end{equation}
where the circular multiplication sign involves convolution in the
time domain along with the $2\times 2$ matrix multiplication. The
argument $(t,t)$ states that the first time argument of $G^{(1)}$
and the last argument of $G^{(l)}$ are the same.

Note that the fermionic Green's function has a different structure
compared to its bosonic counterpart~\eqref{boson-G-fun-2}: the
positions of the $R,A$ and $K$ components in the matrix are
exchanged. The reason, of course, is  the different convention for
transformation of the \textit{bar} fields. One could choose the
fermionic convention to be the same as the bosonic (but {\em not}
the other way around), thus having the same
structure~\eqref{boson-G-fun-2} for the fermions as for the bosons.
The rationale for the Larkin--Ovchinnikov
choice~\eqref{fermion-G-fun} is that the inverse Green's function,
$\hat{G}^{-1}$ and fermionic self--energy $\hat{\Sigma}_F$ have the
same appearance as $\hat{G}$, namely
\begin{equation}\label{fermion-G-inverse}
\hat{G}^{-1}=\left(\begin{array}{cc}
\big[G^{R}\big]^{-1} & \big[G^{-1}\big]^{K} \\
0                & \big[G^{A}\big]^{-1}
\end{array}\right),\qquad
\hat{\Sigma}_F = \left(\begin{array}{cc}
\Sigma_F^R       &  \Sigma_F^{K} \\
0                &  \Sigma_F^{A}
\end{array}\right)\, ,
\end{equation}
whereas in the case of bosons $\hat{G}^{-1}$ (see \eqref{boson-S-1})
and $\hat{\Sigma}$ (see \eqref{int-bos-Sigma}) look differently from
$\hat{G}$ (see~\eqref{boson-G-fun-2}). This fact gives the
form~\eqref{fermion-G-fun} and \eqref{fermion-G-inverse} a certain
technical advantage.

For the single fermionic state, after the Keldysh rotation, the
correlation functions~\eqref{fermion-corr-fun} allow us to find
components of the matrix~\eqref{fermion-G-fun}
\begin{subequations}\label{fermion-G-fun-RAK}
\begin{equation}
\hskip-1.45cm
G^{R}(t,t\,')=-i\theta(t-t\,')e^{-i\epsilon_{0}(t-t')}\to
(\epsilon-\epsilon_{0}+i0)^{-1}\,,
\end{equation}
\begin{equation}
\hskip-1.7cm G^{A}(t,t\,')=i\theta(t'-t)e^{-i\epsilon_{0}(t-t')}\to
(\epsilon-\epsilon_{0}-i0)^{-1}\,,
\end{equation}
\begin{equation}
G^{K}(t,t\,')=-i(1-2n_{F})e^{-i\epsilon_{0}(t-t')}\to -2\pi
i(1-2n_{F})\delta(\epsilon-\epsilon_{0})\,,
\end{equation}
\end{subequations}
where the right--hand side provides also the Fourier transforms. In
thermal equilibrium, one obtains
\begin{equation}\label{fermion-FDT}
G^{K}(\epsilon)
=\left[G^{R}(\epsilon)-G^{A}(\epsilon)\right]\tanh{\epsilon \over
2\,T} \, .
\end{equation}
This is  the FDT for fermions. As in the case of bosons,  FDT is a
generic feature of an equilibrium system, not restricted to the toy
model. In general, it is convenient to parameterize the
anti--Hermitian  Keldysh Green's function by  a Hermitian matrix
$F=F^{\dagger}$ as
\begin{equation}\label{fermion-FDT-general}
G^{K}=G^{R}\circ F-F\circ G^{A} \, .
\end{equation}
The Wigner transform of $F(t,t')$ plays the role of the fermionic
distribution function.

\subsection{Free fermionic fields and their action}\label{sec_fermion-3}

One may proceed  now to a system with many degrees of freedom,
labeled by an index $\mathbf{k}$. To this end, one changes
$\epsilon_0\to \epsilon_\mathbf{k}$ and performs summations over
$\mathbf{k}$. If $\mathbf{k}$ is a momentum and
$\epsilon_\mathbf{k}=\mathbf{k}^2/(2m)$, it is instructive to
transform to the coordinate space representation
$\psi(\mathbf{k},t)\to \psi(\mathbf{r},t)$, while
$\epsilon_\mathbf{k}=\mathbf{k}^2/(2m)\to
-\partial^2_\mathbf{r}/(2m)$. Finally, the Keldysh action for a
non--interacting gas of fermions takes the form
\begin{equation}\label{fermion-S-2}
S_{0}[\bar{\psi},\psi]=\iint \d x\, \d x'\sum\limits_{a,b=1}^2
\bar{\psi}_a(x) \big[\hat G^{-1}(x,x')\big]_{ab} \,\psi_{b}(x')\, ,
\end{equation}
where $x=(\mathbf{r},t)$ and the matrix correlator $[\hat
G^{-1}]_{ab}$ has the structure of~\eqref{fermion-G-inverse} with
\begin{equation}\label{fermion-G-inverse-1}
\big[G^{R(A)}(x,x')\big]^{-1} =\delta(x-x')\left(i\partial_{t} +
{1\over 2m}
\partial_{\mathbf{r}}^2 + \mu \right)\, .
\end{equation}
Although in continuum  notation the $R$ and the $A$ components look
seemingly the same, one has to remember that in the discrete time
representation, they are matrices with the structure below and above
the main diagonal, respectively. The Keldysh component is a pure
regularization, in the sense that it does not have a continuum limit
(the self--energy Keldysh component does have a non--zero continuum
representation). All of this information is already properly taken
into account, however, in the structure of the Green's
function~\eqref{fermion-G-fun}.

\subsection{External fields and sources}\label{sec_fermion-4}

According to the basic idea of the Keldysh technique, the partition
function $Z=1$ is normalized  by construction,
see~\eqref{fermion-Z-unity}. To make the entire theory meaningful
one should introduce auxiliary source fields, which enable one to
compute various observable quantities: density of particles,
currents, etc. For example, one may introduce an external
time--dependent scalar potential $V\rt$ defined along the contour
$\mathcal{C}$. It interacts with the fermions as
$S_{V}=\int\d\mathbf{r}\int_{\mathcal{C}}\d t\,
V\rt\bar{\psi}\rt\psi\rt$. Expressing it via the field components
residing on the forward and backward contour branches, one finds
\begin{eqnarray}\label{fermion-SV}
S_{V}&=&\int\d\mathbf{r}\int^{+\infty}_{-\infty}\d
t\,\big[V_{+}\bar{\psi}_{+}\psi_{+}-V_{-}\bar{\psi}_{-}\psi_{-}\big]\nonumber
\\
&=&\int\d\mathbf{r} \int^{+\infty}_{-\infty}\d
t\,\big[V^{cl}(\bar{\psi}_{+}\psi_{+} - \bar{\psi}_{-}\psi_{-})+
V^{q}(\bar{\psi}_{+}\psi_{+} + \bar{\psi}_{-}\psi_{-})\big]\nonumber
\\
&=&\int\d\mathbf{r}\int^{+\infty}_{-\infty}\d t\
[V^{cl}(\bar{\psi}_{1}\psi_{1}+\bar{\psi}_{2}\psi_{2})+
V^{q}(\bar{\psi}_{1}\psi_{2}+\bar{\psi}_{2}\psi_{1})]\,,
\end{eqnarray}
where the $V^{cl(q)}$ components are defined in the standard way for
real boson fields, $V^{cl(q)} = (V_{+}\pm V_{-})/2$, way. We also
performed rotation from $\psi_{\pm}$ to $\psi_{1(2)}$ according
to~\eqref{fermion-rotation} and \eqref{fermion-rotation-1}. Note
that the physical fermionic density (symmetrized over the two
branches of the Keldysh contour) $\varrho = {1\over 2} \big(
\bar\psi_{+}\psi_{+}+ \bar\psi_{-}\psi_{-} \big)$ is coupled to the
quantum component of the source field, $V^q$. On the other hand, the
classical source component, $V^{cl}$, is nothing but an external
physical scalar potential, the same at the two branches.

Notation may be substantially compactified by introducing two vertex
$\hat \gamma$--matrices:
\begin{equation}\label{fermion-gammas}
\hat{\gamma}^{cl} \equiv \left( \begin{array}{cc}
 1 & 0   \\
 0 & 1
 \end{array} \right)\, , \hskip 1cm
\hat{\gamma}^{q} \equiv \left( \begin{array}{cc}
 0 & 1   \\
 1 & 0
 \end{array} \right)\, .
\end{equation}
With the help of these definitions, the source action
\eqref{fermion-SV} may be written as
\begin{eqnarray}\label{fermion-SV-1}
S_{V}=\int\d\mathbf{r}\int^{+\infty}_{-\infty}\d
t\sum^{2}_{a,b=1}\left[V^{cl}\bar{\psi}_{a}\gamma^{cl}_{ab}\psi_{b}+
V^{q}\bar{\psi}_{a}\gamma^{q}_{ab}\psi_{b}\right]
=\Tr\big\{\vec{\bar{\Psi}}\hat{V}\vec{\Psi}\big\},
\end{eqnarray}
where we have introduced Keldysh doublet $\vec{\Psi}$ and matrix
$\hat{V}$, defined as
\begin{equation}
\vec{\Psi}=\left(\begin{array}{c}\psi_{1} \\
\psi_{2}\end{array}\right)\,,\hskip 1cm
\hat{V}=V^{\alpha}\hat\gamma^{\alpha}= \left(\begin{array}{cc}V^{cl} & V^{q} \\
V^{q} & V^{cl}\end{array}\right)\,,
\end{equation}
where $\alpha=(cl,q)$.

In a similar way one may introduce external vector potential into
the formalism. The corresponding part of the action\footnote{The
vector source $\mathbf{A}\rt$ that we are using here differs from
the actual vector potential by the factor of $e/c$. However, we
refer to it as the vector potential and restore electron charge in
the final expressions.} $S_{A}=\int\d\mathbf{r}\int_{\mathcal{C}}\d
t\, \mathbf{A}\rt\mathbf{j}\rt$ represents the coupling between
$\mathbf{A}\rt$ and the fermion current
$\mathbf{j}\rt=\frac{1}{2mi}[\bar{\psi}\rt\partial_{\mathbf{r}}\psi\rt-
\partial_{\mathbf{r}}\bar{\psi}\rt\psi\rt]$. By splitting
$\int_{\mathcal{C}}\d t$ into forward and backward parts, performing
Keldysh rotation, one finds by analogy with the scalar potential
case~\eqref{fermion-SV} that
\begin{equation}\label{fermion-SA}
S_{A}=\Tr\big\{\vec{\bar{\Psi}} \hat{\mathbf{A}}\mathbf{v}_{F}
 \vec{\Psi}\big\}\,,\qquad
\hat{\mathbf{A}}=\mathbf{A}^{\alpha}\hat{\gamma}^{\alpha}=
\left(\begin{array}{cc}\mathbf{A}^{cl} & \mathbf{A}^{q} \\
\mathbf{A}^{q} & \mathbf{A}^{cl}\end{array}\right)\,.
\end{equation}
We have linearized the fermionic dispersion relation near the Fermi energy and employed that
$-i\partial_{\mathbf{r}}\approx \mathbf{p}_{F}$ and $\mathbf{v}_{F}=\mathbf{p}_{F}/m$.

Let us now define the generating function as
\begin{equation}\label{fermion-ZV}
Z\big[V^{cl},V^{q}\big]\equiv\left\langle\exp\big(iS_{V}\big)\right\rangle\,,
\end{equation}
where the angular brackets denote the functional integration over
the Grassmann fields $\bar{\psi}$ and $\psi$ with the weight
$\exp(iS_{0})$, specified by the fermionic action
\eqref{fermion-S-2}. In the absence of the quantum component,
$V^{q}=0$, the source field is the same at both branches of the time
contour. Therefore, the evolution along the contour  brings the
system back to its exact original state. Thus, one expects that the
classical component alone does not change the fundamental
normalization, $Z=1$. As a result,
\begin{equation}\label{fermion-Z-normalization}
Z[V^{cl}, 0]\equiv1\, ,
\end{equation}
as we already discussed in Section~\ref{sec_bosons},
see~\eqref{boson-causality}. Indeed, one may verify this statement
explicitly by expanding the partition function \eqref{fermion-ZV} in
powers of $V^{cl}$ and employing the Wick theorem. For example, in
the first order, one finds $Z[V^{cl},0]=1+\int \d t\,
\Tr\big[\hat{\gamma}^{cl}\hat{G}(t,t)\big]=1$, where one uses that
$\hat{\gamma}^{cl}=\hat 1$ along with~\eqref{fermion-traces}. It is
straightforward to see that for exactly the same reason all
higher--order terms in $V^{cl}$ vanish as well.

A lesson from~\eqref{fermion-Z-normalization} is that  one
necessarily has  to introduce {\em quantum} sources (which change
sign between the forward and the backward branches of the contour).
The presence of such source fields explicitly violates causality,
and thus changes the generating function. On the other hand, these
fields usually do not have a physical meaning and play only an
auxiliary role. In most cases one uses them only to generate
observables by an appropriate differentiation. Indeed, as was
mentioned above, the physical density is coupled to the quantum
component of the source. In the end, one takes the quantum sources
to be zero, restoring the causality of the action. Note that the
classical component, $V^{cl}$, does {\em not} have to be taken to
zero.

\begin{figure}
  \begin{center}\includegraphics[width=10cm]{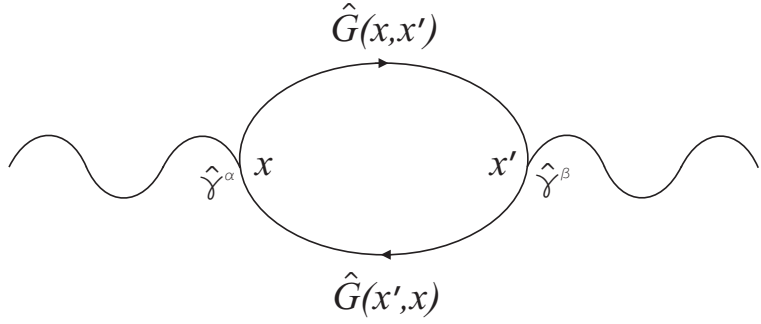}\end{center}
  \caption{Polarization operator $\hat \Pi^{\alpha\beta}(x,x\,')$:
  each solid line stands for the fermion matrix Green
  function \eqref{fermion-G-fun}, wavy lines represent external classical or quantum
  potentials
  $V^{cl(q)}$, and $x=(\mathbf{r},t)$. The loop diagram is a graphic representation
  of the trace in~\eqref{fermion-Pi-matrix}. }\label{Fig-Pi}
\end{figure}

Let us see how it works. Suppose we are interested in the average
fermion density $\varrho$ at time $t$ in the presence of a certain
physical scalar potential $V^{cl}(t)$. According
to~\eqref{fermion-SV} and \eqref{fermion-ZV} it is given by
\begin{equation}\label{fermion-density}
\varrho(x;V^{cl}) = -{i\over 2}\,{\delta \over \delta V^{q}(x)}\,
Z[V^{cl},V^q] \Big|_{V^{q}=0}\, ,
\end{equation}
where $x=(\mathbf{r},t)$. The problem is simplified if the external
field, $V^{cl}$, is weak in some sense. One may then restrict
oneself to the linear response, by defining the  susceptibility
\begin{equation}\label{fermion-Pi-R}
\Pi^{R}(x,x\,') \equiv {\delta \over \delta V^{cl}(x\,')}\,
\varrho(x;V^{cl})\Big|_{V^{cl}=0} = -{i\over 2}\,\left. {\delta^2 \,
Z[V^{cl},V^q] \over \delta V^{cl}(x\,') \delta V^{q}(x)}\, \right|_{
V^q=V^{cl}=0}\, .
\end{equation}
We add the subscript $R$ anticipating on the physical ground that
the response function must be {\em retarded} (causality). We
demonstrate it momentarily. First, let us introduce  the {\em
polarization} matrix as
\begin{equation} \label{fermion-Pi-matrix-def}
\hat \Pi^{\alpha\beta}(x,x\,') \equiv -{i\over 2}\, \left. {\delta^2 \ln
Z[\hat V] \over \delta V^{\beta}(x\,') \delta V^{\alpha}(x)}
\right|_{\hat V=0} = \left(\begin{array}{cc}
0   & \Pi^A(x,x\,')  \\
\Pi^R(x,x\,')  & \Pi^K(x,x\,')
\end{array}\right)\, .
\end{equation}
Owing to the fundamental
normalization,~\eqref{fermion-Z-normalization}, the logarithm is
redundant for the $R$ and the $A$ components and therefore the two
definitions \eqref{fermion-Pi-R} and \eqref{fermion-Pi-matrix-def}
are not in contradiction. The fact that $\Pi^{cl,cl} =0$ is obvious
from~\eqref{fermion-Z-normalization}. To evaluate the polarization
matrix, $\hat{\Pi}$, consider the Gaussian
action~\eqref{fermion-S-2}. Adding the source
term~\eqref{fermion-SV-1}, one finds: $S_0+S_V = \int \d x\,\,
\vec{\bar\Psi}[\hat{G}^{-1} + V^{\alpha}\hat{\gamma^{\alpha}}]
\vec{\Psi}$. Integrating out the fermion fields $\bar\psi$, $\psi$
according to the rules of fermionic Gaussian integration,
Appendix~\ref{app_Gaussian}, one obtains
\begin{equation} \label{fermion-Z-trlog}
 Z[V^{cl},V^{q}]={1\over \Tr{\{\hat\rho_0}\}}\mathrm{Det}
 \left[i\hat{G}^{-1}+ iV^\alpha\hat{\gamma}^\alpha\right]=
\mathrm{Det}\left[\hat 1+\hat{G}\, V^{\alpha}\hat{\gamma}^{\alpha}
\right]=\exp\left\{\Tr\ln[\hat 1 + \hat{G}\,
V^{\alpha}\hat{\gamma}^\alpha]\right\} ,
\end{equation}
where~\eqref{fermion-Z-unity} has been used. Since $Z[0]=1$, the
normalization is exactly right. One may now expand $\ln[\hat 1+
\hat{G}\, V^{\alpha}\hat{\gamma}^\alpha]$ to the second order in
$V^{\alpha}$. As a result, one finds for the polarization matrix
\begin{equation}\label{fermion-Pi-matrix}
\hat \Pi^{\alpha\beta}(x,x\,') = -{i\over 2}\,\, \Tr\left\{\hat{\gamma}^{\alpha}
 \hat{G}(x,x\,')\hat{\gamma}^{\beta}\hat{G}(x\,',x) \right\}\, ,
\end{equation}
which has a transparent diagrammatic representation, see
Figure~\ref{Fig-Pi}.

Substituting the explicit form of the gamma
matrices,~\eqref{fermion-gammas}, and the Green's
functions~\eqref{fermion-G-fun}, one obtains for the {\em response}
and the {\em correlation} components
\begin{subequations}\label{fermion-Pi-RAK}
\begin{equation}
\hskip-3.1cm \Pi^{R(A)}(x,x\,')=-{i\over
2}\left[G^{R(A)}(x,x\,')G^{K}(x\,',x)+
G^{K}(x,x\,')G^{A(R)}(x\,',x)\right]\,,
\end{equation}
\begin{equation}
\Pi^{K}(x,x\,')= -{i\over 2}\left[G^{K}(x,x\,') G^{K}(x\,',x)+
G^{R}(x,x\,')G^{A}(x\,',x)+G^{A}(x,x\,')G^{R}(x\,',x)\right]\,.
\end{equation}
\end{subequations}
From the first line it is obvious that $\Pi^{R(A)}(x,x\,')$ is
indeed a lower (upper) triangular matrix in the time domain,
justifying their superscripts. Moreover, from the symmetry
properties of the fermionic Green's functions one finds:
$\Pi^R=[\Pi^A]^\dagger$ and $\Pi^K=-[\Pi^K]^\dagger$. As a result,
the polarization matrix, $\hat\Pi$, possesses all the symmetry
properties of the bosonic self--energy $\hat\Sigma$,
see~\eqref{int-bos-Sigma}.

Equation \eqref{fermion-Pi-RAK} for $\Pi^R$ constitutes the Kubo
formula~\cite{Mahan,Kubo} for the density--density response
function. In equilibrium it may be derived using the Matsubara
technique. The Matsubara routine involves the analytical
continuation from discrete imaginary frequency $\omega_m=2\pi i m T $ to the real
frequency $\omega$. This procedure may prove to be cumbersome in
specific applications. The purpose of the above discussion is to
demonstrate how the linear response problems may be compactly
formulated in the Keldysh language. The latter allows to circumvent
the analytical continuation and yields results directly in the real
frequency domain.

\subsection{Applications I: Quantum transport}\label{app_Part-I}
\subsubsection{Landauer formula}\label{app_Part-I-1}

\begin{figure}
  \begin{center}\includegraphics[width=10cm]{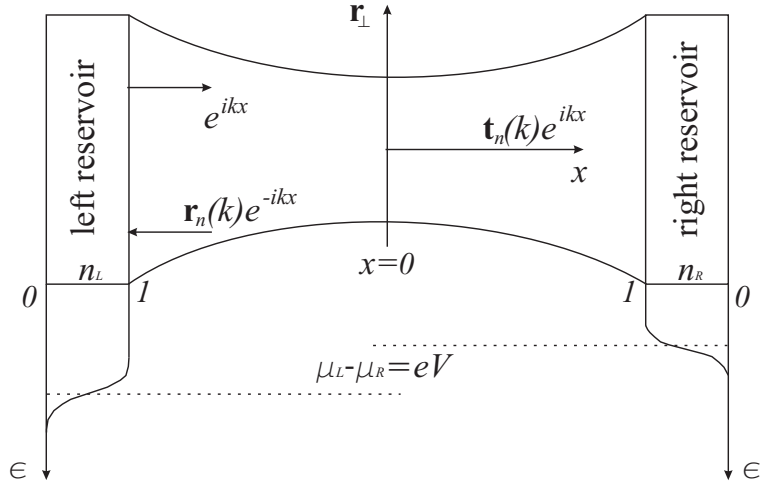}\end{center}
  \caption{Two terminal scattering problem from the
  quantum point contact.}\label{Fig-QPC}
\end{figure}

Let us illustrate how Keldysh technique can be applied to calculate
Landauer conductance~\cite{Landauer} of a quantum point contact
(QPC). For that purpose consider quasi--one--dimensional adiabatic
constriction connected to two reservoirs, to be referred to as left
($L$) and right ($R$). The distribution functions of electrons in
the reservoirs are Fermi distributions
$n_{L(R)}(\epsilon_k)=\big[\exp[(\epsilon_k -
\mu_{L(R)})/T]+1\big]^{-1}$, with electrochemical potentials shifted
by the voltage $\mu_L-\mu_R=eV$. Within QPC electron motion is
separable into transverse and longitudinal components. Owing to the
confinement transverse motion is quantized and we assign quantum
number $n$ to label transverse conduction channels with
$\phi_{n}(\mathbf{r}_{\perp})$ being corresponding transversal wave
functions. The longitudinal motion is described in terms of the
extended scattering states, i.e. normalized electron plane waves
incident from the left
\begin{equation}\label{Part-I-u-L}
u^{L}_{n}(k,\mathbf{r})=\phi_{n}(\mathbf{r}_{\perp})\left\{
\begin{array}{ll}
e^{ikx}+\mathbf{r}_{n}(k)e^{-ikx} & x\to-\infty \\
\mathbf{t}_{n}(k)e^{ikx} & x\to+\infty
\end{array}
\right.\,,
\end{equation}
and the right
\begin{equation}\label{Part-I-u-R}
u^{R}_{n}(k,\mathbf{r})=\phi_{n}(\mathbf{r}_{\perp})\left\{
\begin{array}{ll}
e^{-ikx}+\mathbf{r}_{n}(k)e^{ikx} & x\to+\infty \\
\mathbf{t}_{n}(k)e^{-ikx} & x\to-\infty
\end{array}
\right.\,,
\end{equation}
onto mesoscopic scattering region (Figure~\ref{Fig-QPC}).
Here $k(\epsilon)$ is the electron wave vector and
$\mathbf{t}_{n}(k)$ and $\mathbf{r}_{n}(k)$ are
channel specific transmission and reflection amplitudes. Second
quantized electron field operator is introduced in the standard way
\begin{equation}
\hat{\Psi}(\mathbf{r},t)=\sum_{nk}\left[\hat{\psi}^{L}_{n}(k,t)u^{L}_{n}(k,\mathbf{r})+
\hat{\psi}^{R}_{n}(k,t)u^{R}_{n}(k,\mathbf{r})\right]\,,
\end{equation}
where $\hat{\psi}^{L(R)}_{n}(k,t)$ are fermion destruction operators
in the left and right reservoirs, respectively. For the future use
we define also current operator
\begin{equation}\label{Part-I-Current}
\hat{I}(x,t)=\sum_{nk,n'k'}M^{ab}_{nn'}\hat{\psi}^{\dag
a}_{n}(k,t)\hat{\psi}^{b}_{n'}(k',t)\,,
\end{equation}
with the matrix elements
\begin{equation}
M^{ab}_{nn'}(x;k,k')=\frac{e}{2im}\int
d\mathbf{r}_{\perp}\left[u^{*a}_{n}(k,\mathbf{r})\partial_{x}u^{b}_{n'}(k',\mathbf{r})-
[\partial_{x} u^{*a}_{n}(k,\mathbf{r})]u^{b}_{n'}(k',\mathbf{r})\right]\,
, \quad a=L,R\,,
\end{equation}
which are constructed from the scattering states
\eqref{Part-I-u-L}--\eqref{Part-I-u-R}. Based on the orthogonality
condition $\int
\d\mathbf{r}_{\perp}\phi^{\phantom{*}}_{n}(\mathbf{r}_{\perp})
\phi^{*}_{n'}(\mathbf{r}_{\perp})=\delta_{nn'}$, direct calculation
of $\hat{M}_{nn'}(x;k,k')$ for $x>0$ gives~\footnote{Equation
\eqref{Part-I-M} is obtained as a result of certain approximations.
The exact expression for the current matrix explicitly depends on
coordinate $x$. There are two types of terms. The first depends on
$x$ as $\exp(\pm i(k+k')x)\approx\exp(\pm2ik_{F}x)$, where $k_{F}$
is Fermi momentum, it represents  Friedel oscillations. Their
contribution to the current is small as $(k-k')/k_{F}\ll 1$, and
thus neglected. The second type of terms contains $\exp(\pm
i(k-k')x)\approx 1$, since $|k-k'|\sim L_T^{-1}\ll x^{-1}$, where
$L_{T}=v_{F}/T$ is ballistic thermal length, and the coordinate $x$
is confined by the sample size $L\ll L_T$. See corresponding
discussions in~\cite{Buttiker}.}
\begin{equation}\label{Part-I-M}
\hat{M}_{nn'}(k,k')=ev_{F}\delta_{nn'} \left(\begin{array}{cl}
\mathbf{t}^{*}_{n}(k)\mathbf{t}_{n}(k') &
\mathbf{t}^{*}_{n}(k)\mathbf{r}_{n}(k') \\
\mathbf{r}^{*}_{n}(k)\mathbf{t}_{n}(k') &
\mathbf{r}^{*}_{n}(k)\mathbf{r}_{n}(k')-1
\end{array}\right)\approx ev_{F}\delta_{nn'}
\left(\begin{array}{lr}|\mathbf{t}_{n}|^{2} &
\mathbf{t}^{*}_{n}\mathbf{r}_{n}\\ \mathbf{r}^{*}_{n}\mathbf{t}_{n}
& -|\mathbf{t}_n|^{2}
\end{array}\right) \,,
\end{equation}
where $v_{F}=k_{F}/m$ is Fermi velocity. For $x<0$ the expression
for $\hat{M}$ is similar and different from~\eqref{Part-I-M} by an
overall sign and complex conjugation. The second approximate
relation on the right--hand side is written for the case when the
transmission amplitudes depend weakly on the  wavenumber $k$ on the
scale dictated by temperature or the applied bias, and thus their
momentum dependence may be disregarded.

One can set up now the partition function for this transport problem as
\begin{equation}\label{Part-I-Z}
Z[A]=\frac{1}{\Tr\{\hat{\rho}_{0}\}}\int\D[\bar{\psi}\psi]
\exp\left\{i\,  \vec{\bar
\Psi}[\hat{\mathbf{G}}^{-1}+\hat{A}\hat{M}]\vec{\Psi}\right\}\,,
\end{equation}
here $\vec{\bar \Psi}=(\bar{\psi}^{L},\bar{\psi}^{R})$,
$\hat{\mathbf{G}}=\mathrm{diag}\{\hat{G}_{L},\hat{G}_{R}\}$ is
$4\times4$ Green's function matrix, whereas $\hat{G}_{a}$ is
$2\times2$ matrix in the Keldysh space, and $\hat{A}$ is auxiliary
vector potential, c.f.~\eqref{fermion-SA}. Since the functional
integral over fermionic fields in~\eqref{Part-I-Z} is quadratic, one
finds upon Gaussian integration
\begin{equation}\label{Part-I-logZ}
\ln Z[A]=\Tr\ln\big[\hat{1}+\hat{\mathbf{G}}\hat{A}\hat{M}\big]\,.
\end{equation}
In analogy with~\eqref{fermion-density} the average current is
generated from $Z[A]$ via its functional differentiation with
respect to the quantum component of the vector potential $\langle
I\rangle=-(i/2)\delta\ln Z[A]/\delta A^{q}(t)|_{A^{q}=0}$. By
expanding trace of the logarithm to the linear order in $\hat{A}$,
as $\Tr\ln[\hat{1}+\hat{\mathbf{G}}\hat{A}\hat{M}]\approx
\Tr[\hat{\mathbf{G}}\hat{A}\hat{M}]$, one finds for the current
\begin{equation}
\langle I\rangle=-\frac{iev_{F}}{2}\Tr\left\{
\left(\begin{array}{cc}\hat{G}_{L}\hat{\gamma^{q}} & 0
\\ 0 & \hat{G}_{R}\hat{\gamma}^{q}\end{array}\right)
\left(\begin{array}{lr}|\mathbf{t}_{n}|^{2}
& \mathbf{t}^{*}_{n}\mathbf{r}_{n}\\
\mathbf{r}^{*}_{n}\mathbf{t}_{n} & -|\mathbf{t}_n|^{2}
\end{array}\right)\right\}=-\frac{iev_{F}}{2}\sum_{nk}T_{n}(\epsilon_{k})
\int\frac{\d \epsilon}{2\pi}\,
[G^{K}_{L}(\epsilon,k)-G^{K}_{R}(\epsilon,k)]\,,
\end{equation}
where we used Keldysh trace
$$\Tr\{\hat{G}_{a}\hat{\gamma}^{q}\}=G^{K}_a(t,t,k)=\int
\frac{\d\epsilon}{2\pi}G^{K}_a(\epsilon,k)\,,$$ and introduced QPC
transmission probability
$T_{n}(\epsilon_k)=|\mathbf{t}_{n}(k)|^{2}$. The last step is to
take Keldysh component of the Green's function
$G^{K}_{a}(\epsilon,k)=-2\pi i\delta(\epsilon-\epsilon_{k}+
\mu_{a})[1-2n_F(\epsilon)]$, with $\epsilon_{k}=v_{F}k$
[see~\eqref{fermion-G-fun-RAK}], and to perform momentum integration
which is straightforward owing to the delta function in $G^{K}$. The
result is
\begin{equation}\label{Part-I-I}
\langle I\rangle=\frac{e}{2\pi}\sum_{n}\int\d\epsilon\,
T_{n}(\epsilon)[n_L(\epsilon)-n_{R}(\epsilon)]\,.
\end{equation}
For a small temperature and applied voltage~\eqref{Part-I-I} gives a
conductance $\langle I\rangle=\mathrm{g}V$, where
\begin{equation}\label{Part-I-g}
\mathrm{g}=\frac{e^{2}}{2\pi\hbar}\sum_{n}T_{n}\, ,
\end{equation}
and all transmissions are taken at the Fermi energy
$T_{n}=T_{n}(\epsilon_{F})$ (note that we restored Planck constant
$\hbar$ in the final expression for the conductance). Equation
\eqref{Part-I-g} is known as a multi--channel Landauer formula
(see~\cite{Reviews-on-Landauer-1,Reviews-on-Landauer-2} for detailed
reviews on this subject).

\subsubsection{Shot noise}\label{app_Part-I-2}

Based on the previous example we can make one step forward and
calculate the second moment of the current fluctuations, so--called
noise power, defined as the Fourier transform of current
correlations
\begin{equation}\label{Part-I-noise-def}
\mathcal{S}(\omega,V)=\int \d t\, e^{i\omega t} \langle\delta
\hat{I}(t)\delta\hat{I}(0)+\delta\hat{I}(0)\delta\hat{I}(t)\rangle,\quad
\delta\hat{I}(t)=\hat{I}(t)-\langle I\rangle\,.
\end{equation}
Within Keldysh technique this correlator may be deduced from $Z[A]$
(see~\eqref{Part-I-logZ}). Indeed, one needs now to expand trace of
the logarithm in~\eqref{Part-I-logZ} to the second order in
auxiliary vector potential $\hat{A}$ and differentiate $\ln
Z[A]\propto\Tr\big[\hat{\mathbf{G}}\hat{A}\hat{M}\hat{\mathbf{G}}\hat{A}\hat{M}\big]$
twice over the quantum component, $A^q$:
\begin{equation}
\mathcal{S}(\omega,V)=-\frac{1}{2}\left.\frac{\delta^{2}\ln
Z[A]}{\delta A^{q}(\omega)\delta A^{q}(-\omega)}\right|_{A^q=0}\,.
\end{equation}
This expression automatically gives properly symmetrized noise
power~\eqref{Part-I-noise-def}. As a result of the differentiation
one finds
\begin{eqnarray}
&&\hskip-1cm \mathcal{S}(\omega,V)=\frac{1}{2}
\mathrm{Tr}\left\{\hat{\mathbf{G}}(\epsilon_{+})\hat{\gamma}^{q}
\hat{M}\hat{\mathbf{G}}(\epsilon_{-})\hat{\gamma}^{q}
\hat{M}\right\}=
\frac{e^{2}v^{2}_{F}}{2}\sum_{nkk'}\int\frac{\d\epsilon}{2\pi}\,
\left[T^{2}_{n}\Tr\{\hat{G}_{L}(\epsilon_{+})\hat{\gamma}^{q}
\hat{G}_{L}(\epsilon_{-})\hat{\gamma}^{q}\}\right.\nonumber
\\
&&\hskip-1cm\left.+
T_{n}R_{n}\Tr\{\hat{G}_{L}(\epsilon_{+})\hat{\gamma}^{q}\hat{G}_{R}(\epsilon_{-})\hat{\gamma}^{q}\}
+T_{n}R_{n}\Tr\{\hat{G}_{R}(\epsilon_{+})\hat{\gamma}^{q}\hat{G}_{L}(\epsilon_{-})\hat{\gamma}^{q}\}
+T^{2}_{n}\Tr\{\hat{G}_{R}(\epsilon_{+})\hat{\gamma}^{q}\hat{G}_{R}(\epsilon_{-})\hat{\gamma}^{q}\}
\right]\,,
\end{eqnarray}
where we already calculated partial trace over the left/right
subspace, assuming that transmissions are energy independent, and
used notations $\epsilon_{\pm}=\epsilon\pm\omega/2$ and
$R_{n}=1-T_n$. Calculation of  Keldysh traces
requires~\eqref{fermion-G-fun} and \eqref{fermion-gammas} and gives
\begin{equation}
\Tr\{\hat{G}_{a}\hat{\gamma}^{q}\hat{G}_{b}\hat{\gamma}^{q}\} =
G^{K}_{a}G^{K}_{b} + G^{R}_{a}G^{A}_{b} + G^{A}_{a}G^{R}_{b}\,.
\end{equation}
The remaining step is the momentum integration. One uses
$G^{R(A)}_{a}(\epsilon,k)=(\epsilon-v_{F}k + \mu_{a}\pm i0)^{-1}$
and $G^{K}_{a}(\epsilon,k)=-2\pi
i\delta(\epsilon-v_{F}k+\mu_{a})[1-2n_F(\epsilon)]$
from~\eqref{fermion-G-fun-RAK}, and finds that $\sum_{kk'}\int\d
\epsilon
\Tr\{\hat{G}_{a}\hat{\gamma}^{q}\hat{G}_{b}\hat{\gamma}^{q}\}=v^{-2}_{F}\int\d
\epsilon\, [1-(1-2n_{a})(1-2n_{b})]$. As a result, the final
expression for the noise power obtained by Lesovik~\cite{Lesovik}
reads as
\begin{equation}\label{Part-I-noise}
\mathcal{S}(\omega,V)=\frac{e^{2}}{2\pi\hbar}\sum_{n}\int\d
\epsilon\left[T^{2}_{n}B_{LL}(\epsilon)+T_{n}R_{n}B_{LR}(\epsilon)+
T_{n}R_{n}B_{RL}(\epsilon)+T^{2}_{n}B_{RR}(\epsilon)\right]\,,
\end{equation}
where statistical factors are
$B_{ab}(\epsilon)=n_{a}(\epsilon_{+})[1-n_{b}(\epsilon_{-})]+
n_{b}(\epsilon_{-})[1-n_{a}(\epsilon_{+})]$ and we again restored
$\hbar$ in the end. Despite its complicated appearance, $\epsilon$
integration in~\eqref{Part-I-noise} can be performed in the closed
form~\footnote{Deriving~\eqref{Part-I-noise-general} one writes
statistical factors as
$B_{ab}(\epsilon)=\frac{1}{2}\big[1-\tanh[(\epsilon_{+} - \mu_a)/2T]
\tanh[(\epsilon_{-} - \mu_b)/2T]\big]$ and uses the integral
$\int^{+\infty}_{-\infty}\d x\,
[1-\tanh(x+y)\tanh(x-y)]=4y\coth(2y)$.}
\begin{equation}\label{Part-I-noise-general}
\mathcal{S}(\omega,V)=\frac{e^2}{2\pi\hbar}\sum_{n}
\left[T^{2}_{n}\omega\coth\left(\frac{\omega}{2T}\right)+
T_{n}(1-T_{n})(eV+\omega)
\coth\left(\frac{eV+\omega}{2T}\right)+\{\omega\to-\omega\}\right]\,.
\end{equation}

There are two limiting cases of interest, which can be easily
extracted from~\eqref{Part-I-noise-general}. The first one
corresponds to the thermally equilibrium current fluctuations,
$V\to0$. In this case
\begin{equation}\label{Part-I-noise-FDT}
\mathcal{S}(\omega,0)=2\mathrm{g}\omega\coth\left(\frac{\omega}{2T}\right)\,,
\end{equation}
where we used~\eqref{Part-I-g} for conductance $\mathrm{g}$. This
result is nothing  but familiar fluctuation--dissipation relation
for the current fluctuations. Note, that despite the complicated
dependence on transmission amplitudes in~\eqref{Part-I-noise} the
equilibrium noise power \eqref{Part-I-noise-FDT} is written in terms
of conductance \eqref{Part-I-g} only. The other limiting case is
fully non--equilibrium noise at zero temperature $T\to0$ and a
finite bias $V$. For such a case one finds
from~\eqref{Part-I-noise-general} for the excess part of the noise
\begin{equation}\label{Part-I-noise-shot}
\mathcal{S}(\omega,V) - \mathcal{S}(\omega,0) =
\frac{e^2}{2\pi\hbar} \Big(|eV+\omega|+
|eV-\omega| -2|\omega| \Big)  \sum_{n}T_{n}(1-T_n)\,,
\end{equation}
which is called  the \textit{shot} noise. An important observation
here is that in contrast to equilibrium
noise~\eqref{Part-I-noise-FDT}, shot noise can not be written solely
in terms of the conductance $\mathrm{g}$. Only for the case of
tunnel junction, where all transmissions are small, $T_n\ll 1$,
Equation \eqref{Part-I-noise-shot} reduces to
$\mathcal{S}(0,V)=2eV\mathrm{g}=2e\langle I\rangle$, which is known
as Schottky formula (for a review of shot noise in various systems
see e.g.~\cite{Kogan,Blanter-Buttiker,Martin}).

\subsubsection{Coulomb drag}\label{app_Part-I-3}

\begin{figure}
\begin{center}\includegraphics[width=8cm]{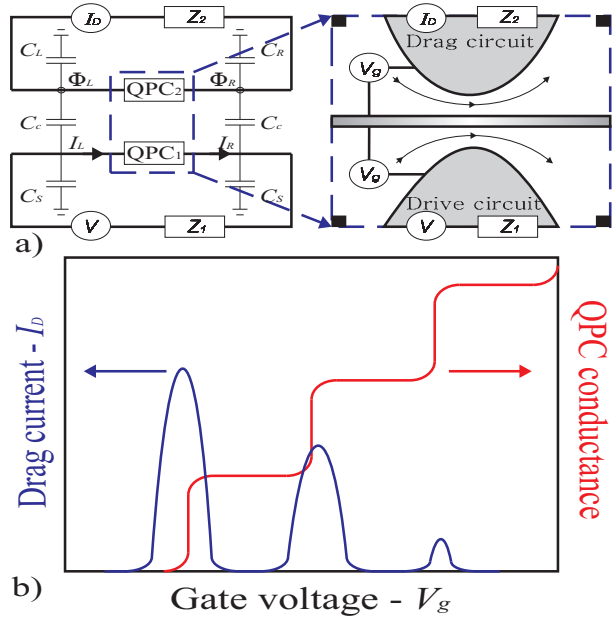}\end{center}
\caption{a) Two coupled QPCs and surrounding electric circuitry. The
Coulomb coupling is due to mutual capacitances $C_c$. Gate voltage
$V_{g}$  control transmission of, e.g., drive  QPC. b) Schematic
representation of conductance of the drive QPC along with the drag
current as a function of the gate voltage. }\label{Fig-drag}
\end{figure}

Drag effect proposed by Pogrebinskii~\cite{Pogrebinskii} and
Price~\cite{Price} by now is one of the standard ways to access and
measure electron--electron scattering. In bulk two--dimensional
systems (two parallel two--dimensional electron gases, separated by
an insulator) the drag effect is well established
experimentally~\cite{Solomon,Gramila,Sivan,Lilly,Pillarisetty} and
studied theoretically~\cite{Smith,MacDonald,Kamenev-Oreg,Flensberg}.
Recently a number of experiments were performed to study Coulomb
drag in quantum confined geometries such as quantum
wires~\cite{Debray-1,Debray-2,Morimoto,Yamamoto}, quantum
dots~\cite{Aguado,Kouwenhoven} or QPCs~\cite{Khrapai}. In  these
systems a source--drain voltage $V$ is applied to generate current
in the \textit{drive  circuit} while an induced current (or voltage)
is measured in the \textit{drag circuit}. Such a drag current  is a
function of the drive voltage $V$ as well as gate voltages, $V_{g}$,
which control transmission of one or both circuits.
Figure~\ref{Fig-drag}a shows  an example of such a setup, where both
drive and drag circuits are represented by two QPCs.

The Keldsyh technique is an efficient way to tackle the drag problem
both in linear response regime and away from the equilibrium, when a
relatively large bias is applied to the drive circuit. Within each
QPC electrons are assumed to be non--interacting and their motion is
separated into quantized--transversal, and extended--longitudinal,
see Section~\ref{app_Part-I-1}. The action describing
non--interacting point contacts is
\begin{equation}\label{Part-I-Action-QPC}
iS_{\mathrm{QPC}}=i\, \Tr\big\{\vec{\bar \Psi}
\hat{\mathbf{G}}^{-1}\vec{\Psi}\big\}\,,
\end{equation}
where $\vec{\bar \Psi}=(\bar{\psi}^{L}_{jn},\bar{\psi}^{R}_{jn})$
and $\hat{\mathbf{G}}=\delta_{jj'}
\mathrm{diag}\{\hat{G}_{L},\hat{G}_{R}\}$. Index $j=1,2$ labels
QPC$_{1(2)}$ respectively, $n$ is the transverse channel index
within each QPC, and $\hat{G}_{L(R)}$ is a $2\times2$ Keldysh
matrix~\eqref{fermion-G-fun}.

The interaction term between the two QPC is
\begin{equation}\label{Part-I-Action-int}
iS_{\mathrm{int}}=\sum_{ab\alpha\beta}\iint^{+\infty}_{-\infty}dtdt'\,
I^{\alpha}_{1a}(t)\mathbf{K}^{\alpha\beta}_{ab}(t-t')I^{\beta}_{2b}(t')\,,
\end{equation}
where $I_{jR(L)}(t)$ are current operators, on the right (left) of
QPC$_j$, coupled by the kernel $\hat{\mathbf{K}}_{ab}(t-t')$, which
encodes electromagnetic environment of the circuit. The retarded and
advanced components of the interaction kernel are related to the
trans--impedance matrix
$\mathbf{K}^{R(A)}_{ab}(\omega)=\mathbf{Z}^{R(A)}_{ab}(\omega)/(\omega\pm
i0)$. The latter is defined as
$\mathbf{Z}^{R(A)}_{ab}(\omega)=\partial\Phi_{a}(\pm
\omega)/\partial I_{b}(\mp \omega)$, where the corresponding local
fluctuating currents $I_a$ and voltages $\Phi_a$ are indicated in
Figure~\ref{Fig-drag}a. The Keldysh component of the interaction
kernel is dictated by the fluctuation--dissipation theorem:
$\mathbf{K}^{K}_{ab}(\omega)=[\mathbf{K}^{R}_{ab}(\omega)-
\mathbf{K}^{A}_{ab}(\omega)]\coth(\omega/2T)$, i.e. we assume that
the surrounding electric environment is close to equilibrium.
Finally the current operators are given by~\eqref{Part-I-Current}
and \eqref{Part-I-M}.

The  drag current is found by averaging $I_{2}$
over the fermionic degrees of freedom
\begin{equation}\label{Part-I-ID-def}
I_{D}=
\int\mathbf{D}[\psi\bar{\psi}]\,\mathrm{Tr}\left[\bar{\psi}_{2}M\psi_{2}\right]\,
\exp\big(iS_{\mathrm{QPC}}[\bar{\psi}\psi]+iS_{\mathrm{int}}[\bar{\psi}\psi]\big)\,.
\end{equation}
Expanding the exponent  to the second order in the interaction term
$S_{\mathrm{int}}$,  one obtains
\begin{equation}\label{Part-I-ID-def-perturb}
I_{D}=\frac{1}{2}\int\mathbf{D}[\psi\bar{\psi}]\,
\mathrm{Tr}\left[\bar{\psi}_{2}M\psi_{2}\right]
\mathrm{Tr}\left[I_{1}\mathbf{K}I_{2}\right]
\mathrm{Tr}\left[I_{1}\mathbf{K}I_{2}\right]
\exp\big(iS_{\mathrm{QPC}}[\bar{\psi}\psi]\big)\,.
\end{equation}
The remaining Gaussian integral over the fermionic fields is
calculated using the Wick's theorem. One employs expression
\eqref{Part-I-Current} for the current operators with the
$M$--matrix given by~\eqref{Part-I-M} and takes into the account all
possible Wick's contractions between the $\psi$--fields. The latter
are given by the Green's functions~\eqref{fermion-G-fun}. This way
one finds  for the drag current
\begin{equation}\label{Part-I-ID-result}
I_D(V)=
\int\frac{d\omega}{4\pi\omega^{2}}\mathrm{Tr}\left[\hat{\mathbf{Z}}(\omega)
\hat{\mathcal{S}}_{1}(\omega,V)\hat{\mathbf{Z}}(-\omega)
\hat{\Gamma}_{2}(\omega)\right]\,.
\end{equation}
The drive circuit is characterized by the \textit{excess} part
$\mathcal{S}^{ab}_{1}(\omega,V)=\mathcal{S}_{ab}(\omega,V)-\mathcal{S}_{ab}(\omega,0)$
of the current--current correlation matrix
$\mathcal{S}_{ab}(\omega,V)=\int dt\, e^{i\omega
t}\big\langle\big\langle\delta\hat{I}_{a}(t)\delta\hat{I}_{b}(0)+
\delta\hat{I}_{b}(0)\delta\hat{I}_{a}(t)\big\rangle\big\rangle$,
given by, e.g.,
\begin{eqnarray}\label{Part-I-noise-Srr}
\mathcal{S}_{RR}(\omega,V)=\frac{2}{R_{Q}}\sum_{n}\int\d\epsilon\,
\big[B_{LL}(\epsilon)|\mathbf{t}^{L}_{n}(\epsilon_{+})|^{2}
|\mathbf{t}^{L}_{n}(\epsilon_{-})|^{2}+B_{LR}(\epsilon)
|\mathbf{t}^{L}_{n}(\epsilon_{+})|^{2}|\mathbf{r}^{R}_{n}(\epsilon_{-})|^{2}\nonumber\\
+B_{RL}(\epsilon)|\mathbf{r}^{R}_{n}(\epsilon_{+})|^{2}
|\mathbf{t}^{L}_{n}(\epsilon_{-})|^{2}+B_{RR}(\epsilon)
[1-\mathbf{r}^{*R}_{n}(\epsilon_{+})\mathbf{r}^{R}_{n}(\epsilon_{-})]
[1-\mathbf{r}^{R}_{n}(\epsilon_{+})\mathbf{r}^{*R}_{n}(\epsilon_{-})]\big]\,,
\end{eqnarray}
where $\epsilon_\pm = \epsilon \pm \omega/2\,\,$,
$\mathbf{t}^{L(R)}_{n}(\epsilon_\pm)=\mathbf{t}^{L(R)}_{n}(\epsilon_\pm+eV_{L(R)})$
and
$\mathbf{r}^{L(R)}_{n}(\epsilon_\pm)=\mathbf{r}^{L(R)}_{n}(\epsilon_\pm+eV_{L(R)})$,
while $R_{Q}=2\pi\hbar/e^2$ is quantum resistance, and statistical
occupation form--factors $B_{ab}(\epsilon)$ are given
by~\eqref{Part-I-noise}. Here $\mathcal{S}_{RR}(\omega,V)$
generalizes~\eqref{Part-I-noise} to the case of energy dependent
transmissions~\cite{Buttiker}. Expressions for other components of
the noise matrix $\mathcal{S}_{LL}$, $\mathcal{S}_{LR}$, and
$\mathcal{S}_{RL}$ are similar, see
Refs.~\cite{Buttiker,LevchenkoKamenev-QPC}.

\begin{figure}
 \begin{center}\includegraphics[width=8cm]{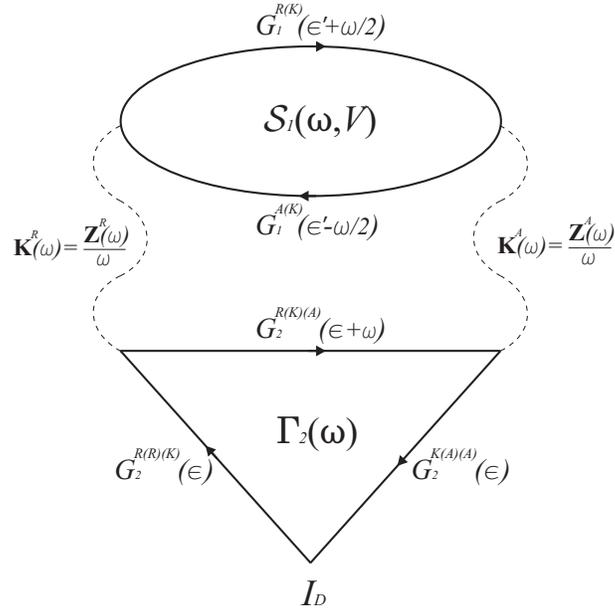}\end{center}
  \caption{Drag current $I_D$ in the second order in
  inter--circuit interactions $\mathbf{K}=\mathbf{Z}/\omega$ (wavy lines). The drag circuit
  is represented by triangular rectification vertex $\Gamma_2(\omega)$, while the drive circuit by the
  non--equilibrium current--current correlator $\mathcal{S}_1(\omega, V)$ (loop).}\label{Fig-drag-diagram}
\end{figure}

The drag circuit in~\eqref{Part-I-ID-result} is characterized by the
rectification coefficient
$\hat{\Gamma}_{2}(\omega)=\Gamma_{2}(\omega)\hat{\varsigma}_{z}$ of
ac voltage fluctuations applied to the (near--equilibrium) drag
QPC$_2$, where $\hat{\varsigma}_{z}$ is the third Pauli matrix
acting in the left--right subspace. Rectification is given
by~\footnote{In terms of the Keldysh matrices the rectification
coefficient is given by the following trace
$\Gamma_{2}(\omega)=\Tr\big[\hat{\mathbf{G}}\hat{\gamma}^{q}\hat{M}
\hat{\mathbf{G}}\hat{\gamma}^{cl}\hat{M}\hat{\mathbf{G}}\hat{\gamma}^{cl}\hat{M}\big]$.
Finding $\Gamma_{2}(\omega)$ in the form of~\eqref{Part-I-Gamma} one
uses Keldysh trace
$\mathrm{Tr}\left[\hat{G}\hat{\gamma}^{q}\hat{G}\hat{\gamma}^{cl}
\hat{G}\hat{\gamma}^{cl}\right]
=\sum_{\pm}\left[G^{R}(\epsilon)G^{R}(\epsilon\pm\omega)G^{K}(\epsilon)+
G^{R}(\epsilon)G^{K}(\epsilon\pm\omega)G^{A}(\epsilon)+
G^{K}(\epsilon)G^{A}(\epsilon\pm\omega)G^{A}(\epsilon)\right]$. To
simplify this expression further one should decompose each Keldysh
component of the Green's function using fluctuation--dissipation
relation
$G^{K}(\epsilon)=\big[G^{R}(\epsilon)-G^{A}(\epsilon)\big][1-2n(\epsilon)]$
and keep in the resulting expression only those terms, which have a
proper causality, i.e. combinations having three Green's functions
of the same kind, like $G^{A}G^{A}G^{A}$ and $G^{R}G^{R}G^{R}$, do
not contribute. In this way, one finds for the Keldysh trace
$\mathrm{Tr}\left[\hat{G}\hat{\gamma}^{q}\hat{G}
\hat{\gamma}^{cl}\hat{G}\hat{\gamma}^{cl}\right]
\propto\big[n_F(\epsilon_{-})-n_F(\epsilon_{+})\big]$. Remaining
trace in the left--right subspace over the current vertex matrices
$\hat{M}$ reduces to the transmission probabilities at shifted
energies, namely $\mathrm{Tr}\big[\hat{M}\hat{M}\hat{M}\big]\propto
|\mathbf{t}_{n}(\epsilon_{+})|^{2}-
|\mathbf{t}_{n}(\epsilon_{-})|^{2}$, leading
to~\eqref{Part-I-Gamma}.}
\begin{equation}\label{Part-I-Gamma}
\Gamma_{2}(\omega)=\frac{2e}{R_{Q}}\sum_{n}\int d\epsilon\,
\big[n_F(\epsilon_{-})-n_F(\epsilon_{+})\big]
\Big[|\mathbf{t}_{n}(\epsilon_{+})|^{2}-|\mathbf{t}_{n}(\epsilon_{-})|^{2}\Big]\,.
\end{equation}
Characteristics of the QPC$_2$ enter through its  energy--dependent
transmission probabilities $|\mathbf{t}_{n}(\epsilon)|^{2}$. This
expression admits a transparent interpretation: potential
fluctuations with frequency $\omega$, say on the left of the QPC,
create electron--hole pairs with energies $\epsilon_\pm$  on the
branch of right moving particles. Consequently the electrons can
pass through the QPC with the probability
$|\mathbf{t}_{n}(\epsilon_{+})|^{2}$, while the holes with the
probability $|\mathbf{t}_{n}(\epsilon_{-})|^{2}$. The difference
between the two gives the dc current flowing across the QPC. Note
that the energy dependence of the transmission probabilities in the
drag QPC is crucial in order to have the asymmetry between electrons
and holes, and thus non--zero rectification $\Gamma_{2}(\omega)$. At
the diagrammatic level~\eqref{Part-I-ID-result} has transparent
representation shown in Figure~\ref{Fig-drag-diagram}.

Focusing on a single partially open channel in a smooth QPC, one may
think of the potential barrier across it as being practically
parabolic. In such a case its transmission probability is given by
\begin{equation}\label{Part-I-t}
|\mathbf{t}(\epsilon)|^2=\left(\exp\{(eV_{g}-\epsilon)/\Delta_j\}+1\right)^{-1}\,,
\end{equation}
where $\Delta_j$ is an energy scale associated with the curvature of
the parabolic barrier in QPC$_j$ and gate voltage $V_g$ shifts the
top of the barrier relative to the Fermi energy. This form of
transmission was used to explain QPC conductance
quantization~\cite{Glazman-QPC} and it turns out to be useful in
application to the Coulomb drag problem. Inserting~\eqref{Part-I-t}
into Eq.~\eqref{Part-I-Gamma} and carrying out the energy
integration, one finds
\begin{equation}\label{Part-I-Gamma-1}
\Gamma_2(\omega)=\frac{2e\Delta_2}{R_{Q}}\,
\ln\left(1+\frac{\sinh^2(\omega/2\Delta_2)}
{\cosh^2(eV_{g}/2\Delta_2)}\right)
\end{equation}
for $T\ll\Delta_2$. In the other limit, $T\gg\Delta_2$, one should
replace $\Delta_2\to T$ in~\eqref{Part-I-Gamma-1}. Note that for
small frequency $\omega\ll \Delta_2$ one has $\Gamma_2\sim
\omega^2$, thus making the integral in~(\ref{Part-I-ID-result})
convergent in $\omega\to 0$ region.

\textbf{\textit{Linear drag regime.}} For small applied voltages $V$
one expects the response current $I_D$ to be linear in $V$.
Expanding $\hat{\mathcal{S}}_{1}(\omega,V)$ to the linear order in
$V$, one finds that only diagonal components of the current--current
correlation matrix contribute to the linear response and as a result,
\begin{equation}\label{Part-I-noise-S-linear}
\hat{\mathcal{S}}_{1}(\omega,V)=V\, \frac{\partial}{\partial\omega}
\left[\coth\frac{\omega}{2T}\right]\Gamma_{1}(\omega)\hat{\varsigma}_{z} +O(V^3)\,,
\end{equation}
where $\Gamma_{1}(\omega)$ is obtained from~\eqref{Part-I-Gamma} by
substituting transmission probabilities of QPC$_2$, by that of
QPC$_1$. Inserting~\eqref{Part-I-noise-S-linear}
into~\eqref{Part-I-ID-result} one finds
\begin{equation}\label{Part-I-ID-linear}
I_{D}=V\, \frac{R^{2}_{Q}}{4\pi}\int
d\omega\,\frac{\alpha_{+}(\omega)}{\omega^{2}}\,
\frac{\partial}{\partial\omega}\left[\coth\frac{\omega}{2T}\right]
\Gamma_{1}(\omega)\, \Gamma_{2}(\omega)\,,
\end{equation}
where dimensionless interaction kernel $\alpha_{+}(\omega)$ is
expressed through the trans--impedance matrix as
$$\alpha_{+}(\omega)=\frac{1}{2R^{2}_{Q}}\Tr\big[
\hat{\mathbf{Z}}(\omega)\hat{\varsigma}_{z}
\hat{\mathbf{Z}}(\omega)\hat{\varsigma}_{z}\big]\,.$$ Equation
\eqref{Part-I-ID-linear} has the same general structure as the one
for the drag current in bulk two--dimensional
systems~\cite{Kamenev-Oreg,Flensberg}. Being symmetric with respect
$1\leftrightarrow 2$ permutation, it satisfies Onsager relation for
the linear response coefficient. Performing remaining frequency
integration in~\eqref{Part-I-ID-linear}, it is sufficient to take
the interaction  kernel at zero frequency. Indeed, frequency scale
at which $\alpha_{+}(\omega)$ changes is set by inverse $RC$--time
of the circuit. If load impedance of the drag circuit is large
compared to that of the drive circuit $Z_{1}\ll Z_2\ll R_Q$, which
is the case for most experiments, and the mutual capacitance of the
two circuits is small $C_c\ll C_{R,L,s}$, see
Figure~\ref{Fig-drag}a, one finds $\tau^{-1}_{RC}=(Z_1 C_s)^{-1}\gg
T$. Since $I_{D}$ in~\eqref{Part-I-ID-linear} is determined by
$\omega\lesssim T$, it is justified to approximate\footnote{For the
circuit shown in the Figure~\ref{Fig-drag} one finds for the
low--frequency limit of the trans--impedance kernel
$$\alpha_{\pm}(0)=\frac{Z^{2}_{1}}{8R^{2}_{Q}}\frac{C^{2}_{c}}{C^{2}_{L}C^{2}_{R}}
\left\{\begin{array}{l}2C^{2}_{L}+2C_{L}C_{R}+2C^{2}_{R}\\
C^{2}_{L}-C^{2}_{R}
\end{array}\right.$$.}
$\alpha_{+}(\omega)\approx\alpha_{+}(0)$.
Substituting~\eqref{Part-I-Gamma-1} into \eqref{Part-I-ID-linear},
one finds for, e.g., low--temperature regime $T\ll \Delta_{1,2}$
\begin{equation}\label{Part-I-ID-linear-1}
I_D= \frac{V}{R_{Q}}\,   \frac{\alpha_{+}(0)\pi^2}{6}\, {T^{2}\over
\Delta_1\Delta_2}\,  {1\over \cosh^{2}(eV_{g}/2\Delta_1)}\,,
\end{equation}
where we assumed that the gate voltage of QPC$_2$ is tuned to adjust
the top of its barrier with the Fermi energy and wrote $I_D$ as a
function of the gate voltage in QPC$_1$. The resulting expression
exhibits a peak at $V_g=0$ similar to that depicted in
Fig.~\ref{Fig-drag}b. This expression describes rectification of
near--equilibrium thermal fluctuations (hence the factor $T^2$),
which is due to the electron--hole asymmetry (hence, non--monotonous
dependence on $V_g$).

\textbf{\textit{Nonlinear regime.}} At larger drive voltages drag
current ceases to be linear in $V$. Furthermore, contrary to the
linear response case, $\hat{\mathcal{S}}_{1}(\omega,V)$ does not
require energy dependence of the transmission probabilities and
could be evaluated for energy independent $|\mathbf{t}_n|^2$ (this
is a fare assumption for $T,eV\ll \Delta_1$). Assuming in addition
$T\ll eV$, one finds $\hat{\mathcal{S}}^{ab}_{1}(\omega,V)=
\big[\mathcal{S}_{ab}(\omega,V)-\mathcal{S}_{ab}(\omega,0)\big]\hat{\varsigma}_{0}$,
where $\mathcal{S}_1(\omega,V)$ is given
by~\eqref{Part-I-noise-shot} (recall that
$T_{n}\equiv|\mathbf{t}_{n}|^{2}$). Inserting it
into~\eqref{Part-I-ID-result}, after the frequency integration
bounded by the voltage, one finds for the drag current
\begin{equation}\label{Part-I-ID-nonlinear}
I_{D}=\frac{eV^2}{\Delta_{2}R_{Q}}\,  \alpha_{-}(0) \,
\sum_{n}T_{n}(1-T_{n})\,.
\end{equation}
Here again we assumed that the detector QPC$_2$ is tuned to the
transition between the plateaus. We also assumed $eV\ll
(Z_1C_s)^{-1}$ to substitute
$$\alpha_{-}(\omega)=\frac{1}{2R^{2}_{Q}}\Tr\big[
\hat{\mathbf{Z}}(\omega)\hat{\varsigma}_{0}
\hat{\mathbf{Z}}(\omega)\hat{\varsigma}_{z}\big]$$ by its dc value,
$\alpha_{-}(0)$. One should notice that while $\alpha_{+}>0$, the
sign of $\alpha_{-}$ is arbitrary, since $\alpha_{-}\propto
C^{2}_{L}-C^{2}_{R}$, see Figure~\ref{Fig-drag}a and Note 12. For a
completely symmetric circuit $\alpha_{-}=0$, while for extremely
asymmetric one $|\alpha_{-}|\approx \alpha_{+}/2$. Although we
presented derivation of~\eqref{Part-I-ID-nonlinear} for $T\ll eV$,
one may show that it remains valid at any temperature as long as
$T\ll \mathrm{min}\{\Delta_1, (Z_1C_s)^{-1}\}$.

Equation \eqref{Part-I-ID-nonlinear} shows that the drag current is
the result of the rectification of the quantum shot noise and is
hence proportional to the Fano factor~\cite{Lesovik} of the drive
circuit. It exhibits the generic behavior depicted in
Figure~\ref{Fig-drag}b, but the reason is rather different from the
similar behavior in the linear regime. The direction of the
non--linear drag current is determined by the inversion asymmetry of
the circuit (through the sign of $\alpha_{-}$) rather than the
direction of the drive current. As a result, for a certain polarity
of the drive voltage, the drag current appears to be {\em negative}.
Finally, assuming that for a generic circuit $\alpha_+\sim \alpha_-$
and comparing~\eqref{Part-I-ID-linear-1} and
\eqref{Part-I-ID-nonlinear} one concludes that the transition from
the linear to the non--linear regime takes place at $V\approx V^*$
with $eV^*=T^2/\Delta_1\ll T$, for $T\ll \Delta_1$. In the opposite
limit, $T>\Delta_1$, the crossover voltage is given by the
temperature $eV^*=T$.  Further details and discussions can be found
in~\cite{LevchenkoKamenev-QPC}.

\section{Disordered fermionic systems}\label{sec_NLSM}

One is often interested in calculating, say,  density--density or
current--current response functions, in the presence of static
(quenched) space--dependent disorder potential
$U_{\mathrm{dis}}(\mathbf{r})$. Moreover, one wants to know their
averages taken over an ensemble of  realizations of
$U_{\mathrm{dis}}(\mathbf{r})$, since the exact form of the disorder
potential is, in general, not known. The response function in the
Keldysh formulation, may be defined as variation of the generating
function and {\em not the logarithm} of the generating function.
More precisely, the two definitions with, and without the logarithm
coincide owing to the fundamental normalization $Z=1$. This is not
the case in the equilibrium formalism, where the presence of the
logarithm (leading to the factor $Z^{-1}$ after differentiation) is
unavoidable in order to have the correct normalization. Such a
disorder--dependent factor $Z^{-1}=Z^{-1}[U_{\mathrm{dis}}]$
formidably complicates the averaging over $U_{\mathrm{dis}}$. Two
techniques were invented to perform the averaging: the replica
trick~\cite{Edwards,Wegner,EfetovLarkinKhmelnitskii,Finkel'stein,BelitzKirkpatrick}
and the supersymmetry~\cite{Efetov,Efetov-book}. The first one
utilizes the observation that $\ln Z=\lim_{n\to 0}(Z^n-1)/n$, to
perform calculations for an integer number, $n$, of replicas of the
same system and take $n\to 0$ in the end of the calculations. The
second one is based on the fact that $Z^{-1}$ of the
non--interacting fermionic system equals to $Z$ of a bosonic system
in the same random potential. One thus introduces an additional
bosonic replica of the fermionic system at hand. The Keldysh
formalism provides an alternative to these two methods ensuring that
$Z=1$ by
construction~\cite{HorbachSchon,KamenevAndreev,ChamonLudwigNayak}.
The purpose of this section is to show how the effective field
theory of disordered electron gas, known as the  non--linear
$\sigma$--model (NLSM), is constructed within Keldysh formalism.

\subsection{Disorder averaging}\label{sec_NLSM-1}

We add the disorder dependent term to the fermionic action
$S_{\mathrm{dis}}[\bar{\psi},\psi]=\int_{\mathcal{C}}\d
t\int\d\mathbf{r}U_{\mathrm{dis}}(\mathbf{r}) \bar{\psi}\rt\psi\rt$,
where $U_{\mathrm{dis}}(\mathbf{r})$ is a static scalar potential,
created by a random configuration of impurities. It is usually
reasonable to assume that impurities are short--ranged and
distributed uniformly over the system, thus having the correlation
function of the form $\langle
U_{\mathrm{dis}}(\mathbf{r})U_{\mathrm{dis}}(\mathbf{r}')\rangle\sim
\delta(\mathbf{r}-\mathbf{r}')$. Assuming, in addition,  Gaussian
distribution of the impurity potential, one ends up with  the
disorder averaging performed with the help of the following
functional integral:
\begin{equation}\label{NLSM-dis-average-def}
\langle\ldots\rangle_{\mathrm{dis}}=\int\D[U_{\mathrm{dis}}] \ldots
\exp\left\{-\pi\nu\tauel\int\d\mathbf{r}\,U^{2}_{\mathrm{dis}}(\mathbf{r})\right\},
\end{equation}
where the disorder strength is characterized by the elastic mean
free time $\tauel$, and $\nu$ is the electronic density of states at
the Fermi energy. Since the disorder potential possesses only the
classical component, it is exactly the same on both branches of the
Keldysh contour. Thus, it is coupled only to $\hat{\gamma}^{cl}=\hat
1$ vertex matrix. Next, we perform the Gaussian integration over
$U_{\mathrm{dis}}$ of the disorder--dependent term of the partition
function (at this step we crucially use the absence of the
normalization factor) and find
\begin{eqnarray}\label{NLSM-dis-average}
\int\D[U_{\mathrm{dis}}]\exp\left(-\int\d\mathbf{r}\left[\pi\nu\tauel
U^{2}_{\mathrm{dis}}(\mathbf{r})-iU_{\mathrm{dis}}(\mathbf{r})\int^{+\infty}_{-\infty}
\d t\, \bar{\psi}^{a}\rt\hat
\gamma^{cl}_{ab}\psi^{b}\rt\right]\right)\nonumber
\\ =\exp\left(-\frac{1}{4\pi\nu\tauel}\int\d\mathbf{r}
\iint^{+\infty}_{-\infty}\d t\d t'
\big[\bar{\psi}^{a}\rt\psi^{a}\rt\big]
\big[\bar{\psi}^{b}\rtp\psi^{b}\rtp\big]\right)\,,
\end{eqnarray}
where $a,b=1,2$, and summations over all repeated indices are
assumed. One can rearrange $[\bar{\psi}^a\rt\psi^{a}\rt][
\bar{\psi}^{b}\rtp\psi^{b}\rtp]=-[\bar{\psi}^{a}\rt
\psi^{b}\rtp][\bar{\psi}^{b}\rtp\psi^{a}\rt]$  in the exponent on
the right--hand side of the last equation (the minus sign originates
from anti--commuting property of the Grassmann
 numbers) and then use  Hubbard--Stratonovich matrix--valued field,
$\hat{Q}=Q^{ab}_{tt'}(\mathbf{r})$ to decouple (time non--local)
four--fermion term as \footnote{Since we do not keep track of the
time--reversal symmetry, i.e. the fact that the Hamiltonian is a
real operator, the following considerations are restricted to the
case, where the time--reversal invariance is broken by, e.g., an
external magnetic field (complex Hermitian Hamiltonian). This is the
so--called {\em unitary} NLSM. The {\em orthogonal} NLSM, i.e. the
one where the time--reversal symmetry is restored is considered in
Section~\ref{sec_SuperCond}, devoted to disordered superconductors.}
\begin{eqnarray}\label{NLSM-HS-transform}
\exp\left(\frac{1}{4\pi\nu\tauel}\int\d\mathbf{r}
\iint^{+\infty}_{-\infty}\d t\d t'
[\bar{\psi}^{a}\rt\psi^{b}\rtp][\bar{\psi}^{b}\rtp\psi^{a}\rt]\right)\nonumber
\\
=\int\D[\hat{Q}]\,\exp\left(-\frac{\pi\nu}{4\tauel}\Tr\{
\hat{Q}^{2}\}-\frac{1}{2\tauel}\int\d\mathbf{r}\iint^{+\infty}_{-\infty}\d
t \d t'
Q^{ab}_{tt'}(\mathbf{r})\bar{\psi}^{b}\rtp\psi^{a}\rt\right)\,.
\end{eqnarray}
Here we have introduced that the trace of the $\hat{Q}^{2}$ implies
summation over the matrix indices as well as time and spatial
integrations
\begin{equation}
\Tr\big\{\hat{Q}^{2}\big\}=\int\d\mathbf{r}\iint^{+\infty}_{-\infty} \d t\d
t'\sum^{2}_{a,b=1}Q^{ab}_{tt'}(\mathbf{r})Q^{ba}_{t't}(\mathbf{r}).
\end{equation}
Now the \textit{averaged} action is quadratic in the Grassmann
variables
$S[\Psi,\hat{Q}]=\Tr\big\{\vec{\bar{\Psi}}\big[\hat{G}^{-1}+
\frac{i}{2\tau_{\mathrm{el}}}\hat{Q}]\vec{\Psi}\big\}$, and they may
be integrated out explicitly, leading to the determinant of the
corresponding quadratic form:
$\hat{G}^{-1}+\frac{i}{2\tau_{\mathrm{el}}}\hat{Q}$. All of the
matrices here should be understood as having  $2\times 2$ Keldysh
structure along with the $N\times N$ structure in  the discrete
time. One thus finds for the disorder averaged generating function
$\mathcal{Z}=\langle Z\rangle_{\mathrm{dis}}$:
\begin{eqnarray}\label{NLSM-action}
&&\mathcal{Z}=\int\D[\hat Q]\,\exp\big(iS[\hat Q]\big)\,,\nonumber\\
&&iS[\hat Q]=-\frac{\pi\nu}{4\tauel}\Tr\big\{\hat{Q}^{2}\big\}+
\Tr\ln\left[\hat{G}^{-1}+\frac{i}{2\tauel}\hat{Q}\right]\,.
\end{eqnarray}
As a result, one has traded the initial functional integral over the
static field $U_{\mathrm{dis}}(\mathbf{r})$ for the functional
integral over the dynamic matrix field $\hat{Q}_{tt'}(\mathbf{r})$.
At a first glance, it does not strike as a terribly bright idea.
Nevertheless, there is a great simplification hidden in this
procedure. The point is that the disorder potential, being
$\delta$--correlated, is a rapidly oscillating function. On the
other hand, as shown below, the $\hat Q$ matrix field is a slow
(both in space and time) function. Thus, it represents  true
macroscopic (or hydrodynamic) degrees of freedom of the system,
which are diffusively propagating modes.

\subsection{Non--linear $\sigma$--model}\label{sec_NLSM-2}

To proceed we  look for stationary configurations of the action
$S[\hat{Q}]$ in~\eqref{NLSM-action}. Taking the variation over
$\hat{Q}_{tt'}(\mathbf{r})$, one obtains the saddle point equation
\begin{equation}\label{NLSM-saddle-point-eq}
\underline{\hat{Q}}_{tt'}(\mathbf{r})= \frac{i}{\pi\nu}
\left(\hat{G}^{-1}+\frac{i}{2\tauel}\underline{\hat{Q}}\right)^{-1}_{tt',\mathbf{r}\mathbf{r}}
\,,
\end{equation}
where $\underline{\hat{Q}}_{tt'}(\mathbf{r})$ denotes a stationary
configuration of the fluctuating field $\hat{Q}_{tt'}(\mathbf{r})$.
The strategy is to find first a spatially uniform and
time--translationally invariant solution $\underline{\hat{Q}}_{\,
t-t'} $ of~\eqref{NLSM-saddle-point-eq} and then consider space--
and time--dependent deviations from such a solution. This strategy
is adopted from the theory of magnetic systems, where one first
finds a uniform static magnetized configurations and then treats
spin--waves as smooth perturbations on top of such a static uniform
solution. From the structure of~\eqref{NLSM-saddle-point-eq} one
expects that the stationary  configuration $\underline{\hat{Q}}$
possesses the same form as the fermionic
self--energy~\eqref{fermion-G-inverse} (more accurately, one expects
that among possible stationary configurations there is a
\textit{classical} configuration that admits  the causality
structure~\eqref{fermion-G-inverse}). One looks, therefore, for a
solution of~\eqref{NLSM-saddle-point-eq} in the form of the matrix
\begin{equation}\label{NLSM-Lambda-trial}
\underline{\hat{Q}}_{\, t-t'}=\hat{\Lambda}_{t-t'}=\left(
\begin{array}{cc}\Lambda^{R}_{t-t'} & \Lambda^{K}_{t-t'}\\
0 & \Lambda^{A}_{t-t'}\end{array}\right)\,.
\end{equation}
Substituting this expression into~\eqref{NLSM-saddle-point-eq},
which in the energy/momentum representation reads as
$\hat{\Lambda}_{\epsilon}=\frac{i}{\pi\nu}\sum_{p}\big(
\epsilon-\epsilon_{p}+\frac{i}{2\tauel}\hat{\Lambda}_{\epsilon}\big)^{-1}$,
with $\epsilon_{p}\equiv p^{2}/2m-\epsilon_{F}$, one finds
\begin{equation}
\Lambda^{R(A)}_{\epsilon}=\frac{i}{\pi\nu}\sum_{p}
\frac{1}{\epsilon-\epsilon_{p}+\frac{i}{2\tauel}
\Lambda^{R(A)}_{\epsilon}}=\pm1\,,
\end{equation}
where one adopts the convention
$\sum_{p}\ldots\to\nu\int\d\epsilon_{p}$. The signs on the
righ--hand side are chosen so as to respect causality: the retarded
(advanced) Green's function is analytic in the entire upper (lower)
half--plane of complex energy $\epsilon$. One has also assumed that
$1/\tauel\ll\epsilon_{F}$ to extend  the energy integration to minus
infinity, while using  constant density of states $\nu$. The Keldysh
component, as always, may be parameterized through a Hermitian
distribution function: $\Lambda^{K}=\Lambda^{R}\circ
F-F\circ\Lambda^{A}$, where the distribution function $F$ is not
fixed by the saddle point equation \eqref{NLSM-saddle-point-eq} and
must be determined through the boundary conditions. In  equilibrium,
however, $F$ is nothing but the thermal fermionic distribution
function $F^{eq}_{\epsilon}=\tanh\frac{\epsilon}{2T}$, thus
$\Lambda^{K}_{\epsilon}=(\Lambda^{R}_{\epsilon}-\Lambda^{A}_{\epsilon})
F^{eq}_{\epsilon}= 2F^{eq}_{\epsilon}$. Finally we have for the
stationary $\underline{\hat{Q}}$ matrix configuration
\begin{equation}\label{NLSM-Lambda}
\hat{\Lambda}_{\epsilon}=\left(
\begin{array}{cc} 1^{R}_{\epsilon} & 2F_{\epsilon}\\
0 & -1^{A}_{\epsilon}\end{array}\right)\,,
\end{equation}
where we have introduced the retarded and advanced unit matrices to
remind about causality structure and the superscript "$eq$" in the
distribution  $F$ was suppressed for brevity. Transforming back to
the time representation, one finds
$\Lambda^{R(A)}_{t-t'}=\pm\delta(t-t'\mp0)$, where $\mp0$ indicates
that $\delta$--function is shifted below (above) the main diagonal,
$t=t'$. As a result, $\Tr\{\hat{\Lambda}\}=0$ and $S[\hat
\Lambda]=0$, as it should be, of course, for any purely classical
field configuration~\eqref{NLSM-Lambda-trial}. One should note,
however, that this particular form of the saddle point
solution~\eqref{NLSM-Lambda} is a result of the approximation that
the single--particle density of states $\nu$ is independent of
energy. In general, it does depend on $\epsilon$ and thus retarded
(advanced) components of $\hat{\Lambda}_{\epsilon}$ are analytic
functions of energy in the upper (lower) half--plane, which do
depend on energy on the scale of order of the Fermi energy
$\epsilon_{F}$. Therefore, the infinitesimally shifted
$\delta$-functions in $\Lambda^{R(A)}_{t-t'}=\pm\delta(t-t'\mp0)$
should be understood as $\delta_{t\mp0}=f_{\pm}(t)\theta(\pm t)$,
where $\theta(\pm t)$ is the Heaviside step function, and
$f_{\pm}(t)$ are functions that are highly peaked for
$|t|\lesssim\epsilon^{-1}_{F}$ and satisfy the normalization
$\int^{\pm\infty}_{0}\d t f_{\pm}(t)=\pm 1$. This high--energy
regularization is important to remember in calculations to avoid
spurious unphysical constants. In particular, for this reasons
$1^{R}_{t-t'}M^{R}_{t',t}=0$, and $1^{A}_{t-t'}M^{A}_{t',t}=0$,
where $M^{R(A)}_{t',t}$ is an arbitrary retarded (advanced) matrix
in the time space.

Now we are on a position to examine the fluctuations around the
saddle point~\eqref{NLSM-Lambda}. The fluctuations of $\hat{Q}$ fall
into two general classes: (i) massive, with the mass $\propto
\nu/\tauel$ and (ii) massless, i.e. such that the action depends
only on gradients or time derivatives of these degrees of freedom.
The fluctuations along the massive modes can be integrated out in
the  Gaussian approximation and lead to insignificant
renormalization of the parameters in the action. The massless, or
Goldstone, modes describe diffusive motion of the electrons. The
fluctuations of $\hat{Q}$ matrix along these massless modes are not
small and should be parameterized by the matrices satisfying a
certain non--linear constraint. To identify the relevant Goldstone
modes consider the first term in the action $S[\hat Q]$
of~\eqref{NLSM-action}. The stationary configuration given
by~\eqref{NLSM-Lambda} satisfies
\begin{equation}\label{NLSM-Q-constraint}
\hat{Q}^{2}=\left(\begin{array}{cc}1^{R}_{\epsilon} & 0
\\ 0 & 1^{A}_{\epsilon}\end{array}\right)=\hat{1}\,.
\end{equation}
Note that $\Tr\big\{\hat{Q}^{2}\big\}=\Tr\,\{\hat{1}^R\}+
\Tr\,\{\hat{1}^A\}=0$, owing to the definition of the
retarded/advanced unit matrices. The fluctuations of $\hat{Q}$ which
do not satisfy~\eqref{NLSM-Q-constraint} are massive. The  class of
$\hat{Q}$ matrix configurations, that obeys the
constraint~\eqref{NLSM-Q-constraint}, is generated by rotations of
the stationary matrix $\hat{\Lambda}_{\epsilon}$ and may be
parameterized as follows
\begin{equation}\label{NLSM-Q-Rotated}
\hat{Q}=\hat{\mathcal{R}}^{-1}\circ
\hat{\Lambda}\circ\hat{\mathcal{R}}\,.
\end{equation}
The specific  form of $\hat{\mathcal{R}}$ is not important at the
moment and will be chosen later. The massless modes, or spin waves,
if one adopts magnetic analogy, which are associated with
$\hat{\mathcal{R}}_{tt'}(\mathbf{r})$ are slow functions of $t+t'$
and $\mathbf{r}$ and their gradients are small. Our goal now is to
derive an action for  soft--mode $\hat{Q}$ field configurations
given by~\eqref{NLSM-Q-constraint} and \eqref{NLSM-Q-Rotated}.

To this end, one substitutes~\eqref{NLSM-Q-Rotated}
into~\eqref{NLSM-action} and cyclically permutes $\hat{\mathcal{R}}$
matrices under the trace. This way one arrives at
$\hat{\mathcal{R}}\circ
\hat{G}^{-1}\circ\hat{\mathcal{R}}^{-1}=\hat{G}^{-1}+\hat{\mathcal{R}}
\circ[\hat{G}^{-1}\stackrel{\circ}{,}\hat{\mathcal{R}}^{-1}]
=\hat{G}^{-1}+i\hat{\mathcal{R}}\partial_{t}\hat{\mathcal{R}}^{-1}+
i\hat{\mathcal{R}}\mathbf{v}_{F}\partial_{\mathbf{r}}\hat{\mathcal{R}}^{-1}$,
where one has linearized the dispersion relation near the Fermi
surface
$\epsilon_{p}=p^{2}/2m-\epsilon_{F}\approx\mathbf{v}_{F}\mathbf{p}\to
-i\mathbf{v}_{F}\partial_{\mathbf{r}}$. As a result, the desired
action has the form
\begin{equation}\label{NLSM-action-trlog-R}
iS[\hat Q]=\Tr\ln\left[\hat{1}+i\hat{\mathcal{G}}\hat{\mathcal{R}}
\partial_{t}\hat{\mathcal{R}}^{-1}+
i\hat{\mathcal{G}}\hat{\mathcal{R}}\mathbf{v}_{F}
\partial_{\mathbf{r}}\hat{\mathcal{R}}^{-1}\right]\,,
\end{equation}
where we omit circular multiplication sign for brevity. Here
$\hat{\mathcal{G}}$ is the \textit{impurity dressed} Green's
function matrix, defined through the Dyson equation
$\big(\hat{G}^{-1}+\frac{i}{2\tauel}\hat{\Lambda}\big)\hat{\mathcal{G}}=\hat{1}$.
For  practical calculations it is convenient to write
$\hat{\mathcal{G}}$ in the form
\begin{equation}\label{NLSM-G-impurity-dressed}
\hat{\mathcal{G}}=\left(\begin{array}{cc}\mathcal{G}^{R} &
\mathcal{G}^{K}\\ 0 &
\mathcal{G}^{A}\end{array}\right)=\frac{1}{2}\mathcal{G}^{R}[\hat{1}+\hat{\Lambda}]+
\frac{1}{2}\mathcal{G}^{A}[\hat{1}-\hat{\Lambda}]\,,
\end{equation}
with retarded, advanced and Keldysh components given by
\begin{equation}\label{NLSM-G-impurity-dressed-RAK}
\mathcal{G}^{R(A)}(\mathbf{p},\epsilon)=\big[\epsilon-\epsilon_{p}\pm
i/2\tauel\big]^{-1},\qquad
\mathcal{G}^{K}(\mathbf{p},\epsilon)=\mathcal{G}^{R}(\mathbf{p},\epsilon)
F_{\epsilon}-F_{\epsilon}\mathcal{G}^{A}(\mathbf{p},\epsilon)\,.
\end{equation}
One may now expand the logarithm in~\eqref{NLSM-action-trlog-R} in
gradients of the rotation matrices $\hat{\mathcal{R}}$ to the linear
order in $\partial_{t}\hat{\mathcal{R}}^{-1}$ and to the quadratic
order in $\partial_{\mathbf{r}}\hat{\mathcal{R}}^{-1}$ terms
(contribution, linear in the spatial gradient, vanishes owing to the
angular integration). As a result
\begin{equation}\label{NLSM-action-trlog-Approx}
iS[\hat Q]\approx
i\Tr\big\{\hat{\mathcal{G}}\hat{\mathcal{R}}\partial_{t}\hat{\mathcal{R}}^{-1}\big\}
+\frac{1}{2}\Tr\big\{
\hat{\mathcal{G}}\big(\hat{\mathcal{R}}\mathbf{v}_{F}
\partial_{\mathbf{r}}\hat{\mathcal{R}}^{-1}\big)
\hat{\mathcal{G}}\big(\hat{\mathcal{R}}\mathbf{v}_{F}
\partial_{\mathbf{r}}\hat{\mathcal{R}}^{-1}\big)\big\}\,.
\end{equation}
Since
$\sum_{p}\hat{\mathcal{G}}(\mathbf{p},\epsilon)=-i\pi\nu\hat{\Lambda}_{\epsilon}$,
which directly follows from the saddle point Equation
\eqref{NLSM-saddle-point-eq}, one finds for the $\partial_{t}$ term
in the action
$i\Tr\{\hat{\mathcal{G}}\hat{\mathcal{R}}\partial_{t}\hat{\mathcal{R}}^{-1}\}=
\pi\nu\Tr\{\partial_{t}\hat{Q}\}$. For the $\partial_{\mathbf{r}}$
term, one finds $-\frac{1}{4}\pi\nu D\Tr\big\{(\partial_{\mathbf{r}}
Q)^{2}\big\}$, where $D=v^{2}_{F}\tauel/d$ is the diffusion constant
and $d$ is the spatial dimensionality. Indeed, for the product of
the Green's functions one uses
$\sum_{p}\mathcal{G}^{R}(\mathbf{p},\epsilon)\mathbf{v}_{F}
\mathcal{G}^{A}(\mathbf{p},\epsilon)\mathbf{v}_{F}=2\pi\nu\tauel
v^{2}_{F}/d=2\pi\nu D$, while the corresponding $R--R$ and $A--A$
terms vanish upon performing $\epsilon_{p}$ integration. Employing
then~\eqref{NLSM-G-impurity-dressed}, one arrives at
$\Tr\big\{[\hat{1}+\hat{\Lambda}]
(\hat{\mathcal{R}}\partial_{\mathbf{r}} \hat{\mathcal{R}}^{-1})
[\hat{1}-\hat{\Lambda}](\hat{\mathcal{R}}
\partial_{\mathbf{r}}\hat{\mathcal{R}}^{-1})\big\}=-\frac{1}{2}
\Tr\big\{\big(\partial_{\mathbf{r}}(\hat{\mathcal{R}}^{-1}
\hat{\Lambda}\hat{\mathcal{R}})\big)^{2}\big\}=
-\frac{1}{2}\Tr\big\{(\partial_{\mathbf{r}}\hat{Q})^{2}\big\}$.
Finally, one finds for the action of the soft--mode
configurations~\cite{HorbachSchon,KamenevAndreev,ChamonLudwigNayak}
\begin{equation}\label{NLSM-action-noninteracting}
iS[\hat Q]=-\frac{\pi\nu}{4}\,\Tr\Big\{D(\partial_{\mathbf{r}}
\hat{Q})^{2}-4\partial_{t}\hat{Q}\Big\}\,.
\end{equation}
Despite of its simple appearance, the action
\eqref{NLSM-action-noninteracting} is highly non--linear owing to
the constraint $\hat{Q}^{2}=\hat{1}$. The theory specified
by~\eqref{NLSM-Q-constraint} and \eqref{NLSM-action-noninteracting}
is called the \textit{matrix non--linear $\sigma$--model}. The name
came from the theory of magnetism, where the unit--length vector
$\vec \sigma(\mathbf{r})$, represents a local (classical) spin, that
may rotate over the sphere $\vec \sigma^{2}=1$.

One may now incorporate source terms $S_{V}$ and $S_{A}$
(see~\eqref{fermion-SV} and \eqref{fermion-SA}) into the fermionic
part of the action:
$$\Tr\left\{\vec{\bar{\Psi}}\left[\hat{G}^{-1}+\frac{i}{2\tauel}\hat{Q}+
\hat{V}+\mathbf{v}_{F}\hat{\mathbf{A}}\right]\vec{\Psi}\right\}\,.$$
After Gaussian integration over $\bar{\Psi}$ and $\Psi$, one finds
for the source--fields--dependent partition function, compare with
(cf.~\eqref{NLSM-action})
\begin{eqnarray}\label{NLSM-action-SAV}
&&\mathcal{Z}[\mathbf{A},V]=\int\D[\hat Q]
\exp\big(iS[\hat Q,\mathbf{A},V]\big)\,,\nonumber\\
&& iS[\hat
Q,\mathbf{A},V]=-\frac{\pi\nu}{4\tauel}\Tr\{\hat{Q}^{2}\}+
\Tr\ln\left[\hat{G}^{-1}+\frac{i}{2\tauel}\hat{Q}+
\hat{V}+\mathbf{v}_{F}\hat{\mathbf{A}}\right]\,.
\end{eqnarray}
Expanding trace of the logarithm in gradients of $\hat{Q}$ with the
help of~\eqref{NLSM-Q-Rotated}, one assumes that source fields
$\hat{V}$ and $\hat{\mathbf{A}}$ are small in some sense and do not
disturb the stationary configuration \eqref{NLSM-Lambda} (see
Section~\ref{sec_int_ferm} for discussions of this point). Then,
similarly to~\eqref{NLSM-action-noninteracting}, one finds
from~\eqref{NLSM-action-SAV}
\begin{eqnarray}\label{NLSM-action-general}
iS[\hat Q,\mathbf{A},V]=\frac{i\nu}{2}\, \Tr\big\{\hat{V}\hat{\sigma}_{x}\hat{V}\big\}
-\frac{\pi\nu}{4}\, \Tr\Big\{D(\hat{\bm{\partial}}_{\mathbf{r}}\hat{Q})^{2}-
4\partial_{t}\hat{Q}+4i\hat{V}\hat{Q}\Big\}\,,
\end{eqnarray}
where $\hat{\sigma}_{x}$ is the Pauli matrix acting in the Keldysh
space, and we have introduced covariant derivative
\begin{equation}\label{NLSM-covariant-deriv}
\hat{\bm{\partial}}_{\mathbf{r}}\hat{Q}=
\partial_{\mathbf{r}}\hat{Q}-i[\hat{\mathbf{A}},\hat{Q}]\,.
\end{equation}
A few comments are in order regarding~\eqref{NLSM-action-general}.
First, it is still restricted to the manifold of $\hat{Q}$ matrices
satisfying $\hat{Q}^{2}=\hat{1}$. The second trace on the
right--hand side of~\eqref{NLSM-action-general}, containing
$\hat{Q}$, originates from
$\sum_{p}\mathbf{v}_{F}\mathcal{G}^{R}\mathbf{v}_{F}\mathcal{G}^{A}$
and $\sum_{p}\mathcal{G}^{R(A)}$ combinations in the expansion of
the logarithm. On the other hand, the first term on the right--hand
side of~\eqref{NLSM-action-general} originates from
$\sum_{p}\mathcal{G}^{R}\mathcal{G}^{R}$ and
$\sum_{p}\mathcal{G}^{A}\mathcal{G}^{A}$ combinations. These terms
should be retained since the matrix
$V^{\alpha}(\epsilon-\epsilon')\hat{\gamma}^{\alpha}$ is not
restricted to the $1/\tauel$ shell near the Fermi energy. This is
so, because the scalar potential shifts the entire electronic band
and not only energy strip $|\epsilon|,|\epsilon'|<1/\tauel$. Thus,
it is essential to follow the variations of the electron spectrum
all the way down to the bottom of the band to respect the charge
neutrality. To derive $\Tr\{\hat{V}\hat{\sigma}_{x}\hat{V}\}$ one
has to employ the fact that for any physical fermionic distribution
function $F_{\epsilon\to\pm\infty}\to\pm1$. Equations
\eqref{NLSM-action-general} and \eqref{NLSM-covariant-deriv}
generalize an effective $\sigma$--model action given
by~\eqref{NLSM-action-noninteracting}. Additional technical details
needed to derive~\eqref{NLSM-action-general}
from~\eqref{NLSM-action-SAV} are provided in
Appendix~\ref{app_GradientExpansion}.

\subsection{Tunneling action}\label{sec_NLSM-3}

Consider  two metallic leads  separated by a tunneling barrier,
such that upon applying external voltage a current may flow between
them. In this case one has to add corresponding tunneling term to
the Hamiltonian of the system
$$\hat{H}_{T}=\int_{\mathbf{r}\in
L}\d\mathbf{r}\int_{\mathbf{r}'\in
R}\d\mathbf{r}'\big[T^{\phantom{*}}_{\mathbf{rr}'}
\hat{\psi}^{\dag}_{L}(\mathbf{r})
\hat{\psi}^{\phantom{\dag}}_{R}(\mathbf{r}')+T^{*}_{\mathbf{rr}'}
\hat{\psi}^{\dag}_{R}(\mathbf{r}')\hat{\psi}^{\phantom{\dag}}_{L}(\mathbf{r})\big]\,,$$
where $\hat{\psi}_{L(R)}$ is the electron annihilation operator to
the left(right) from the tunneling barrier. The
$\hat{\psi}^{\dag}_{L(R)}$ is corresponding creation operator. The
$T^{\phantom{*}}_{\mathbf{rr}'}$ and $T^{*}_{\mathbf{rr}'}$ are
tunneling matrix elements whose range is restricted to the vicinity
of the junction, since the overlap of electron wave functions decay
exponentially away from it. Tunneling Hamiltonian translates into
the fermionic tunneling action
$$iS_{T}=\int_{\mathcal{C}}\d
t\iint\d\mathbf{r}\d\mathbf{r}' \big[T^{\phantom{*}}_{\mathbf{rr}'}
\bar{\psi}_{L}(\mathbf{r},t)
\psi_{R}(\mathbf{r}',t)+T^{*}_{\mathbf{rr}'}
\bar{\psi}_{R}(\mathbf{r}',t)\psi_{L}(\mathbf{r},t)\big]\,.$$
Since
$S_{T}$ is still quadratic in fermion fields,  the Gaussian
integration over them is straightforward, leading to the disorder
averaged action in the form
\begin{eqnarray}\label{NLSM-tunnel-Z}
&&\mathcal{Z}=\int\D[\hat Q_{L}, \hat Q_{R}]\,\exp\big(iS[\hat Q_{L},\hat Q_{R}]\big)\,,\nonumber
\\
&&iS[\hat Q_{L}, \hat
Q_{R}]=-\frac{\pi\nu}{4\tauel}\sum_{a=L,R}\Tr\big\{\hat{Q}^{2}_{a}\big\}
+\Tr\ln\left(\begin{array}{cc}
\hat{G}_L^{-1}+\frac{i}{2\tauel}\hat{Q}_{L} & \hat{T} \\
\hat{T}^{\dag} & \hat{G}_R^{-1}+\frac{i}{2\tauel}\hat{Q}_{R}
\end{array}\right)\,.
\end{eqnarray}
Deriving~\eqref{NLSM-tunnel-Z} one has to introduce two $\hat Q$
matrices to decouple disorder mediated four--fermion
term~\eqref{NLSM-HS-transform} in each of the two leads
independently. In doing so it was assumed for simplicity that both
disordered samples are characterized by  equal mean free times and
bare electronic densities  of states. Equation~\eqref{NLSM-tunnel-Z}
contains an additional $2\times2$ matrix structure in the space of
left--right electronic subsystems, described by $\hat{Q}_{L(R)}$,
respectively. Note also that the tunneling matrix elements entering
$S[\hat{Q}_{L},\hat{Q}_{R}]$ are unit matrices in the Keldysh
subspace $\hat{T}_{\mathbf{rr}'}=T_{\mathbf{rr}'}\hat{\sigma}_{0}$.

Introducing the notation $\hat{\mathbf{G}}^{-1}_{a}=\hat{G}^{-1}_a +
\frac{i}{2\tauel}\hat{Q}_{a}$, one identically rewrites the last
term of the action $S[\hat{Q}_{L},\hat{Q}_{R}]$
in~\eqref{NLSM-tunnel-Z} as
\begin{equation}\label{NLSM-trlog-transform}
\Tr\ln\left(\begin{array}{cc}
\hat{\mathbf{G}}^{-1}_{L} & \hat{T} \\
\hat{T}^{\dag} & \hat{\mathbf{G}}^{-1}_{R}
\end{array}\right)=\Tr\ln\left(\begin{array}{cc}
\hat{\mathbf{G}}^{-1}_{L} & 0 \\
0 & \hat{\mathbf{G}}^{-1}_{R}\end{array}\right)
+\Tr\ln\left[\hat 1+\left(\begin{array}{cc}
0 & \hat{\mathbf{G}}_{L}\hat{T} \\
\hat{\mathbf{G}}_{R}\hat{T}^{\dag} & 0
\end{array}\right)\right]\,.
\end{equation}
Expanding now $\Tr\ln\hat{\mathbf{G}}^{-1}_{a}$ in gradients of
$\hat{Q}_{a}$ matrix around the saddle point $\hat{\Lambda}_{a}$,
one obtains sigma model action~\eqref{NLSM-action-noninteracting},
for each of the two leads independently. The coupling between them
is described by the second term on the right--hand side
of~\eqref{NLSM-trlog-transform}, which defines tunneling action
$S_{T}[\hat{Q}_{L},\hat{Q}_{R}]$. For a small transparency tunneling
junction, one may expand trace of the logarithm to the leading
(second) order in $\hat{T}$ and obtain
\begin{equation}\label{NLSM-action-S-T-log-expansion}
iS_{T}[\hat Q_{L},\hat Q_{R}]=\Tr\ln\left[\hat 1+\left(\begin{array}{cc}
0 & \hat{\mathbf{G}}_{L}\hat{T} \\
\hat{\mathbf{G}}_{R}\hat{T}^{\dag} & 0
\end{array}\right)\right]
\approx-\Tr\big\{\hat{\mathbf{G}}_{L}\hat{T}\hat{\mathbf{G}}_{R}\hat{T}^{\dag}\big\}+\ldots\,.
\end{equation}
Employing the local nature of  matrix elements $T_{{\bf r r}'}$ and
the fact that at the soft--mode manifold
$\hat{Q}_{a}=\frac{i}{\pi\nu}\hat{\mathbf{G}}_{a}({\bf r},{\bf r})$,
see~\eqref{NLSM-saddle-point-eq}, one finds for the tunneling part
of the action
\begin{equation}\label{NLSM-action-S-T}
iS_{T}[\hat Q_{L},\hat Q_{R}]=\frac{
\mathrm{g}_{T}}{4\mathrm{g}_{Q}}\Tr\big\{\hat{Q}_{L}\hat{Q}_{R}\big\}
= - \frac{\mathrm{g}_{T}}{8\mathrm{g}_{Q}}
\Tr\big\{(\hat{Q}_{L} - \hat{Q}_{R})^2 \big\}\,.
\end{equation}
Here we approximated the tunneling  matrices as
$T_{\mathbf{rr}'}=T_{0}\delta(\mathbf{r}-\mathbf{r}')$ and
introduced the tunneling conductance $\mathrm{g}_{T}=4\pi^2
e^{2}|T_{0}|^{2}\nu^2$, and the quantum conductance
$\mathrm{g}_Q=e^2/(2\pi \hbar )$. The tunneling action (\ref{NLSM-action-S-T}) is
a generalization of the $\Tr\big\{D(\partial_{\bf r} Q)^2\big\}$ term
of the NLSM action~(\ref{NLSM-action-noninteracting}) for the
tunneling geometry.

If the tunneling amplitudes  $T_{\mathbf{rr}'}$ are not small one
needs to keep higher orders in the expansion of the logarithm
in~(\ref{NLSM-action-S-T-log-expansion}). It is convenient to
express products of the even number of the tunneling amplitudes
$T_{\mathbf{rr}'}$ through the transmission probabilities of
individual transverse channels $T_n$ (see, for example, Appendix C
of~\cite{ABG}). With the help of~\eqref{NLSM-saddle-point-eq}, one
may show that expansion of the logarithm
in~(\ref{NLSM-action-S-T-log-expansion}) is order by order
equivalent to the expansion of the following action
\cite{Nazarov-TunnelingAction,BelzigNazarov,FKLS}
\begin{equation}\label{NLSM-action-S-T-general}
iS_{T}[\hat Q_{L},\hat Q_{R}]=\frac{
1}{2}\sum_{n}\Tr\ln\left[\hat 1 - \frac{T_{n}}{4}
\left(\hat{Q}_{L}-\hat{Q}_{R}\right)^2 \right]\,.
\end{equation}
If all transmissions are small, $T_{n}\ll 1$,  one may
expand~\eqref{NLSM-action-S-T-general} to the leading order in
$T_{n}$ and recover~\eqref{NLSM-action-S-T}, identifying the
tunneling conductance as $\mathrm{g}_{T}=\mathrm{g}_{Q}\sum_n T_n$,
c.f.~(\ref{Part-I-g}). Equation (\ref{NLSM-action-S-T-general}) goes
beyond this limit and allows  mesoscopic transport to be treat in
arbitrary two--terminal geometries. Its generalization for
multi--terminal case was also developed by Nazarov {\em et.
al.}~\cite{Nazarov-TunnelingAction,Nazarov-CBWTJ,SnymanNazarov}.

\subsection{Usadel equation}\label{sec_NLSM-4}

Let us return to the action specified
by~\eqref{NLSM-action-noninteracting}. Our  goal is to investigate
the physical consequences of  NLSM. As a first step, one needs to
determine the most probable (stationary) configuration,
$\underline{\hat{Q}}_{tt'}(\mathbf{r})$, on the soft--mode
manifold~\eqref{NLSM-Q-constraint}. To this end, one parameterizes
deviations from $\underline{\hat{Q}}_{tt'}(\mathbf{r})$ as
$\hat{Q}=\hat{\mathcal{R}}^{-1}\circ\underline{\hat{Q}}\circ\hat{\mathcal{R}}$
and chooses $\hat{\mathcal{R}}=\exp(\hat{\mathcal{W}}/2)$, where
$\hat{\mathcal{W}}_{tt'}(\mathbf{r})$ is the generator of rotations.
Expanding to the first order in $\hat{\mathcal{W}}$, one finds
$\hat{Q}=\underline{\hat{Q}}-[\hat{\mathcal{W}}\stackrel{\circ}{,}\underline{\hat{Q}}]/2$.
One may now substitute such a $\hat Q$ matrix into the action
\eqref{NLSM-action-noninteracting} and require that the terms linear
in $\hat{\mathcal{W}}$ vanish. This leads to the saddle point
equation for $\hat{\underline{Q}}$. For the first term in the curly
brackets on the right--hand side
of~\eqref{NLSM-action-noninteracting} one obtains
$\frac{1}{2}\Tr\big\{\hat{\mathcal{W}}\partial_{\mathbf{r}}
D\,\big[(\partial_{\mathbf{r}}\underline{\hat{Q}})
\underline{\hat{Q}}-\underline{\hat{Q}}
\partial_{\mathbf{r}}\underline{\hat{Q}}\big]\big\}=
-\Tr\big\{\hat{\mathcal{W}}\partial_{\mathbf{r}}
D\,\big(\underline{\hat{Q}}\partial_{\mathbf{r}}
\underline{\hat{Q}}\big)\big\}$, where one has employed
$\partial_{\mathbf{r}}\underline{\hat{Q}}\circ
\underline{\hat{Q}}+\underline{\hat{Q}}\circ
\partial_{\mathbf{r}}\underline{\hat{Q}}=0$, since
$\underline{\hat{Q}}^{2}=\hat{1}$. For the second term one finds
$\Tr\big\{\hat{\mathcal{W}}_{tt'}\big(\partial_{t}+\partial_{t'}\big)
\underline{\hat{Q}}_{t't}\big\}=
\Tr\big\{\hat{\mathcal{W}}\{\partial_{t},\underline{\hat{Q}}\}\big\}$.
Demanding that the linear term in $\hat{\mathcal{W}}$ vanishes, one
obtains
\begin{equation}\label{NLSM-Usadel}
\partial_{\mathbf{r}}
\big(D\,
\underline{\hat{Q}}\circ\partial_{\mathbf{r}}\underline{\hat{Q}}\big)-
\{\partial_{t},\underline{\hat{Q}}\}=0\,.
\end{equation}
This is the Usadel equation~\cite{Usadel} for the stationary
$\underline{\hat{Q}}$--matrix. If one looks for the solution of the
Usadel equation in the subspace of "classical", having causality
structure, configurations, then one takes
$\underline{\hat{Q}}=\hat{\Lambda}$, with  as--yet unspecified
distribution function $F_{tt'}(\mathbf{r})$. Therefore, in this case
the Usadel equation is reduced to the single equation for the
distribution function $F_{tt'}(\mathbf{r})$. Substituting
$\hat{\Lambda}$ from (\ref{NLSM-Lambda}) into~\eqref{NLSM-Usadel}
and performing the Wigner transformation
\begin{equation}\label{NLSM-Wigner}
F_{tt'}(\mathbf{r})=\int\frac{\d\epsilon}{2\pi}\,F_{\epsilon}
\left(\mathbf{r},\tau\right)\,e^{-i\epsilon(t-t')},\quad \quad
\tau=\frac{t+t'}{2}\, ,
\end{equation}
one obtains
\begin{equation}\label{NLSM-Diffusion-Eq}
\partial_{\mathbf{r}}\big[D(\mathbf{r}) \partial_{\mathbf{r}} F_{\epsilon}(\mathbf{r},\tau)\big]-
\partial_{\tau}F_{\epsilon}(\mathbf{r},\tau)=0\,,
\end{equation}
where we allowed for a (smooth) spatial dependence of the diffusion
constant. This is the kinetic equation for the fermionic
distribution function of the disordered system in the
non--interacting limit, which happens to be the diffusion equation.
Note that it is the same equation for any energy $\epsilon$ and
different energies do not "talk" to each other, which is natural for
the non--interacting system. In the presence of interactions, the
equation acquires the collision integral on the right--hand side
that mixes different energies between themselves. It is worth
mentioning that elastic scattering does not show up in the collision
integral. It was already fully taken into account in the derivation
of the Usadel equation and went into the diffusion term.

As an example, let us consider a disordered quasi--one--dimensional wire of
length $L$, attached to two leads, kept at different
voltages~\cite{Pierre}. We look for the space dependent, stationary
function $F_{\epsilon}(x)$ with $x$ being coordinate along the
wire, that satisfies $D\,\partial^{2}_{x}F_{\epsilon}(x)=0$,
supplemented by the boundary conditions
$F_{\epsilon}(x=0)=F_{L}(\epsilon)$ and
$F_{\epsilon}(x=L)=F_{R}(\epsilon)$, where $F_{R(L)}(\epsilon)$ are
the distribution functions of the left and right leads. The proper solution is
\begin{equation}
F_{\epsilon}(x)=F_{L}(\epsilon)+[F_{R}(\epsilon)-F_{L}(\epsilon)]\frac{x}{L}\,.
\end{equation}
The distribution function inside the wire interpolates between the
two distribution linearly. At low temperatures it looks like a
two--step function, where the energy separation between the steps is
the applied voltage, $eV$, while the relative height depends on the
position $x$. Comparing~\eqref{NLSM-Diffusion-Eq} with the
continuity equation, one notes that the current density  (at a given
energy $\epsilon$) is given by $j(\epsilon)=D\,\partial_{x}
F_{\epsilon}(x)=D[F_{R}(\epsilon)-F_{L}(\epsilon)]/L$. The total
electric current, is thus
$$I=e\nu\int\d\epsilon
j(\epsilon)=\frac{e\nu
D}{L}\int\d\,\epsilon[F_{R}(\epsilon)-F_{L}(\epsilon)]=e^{2}\frac{\nu
D}{L}V=\sigma_D V/L\,,$$
where the Drude conductivity of the
diffusive wire is given by $\sigma_{D}=e^{2}\nu D$.

\subsection{Fluctuations}\label{sec_NLSM-5}

Following the discussions in  previous sections we consider
fluctuations near the stationary solution
$\underline{\hat{Q}}_{tt'}(\mathbf{r})=\hat{\Lambda}_{t-t'}$,
see~\eqref{NLSM-Lambda}. We restrict ourselves to the soft--mode
fluctuations that satisfy $\hat{Q}^{2}=\hat{1}$  and neglect all
massive modes that stay outside of this manifold. The massless
fluctuations of the $\hat Q$--matrix may be parameterized as
\begin{equation}\label{NLSM-Q-U-W-parametrization}
\hat{Q}=\hat{\mathcal{U}}\circ
e^{-\hat{\mathcal{W}}/2}\circ\hat{\sigma}_{z}\circ
e^{\hat{\mathcal{W}}/2}\circ\hat{\mathcal{U}}^{-1}\,,
\end{equation}
where rotation generators are given by
\begin{equation}\label{NLSM-U-W}
\hat{\mathcal{W}}=
\left(\begin{array}{cc}0&d\\
\bar{d}&0\end{array}\right)\,,\; \quad\quad \quad
\hat{\mathcal{U}}=\hat{\mathcal{U}}^{-1}=
\left(\begin{array}{cc}1&F\\0&-1\end{array}\right)\, .
\end{equation}
Here $d_{tt'}(\mathbf{r})$ and $\bar{d}_{tt'}(\mathbf{r})$ are two
independent Hermitian matrices in the time space. One, thus,
understands the functional integration over
$\hat{Q}_{tt'}(\mathbf{r})$ in~\eqref{NLSM-action-SAV} as an
integration over two mutually independent Hermitian matrices in the
time domain, $d_{tt'}(\mathbf{r})$ and $\bar{d}_{tt'}(\mathbf{r})$.
The physical meaning of $d_{tt'}(\mathbf{r})$ is a deviation of the
fermionic distribution function $F_{tt'}(\mathbf{r})$ from its
stationary value. At the same time, $\bar{d}_{tt'}(\mathbf{r})$ has
no classical interpretation. To a large extent, it plays the role of
the quantum counterpart of $d_{tt'}(\mathbf{r})$, that appears only
as the internal line in the diagrams. The reason for choosing
$\hat{Q}$ in the form of~\eqref{NLSM-Q-U-W-parametrization} can be
justified as follows. First, one notes that
$\underline{\hat{Q}}\equiv\hat{\Lambda}=
\hat{\mathcal{U}}\,\hat{\sigma}_{z}\,\hat{\mathcal{U}}^{-1}$.
Second, one should realize that the part of $\hat{\mathcal{W}}$ that
commutes with $\underline{\hat{Q}}$ does not generate any
fluctuations, therefore, one restricts $\hat{\mathcal{W}}$ to
satisfy $\hat{\mathcal{W}}\,\hat{\sigma}_{z}+
\hat{\sigma}_{z}\,\hat{\mathcal{W}}=0$. Thus, $\hat{\mathcal{W}}$
has to be off--diagonal and most generally parameterized by two
independent fields, $d$ and $\bar{d}$, see~\eqref{NLSM-U-W}.

One may expand now the action~\eqref{NLSM-action-noninteracting} in
powers of $\bar{d}_{tt'}(\mathbf{r})$ and $d_{tt'}(\mathbf{r})$.
Since $\underline{\hat{Q}}_{tt'}$ was chosen to be a stationary
point, the expansion starts from the second order. If stationary
$F_{t,t'}({\bf r})$ is  spatially uniform, one obtains
\begin{equation}\label{NLSM-S-W}
iS[\hat{\mathcal{W}}]=-\frac{\pi\nu}{2}\int\d\mathbf{r} \iint\d t\d t'\,
\bar{d}_{tt'}(\mathbf{r})
\left[-D\,\partial^{2}_{\mathbf{r}}+\partial_{t}+\partial_{t\,'}\right]
d_{t't}(\mathbf{r})\,.
\end{equation}
The quadratic form may be  diagonalized by transforming to the
energy/momentum representation
$$\hat{\mathcal{W}}_{\epsilon\epsilon'}(\mathbf{q})=\int\d\mathbf{r}
\iint\d t\d t'\hat{\mathcal{W}}_{tt'}(\mathbf{r})\exp(i\epsilon
t-i\epsilon't')\exp(-i\mathbf{qr})\,.$$ As a result, the propagator
of small $\hat Q$ matrix fluctuations is
\begin{equation}\label{NLSM-d-d}
\langle d_{\epsilon_{2}\epsilon_{1}}(\mathbf{q})
\bar{d}_{\epsilon_{3}\epsilon_{4}}(-\mathbf{q})\rangle_{\mathcal{W}}=
-\frac{2}{\pi\nu}\,\frac{\delta_{\epsilon_{1}\epsilon_{3}}\delta_{\epsilon_{2}\epsilon_{4}}}
{Dq^2+i\omega}\equiv -\frac{2}{\pi\nu}\,
\delta_{\epsilon_{1}\epsilon_{3}}\delta_{\epsilon_{2}\epsilon_{4}}\,
\mathcal{D}^A(\mathbf{q},\omega)\,,
\end{equation}
where $\omega\equiv\epsilon_{1}-\epsilon_{2}=\epsilon_3-\epsilon_4$
and  object
$\mathcal{D}^{R(A)}(\mathbf{q},\omega)=\mathcal{D}^{R(A)}(\mathbf{q},\epsilon_{1}-\epsilon_{2})=
\big[Dq^{2}\mp i(\epsilon_{1}-\epsilon_{2})\big]^{-1}$ is called the
\textit{diffuson}. The higher--order terms of the
action~\eqref{NLSM-action-noninteracting} expansion over
$d_{tt'}(\mathbf{r})$ and $\bar{d}_{tt'}(\mathbf{r})$ describe
non--linear interactions of the diffusive modes with the vertices
called \textit{Hikami
boxes}~\cite{Hikami,Gor'kovLarkinKhmelnitskii}. These non--linear
terms are responsible for weak--localization
corrections~\cite{Gor'kovLarkinKhmelnitskii,AAKL,LeeRamakrishnan,ALW-book}.
If the distribution function $F_{tt'}(\mathbf{r})$ is spatially
non--uniform, there is an additional term in the quadratic action
$-(\pi\nu D/2)\Tr\big\{\bar{d}(\partial_{\mathbf{r}}
F)\bar{d}(\partial_{\mathbf{r}} F)\big\}$. This term generates
non--zero correlations of the type $\langle
dd\rangle_{\mathcal{W}}$, which are important for some applications.

\subsection{Applications II: Mesoscopic effects}\label{app_Part-II}
\subsubsection{Kubo formula and linear response}\label{app_Part-II-1}

It was demonstrated in  Section~\ref{sec_fermion-4} how the linear
response theory is formulated in the Keldysh technique. Let us see
now how the polarization operator of the disordered electron gas may
be obtained from  NLSM action. To this end, one uses general
definition of the density response function $\Pi^{R}(x,x\,')$ given
by~\eqref{fermion-Pi-matrix-def} along with the disorder averaged
action~\eqref{NLSM-action-general}, which gives
\begin{equation}\label{Part-II-Pi}
\Pi^{R}(x,x\,')=-\frac{i}{2}\left.\frac{\delta^{2}
\mathcal{Z}[V^{cl},V^{q}]}{\delta V^{cl}(x\,')\delta
V^{q}(x)}\right|_{\hat{V}=0}=\nu\delta(\mathbf{r}-\mathbf{r}')\delta(t-t')
+\frac{i}{2}(\pi\nu)^{2}
\left\langle\Tr\big\{\hat{\gamma}^{q}\hat{Q}_{tt}(\mathbf{r})\big\}
\Tr\big\{\hat{\gamma}^{cl}\hat{Q}_{t\,'t\,'}(\mathbf{r}')\big\}\right\rangle_{Q}\,,
\end{equation}
where $x=(\mathbf{r},t)$ and angular brackets stand for the
averaging over the action~\eqref{NLSM-action-noninteracting}. The
first term on the right--hand side of~\eqref{Part-II-Pi} originates
from the differentiation of
$\Tr\big\{\hat{V}\hat{\sigma}_{x}\hat{V}\big\}$ part of the action
\eqref{NLSM-action-general}, while the second term comes from
differentiation of $\Tr\big\{\hat{V}\hat{Q}\big\}$. Equation
\eqref{Part-II-Pi} represents the $\sigma$--model equivalent of the
Kubo formula for the linear density response.

In the Fourier representation the last equation takes the form
\begin{equation}\label{Part-II-Pi-Fourier}
\Pi^{R}\qo=\nu+\frac{i}{2}(\pi\nu)^{2}\iint
\frac{\d\epsilon\d\epsilon'}{4\pi^{2}}
\left\langle\Tr\big\{\hat{\gamma}^{q}\hat{Q}_{\epsilon+\omega,\epsilon}(\mathbf{q})\big\}
\Tr\{\hat{\gamma}^{cl}\hat{Q}_{\epsilon',\epsilon'+\omega}(-\mathbf{q})\}
\right\rangle_{Q}\,.
\end{equation}
Employing~\eqref{NLSM-Q-U-W-parametrization} and \eqref{NLSM-U-W},
one finds in the liner order in the diffusive fluctuations (the only
contribution in the zeroth order is $\nu$, indeed
$\Tr\{\hat{\gamma}^{cl}\hat{\Lambda}\}=0$)
\begin{eqnarray}
&&\Tr\big\{\hat{\gamma}^{cl}\hat{Q}_{\epsilon',\epsilon'+\omega}(-\mathbf{q})\big\}=
\bar{d}_{\epsilon',\epsilon'+\omega}(-\mathbf{q})
(F_{\epsilon'+\omega}-F_{\epsilon'})\,,\nonumber\\
&&\Tr\big\{\hat{\gamma}^{q}\hat{Q}_{\epsilon+\omega,\epsilon}(\mathbf{q})\big\}=
\bar{d}_{\epsilon+\omega,\epsilon}(\mathbf{q})
(1-F_{\epsilon}F_{\epsilon+\omega})-d_{\epsilon+\omega,\epsilon}(\mathbf{q})\,.
\end{eqnarray}
Since $\langle\bar{d}\bar{d}\rangle_{\mathcal{W}}\equiv0$ only the
last term of the last expression contributes to the average
in~(\ref{Part-II-Pi-Fourier}). The result is
\begin{equation}\label{Part-II-Pi-diffusive}
\Pi^{R}\qo=\nu+\frac{i\pi\nu^{2}}{4}
\int^{+\infty}_{-\infty}\d\epsilon\,\big(F_{\epsilon}-F_{\epsilon+\omega}\big)
\left\langle d_{\epsilon+\omega,\epsilon}(\mathbf{q})
\bar{d}_{\epsilon,\epsilon+\omega}(-\mathbf{q})\right\rangle_{\mathcal{W}}
=\nu\left[1+\frac{i\omega}{Dq^{2}-i\omega}\right]=\frac{\nu
Dq^{2}}{Dq^{2}-i\omega},
\end{equation}
where we have used the propagator of diffusons~\eqref{NLSM-d-d} and
the integral
$\int\d\epsilon\,(F_{\epsilon}-F_{\epsilon+\omega})=-2\omega$. The
fact that $\Pi^{R}(0,\omega)=0$ is a consequence of the particle
number conservation. One has obtained the diffusion form of the
density--density response function. Also note that this function is
indeed retarded (analytic in the upper half--plane of complex
$\omega$), as it should be. The current--current response function,
$K^{R}\qo$, may be obtained in the similar manner. However, more
straightforward way is to use continuity equation
$\mathbf{q}\cdot\mathbf{j}+\omega\varrho=0$, which implies the
following relation between density and current response functions
$K^{R}\qo=\omega^{2}\Pi^{R}\qo/q^{2}$. As a result the conductivity
is given by
\begin{equation}
\sigma\qo=\frac{e^{2}}{i\omega}\,K^{R}\qo =
e^{2}\, \frac{-i\omega}{q^2}\, \Pi^R\qo
= e^{2}\nu D\, \frac{-i\omega}{Dq^{2}-i\omega}\, ,
\end{equation}
which in the uniform limit $\mathbf{q}\to0$ reduces to the Drude
result $\sigma_{D}\equiv\sigma(0,\omega)=e^{2}\nu D$.

\subsubsection{Spectral statistics}\label{app_Part-II-2}

Consider a piece of  disordered metal of size $L$ such that $L\gg
l$, where $l\equiv v_F\tauel$ is the elastic mean free path. The
spectrum of the Schr\"odinger equation consists of a discrete set of
levels, $\epsilon_n$, that may be characterized by the {\em
sample--specific} density of states (DOS), $\nu(\epsilon)=
\sum_n\delta(\epsilon-\epsilon_n)$. This quantity fluctuates
strongly and usually cannot (and does not need to) be calculated
analytically. One may average it over  realizations of disorder to
obtain a mean DOS: $\langle\nu(\epsilon)\rangle_{\mathrm{dis}}$. The
latter is a smooth function of energy on the scale of the Fermi
energy and thus may be taken as a constant
$\langle\nu(\epsilon_F)\rangle_{\mathrm{dis}}\equiv\nu$. This is
exactly the DOS that was used in the previous sections.

One may wonder how to sense fluctuations of the sample--specific DOS
$\nu(\epsilon)$ and, in particular, how a given spectrum at one energy
$\epsilon$ is correlated with itself at another energy $\epsilon'$.
To answer this question one may calculate the spectral correlation
function
\begin{equation}\label{Part-I-R-def}
R(\epsilon,\epsilon')\equiv\langle\nu(\epsilon)\nu(\epsilon')\rangle_{\mathrm{dis}}
-\nu^2\, .
\end{equation}
This function was calculated in the seminal paper of Altshuler
and Shklovskii~\cite{AltshulerShklovskii}. Here we derive it
using the Keldysh NLSM.

\begin{figure}[h!]
\begin{center}\includegraphics[width=8cm]{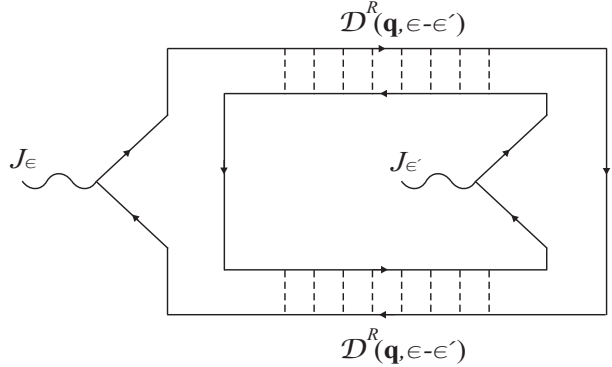}\end{center}
\caption{Diagram for calculation of mesoscopic fluctuations of the
density of states,  $R(\epsilon,\epsilon')$, see~\eqref{Part-I-R}.
It is generated from the Wick contraction $ \langle
d_{\epsilon\epsilon_{1}}\bar{d}_{\epsilon_{1}\epsilon}
\bar{d}_{\epsilon'\epsilon_{2}}d_{\epsilon_{2}\epsilon'}\rangle_{\mathcal{W}}\to
\langle d_{\epsilon\epsilon_{1}}
\bar{d}_{\epsilon'\epsilon_{2}}\rangle_{\mathcal{W}}
\langle\bar{d}_{\epsilon_{1}\epsilon}d_{\epsilon_{2}\epsilon'}
\rangle_{\mathcal{W}}
\propto[\mathcal{D}^{R}(\mathbf{q},\epsilon-\epsilon')]^{2}
\delta_{\epsilon_1\epsilon'}\delta_{\epsilon_2\epsilon}$,
see~\eqref{Part-I-R-dd}. There is also a similar diagram with the
advanced diffusons. \label{Fig-AS-DOS}}
\end{figure}

The DOS is defined as $$\nu(\epsilon) =
i\sum_{k}(G^R(\mathbf{k},\epsilon)-
G^A(\mathbf{k},\epsilon))/(2\pi)=(\langle \psi_1\bar\psi_1\rangle
-\langle \psi_2\bar\psi_2\rangle)/(2\pi) =
-\big\langle\vec{\bar{\Psi}}\hat{\sigma}_z\vec{\Psi}\big\rangle/(2\pi)\,,$$
where the angular brackets denote quantum (as opposed to disorder)
averaging and the indices are in  Keldysh space. To generate the DOS
at any given energy one adds a source term
$$iS_{\mathrm{DOS}}=-\int\frac{\d\epsilon}{2\pi} J_\epsilon\int
\d\mathbf{r}\,\vec{\bar{\Psi}}(\epsilon,\mathbf{r})\hat{\sigma}_z
\vec{\Psi}(\epsilon,\mathbf{r})=-\iint\d t\d t'\int\d\mathbf{r}\,
\vec{\bar{\Psi}}\rt
J_{t-t'}\hat{\sigma}_z\vec{\Psi}(\mathbf{r},t')\,,$$ to the
fermionic action~\eqref{NLSM-action-noninteracting}. After averaging
over disorder and changing to the $\hat Q$ matrix representation the
DOS source term is translated to
$$iS_{\mathrm{DOS}}=\pi\nu\int\frac{\d\epsilon}{2\pi}
J_\epsilon\int\d\mathbf{r}\,\Tr\{\hat{Q}_{\epsilon\epsilon}(\mathbf{r})\hat
\sigma_z\}\,.$$ Then the DOS is generated by
$\nu(\epsilon)=\delta\mathcal{Z}[J]/\delta J_\epsilon$. It is now
clear that $\langle\nu(\epsilon)\rangle_{\mathrm{dis}}={1\over 2}\nu
\langle \Tr\{\hat{Q}_{\epsilon\epsilon}\hat{\sigma}_z\}\rangle_Q$.
Substituting $\hat{Q}_{\epsilon\epsilon}=\hat{\Lambda}_\epsilon$ one
finds $\langle\nu(\epsilon)\rangle_{\mathrm{dis}}=\nu$, as it should
be, of course. It is also easy to check that the fluctuations around
$\hat{\Lambda}$ do not change the result (all the fluctuation
diagrams cancel owing to the causality constraints). We are now in a
position to calculate the correlation function~\eqref{Part-I-R-def},
\begin{equation}\label{Part-I-R}
R(\epsilon,\epsilon')\equiv\frac{\delta^{2}\mathcal{Z}[J]}{\delta
J_\epsilon \delta J_{\epsilon'}} -\nu^2=\nu^{2}\left[\frac{1}{4}
\left\langle \Tr\{\hat{Q}_{\epsilon\epsilon}\hat{\sigma}_{z}\}\Tr\{
\hat{Q}_{\epsilon'\epsilon'}\hat{\sigma}_{z}\}\right\rangle_{Q}-1\right]\,.
\end{equation}
Employing the parametrization of~\eqref{NLSM-Q-U-W-parametrization},
one finds, up to the second order in the diffusive fluctuations
$\hat{\mathcal{W}}$,
\begin{equation}
\Tr\big\{\hat{Q}\hat{\sigma}_{z}\big\}=\frac{1}{2}\,\big[4-2\,F\circ\bar{d}-2\,\bar{d}\circ
F+d\circ\bar{d}+\bar{d}\circ d\big]\,.
\end{equation}
Since $\langle\bar{d}\bar{d}\rangle_{\mathcal{W}}=0$, the only
non--vanishing terms contributing to~\eqref{Part-I-R} are those with
no $d$ and $\bar{d}$ at all (they cancel $\nu^{2}$ term) and those
of the type $\langle d\bar{d}d\bar{d}\rangle_{\mathcal{W}}$.
Collecting the latter terms one finds
\begin{equation}\label{Part-I-R-dd}
R(\epsilon,\epsilon')=\frac{\nu^{2}}{16}\int\d\mathbf{r}
\iint\frac{\d\epsilon_{1}\d\epsilon_{2}}{(2\pi)^{2}}
\,\left\langle\left(d_{\epsilon\epsilon_{1}}\bar{d}_{\epsilon_{1}\epsilon}+
\bar{d}_{\epsilon\epsilon_{1}}d_{\epsilon_{1}\epsilon}\right)
\left(d_{\epsilon'\epsilon_{2}}\bar{d}_{\epsilon_{2}\epsilon'}+
\bar{d}_{\epsilon'\epsilon_{2}}d_{\epsilon_{2}\epsilon'}\right)\right\rangle_{\mathcal{W}}\,.
\end{equation}
Now one has to perform Wick's contractions, using correlation
function $\langle
d_{\epsilon\epsilon'}\bar{d}_{\epsilon'\epsilon}\rangle_{\mathcal{W}}
\propto\mathcal{D}^{R}(\epsilon-\epsilon')$, which follows
from~\eqref{NLSM-d-d}, and also take into account
$\int\d\epsilon_{1}[\mathcal{D}^{R(A)}(\mathbf{q},\epsilon-\epsilon_{1})]^{2}=0$,
owing to the integration of a function which is analytic in the
entire upper (lower) half--plane of $\epsilon_{1}$. As a result,
\begin{equation}
R(\epsilon,\epsilon')=\frac{1}{4\pi^{2}}\sum_{q}
\left[\big(\mathcal{D}^{R}(\mathbf{q},\epsilon-\epsilon')\big)^{2}+
\big(\mathcal{D}^{A}(\mathbf{q},\epsilon-\epsilon')\big)^{2}\right]\,,
\end{equation}
where the momentum summation stands for a summation over the
discrete modes of the diffusion operator
$D\partial^{2}_{\mathbf{r}}$ with the zero current (zero derivative)
at the boundary of the metal. This is the result of Altshuler and
Shklovskii~\cite{AltshulerShklovskii} for the unitary symmetry
class. Note that the correlation function $R(\epsilon,\epsilon')$
depends only on the energy difference $\omega=\epsilon-\epsilon'$.
Diagrammatic representation of $R(\epsilon,\epsilon')$ function is
shown in Figure~\ref{Fig-AS-DOS}. Adopting an explicit form of the
diffusion propagator, we find spectral correlation function in the
form
\begin{equation}
R(\epsilon-\epsilon')=-\frac{1}{2\pi^{2}}\,\mathrm{Re}\sum_{n}\frac{1}
{\big(\epsilon-\epsilon'+iDq^{2}_{n}\big)^{2}}\,,
\end{equation}
where $q^{2}_{n}=\sum_{\mu}\pi^{2}n^{2}_{\mu}/L^{2}_{\mu}$, with
$\mu=x,y,z\,$; $n_{\mu}=0, 1, 2\ldots$ and $L_{\mu}$ are spatial
dimensions of the mesoscopic sample.

For a small energy difference $\omega\ll E_{Th}=D/L^{2}$ only the
lowest homogenous mode, $q_{n}=0$, of the diffusion operator (the
so--called zero mode) may be retained and, thus, $R(\omega)=
-1/(2\pi^2\omega^2)$. This is the universal random matrix result.
The fact that the correlation function is negative means that the
energy levels are \textit{less} likely to be found at a small
distance $\omega$ from each other. This is a manifestation of the
energy levels repulsion. Note that the correlations decay very
slowly --- as the inverse square of the energy distance. One may
note that the random matrix result~\cite{Mehta}
\begin{equation}
R_{RMT}(\omega)=-\frac{1-\cos(2\pi\omega/\delta)}{2\pi^2\omega^2}\,,
\end{equation}
where $\delta$ is the mean level spacing, contains also an
oscillatory function of the energy difference. These oscillations
reflect discreteness of the underlying energy spectrum. They cannot
be found in the perturbation theory in small fluctuations near the
$\hat{\Lambda}$ ``point''. However, they may be recovered once
additional stationary points (not possessing the  causality
structure) are taken into account~\cite{AltlandKamenev}. The saddle
point method and perturbation theory  work as long as $\omega\gg
\delta$. Currently it is not known  how to treat the Keldysh NLSM at
$\omega\lesssim \delta$.

\subsubsection{Universal conductance fluctuations}\label{Part-II-3}

Similarly to the discussions of the previous section consider an
ensemble of small metallic samples with the size $L$ comparable to
the electron phase coherence length, $L\sim L_{\varphi}$. Their
conductances  exhibit sample--to--sample  fluctuations owing to
differences in their specific realizations of disorder potential.
These reproducible fluctuations are known as \textit{universal
conductance fluctuations} (UCFs). Theoretical studies of UCFs were
initiated by Altshuler~\cite{Altshuler-UCF}, and Lee and
Stone~\cite{LeeStone}. More detailed study of UCFs was given later
in~\cite{AltshulerShklovskii,LeeStoneFukuyama}. The technique
developed in this section for treating UCFs in the framework of the
Keldysh non--linear $\sigma$--model is closely parallel to the
theory developed by Altshuler, Kravtsov and Lerner~\cite{AKL-UCF} in
the framework of the zero temperature replica non--linear sigma
model.

\begin{figure}
\begin{center}\includegraphics[width=10cm]{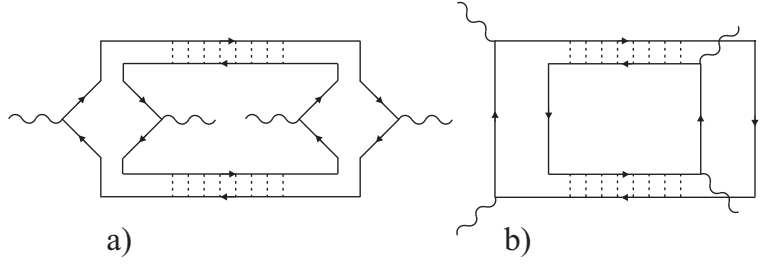}\end{center}
\caption{Diagrams for the variance of conductance
fluctuations.\label{Fig-UCF}}
\end{figure}

Our starting point is the expression for the $\mathrm{dc}$
conductivity within the linear response given by
\begin{equation}\label{Part-II-sigma-def}
\sigma_{\mu\nu}=-\frac{e^2}{2}\lim_{\Omega\to 0}\frac{1}{\Omega}
\left(\frac{\delta^{2}\mathcal{Z}[\mathbf{A}^{cl},\mathbf{A}^{q}]}
{\delta
A^{cl}_{\nu}(\Omega)A^{q}_{\mu}(-\Omega)}\right)_{\mathbf{A}^{cl}=\mathbf{A}^{q}=0}\,,
\end{equation}
where indices $\mu,\nu$ stand for the spatial Cartesian coordinates.
Expanding action~\eqref{NLSM-action-general} to the quadratic order
in the vector potential with the help
of~\eqref{NLSM-covariant-deriv} one finds that corresponding term in
the generating function reads as
$\mathcal{Z}[\mathbf{A}^{cl},\mathbf{A}^{q}]=\frac{\pi\nu
D}{2}\big\langle\Tr\big\{\hat{\mathbf{A}}
\hat{Q}\hat{\mathbf{A}}\hat{Q}\big\}\big\rangle_{Q}$. At the saddle
point $\hat{Q}=\hat{\Lambda}$, after consecutive differentiation
over the vector potential in~\eqref{Part-II-sigma-def} one finds for
the average conductivity
\begin{equation}\label{NLSM-sigma-Drude}
\langle\sigma_{\mu\nu}\rangle_{\mathrm{dis}}=
\delta_{\mu\nu}\lim_{\Omega\to 0}\frac{\pi\sigma_{D}}{4\Omega}
\Tr\left\{\hat{\gamma}^{cl}\hat{\Lambda}_{\epsilon+\Omega}
\hat{\gamma}^{q}\hat{\Lambda}_{\epsilon-\Omega}\right\}=
\delta_{\mu\nu}\frac{\pi\sigma_{D}}{2}\lim_{\Omega\to0}
\frac{1}{\Omega}\int\frac{\d\epsilon}{2\pi}
\big(F_{\epsilon+\Omega}-F_{\epsilon-\Omega}\big)=\sigma_{D}\delta_{\mu\nu}
\end{equation}
where $\sigma_D=e^2\nu D$, as it should be of course. At this level,
retaining fluctuations $\hat{\mathcal{W}}$ of the $\hat Q$ matrix
around the  saddle point $\hat{\Lambda}$, one can calculate
weak--localization
corrections~\cite{Hikami,Gor'kovLarkinKhmelnitskii,AAKL,LeeRamakrishnan,ALW-book}
to the average conductivity. In what follows we are interested in
calculation of the irreducible correlation function for the
conductivity fluctuations which is defined in the following way
$$\langle\delta\sigma_{\mu_{1}\nu_{1}}
\delta\sigma_{\mu_{2}\nu_{2}}\rangle_{\mathrm{dis}}=
\left\langle\big(\sigma_{\mu_{1}\nu_{1}}-\langle\sigma_{\mu_{1}\nu_{1}}\rangle\big)
\big(\sigma_{\mu_{2}\nu_{2}}-
\langle\sigma_{\mu_{2}\nu_{2}}\rangle\big)\right\rangle_{\mathrm{dis}}\,.$$
In  view of~\eqref{Part-II-sigma-def} this irreducible correlator
can be expressed through the $\hat Q$ matrix as
\begin{eqnarray}\label{Part-II-UCF-def}
\langle\delta\sigma_{\mu_{1}\nu_{1}}
\delta\sigma_{\mu_{2}\nu_{2}}\rangle_{\mathrm{dis}}=
\left(\frac{\pi\sigma_{D}}{4}\right)^{2}
\prod^{2}_{i=1}\left(\lim_{\Omega_{i}\to 0}
\frac{1}{\Omega_{i}}\frac{\delta^{2}}{\delta
A^{cl}_{\nu_{i}}(\Omega_{i})\delta
A^{q}_{\mu_{i}}(-\Omega_{i})}\right)
\left\langle\Tr\{\hat{\mathbf{A}}\hat{Q}\hat{\mathbf{A}}\hat{Q}\}
\Tr\{\hat{\mathbf{A}}\hat{Q}\hat{\mathbf{A}}\hat{Q}\}\right\rangle_{Q}\nonumber\\
-\sigma^{2}_{D} \delta_{\mu_{1}\nu_{1}}\delta_{\mu_{2}\nu_{2}}\,,
\end{eqnarray}
where we have used~\eqref{NLSM-action-general} and expanded
$\exp(iS[\hat{Q},\mathbf{A}])$ up to the forth order in the vector
potential. Now one has to account for fluctuations of the $\hat Q$
matrix up to the second order in generators $\hat{\mathcal{W}}$.
There are two possibilities here: within each trace on the
right--hand side of~\eqref{Part-II-UCF-def} one may expand each
$\hat Q$ matrix either to the linear order in $\hat{\mathcal{W}}$
resulting in
$\mathcal{T}_{1}[\hat{\mathcal{W}}]=\Tr\{\hat{\mathbb{A}}\hat{\sigma}_{z}
\hat{\mathcal{W}}\hat{\mathbb{A}}\hat{\sigma}_{z}\hat{\mathcal{W}}\}$;
or alternatively set one of $\hat Q$ matrices to be $\hat{\Lambda}$,
while expanding the other one to the second order, resulting in
$\mathcal{T}_{2}[\hat{\mathcal{W}}]=
\Tr\{\hat{\mathbb{A}}\hat{\sigma}_{z}\hat{\mathbb{A}}\hat{\sigma}_{z}
\hat{\mathcal{W}}^{2}\}$, where
$\hat{\mathbb{A}}=\hat{\mathcal{U}}^{-1}
\hat{\mathbf{A}}\hat{\mathcal{U}}$. As a result,
\eqref{Part-II-UCF-def} takes the form
\begin{eqnarray}\label{Part-II-UCF-W}
&&\langle\delta\sigma_{\mu_{1}\nu_{1}}
\delta\sigma_{\mu_{2}\nu_{2}}\rangle_{\mathrm{dis}}=
\left(\frac{\pi\sigma_{D}}{4}\right)^{2}
\prod^{2}_{i=1}\left(\lim_{\Omega_{i}\to 0}
\frac{1}{\Omega_{i}}\frac{\delta^{2}}{\delta
A^{cl}_{\nu_{i}}(\Omega_{i})\delta
A^{q}_{\mu_{i}}(-\Omega_{i})}\right) \nonumber \\
&&\left[\big\langle\mathcal{T}_{1}[\hat{\mathcal{W}}]
\mathcal{T}_{1}[\hat{\mathcal{W}}]\big\rangle_{\mathcal{W}}+
\big\langle\mathcal{T}_{2}[\hat{\mathcal{W}}]
\mathcal{T}_{2}[\hat{\mathcal{W}}]\big\rangle_{\mathcal{W}}\right]-
\sigma^{2}_{D} \delta_{\mu_{1}\nu_{1}}\delta_{\mu_{2}\nu_{2}}\,.
\end{eqnarray}
It is convenient to represent each average here diagrammatically,
see Figure~\ref{Fig-UCF}. A rhombus in Figure~\ref{Fig-UCF}a
correspond to the term with $\mathcal{T}_{1}[\hat{\mathcal{W}}]$,
where the opposite vertices represent matrices $\hat{\mathbb{A}}$,
while rectangles with adjacent vertices in Figure~\ref{Fig-UCF}b
correspond to the term with $\mathcal{T}_{2}[\hat{\mathcal{W}}]$.
The vertices are connected by the diffuson propagators of the field
$\hat{\mathcal{W}}$. Equation \eqref{Part-II-UCF-W} should also
contain the cross--contribution
$2\langle\mathcal{T}_{1}[\hat{\mathcal{W}}]\mathcal{T}_{2}[\hat{\mathcal{W}}]
\rangle_{\mathcal{W}}$, which vanishes, however, upon
$\hat{\mathcal{W}}$ averaging.

Differentiating each term of the~\eqref{Part-II-UCF-W} individually,
multiplying matrices and using diffuson propagators
from~\eqref{NLSM-d-d}, one finds for~\eqref{Part-II-UCF-W}
\begin{eqnarray}\label{Part-II-UCF-AS}
&& \langle\delta\sigma_{\mu_{1}\nu_{1}}
\delta\sigma_{\mu_{2}\nu_{2}}\rangle_{\mathrm{dis}}=
\left(\frac{4\sigma_{D}}{\pi\nu}\right)^{2}\!\!
\iint^{+\infty}_{-\infty}\frac{\d\epsilon_{1}\d\epsilon_{2}}
{\big[2T\cosh(\epsilon_{1}/2T)\cosh(\epsilon_{2}/2T)\big]^{2}}\nonumber
\\
&&\sum_{q}\left[
|\mathcal{D}^{R}(\mathbf{q},\epsilon_{1}-\epsilon_{2})|^{2}
\big(\delta_{\mu_{1}\mu_{2}}\delta_{\nu_{1}\nu_{2}}+
\delta_{\mu_{1}\nu_{2}}\delta_{\nu_{1}\mu_{2}}\big)+\mathrm{Re}\,
\big[\mathcal{D}^{R}(\mathbf{q},\epsilon_{1}-\epsilon_{2})\big]^{2}
\delta_{\mu_{1}\nu_{1}}\delta_{\mu_{2}\nu_{2}}\right]\,.
\end{eqnarray}
The first term in the square  brackets of~\eqref{Part-II-UCF-AS}
corresponds to Figure~\ref{Fig-UCF}a and the second one to
Figure~\ref{Fig-UCF}b. Introducing
$\epsilon_{1}-\epsilon_{2}=\omega$ and
$\epsilon_{1}+\epsilon_{2}=2\epsilon$, and integrating over
$\epsilon$,  Equation~\eqref{Part-II-UCF-AS} may be cast into the
form
\begin{equation}
\langle\delta\sigma_{\mu_{1}\nu_{1}}
\delta\sigma_{\mu_{2}\nu_{2}}\rangle_{\mathrm{dis}}=
\sigma^{2}_{1}\big(\delta_{\mu_{1}\mu_{2}}\delta_{\nu_{1}\nu_{2}}+
\delta_{\mu_{1}\nu_{2}}\delta_{\nu_{1}\mu_{2}}\big)+
\sigma^{2}_{2}\,\delta_{\mu_{1}\nu_{1}}\delta_{\mu_{2}\nu_{2}},
\end{equation}
where
\begin{subequations}
\begin{equation}
\sigma^{2}_{1}=\left(\frac{4\sigma_{D}}{\pi\nu}\right)^{2}\int^{+\infty}_{-\infty}\
\frac{\d\omega}{2T}\
\mathcal{F}\left(\frac{\omega}{2T}\right)\sum_{q}
\frac{1}{\big(Dq^{2}\big)^{2}+\omega^{2}}\,,
\end{equation}
\begin{equation}
\sigma^{2}_{2}=\left(\frac{4\sigma_{D}}{\pi\nu}\right)^{2}\int^{+\infty}_{-\infty}\
\frac{\d\omega}{2T}\
\mathcal{F}\left(\frac{\omega}{2T}\right)\mathrm{Re}\sum_{q}
\frac{1}{\big(Dq^{2}-i\omega\big)^{2}}\,,
\end{equation}
\end{subequations}
and dimensionless function is given by $
\mathcal{F}(x)=[x\coth(x)-1]/\sinh^{2}(x)$. Here $\sigma^{2}_{1}$
may be regarded as contribution from the mesoscopic fluctuations of
the diffusion coefficient, Figure~\ref{Fig-UCF}a, while
$\sigma^{2}_{2}$ as  the corresponding contribution from the
fluctuations of the density of states, Figure~\ref{Fig-UCF}b. The
fact that
$\langle\mathcal{T}_{1}[\hat{\mathcal{W}}]\mathcal{T}_{2}[\hat{\mathcal{W}}]
\rangle_{\mathcal{W}} = 0$ implies that mesoscopic fluctuations of
the diffusion coefficient and density of states are statistically
independent. In general, $\sigma^{2}_{1}$ and $\sigma^{2}_{2}$
contributions are distinct. At zero temperature $\omega\to 0$,
however, they are equal, resulting in
\begin{equation}
\langle\delta\sigma_{\mu_{1}\nu_{1}}
\delta\sigma_{\mu_{2}\nu_{2}}\rangle=c_{d}
\left(\frac{e^{2}}{2\pi\hbar}\right)^{2}
\big(\delta_{\mu_{1}\mu_{2}}\delta_{\nu_{1}\nu_{2}}+
\delta_{\mu_{1}\nu_{2}}\delta_{\nu_{1}\mu_{2}}+
\delta_{\mu_{1}\nu_{1}}\delta_{\mu_{2}\nu_{2}}\big)\,,
\end{equation}
where $c_{d}=(4/\pi)^2\sum_{n_{\mu}}(\pi n_{\mu}n_{\mu})^{-2}$ is
dimensionality-- and geometry--dependent coefficient (note that in
the final answer we have restored Planck's constant). This
expression reflects the universality of conductance fluctuations
and, of course, coincides with the result obtained originally from
the impurity diagram technique~\cite{AltshulerShklovskii,LeeStone},
for review see~\cite{ALW-book,Beenakker}.

\subsubsection{Full counting statistics}\label{Part-II-4}

When current $I(t)$ flows in a conductor it generally fluctuates
around its average value $\langle I\rangle$. One is often interested
in calculation of the second, or even higher moments of current
fluctuations. An example of this sort was already considered in the
Section~\ref{app_Part-I-2}. Remarkably, in certain cases one may
calculate not only a given moment of the fluctuating  current, but
rather restore full distribution function of current fluctuations.
Theoretical approach, utilizing Keldysh technique, to the full
counting statistics (FCS) of electron transport was pioneered by
Levitov and Lesovik and
coworkers~\cite{LevitovLesovik-1,LevitovLesovik-2,LevitovLee}. In
the following we consider its application to the diffusive
electronic transport developed by Nazarov~\cite{Nazarov-FCS}.

Consider two reservoirs, with the chemical potentials shifted by
externally applied voltage $V$. It is assumed that reservoirs are
connected to each other by  diffusive quasi--one--dimensional wire
of  length $L$. The wire conductance is
$\mathrm{g}_{D}=\sigma_{D}\mathcal{A}/L$, with $\mathcal{A}$ being
wire cross section. Describing diffusive electron transport across
the wire one starts from the disorder averaged generating function
$\mathcal{Z}[\chi]=\int\D[\hat Q]\exp(iS[\hat Q,A_{\chi}])$. The
action is given by~\eqref{NLSM-action-general}, while the auxiliary
vector potential $\hat{A}_{\chi}$ enters the problem through the
covariant derivative~\eqref{NLSM-covariant-deriv}. We choose
$\hat{A}_{\chi}$ to be purely quantum, without classical component,
as
\begin{equation}\label{Part-II-FCS-A}
\hat{A}_{\chi}(t)=\frac{\hat{\gamma}^{q}}{2L}\left\{\begin{array}{cc}\chi\quad
& 0<t<t_{0}
\\ 0\quad & \text{otherwise}\end{array}\right.\,.
\end{equation}
Here the quantum Keldysh matrix $\hat{\gamma}^{q}$ is given
by~\eqref{fermion-gammas} and $\chi$ is called \textit{counting
field}. The action $S[\hat{Q},A_{\chi}]$ is accompanied by the
boundary conditions on  $\hat Q(x)$ matrix at the ends of the wire:
\begin{eqnarray}\label{Part-II-Q-boundary}
\hat Q(0)=\left(\begin{array}{cc} 1\quad & 2F_{\epsilon}
\\ 0\quad & -1
\end{array}\right)\,, \qquad
\hat Q(L)=
\left(\begin{array}{cc} 1\quad & 2F_{\epsilon-eV}
\\ 0\quad & -1
\end{array}\right)\,.
\end{eqnarray}
Knowing $\mathcal{Z}[\chi]$ one can find then any moment $\langle
q^n\rangle$ of the number of electrons transferred between reservoirs during the
time of measurement $t_0$ via differentiation of $\mathcal{Z}[\chi]$
with respect to the counting field $\chi$. The irreducible
correlators are defined as $\mathcal{C}_{1}=\langle q\rangle=q_0$ and $\mathcal{C}_{n}=\langle
(q-q_{0})^n\rangle$ with $n=2,3,\ldots$, where $q=\frac{1}{e}\int_0^{t_0}I(t)dt$ and
$q_0=t_{0}\mathrm{g}_{D}V/e=t_{0}\langle I\rangle/e$, where
$\mathrm{g}_D$ is the average diffusive conductance. They may be found
through the expansion of the logarithm of $\mathcal{Z}[\chi]$ in powers
of the counting field
\begin{equation}
\label{Part-II-log-Z-chi}
\ln\mathcal{Z}[\chi]=\sum_{n=0}^{\infty}\frac{(i\chi)^n}{n!}\mathcal{C}_n\,.
\end{equation}

One calculates $\mathcal{Z}[\chi]$, by taking the action at the
saddle point $\underline{\hat{Q}}=\hat{\Lambda}_{\chi}$ which
extremizes $S[\hat Q,A_{\chi}]$. The difficulty  is that the action
$S[\hat Q,A_{\chi}]$ depends explicitly on the counting field $\chi$
and solution of the corresponding saddle point equation is  not know
for an arbitrary $A_{\chi}$. This obstacle can be overcame by
realizing that vector potential~\eqref{Part-II-FCS-A} is a pure
gauge and it can be gauged away from the action $S[\hat
Q,A_{\chi}]\to S[\hat Q_{\chi}]$ by the transformation
\begin{equation}
\hat{Q}(x\,;t,t')=\exp\big\{ix\hat{A}_{\chi}(t)\big\}\,
\hat{Q}_{\chi}(x\,;t,t')\,\exp\big\{-ix\hat{A}_{\chi}(t')\big\}\,.
\end{equation}
It comes with the price though: the boundary
conditions~\eqref{Part-II-Q-boundary} change accordingly
\begin{equation}\label{Part-II-Q-boundary-twisted}
\hat{Q}_{\chi}(0)=\hat{Q}(0), \qquad
\hat{Q}_{\chi}(L)=\exp\big(-i\chi\hat{\gamma}^{q}/2\big)\hat{Q}(L)
\exp\big(i\chi\hat{\gamma}^{q}/2\big)\,.
\end{equation}
The advantage of this transformation is that the saddle point equation
for $\hat{Q}_{\chi}$, which is nothing else but the Usadel equation~\eqref{NLSM-Usadel}
\begin{equation}
D\frac{\partial}{\partial
x}\left(\hat{Q}_{\chi}\circ\frac{\partial\hat{Q}_{\chi}}{\partial
x}\right)=0\,,
\end{equation}
can be solved explicitly now. To this end, notice that
$\hat{Q}_{\chi}\circ\partial_x\hat{Q}_{\chi}=-
\partial_x\hat{Q}_{\chi}\circ \hat{Q}_{\chi} =\hat{J}$ is a
constant, i.e. $x$--independent, matrix. Since
$\hat{Q}_{\chi}^2=\hat{1}$, $\hat J$   anti--commutes with
$\hat{Q}_{\chi}$, i.e.
$\hat{Q}_{\chi}\circ\hat{J}+\hat{J}\circ\hat{Q}_{\chi}=0$. As a
result one finds
$\hat{Q}_{\chi}(x)=\hat{Q}_{\chi}(0)\exp\big(\hat{J}x\big)$. Putting
$x=L$ and multiplying by $\hat{Q}_{\chi}(0)$ from the left, one
expresses  as--yet unknown matrix $\hat J$ through the boundary
conditions~\eqref{Part-II-Q-boundary-twisted}:
$\hat{J}=\frac{1}{L}\ln\big[\hat{Q}_{\chi}(0)\hat{Q}_{\chi}(L)\big]$.

Having determined the saddle point configuration of the
$\hat{Q}_{\chi}$ matrix, for an arbitrary choice of the counting
field $\chi$, one substitutes  it back into the action $S[\hat
Q_{\chi}]$ to find the generating function
$$\ln\mathcal{Z}[\chi]=iS[\hat{Q}_{\chi}]=-\frac{\pi\nu D}{4}\,
\Tr\{(\partial_x\hat{Q}_{\chi})^{2}\}= \frac{\pi \nu D}{4}\,
\Tr\{\hat{J}^2\}\,,$$ where we used anti--commutativity relation
$\{\hat{Q}_{\chi}(0),\hat{J}\}=0$. Calculating time integrals one
passes to the Wigner transform $\iint\d t\d t'\to
t_{0}\int\frac{\d\epsilon}{2\pi}$, where $t_{0}$ emerges from the
integral over the central time, and finds
\begin{equation}\label{Part-II-FCS-Z}
\ln\mathcal{Z}[\chi]=\frac{t_{0}\mathrm{g}_{D}}{8e^{2}}\int\d\epsilon\
\mathrm{Tr}\ln^{2}\left[\hat{Q}(0)\exp\big(-i\chi\hat{\gamma}^{q}/2\big)
\hat{Q}(L)\exp\big(i\chi\hat{\gamma}^{q}/2\big)\right]\,.
\end{equation}

In the following we analyze~\eqref{Part-II-FCS-Z} in the
zero--temperature limit, $T=0$, where
$F_{\epsilon}=\tanh(\epsilon/2T)\to \mbox{sign}(\epsilon)$. The
algebra can be further significantly shortened by performing the
rotation
$\hat{\mathcal{Q}}=\hat{\mathcal{O}}^{-1}\hat{Q}\hat{\mathcal{O}}$
with the help of the matrix
\begin{equation}
\hat{\mathcal{O}}=\frac{1}{\sqrt{2}}\left(\begin{array}{cr}1 & -1
\\ 1 & 1\end{array}\right)\,.
\end{equation}
One should note also that $\hat{\mathcal{O}}^{-1}\exp(\pm
i\chi\hat{\gamma^{q}}/2)\hat{\mathcal{O}}=\exp(\pm
i\chi\hat{\sigma}_{z}/2)$. It is not difficult to show that for
$T=0$ the only energy interval that contributes to the trace
in~\eqref{Part-II-FCS-Z} is that where $0<\epsilon<eV$. Furthermore,
at such energies rotated $\mathcal{Q}$--matrices are energy
independent and given by
\begin{equation}
\hat{\mathcal{Q}}(0)=\left(\begin{array}{cc}-1 & -2 \\ \phantom{-}0
& \phantom{-}1\end{array}\right)\,,\qquad
\hat{\mathcal{Q}}(L)=\left(\begin{array}{cc}\phantom{-}1 &
\phantom{-}0
\\ -2 & -1\end{array}\right)\,.
\end{equation}
As a result, the $\epsilon$ integration in Eq.~\eqref{Part-II-FCS-Z}
gives  a factor $eV$ and inserting $\hat{\mathcal{Q}}$ into
$\ln\mathcal{Z}[\chi]$ the latter reduces to
\begin{equation}\label{Part-II-FCS-Z-T-0}
\ln\mathcal{Z}[\chi]=\frac{t_{0}\mathrm{g}_{D}V}{8e}\,\Tr
\ln^{2}\left(\begin{array}{cc} -1+4e^{i\chi} & \phantom{-}2 \\
-2e^{i\chi} & -1
\end{array}\right)\,.
\end{equation}
Since the trace is invariant with respect to the choice of the
basis, it is convenient to evaluate it in the basis where matrix
under the logarithm in~\eqref{Part-II-FCS-Z-T-0} is diagonal.
Solving the eigenvalue problem and calculating the trace, as the
final result one finds
\begin{equation}
\ln\mathcal{Z}[\chi]=\frac{t_{0}\mathrm{g}_{D}V}{4e}
\ln^{2}\left[p_{\chi}+\sqrt{p^{2}_{\chi}-1}\right]\,,\quad
p_{\chi}=2e^{i\chi}-1\,.
\end{equation}
Knowing $\ln \mathcal{Z}[\chi]$ one can extract now all the
cummulants of interest by expanding in powers of $\chi$ and
employing~(\ref{Part-II-log-Z-chi}), for example,
$\mathcal{C}_{1}=q_{0}$, $\mathcal{C}_{2}=q_0/3$,
$\mathcal{C}_{3}=q_0/15$, {\em etc}. For a review devoted to FCS
see~\cite{Levitov-FCS}.

\section{Interactions and kinetic equation for
fermions}\label{sec_int_ferm}
\subsection{Interactions}\label{sec_int_ferm-1}

Consider a liquid of electrons that interact through the
instantaneous density--density interactions
$$\hat{H}_{\mathrm{int}}=-{1\over 2} \iint\d\mathbf{r}\d\mathbf{r}'
:\hat\varrho(\mathbf{r}) U_{0}(\mathbf{r}-\mathbf{r}')
\hat{\varrho}(\mathbf{r}'):\,,$$ where $\hat{\varrho}(\mathbf{r})
=\hat{\psi}^\dagger(\mathbf{r})\hat{\psi}(\mathbf{r})$ is the local
density operator, $U_{0}(\mathbf{r}-\mathbf{r}')$ is the bare
Coulomb interaction potential and $:\ldots:$ stands for normal
ordering. The corresponding Keldysh contour action has the form
\begin{equation}\label{int-ferm-S-int}
S_{\mathrm{int}}[\bar{\psi},\psi]=-{1\over 2}\int_{\cal C} \d t
\iint\d\mathbf{r}\d\mathbf{r}'\,\bar{\psi}(\mathbf{r},t)
\bar{\psi}(\mathbf{r}',t)U_{0}(\mathbf{r}-\mathbf{r}')
\psi(\mathbf{r}',t)\psi(\mathbf{r},t)\,.
\end{equation}
One may now perform the Hubbard--Stratonovich transformation with
the help of a real boson field $\phi(\mathbf{r},t)$, defined along
the contour, to decouple the interaction term
\begin{eqnarray}\label{int-ferm-HS}
\exp\big(iS_{\mathrm{int}}[\bar{\psi},\psi]\big)=\int\D[\phi]\!\!\!\!\!\!\!\!&&\!\!
\exp\left(\frac{i}{2}\int_{\mathcal{C}}\d t
\iint\d\mathbf{r}\d\mathbf{r}'\phi\rt\,
U^{-1}_{0}(\mathbf{r}-\mathbf{r}')\,\phi(\mathbf{r}',t)\right)\nonumber \\
\times \!\!&&\!\!\!\!\!\!\! \exp\left(i\int_{\mathcal{C}}\d
t\int\d\mathbf{r}\ \phi\rt \bar{\psi}\rt\psi\rt \right)\, ,
\end{eqnarray}
where $U^{-1}_{0}$ is an inverse interaction kernel, i.e.
$\int\d\mathbf{r}'' U_{0}(\mathbf{r}-\mathbf{r}'')
U^{-1}_{0}(\mathbf{r}''-\mathbf{r}')=\delta(\mathbf{r}-\mathbf{r}')$.
One may notice that the auxiliary bosonic field, $\phi\rt$, enters
the fermionic action in exactly the same manner as a scalar source
field~\eqref{fermion-SV}. Following~\eqref{fermion-SV-1}, one
introduces $\phi^{cl(q)}\equiv(\phi_{+}\pm\phi_{-})/2$ and rewrites
the fermion--boson interaction term as
$\bar\psi_{a}\phi^{\alpha}\hat \gamma^\alpha_{ab}\psi_b\,$, where
summations over $a,b=(1,2)$ and $\alpha=(cl,q)$ are assumed and
gamma matrices $\hat{\gamma}^{\alpha}$ are defined
by~\eqref{fermion-gammas}. The free bosonic term takes the form
${1\over 2} \phi U^{-1}_{0}\phi\to \phi^\alpha U^{-1}\hat
\sigma^{\alpha\beta}_{x}\phi^\beta$. Following~\eqref{int-ferm-HS}
one may integrate fermions explicitly to obtain the partition
function for the interacting disordered electron liquid
\begin{eqnarray}\label{int-ferm-NLSM-action}
&&\mathcal{Z}=\int\D[\Phi]\,\exp
\left(i\Tr\{\vec{\Phi}^{T}U^{-1}_{0}\hat{\sigma}_{x}\vec{\Phi}\}\right)
\int\D[\hat Q]\,\exp\left(iS[\hat Q,\Phi]\right)\,,\nonumber\\
&&iS[\hat Q,\Phi]=-\frac{\pi\nu}{4\tauel}\Tr\{\hat{Q}^{2}\}+
\Tr\ln\left[\hat{G}^{-1}+\frac{i}{2\tauel}\hat{Q}+
\hat{\Phi}+\mathbf{v}_{F}\hat{\mathbf{A}}\right]\,,
\end{eqnarray}
where we introduced doublet $\vec{\Phi}^{T}=(\phi^{cl},\phi^{q})$
and matrix $\hat{\Phi}=\phi^{\alpha}\hat{\gamma}^{\alpha}$. This
should be compared to the non--interacting version of the action
given by~\eqref{NLSM-action-general}. An extra complication, which
stems from interactions, is an additional functional integral over
the dynamic bosonic field $\hat{\Phi}$
entering~\eqref{int-ferm-NLSM-action}.

Varying the action in~\eqref{int-ferm-NLSM-action} over the $\hat Q$
matrix $\delta S[\hat Q,\Phi]/\delta \hat{Q}=0$, at zero external
vector potential $\hat{\mathbf{A}}=0$, one obtains the following
equation for the saddle point matrix
$\underline{\hat{Q}}=\underline{\hat{Q}}[\Phi]$:
\begin{equation}\label{int-ferm-NLSM-saddle-point-eq}
\underline{\hat{Q}}_{tt'}(\mathbf{r})= \frac{i}{\pi\nu}
\left(\hat{G}^{-1}+\frac{i}{2\tauel}\underline{\hat{Q}}+
\hat{\Phi}\right)^{-1}_{tt',\mathbf{r}\mathbf{r}} \,,
\end{equation}
which is a generalization of~\eqref{NLSM-saddle-point-eq} for the
interacting case. Our strategy will be to find a stationary solution
of~\eqref{int-ferm-NLSM-saddle-point-eq} for a given realization of
the fluctuating bosonic field $\hat{\Phi}$, and then consider
space-- and time--dependent deviations from such a solution.

The conceptual  problem here is that the saddle point Equation
(\ref{int-ferm-NLSM-saddle-point-eq}) can not be solved exactly for
an arbitrary $\hat{\Phi}\rt$. Note, however, that
\eqref{int-ferm-NLSM-saddle-point-eq} can be solved for a particular
case of spatially uniform realization of the boson field,
$\hat{\Phi}=\hat{\Phi}(t)$. This is achieved with the help of the
gauge transformation of the non--interacting saddle point
\begin{equation}\label{int-ferm-NLSM-Q-gauge}
\underline{\hat{Q}}_{tt'}[\Phi(t)]=\exp\left(i\int^{\,t}\d
t\,\hat{\Phi}(t)\right)
\hat{\Lambda}_{t-t'}\,\exp\left(-i\int^{\,t'}\d
t\,\hat{\Phi}(t)\right)\,.
\end{equation}
The validity of this solution can be verified by acting with the
operator $\hat{G}^{-1}+i/(2\tauel)\underline{\hat{Q}}+\hat{\Phi}$ on
both sides of~\eqref{int-ferm-NLSM-saddle-point-eq}, and utilizing
the fact that $\hat{\Lambda}_{t-t'}$
solves~\eqref{int-ferm-NLSM-saddle-point-eq} with $\hat{\Phi}=0$. We
also rely on the commutativity of the vertex matrices
$[\hat{\gamma}^{cl},\hat{\gamma}^{q}]=0$, in writing the solution in
the form of~\eqref{int-ferm-NLSM-Q-gauge}. This example shows that a
properly chosen gauge may considerably simplify the task of finding
the saddle point and performing perturbative  expansion around it.
We show in the following that there is a particularly convenient
gauge (the $\EuScript{K}$ gauge) suited for calculations of
interaction effects.

\subsection{$\EuScript{K}$ Gauge }\label{sec_int_ferm-2}

Let us perform a gauge transformation from  the old $\hat Q$ matrix
to a new one, which we call $\Qk$ matrix. It is defined as
\begin{equation}\label{int-ferm-NLSM-Q-K-gauge}
\hat{Q}_{\EuScript{K}}(\mathbf{r};t,t')=\exp\left(-i\hat{\EuScript{K}}\rt\right)
\,\hat{Q}_{tt'}(\mathbf{r})\,
\exp\left(i\hat{\EuScript{K}}\rtp\right),
\end{equation}
where the matrix
$\hat{\EuScript{K}}\rt=\EuScript{K}^{\alpha}\rt\hat{\gamma}^{\alpha}$
is defined through  two scalar fields $\EuScript{K}^{\alpha}\rt$
with $\alpha=(cl,q)$, which are specified below. Substituting $\hat
Q =e^{i \hat{\EuScript{K}}} \hat{Q}_{\EuScript{K}}
e^{-i\hat{\EuScript{K}}}$ into the
action~\eqref{int-ferm-NLSM-action} and using the invariance of the
trace under a cyclic permutations, we can rewrite the action
as~\footnote{Deriving~\eqref{int-ferm-NLSM-action-K-gauge} one uses
obvious equality between the traces
$\Tr\{\Qk^{2}\}=\Tr\{\hat{Q}^{2}\}$. As to the logarithm term, one
writes $\Tr\left\{e^{-i\hat{\EuScript{K}}}
\ln\left[\hat{G}^{-1}+\hat{\Phi}+\mathbf{v}_{F}\hat{\mathbf{A}}+\frac{i}{2\tauel}
e^{i\hat{\EuScript{K}}}\Qk
e^{-i\hat{\EuScript{K}}}\right]e^{i\hat{\EuScript{K}}}\right\}= \Tr
\ln\left[e^{-i\hat{\EuScript{K}}}\hat{G}^{-1}e^{i\hat{\EuScript{K}}}+
\hat{\Phi}+\mathbf{v}_{F}\hat{\mathbf{A}}+\frac{i}{2\tauel}
\Qk\right]$, where familiar algebraic identity
$\Tr\{\hat{L}f(\hat{A})\hat{L}\}=\Tr\{
f(\hat{L}\hat{A}\hat{L}^{-1})\}$ was used, which holds for any
analytic function $f$ of  matrix $\hat{A}$. Finally, one rewrites
$e^{-i\hat{\EuScript{K}}}\hat{G}^{-1}e^{i\hat{\EuScript{K}}}=\hat{G}^{-1}+
e^{-i\hat{\EuScript{K}}}[\hat{G}^{-1},e^{i\hat{\EuScript{K}}}]$ and
calculates the commutator $[\hat{G}^{-1},e^{i\hat{\EuScript{K}}}]=
e^{i\hat{\EuScript{K}}} \left(-\partial_{t}
\hat{\EuScript{K}}-\mathbf{v}_{F}\partial_{\mathbf{r}}\hat{\EuScript{K}}
-\frac{1}{2m}
(\partial_{\mathbf{r}}\hat{\EuScript{K}})^{2}\right)$.}
\begin{equation}\label{int-ferm-NLSM-action-K-gauge}
iS[\hat
Q_{\EuScript{K}},\Phi]=-\frac{\pi\nu}{4\tauel}\Tr\big\{\Qk^{2}\big\}+
\Tr\ln\left[\hat{G}^{-1}+\hat{C}+\frac{i}{2\tauel}\Qk-\frac{1}{2m}(\partial_{\mathbf{r}}
\hat{\EuScript{K}})^{2}\right]\,,
\end{equation}
where we have introduced the notation $\hat{C}\rt = \hat{\Phi}_{\EuScript{K}}\rt+
\mathbf{v}_{F}\hat{\mathbf{A}}_{\EuScript{K}}\rt $ along with the gauge transformed
electromagnetic potentials
\begin{equation}\label{int-ferm-NLSM-C}
\hat{\Phi}_{\EuScript{K}}\rt=\hat{\Phi}\rt-\partial_{t}\hat{\EuScript{K}}\rt\,,\quad
\hat{\mathbf{A}}_{\EuScript{K}}\rt=
\hat{\mathbf{A}}\rt-\partial_{\mathbf{r}}\hat{\EuScript{K}}\rt\,.
\end{equation}

We assume now that the saddle point of the new field $\Qk$ is close
to the non--interacting saddle point $\hat \Lambda$,
(\ref{NLSM-Lambda}), and use the freedom of choosing  two fields
$\hat{\EuScript{K}}^\alpha$  to enforce it. To this end, we
substitute $\Qk=\hat{\Lambda}+\delta\Qk$
into~\eqref{int-ferm-NLSM-action-K-gauge} and expand it in powers of
the deviation $\delta\Qk$ as well as the electromagnetic potentials,
encapsulated in $\hat C$. The first non--trivial term of such an
expansion is
\begin{equation}\label{int-ferm-NLSM-action-Q-linear}
iS[\delta \hat Q_{\EuScript{K}},\Phi]= -\frac{i}{2\tauel}
\Tr\big\{\hat{\mathcal{G}}\hat{C}\hat{\mathcal{G}}\delta\Qk\big\}+\ldots\,,
\end{equation}
where we have employed the fact that $\hat{\Lambda}$ is the saddle
point of the non--interacting model and, thus, in the absence of the
electromagnetic potentials, there are no linear terms in deviations
$\delta\Qk$. We have also neglected the diamagnetic
$(\partial_{\mathbf{r}}\hat{\EuScript{K}})^{2}/2m$ term, since it is
quadratic in $\hat{\EuScript{K}}$, and hence (as shown below) in
$\hat{\Phi}$.

We now demand that this linear in $\delta \hat Q_{\EuScript{K},tt'}(\mathbf{r})$ term vanishes.
Performing the Fourier transform, one notices that this takes place for an arbitrary
$\delta \hat Q_{\EuScript{K},\epsilon_- \epsilon_+}({\bf q})$,
if the following matrix identity holds for any $\epsilon$, $\omega$ and $\mathbf{q}$
\begin{equation}\label{int-ferm-NLSM-Phi-K}
\sum_{p}\hat{\mathcal{G}}(\mathbf{p}_{+},\epsilon_{+})
\hat{C}(\mathbf{q},\omega)
\hat{\mathcal{G}}(\mathbf{p}_{-},\epsilon_{-})=0\,,
\end{equation}
where $\mathbf{p}_{\pm}=\mathbf{p}\pm\mathbf{q}/2\,$ and
$\epsilon_{\pm}=\epsilon\pm\omega/2$.
Condition~\eqref{int-ferm-NLSM-Phi-K} represents matrix equation,
which expresses as--yet unspecified gauge fields
$\EuScript{K}^{\alpha}$ through $\Phi^{\alpha}$ and ${\bf
A}^\alpha$. Employing~\eqref{NLSM-G-impurity-dressed}, and the
following identities
\begin{subequations}\label{int-ferm-NLSM-integrals}
\begin{equation}
\hskip-1.2cm
\sum_{p}\mathcal{G}^{R}(\mathbf{p}_{\pm},\epsilon_{\pm})
\mathcal{G}^{A}(\mathbf{p}_{\mp},\epsilon_{\mp})\approx2\pi\nu\tauel\,,
\end{equation}
\begin{equation}
\!\!\! \sum_{p}\mathbf{v}_{F}\,
\mathcal{G}^{R}(\mathbf{p}_{\pm},\epsilon_{\pm})
\mathcal{G}^{A}(\mathbf{p}_{\mp},\epsilon_{\mp})\approx\mp2\pi
i\nu\tauel D \mathbf{q}\,,
\end{equation}
\end{subequations}
one may transform~\eqref{int-ferm-NLSM-Phi-K} into
\begin{equation}\label{int-ferm-NLSM-Phi-K-1}
\frac{1}{\pi\nu\tauel}
\sum_{p}\hat{\mathcal{G}}(\mathbf{p}_{+},\epsilon_{+})
\hat{C}(\mathbf{q},\omega)
\hat{\mathcal{G}}(\mathbf{p}_{-},\epsilon_{-})=
\big(\hat{\gamma}^{\alpha}-\hat{\Lambda}_{\epsilon_{+}}\hat{\gamma}^{\alpha}
\hat{\Lambda}_{\epsilon_{-}}\big)\Phi^{\alpha}_{\EuScript{K}}-
\big(\hat{\Lambda}_{\epsilon_{+}}\hat{\gamma}^{\alpha}-
\hat{\gamma}^{\alpha}\hat{\Lambda}_{\epsilon_{-}}\big)
D\,\mathrm{div}\mathbf{A}_{\EuScript{K}}^{\alpha}=0\,.
\end{equation}
It is in general impossible to satisfy this condition for any
$\epsilon$ and $\omega$ by a choice of two fields
$\EuScript{K}^\alpha(\mathbf{r},\omega)$. In  thermal
equilibrium, however, there is a ``magic'' fact that
\begin{equation}\label{int-ferm-F-B}
\frac{1-F_{\epsilon_{+}}F_{\epsilon_{-}}}
{F_{\epsilon_{+}}-F_{\epsilon_{-}}}=\coth\frac{\omega}{2T}\equiv
B_{\omega}\, ,
\end{equation}
which depends on $\omega$ only, but {\em not} on $\epsilon$. This
allows for the condition (\ref{int-ferm-NLSM-Phi-K-1}) to be
satisfied if the following vector relation between the gauge
transformed potentials~(\ref{int-ferm-NLSM-C}) holds:
\begin{equation}\label{int-ferm-NLSM-K-functional}
\vec{\Phi}_{\EuScript{K}}(\mathbf{r},\omega)=
\left(\begin{array}{cc} 1& 2B_{\omega} \\ 0 &
-1\end{array}\right)D\,\mathrm{div}
\vec{\mathbf{A}}_{\EuScript{K}}(\mathbf{r},\omega)\, .
\end{equation}
This equation specifies the $\EuScript{K}$--gauge for both classical
and quantum components of the electromagnetic potentials.

The advantage of the $\EuScript{K}$ gauge is that the action does
not contain  terms linear in the deviations of the $\Qk$ matrix from
its saddle point  $\hat \Lambda$ {\em and linear} in the
electromagnetic potentials. Note that there are still terms which
are linear in $\delta \Qk$ and quadratic in electromagnetic
potentials. This means that, strictly speaking, $\hat \Lambda$ is
not the exact saddle point on the $\Qk$ manifold for any realization
of the electromagnetic potentials. However, since the deviations
from the true saddle point are pushed to the second order in
potentials, the $\EuScript{K}$ gauge substantially simplifies the
structure of the perturbation theory. Moreover, this state of
affairs holds only in equilibrium. For out--of--equilibrium
situations condition (\ref{int-ferm-NLSM-Phi-K-1}) cannot be
identically satisfied and terms linear in $\delta\Qk$ and
electromagnetic fields appear in the action. As we explain below, it
is precisely these terms which are responsible for the collision
integral in the kinetic equation. Still the $\EuScript{K}$ gauge is
a useful concept in the out--of--equilibrium context as well. In
such a case one should define the bosonic distribution function
$B_{\omega}$ in~(\ref{int-ferm-NLSM-K-functional}) as
\begin{equation}\label{int-ferm-NLSM-bosonic-distribution}
B_{\omega}(\mathbf{r},\tau)=\frac{1}{2\omega}\int^{+\infty}_{-\infty}
\d\epsilon\,\left[1-F_{\epsilon+\omega/2}(\mathbf{r},\tau)
F_{\epsilon-\omega/2}(\mathbf{r},\tau)\right]\,,
\end{equation}
where $F_{\epsilon}(\mathbf{r},\tau)$ is WT of the fermionic matrix
$F_{t,t'}(\mathbf{r})$.

With the help of~(\ref{int-ferm-NLSM-C}) the definition of the
$\EuScript{K}$ gauge~\eqref{int-ferm-NLSM-K-functional} may be
viewed as an explicit relation determining the gauge fields
$\EuScript{K}^\alpha$ through the electromagnetic potentials
$\Phi^\alpha$ and ${\bf A}^\alpha$. Taking $\hat{\mathbf{A}}=0$ for
simplicity, one finds for the quantum and classical components of
the gauge field
\begin{subequations}\label{int-ferm-NLSM-K-q-cl}
\begin{equation}
\hskip-3cm
(D\partial^{2}_{\mathbf{r}}-i\omega)\EuScript{K}^{q}(\mathbf{r},\omega)=
\Phi^{q}(\mathbf{r},\omega)\,,
\end{equation}
\begin{equation}
(D\partial^{2}_{\mathbf{r}}+i\omega)\EuScript{K}^{cl}(\mathbf{r},\omega)+
2B_{\omega}D\partial^{2}_{\mathbf{r}}\EuScript{K}^{q}(\mathbf{r},\omega)
=-\Phi^{cl}(\mathbf{r},\omega)\,.
\end{equation}
\end{subequations}
In  general case it is convenient to cast
these relations  into the matrix form
\begin{equation}\label{int-ferm-NLSM-Phi-K-matrix}
\vec{\EuScript{K}}\qo =  \hat{\mathcal{D}}^{-1}\qo \Big(
\hat{\EuScript{B}}^{-1}_{\omega}\vec{\Phi}\qo-
D\,\hat{\sigma}_{x}\,\mathbf{q}\cdot\vec{\mathbf{A}}\qo \Big)\,,
\end{equation}
with the vector
$\vec{\EuScript{K}}^{T}=(\EuScript{K}^{cl},\EuScript{K}^{q})$. Here
we have introduced diffuson bosonic matrix propagator
\begin{equation}\label{int-ferm-NLSM-D}
\hat{\mathcal{D}}\qo=\left(\begin{array}{cc}\mathcal{D}^{K}\qo &
\mathcal{D}^{R}\qo \\ \mathcal{D}^{A}\qo & 0\end{array}\!\right),\,
\end{equation}
having matrix components
\begin{equation}\label{int-ferm-NLSM-D-RAK}
\mathcal{D}^{R(A)}\qo=\big(Dq^{2}\mp i\omega\big)^{-1},\quad
\mathcal{D}^{K}\qo=B_{\omega}\big[\mathcal{D}^{R}\qo-\mathcal{D}^{A}\qo\big]\,,
\end{equation}
and
\begin{equation}\label{int-ferm-NLSM-B}
\hat{\EuScript{B}}_{\omega}=\left(\begin{array}{lr}2B_{\omega} &
1^{R}_{\omega} \\ -1^{A}_{\omega} & 0\end{array}\right)\,.
\end{equation}

Equation \eqref{int-ferm-NLSM-Phi-K-matrix} provides an explicit
{\em linear} relation between the gauge fields $\EuScript{K}^\alpha$
and the electromagnetic potentials. It thus gives an explicit
definition of the gauge transformed field $\Qk$,
cf.~(\ref{int-ferm-NLSM-action-K-gauge}). The latter has the saddle
point which is rather close to the non--interacting saddle point
$\hat \Lambda$ (with deviations being quadratic in electromagnetic
fields). Returning to the original gauge, one realizes that the
following $\hat Q$ matrix
\begin{equation}\label{int-ferm-NLSM-Q-gauge-general}
\underline{\hat{Q}}_{tt'}(\mathbf{r})=
\exp\left(i\EuScript{K}^{\alpha}\rt\hat{\gamma}^{\alpha}\phantom{^{'}}\right)
\hat{\Lambda}_{t-t'}\,
\exp\left(-i\EuScript{K}^{\beta}\rtp\hat{\gamma}^{\beta}\right),
\end{equation}
provides a  good approximation for the solution of the generic
saddle point Equation (\ref{int-ferm-NLSM-saddle-point-eq}) for any
given realization of the fluctuating potentials. This statement
holds only for the equilibrium conditions. Away from equilibrium,
$\hat{\Phi}\delta\Qk$ terms reappear and have to be taken into the
account to obtain the proper form of the kinetic equation (see
further discussions in  Section~\ref{sec_int_ferm-5}). In addition,
terms  $\sim\hat{\Phi}^{2}\delta\Qk$ exist even in equilibrium. They
lead to  interaction corrections to the transport coefficients
(details are given in  Section~\ref{app_Part-III}).

\subsection{Non--linear $\sigma$--model  for interacting systems}\label{sec_int_ferm-3}

Performing gradient expansion for the trace of the logarithm term
in~\eqref{int-ferm-NLSM-action-K-gauge} (this procedure is closely
analogous to that presented in Section~\ref{sec_NLSM-2}),
 one obtains an
effective action written in terms of $\Qk$ matrix field and
electromagnetic potentials in the $\EuScript{K}$ gauge
\begin{eqnarray}\label{int-ferm-action-interacting}
iS[\hat Q_{\EuScript{K}},\Phi]=\frac{i\nu}{2}
\Tr\left\{\hat{\Phi}_{\EuScript{K}}
\hat{\sigma}_{x}\hat{\Phi}_{\EuScript{K}}\right\}-\frac{\pi\nu}{4}
\Tr\left\{D(\hat{\bm{\partial}}_{\mathbf{r}}\Qk)^{2}-
4\partial_{t}\Qk+4i\hat{\Phi}_{\EuScript{K}}\Qk\right\}\,,
\end{eqnarray}
where
\begin{eqnarray}\label{int-ferm-covariant-derivative}
\hat{\bm{\partial}}_{\mathbf{r}}\Qk=\partial_{\mathbf{r}}\Qk-
i\big[\hat{\mathbf{A}}_{\EuScript{K}},\Qk\big]\,.
\end{eqnarray}
Equation \eqref{int-ferm-action-interacting}, together with the
saddle point
condition~\eqref{int-ferm-NLSM-Phi-K-matrix}--\eqref{int-ferm-NLSM-B},
generalizes  the effective $\sigma$--model
action~\eqref{NLSM-action-noninteracting}  to include Coulomb
interaction effects. Employing the explicit form of the long
covariant derivative~\eqref{int-ferm-covariant-derivative}, and the
relation between the $\hat{\EuScript{K}}$ and $\hat{\Phi}$ fields at
$\hat{\mathbf{A}}=0$ (see~\eqref{int-ferm-NLSM-K-q-cl}), one finds
for the partition function
\begin{equation}\label{int-ferm-Z}
\mathcal{Z}=\int\D[\Phi]
\exp\left(i\Tr\{\vec{\Phi}^{T}\hat{U}^{-1}_{RPA}\vec{\Phi}\}\right)
\int\D[\hat Q_{\EuScript{K}}] \exp\left(iS_{0}[\hat
Q_{\EuScript{K}}]+ iS_{1}[\hat
Q_{\EuScript{K}},\partial_{\mathbf{r}}\EuScript{K}]+ iS_{2}[\hat
Q_{\EuScript{K}},\partial_{\mathbf{r}}\EuScript{K}]\right),
\end{equation}
where $S_{l}$, with $l=0,1,2$  contain  the $l$-th power of the
electromagnetic potentials and are given
by
\begin{subequations}\label{int-ferm-S-012}
\begin{equation}
\hskip-3cm iS_{0}[\hat Q_{\EuScript{K}}]=-\frac{\pi\nu}{4}
\Tr\left\{D(\partial_{\mathbf{r}}\Qk)^{2}-4i\partial_{t}\Qk\right\}\,,
\end{equation}
\begin{equation}
\hskip-1cm
iS_{1}[\hat Q_{\EuScript{K}},\partial_{\mathbf{r}}\EuScript{K}]=
-i\pi\nu\Tr\left\{D(\partial_{\mathbf{r}}\hat{\EuScript{K}})\Qk(\partial_{\mathbf{r}}\Qk)+
\hat{\Phi}_{\EuScript{K}}\Qk\right\}\,,
\end{equation}
\begin{equation}
iS_{2}[\hat Q_{\EuScript{K}},\partial_{\mathbf{r}}\EuScript{K}]=\frac{\pi\nu
D}{2}\Tr\left\{(\partial_{\mathbf{r}}\hat{\EuScript{K}})\Qk
(\partial_{\mathbf{r}}\hat{\EuScript{K}})\Qk-
(\partial_{\mathbf{r}}\hat{\EuScript{K}})\hat{\Lambda}
(\partial_{\mathbf{r}}\hat{\EuScript{K}})\hat{\Lambda}\right\}\,.
\end{equation}
\end{subequations}
The effective interaction matrix $\hat{U}_{RPA}$ is nothing but the
screened interaction in the random--phase approximation (RPA)
\begin{equation}\label{int-ferm-U-RPA}
\hat{U}_{RPA}\qo=\big[U^{-1}_{0}\hat{\sigma}_{x}+\hat{\Pi}\qo\big]^{-1}\,,
\end{equation}
where $\hat{\Pi}\qo$ is the  density--density correlator. According
to~(\ref{fermion-Pi-matrix-def}) and (\ref{Part-II-Pi-diffusive}) it
has a typical form of a bosonic propagator in the Keldysh space
\begin{equation}\label{int-ferm-Pi-matrix}
\hat{\Pi}\qo=\left(\begin{array}{cc}0& \Pi^{A}\qo \\
\Pi^{R}\qo & \Pi^{K}\qo\end{array}\right)\,,
\end{equation}
with the components
\begin{equation}\label{int-ferm-Pi-RAK}
\Pi^{R(A)}\qo=\frac{\nu Dq^{2}}{Dq^{2}\mp i\omega}\,,\,
\qquad\Pi^{K}\qo=B_{\omega}\big[\Pi^{R}\qo-\Pi^{A}\qo\big]\,.
\end{equation}
To derive~\eqref{int-ferm-Z}--\eqref{int-ferm-Pi-RAK} one has to add
and subtract the term
$\Tr\big\{(\partial_{\mathbf{r}}\hat{\EuScript{K}})\hat{\Lambda}
(\partial_{\mathbf{r}}\hat{\EuScript{K}})\hat{\Lambda}\big\}$, and
employ the equation
\begin{equation}\label{int-ferm-integral-1}
\int^{+\infty}_{-\infty}\d\epsilon\,\Tr\left\{\hat{\gamma}^{\alpha}
\hat{\gamma}^{\beta}-
\hat{\gamma}^{\alpha}\hat{\Lambda}_{\epsilon_{+}}
\hat{\gamma}^{\beta}\hat{\Lambda}_{\epsilon_{-}}\right\}=
4\omega\big(\hat{\EuScript{B}}^{-1}_{\omega}\big)^{\alpha\beta}\,,
\end{equation}
where $\epsilon_{\pm}=\epsilon\pm\omega/2$, and matrices
$\hat{\Lambda}$ and $\hat{\EuScript{B}}$ are defined
by~\eqref{NLSM-Lambda} and \eqref{int-ferm-NLSM-B} correspondingly.
Equation \eqref{int-ferm-integral-1} is a consequence of the
following integral relations between bosonic and fermionic
distribution functions
\begin{equation}\label{int-ferm-integral-2}
\int^{+\infty}_{-\infty}\d\epsilon\,\big(F_{\epsilon_{+}}-F_{\epsilon_{-}}\big)=2\omega\,,\qquad
\int^{+\infty}_{-\infty}\d\epsilon\,\big(1-F_{\epsilon_{+}}F_{\epsilon_{-}}\big)=2\omega
B_{\omega}\,.
\end{equation}

Equations \eqref{int-ferm-Z}--\eqref{int-ferm-Pi-RAK} constitute an
effective non--linear $\sigma$--model for interacting disordered
Fermi liquid. The model consists of two interacting fields: the
matrix field $\Qk$, obeying non--linear constraint
$\Qk^{2}=\hat{1}$, and the bosonic longitudinal  field
$\partial_{\mathbf{r}}\hat{\EuScript{K}}$ (or equivalently
$\hat{\Phi}$). As will be apparent later, $\Qk$ field describes
fluctuations of the quasi--particle distribution function, whereas
$\hat{\Phi}$ (or $\hat{\EuScript{K}}$) represents propagation of
electromagnetic modes through the media.

\subsection{Interaction propagators}\label{sec_int_ferm-4}

\begin{figure}
\begin{center}\includegraphics[width=10cm]{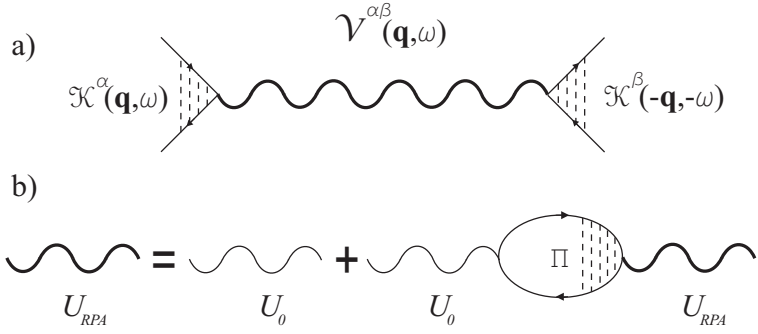}\end{center}
\caption{ a) Diagrammatic representation of the gauge field
propagator $\hat{\mathcal{V}}\qo$: wavy line represents Coulomb
interaction. Vertices dressed by the diffusons are shown by the
ladders of dashed lines. b) Screened Coulomb interaction in  RPA,
$\hat{U}_{RPA}\qo$. Bold and thin wavy lines represent    screened
and bare interactions correspondingly, the loop represents
polarization operator dressed by the diffusion ladder.
\label{Fig-K-corr-fun}}
\end{figure}

For  future applications we introduce correlation function
\begin{equation}\label{int-ferm-V-def}
\mathcal{V}^{\alpha\beta}(\mathbf{r}-\mathbf{r}',t-t')\! = -2i
\big\langle\EuScript{K}^{\alpha}(\mathbf{r},t)\EuScript{K}^{\beta}(\mathbf{r}',t')\big\rangle\!
= -2i\!\int\! \D[\hat \Phi]\,\EuScript{K}^{\alpha}\rt
\EuScript{K}^{\beta}(\mathbf{r}',t')\,
\exp\left(i\Tr\{\vec{\Phi}^{T}\hat{U}^{-1}_{RPA}\vec{\Phi}\}\right)\,,
\end{equation}
where factor $-2i$ is used for convenience. Since $\hat{\Phi}$ and
$\hat{\EuScript{K}}$ are linearly related
through~\eqref{int-ferm-NLSM-Phi-K-matrix},  one may evaluate this
Gaussian integral and find for the gauge field correlation function
\begin{equation} \label{int-ferm-K-corr-fun}
\hat{\mathcal{V}}\qo=\hat{\mathcal{D}}\qo\hat{\EuScript{B}}^{-1}_{\omega}\hat{U}_{RPA}\qo
\big(\hat{\EuScript{B}}^{-1}_{-\omega}\big)^{T}\hat{\mathcal{D}}^{T}\qom\,.
\end{equation}
The bosonic correlation matrix $\hat{\mathcal{V}}\qo$ has the
standard Keldysh structure
\begin{equation}\label{int-ferm-V-matrix}
\hat{\mathcal{V}}\qo=\left(\begin{array}{cc}\mathcal{V}^{K}\qo &
\mathcal{V}^{R}\qo \\ \mathcal{V}^{A}\qo & 0\end{array}\right),
\end{equation}
with the elements
\begin{subequations}\label{int-ferm-V-RAK}
\begin{equation}
\mathcal{V}^{R(A)}\qo=-\frac{1}{(Dq^{2}\mp
i\omega)^{2}}\left(U^{-1}_{0}+\frac{\nu Dq^{2}}{Dq^{2}\mp
i\omega}\right)^{-1}\,,
\end{equation}
\begin{equation}
\hskip-1.5cm \mathcal{V}^{K}\qo=B_{\omega}
\big[\mathcal{V}^{R}\qo-\mathcal{V}^{A}\qo\big]\,.
\end{equation}
\end{subequations}
This propagator  corresponds to the screened dynamic Coulomb
interaction, dressed by the two diffusons at the vertices,
Figure~\ref{Fig-K-corr-fun}a. Thus, the role of the gauge field
$\EuScript{K}$ is to take into account automatically both the
RPA--screened interactions, Figure~\ref{Fig-K-corr-fun}b, and its
vertex renormalization by the diffusons. Owing to the linear
dependence between $\hat{\Phi}$ and $\hat{\EuScript{K}}$,
(see~\eqref{int-ferm-NLSM-Phi-K-matrix}), we use averaging over
$\hat{\Phi}$ or $\hat{\EuScript{K}}$ fields interchangeably. The
essence is that the correlator of two $\hat{\EuScript{K}}^\alpha$
fields is given by~(\ref{int-ferm-V-def})--(\ref{int-ferm-V-RAK}).

\subsection{Kinetic equation}\label{sec_int_ferm-5}

The aim of this section is to show how the kinetic equation for the
distribution function $F$ appears naturally in the framework of the
Keldysh formulation. In  Section~\ref{sec_NLSM-4} it was
demonstrated that the kinetic equation for non--interacting fermions
is nothing but the saddle point equation for the effective action of
the $\hat Q$ matrix. In the case of interacting electrons it is
obtained from the action $S[\hat Q_{\EuScript{K}},\Phi]$,
(see~\eqref{int-ferm-action-interacting}), by first integrating out
fast degrees of freedom: diffusive, $\hat{\mathcal{W}}$, and
electromagnetic, $\hat{\EuScript{K}}$ (or, equivalently,
$\hat{\Phi}$).

Let us outline the logic of the entire procedure, which leads from
the partition function~\eqref{int-ferm-Z} and \eqref{int-ferm-S-012}
to the kinetic equation. As the first step we separate slow and fast
degrees of freedom in the action $S_{l}[\hat
Q_{\EuScript{K}},\partial_{\mathbf{r}}\EuScript{K}]$, where
$l=0,1,2$ (see~\eqref{int-ferm-S-012}). The former are encoded in
the distribution function $F_{tt'}(\mathbf{r})$, while the latter
are carried by diffusons $\hat{\mathcal{W}}_{tt'}(\mathbf{r})$ and
electromagnetic modes $\hat{\EuScript{K}}\rt$. This separation is
achieved by an appropriate parametrization of the $\Qk$--matrix. One
convenient choice is
$\Qk=\hat{\mathcal{U}}_{z}\circ\hat{Q}_{\mathrm{fast}}\circ\hat{\mathcal{U}}^{-1}_{z}$,
where rotation matrices
\begin{equation}\label{int-ferm-R}
\hat{\mathcal{U}}_{z}=\left(\begin{array}{cr}1-F\circ Z & F \\ Z &
-1
\end{array}\right)\,,\qquad \hat{\mathcal{U}}^{-1}_{z}=
\left(\begin{array}{cc} 1 & F \\ Z & -1+Z\circ F
\end{array}\right)\,,
\end{equation}
with $A\circ B=\int\d t' A_{tt'}B_{t't''}$ carry information about
slow degrees of freedom, and the fast part of $\Qk$ matrix is
parameterized by the diffuson  fields
$\hat{Q}_{\mathrm{fast}}=\exp\{-\hat{\mathcal{W}}/2\}\circ
\hat{\sigma}_{z}\circ\exp\{\hat{\mathcal{W}}/2\}$ (compare this
parametrization with that given
by~\eqref{NLSM-Q-U-W-parametrization}). In the last equation
$Z_{tt'}(\mathbf{r})$ (not to be confused with the partition
function) may be thought of as the \textit{quantum} component of the
distribution function $F_{tt'}(\mathbf{r})$. Although
$Z_{tt'}(\mathbf{r})$ is put to  zero in the end of the
calculations, it was emphasized in the~\cite{Zala} that
$Z_{tt'}(\mathbf{r})$ must be kept explicitly in $\hat Q$
parametrization to obtain the proper form of the collision integral
in the kinetic equation.

As the second step, one performs integrations over $\hat{\Phi}$ (or
equivalently $\hat{\EuScript{K}}$, since the relation between them
is fixed by~(\ref{int-ferm-NLSM-Phi-K-matrix})), and over
$\hat{\mathcal{W}}$ fields  in the partition
function~\eqref{int-ferm-Z}, to arrive at the effective action
\begin{equation}\label{int-ferm-S-eff-def}
\mathcal{Z}=\int\D[\hat Q_{\EuScript{K}},\Phi]\,
\exp\big(iS[\hat{\mathcal{W}},\partial_{\mathbf{r}}\EuScript{K}]\big)=\int\D[F,Z]\,
\exp\big(iS_{\mathrm{eff}}[F,Z]\big)\,.
\end{equation}
Note  that after the decomposition given
by~\eqref{NLSM-Q-U-W-parametrization}, with the
$\hat{\mathcal{U}}_{z}$ and $\hat{\mathcal{U}}^{-1}_{z}$ matrices in
the form of~\eqref{int-ferm-R}, one understands the functional
integral over $\hat Q_{\EuScript{K}}$ matrix in
the~\eqref{int-ferm-S-eff-def} as taken over the independent matrix
fields $F$, $Z$ and $\hat{\mathcal{W}}$.  As a result, the effective
action $S_{\mathrm{eff}}$ will depend on $F$ and its quantum
component $Z$, and possibly the classical external fields, such as,
e.g., scalar or vector potentials. One then looks for the saddle
point equation for the distribution function $F$:
\begin{equation}\label{int-ferm-KE-def}
\left.\frac{\delta S_{\mathrm{eff}}[F,Z]}{\delta Z}\right|_{Z=0} =
0\,,
\end{equation}
which is  a  desired kinetic equation.

Proceeding along these lines, one expands the action
\eqref{int-ferm-S-012} in terms of $F$, $Z$, $\hat{\mathcal{W}}$,
and electromagnetic potentials $\Phi$ and $\EuScript{K}$. For the
slow part of the action one finds from~(\ref{int-ferm-S-012}a) that
$\Tr\big\{(\partial_{\mathbf{r}}
\Qk)^{2}\big\}=8\tr\big\{\partial_{\mathbf{r}}
F_{tt'}\partial_{\mathbf{r}} Z_{t't}\big\}+O(Z^{2})$ and
$\Tr\{\partial_{t}\Qk\}=2\mathrm{tr}\{\partial_{t}Z_{tt'}F_{t't}-
\partial_{t}F_{tt'}Z_{t't}\}$, where $\tr\{\ldots\}$ denotes
the spatial and time integrations only and Keldysh structure was
traced out explicitly. Passing to the Wigner transform
representation~\eqref{NLSM-Wigner}, one obtains
\begin{equation}\label{int-ferm-S-0-FZ}
iS_{0}[F,Z]=2\pi\nu\,\tr\left\{\big[D\partial^{2}_{\mathbf{r}}F_{\epsilon}(\mathbf{r},\tau)-
\partial_{\tau}F_{\epsilon}(\mathbf{r},\tau)\big]Z_{\epsilon}(\mathbf{r},\tau)\right\}\,,
\end{equation}
where $\tau=(t+t')/2$. Already at this stage, differentiating
$S_{0}[F,Z]$ with respect to $Z$ one recovers
from~\eqref{int-ferm-KE-def} the non--interacting kinetic
Equation~\eqref{NLSM-Diffusion-Eq}. In a similar fashion, one finds
dynamic part of the action for the fast degrees of freedom,
\begin{equation}\label{int-ferm-S-0-W}
iS_{0}[\hat{\mathcal{W}}]=-\frac{\pi\nu}{2}\,\tr\left\{\bar{d}_{\epsilon}(\mathbf{r},\tau)
\big[D\partial^{2}_{\mathbf{r}}-\partial_{\tau}\big]d_{\epsilon}(\mathbf{r},\tau)\right\}\,,
\end{equation}
which is nothing else but Wigner representation of~\eqref{NLSM-S-W}.

We continue now with the coupling terms between the
$\hat{\mathcal{W}}$ and $\Phi$ modes. For
$S_{1}[\hat{\mathcal{W}},F,Z]$ part of the action, which follows
from~(\ref{int-ferm-S-012}b) upon expansion, one obtains
\begin{eqnarray}\label{int-ferm-S-1-WFZ}
iS_{1}[\hat{\mathcal{W}},F,Z]=-i\pi\nu\,\tr\left\{\Big([F,X^{cl}_{+}]+
X^{q}_{+}-FX^{q}_{+}F\Big)\bar{d} +
\Big( [Z,X^{cl}_{-}] - X^{q}_{-} +
ZFX^{q}_{-} + X^{q}_{-}FZ \Big) d \right\}\,,
\end{eqnarray}
where
\begin{equation}\label{int-ferm-X}
X^{\alpha}_{\pm}=\Phi^{\alpha}-\partial_{t}\EuScript{K}^{\alpha}\pm
D\partial^{2}_{\mathbf{r}}\EuScript{K}^{\alpha}.
\end{equation}
Deriving the functional relation between $\hat{\Phi}$ and
$\hat{\EuScript{K}}$ fields, our logic was to nullify $S_{1}$ part
of the action (recall~\eqref{int-ferm-NLSM-action-Q-linear}). This
step turns out to be impossible to implement for the
non--equilibrium situation. However, we may still
satisfy~(\ref{int-ferm-NLSM-K-q-cl}a) by imposing a condition
$X^{q}_{-}=0$. Although the Keldysh component
of~(\ref{int-ferm-NLSM-Phi-K-1}) cannot be satisfied identically, it
still makes sense to demand that $\EuScript{K}^{cl}$ obeys the
following non--equilibrium generalization of
equation~(\ref{int-ferm-NLSM-K-q-cl}b)
\begin{equation}\label{int-ferm-NLSM-K-cl-NonEq}
(D\partial^{2}_{\mathbf{r}}+i\omega)\EuScript{K}^{cl}(\mathbf{r},\omega)+
2B_{\omega}(\mathbf{r},\tau)D\partial^{2}_{\mathbf{r}}\EuScript{K}^{q}(\mathbf{r},\omega)
=-\Phi^{cl}(\mathbf{r},\omega)\,,
\end{equation}
where non--equilibrium bosonic distribution function is defined
by~(\ref{int-ferm-NLSM-bosonic-distribution}). Note, however, that
this generalization \textit{does not} imply that linear in
$\hat{\mathcal{W}}$ (i.e. in $d$ and $\bar d$) terms  vanish
in~\eqref{int-ferm-S-1-WFZ}. Indeed,
using~(\ref{int-ferm-NLSM-K-q-cl}a) which relates quantum components
of $\hat{\Phi}$ and $\hat{\EuScript{K}}$,
and~\eqref{int-ferm-NLSM-K-cl-NonEq}, performing Wigner transform,
one finds that $S_{1}[\hat{\mathcal{W}},F,Z]$ part of the action can
be brought to the form
\begin{equation}\label{int-ferm-S-1-WFZ-1}
iS_{1}[\hat{\mathcal{W}},F,Z]=-i\pi\nu\,
\tr\left\{\mathcal{I}[F]X^{q}_{+}(\mathbf{r},\omega)
\bar{d}_{\epsilon_{-}}(\mathbf{r},\tau)e^{-i\omega\tau}+
Z_{\epsilon}(\mathbf{r},\tau)X^{cl}_{-}(\mathbf{r},\omega)
[d_{\epsilon_{-}}(\mathbf{r},\tau)-
d_{\epsilon_{+}}(\mathbf{r},\tau)] e^{-i\omega\tau}\right\}\,,
\end{equation}
where $\epsilon_{\pm}=\epsilon\pm\omega/2$ and we have introduced
functional
\begin{equation}
\mathcal{I}[F]=B_{\omega}(\mathbf{r},\tau)
\big[F_{\epsilon-\omega}(\mathbf{r},\tau)-F_{\epsilon}(\mathbf{r},\tau)\big]
+ 1-F_{\epsilon-\omega}(\mathbf{r},\tau)F_{\epsilon}(\mathbf{r},\tau) \,.
\end{equation}
Note that, in equilibrium, $\mathcal{I}[F]\equiv 0$.
In~\eqref{int-ferm-S-1-WFZ-1} one keeps an explicit $\omega$
dependence, thus not performing expansion for small $\omega$ as
compared to $\epsilon$ in the conventional Wigner transform sense.
In addition, equation \eqref{int-ferm-S-1-WFZ-1} should also contain
terms proportional to $FZX^{q}_{-}d$, which will not contribute to
the kinetic equation after $\EuScript{K}$ averaging, thus omitted
for brevity.

The remaining  $S_{2}$ part of the action~(\ref{int-ferm-S-012}c) is
already quadratic in the fast degrees of freedom
$S_{2}\propto(\partial_{\mathbf{r}}\EuScript{K})^{2}$, therefore it
can be taken at $\hat{\mathcal{W}}=0$:
\begin{eqnarray}\label{int-ferm-S-2-FZ}
iS_{2}[F,Z]\!\!\!\!&=&\!\!\!\!4\pi\nu D\, \tr \Big\{
(\partial_{\mathbf{r}}\EuScript{K}^{cl})
(\partial_{\mathbf{r}}\EuScript{K}^{q})Z -
(\partial_{\mathbf{r}}\EuScript{K}^{cl}) F
(\partial_{\mathbf{r}}\EuScript{K}^{q})FZ-
(\partial_{\mathbf{r}}\EuScript{K}^{q})
(\partial_{\mathbf{r}}\EuScript{K}^{cl})Z +
(\partial_{\mathbf{r}}\EuScript{K}^{q}) F
(\partial_{\mathbf{r}}\EuScript{K}^{cl})FZ \nonumber
\\
&+&\!\!\! (\partial_{\mathbf{r}}\EuScript{K}^{cl})Z
(\partial_{\mathbf{r}}\EuScript{K}^{cl})F -
\frac{1}{2}(\partial_{\mathbf{r}}\EuScript{K}^{cl})
(\partial_{\mathbf{r}}\EuScript{K}^{cl})FZ  -
\frac{1}{2}(\partial_{\mathbf{r}}\EuScript{K}^{cl})
(\partial_{\mathbf{r}}\EuScript{K}^{cl})ZF \Big\}\,.
\end{eqnarray}

The next step is to perform the Gaussian integration over the fast
degrees of freedom:  diffusons $(d,\bar{d})$ and gauge fields
$(\EuScript{K}^{cl},\EuScript{K}^{q})$. For $S_1$ part of the
action, employing~\eqref{int-ferm-S-0-W} and
\eqref{int-ferm-S-1-WFZ-1} we obtain
\begin{equation}
\left\langle\exp\big(iS_{0}[\hat{\mathcal{W}}]+iS_{1}[\hat{\mathcal{W}},F,Z]\big)
\right\rangle_{\mathcal{W},\EuScript{K}}=
\exp\big(iS^{(1)}_{\mathrm{eff}}[F,Z]\big)\,,
\end{equation}
where
\begin{eqnarray}\label{int-ferm-S-eff-b}
iS^{(1)}_{\mathrm{eff}}[F,Z]=-4i\pi\nu\,
\tr\left\{\big(Dq^{2}\big)^{2}
\big[\mathcal{D}^{A}\qo\mathcal{V}^{R}\qo
-\mathcal{D}^{R}\qo\mathcal{V}^{A}\qo\big]
\mathcal{I}[F]Z\right\}\,.
\end{eqnarray}
To derive $S^{(1)}_{\mathrm{eff}}$ in the form
of~\eqref{int-ferm-S-eff-b}, one observes that upon
$\hat{\mathcal{W}}$ integration the terms
$\tr\{\mathcal{I}[F]X^{q}_{+}\bar{d}\}$ and $\tr\{ZX^{cl}_{-}d\}$
in~\eqref{int-ferm-S-1-WFZ-1} produce an effective interaction
vertex between $F$ and $Z$, namely:
$\big\langle\exp(iS_{1})\big\rangle_{\mathcal{W}}=
\exp\big(\tr\{\mathcal{I}[F]X^{q}_{+}\mathcal{D}^{A}ZX^{cl}_{-}\}\big)$.
The latter has to be averaged over $\EuScript{K}$, which is done
observing that
\begin{eqnarray}
\left\langle
X^{cl}_{-}(\mathbf{q},\omega)X^{q}_{+}(-\mathbf{q},-\omega)\right\rangle_{\EuScript{K}}=
-4D^{2}\left\langle\partial^{2}_{\mathbf{r}}\EuScript{K}^{cl}\qo
\partial^{2}_{\mathbf{r}}\EuScript{K}^{q}\qom\right\rangle_{\EuScript{K}}=
-2i\big(Dq^{2}\big)^{2}\mathcal{V}^{R}\qo\,.
\end{eqnarray}
The last equation is a direct consequence of~\eqref{int-ferm-X} and
\eqref{int-ferm-NLSM-K-cl-NonEq}, and correlator given
by~\eqref{int-ferm-V-def}.

For  $S_2$ part of the action, using~\eqref{int-ferm-S-2-FZ}, one
finds
\begin{equation}
\big\langle\exp\big(iS_{2}[F,Z]\big)\big\rangle_{\EuScript{K}}=
\exp\big(iS^{(2)}_{\mathrm{eff}}[F,Z]\big)\,,
\end{equation}
where
\begin{equation}\label{int-ferm-S-eff-a}
iS^{(2)}_{\mathrm{eff}}[F,Z]=2i\pi\nu\,\tr\left\{Dq^{2}[\mathcal{V}^{R}\qo-
\mathcal{V}^{A}\qo]\mathcal{I}[F]Z\right\}\,.
\end{equation}
To derive equation \eqref{int-ferm-S-eff-a} one has to use
interaction propagators for the gauge fields
\eqref{int-ferm-K-corr-fun}, and adopt \textit{quasi--equilibrium}
FDT relation for the Keldysh component at coinciding arguments
\begin{equation}\label{int-ferm-quasi-FDT}
\mathcal{V}^{K}(\mathbf{r},\mathbf{r},\tau)=B_{\omega}(\mathbf{r},\tau)
\sum_{q}\big[\mathcal{V}^{R}\qo-\mathcal{V}^{A}\qo\big]\,,
\end{equation}
which holds in the non--equilibrium conditions as long as
$F_{\epsilon}(\mathbf{r},\tau)$ changes slowly on the spatial scale
$L_{T}=\sqrt{D/T}$ (this implies that gradient of
$F_{\epsilon}(\mathbf{r},\tau)$ is small). The correction to
the~\eqref{int-ferm-quasi-FDT} is of the form
$\propto\omega\int\d\mathbf{r}'\mathcal{D}^{R}(\mathbf{r}-\mathbf{r}',\omega)
\partial_{\tau}B_{\omega}(\mathbf{r}',\tau)\partial_{\omega}
\mathcal{D}^{A}(\mathbf{r}'-\mathbf{r})$, see~\cite{KamenevAndreev}.

As the final step, one combines $S_{0}[F,Z]$ from
Eq.~\eqref{int-ferm-S-0-FZ}, together with
$S^{(1),(2)}_{\mathrm{eff}}[F,Z]$ parts of the action given by
Eqs.~\eqref{int-ferm-S-eff-b} and \eqref{int-ferm-S-eff-a}, and
employs Eq.~\eqref{int-ferm-KE-def} to arrive at the kinetic
equation
\begin{equation}\label{int-ferm-KE}
D\partial^{2}_{\mathbf{r}}F_{\epsilon}(\mathbf{r},\tau)-\partial_{\tau}
F_{\epsilon}(\mathbf{r},\tau)=\mathcal{I}_{\mathrm{col}}[F]\,,
\end{equation}
where the collision integral is given by
\begin{equation}\label{int-ferm-col-int}
\mathcal{I}_{\mathrm{coll}}[F]=\sum_{q}\int\frac{\d\omega}{2\pi}\mathcal{M}
\qo\Big[1 - F_{\epsilon-\omega}(\mathbf{r},\tau)
F_{\epsilon}(\mathbf{r},\tau) + B_{\omega}(\mathbf{r},\tau)
[F_{\epsilon-\omega}(\mathbf{r},\tau)-F_{\epsilon}(\mathbf{r},\tau)]\Big]\,,
\end{equation}
with the kernel
\begin{equation}
\mathcal{M}\qo=-iDq^2\Big\{\big[\mathcal{V}^{R}\qo-\mathcal{V}^{A}\qo\big]-2Dq^2
\big[\mathcal{D}^{A}\qo\mathcal{V}^{R}\qo-\mathcal{D}^{R}\qo\mathcal{V}^{A}\qo\big]\Big\}\,.
\end{equation}
This equation can be simplified by noticing that the gauge field
propagator $\mathcal{V}^{R(A)}\qo$ may be written in terms of the
diffusons and screened RPA interactions, as
$\mathcal{V}^{R}\qo=-\big[\mathcal{D}^{R}\qo\big]^2U^{R}_{RPA}\qo$
and similarly for the advanced component, which is direct
consequence of~\eqref{int-ferm-U-RPA}. Using this form of
$\mathcal{V}^{R(A)}\qo$, after some algebra the interaction kernel
$\mathcal{M}\qo$ reduces to
\begin{equation}\label{int-ferm-col-int-kernel}
\mathcal{M}\qo=2\,\mathrm{Re}[\mathcal{D}^{R}\qo]\,\mathrm{Im}[U^{R}_{RPA}\qo]\,.
\end{equation}

For the conventional choice of the fermion distribution function
$\mathbf{n}_{\epsilon}(\mathbf{r},\tau)=(1-F_{\epsilon}(\mathbf{r},\tau))/2$,
one can rewrite the collision integral~\eqref{int-ferm-col-int} in
the usual form with "out" and "in" relaxation terms. Indeed,
employing~\eqref{int-ferm-integral-2}, one identically rewrites the
right hand side of~\eqref{int-ferm-KE}
as~\cite{Altshuler,AltshulerAronov-KE}
\begin{equation}\label{int-ferm-col-int-in-out}
\mathcal{I}_{\mathrm{coll}}[\mathbf{n}]=\sum_{q}\iint^{+\infty}_{-\infty}\d\omega\d\epsilon'\,
\mathbb{K}\qo\,\Big[\mathbf{n}_{\epsilon}
\mathbf{n}_{\epsilon'-\omega}(1-\mathbf{n}_{\epsilon'})(1-\mathbf{n}_{\epsilon-\omega})
-\mathbf{n}_{\epsilon'}\mathbf{n}_{\epsilon-\omega}
(1-\mathbf{n}_{\epsilon})(1-\mathbf{n}_{\epsilon'-\omega})\Big]\,,
\end{equation}
where collision kernel is $\mathbb{K}\qo=2\mathcal{M}\qo/\pi\omega$.

There are several important points which has to be discussed
regarding the general structure of the kinetic equation. (i) The
term $\tr\{ZFX^{q}_{-}d\}$, neglected in
the~\eqref{int-ferm-S-1-WFZ-1}, produces an effective vertex of the
type
$\tr\big\{\mathcal{I}[F]X^{q}_{+}\mathcal{D}^{A}ZFX^{q}_{-}\big\}$
after $\hat{\mathcal{W}}$ integration, which indeed vanishes after
$\EuScript{K}$ averaging, since $\langle
X^{q}_{\pm}X^{q}_{\pm}\rangle_{\EuScript{K}}\equiv0$. Thus, it
indeed does not generate any additional terms into the collision
integral. (ii) Throughout the derivation of the collision integral
we persistently neglected all spatial $\partial_{\mathbf{r}}
F_{\epsilon}(\mathbf{r},\tau)$ and time
$\partial_{\tau}F_{\epsilon}(\mathbf{r},\tau)$ derivatives  of the
distribution function, e.g. in~\eqref{int-ferm-quasi-FDT}. This is
justified as long as there is a spatial scale at which
$F_{\epsilon}(\mathbf{r},\tau)$ changes  slowly. In fact, gradients
of the distribution, if kept explicitly, contribute to the elastic
part of the collision integral~\cite{Zala,CatelaniAleiner}. (iii) We
kept in the effective action only terms which are linear in the
quantum component of the distribution function. There are, however,
terms which are quadratic in $Z_{\epsilon}(\mathbf{r},\tau)$. These
terms are responsible for the fluctuations in the distribution
function and lead to the so--called \textit{stochastic kinetic
equation} or, equivalently, Boltzmann--Langevin kinetic
theory~\cite{Kogan,KoganShulman,Nagaev}. It was shown recently that
Keldysh $\sigma$--model with retained
$Z^{2}_{\epsilon}(\mathbf{r},\tau)$ terms is equivalent to the
effective Boltzmann--Langevin description~\cite{Gutman,Bagrets}.
(iv) A collision integral similar to Eq.~\eqref{int-ferm-col-int}
was derived  within Keldysh $\sigma$--model formalism
in~\cite{KamenevAndreev}. However, the $S^{(1)}_{\mathrm{eff}}$ part
of the effective action was overlooked and as a result, the obtained
kernel of the collision integral turns out to be correct only in the
universal limit $U^{-1}_{0}\to0$. One finds
from~\eqref{int-ferm-col-int-kernel} for $U^{-1}_{0}\to0$ that
$\mathcal{M}\qo$ reduces to
$\mathcal{M}\qo=-\frac{2}{\nu}\,\mathrm{Im}\big[\mathcal{D}^{R}\qo\big]$,
which is result of~\cite{KamenevAndreev}. (v) Finally, the present
discussion can be generalized to include a spin degree of freedom.
Corresponding kinetic equation and collision kernel were obtained
in~\cite{DimitrovaKtavtsov,BurmiChelk}.

\subsection{Applications III: Interaction effects in disordered metals}\label{app_Part-III}
\subsubsection{Zero--bias anomaly}\label{app_Part-III-1}

Having discussed in  Section~\ref{app_Part-II} several examples,
where non--interacting version of the $\sigma$--model may be
applied, we turn now to  consideration of  interaction effects. The
first example of interest is the modification of the bare single
particle density of states $\nu$ of free electrons by Coulomb
interactions. The question  was addressed by Altshuler, Aronov and
Lee~\cite{AltshulerAronov,AAL,AltshulerAronov-review}. Although in
their original work only leading order interaction correction was
calculated, one may extent treatment of zero--bias anomaly beyond
the perturbation
theory~\cite{Finkel'stein,Nazarov-ZBA,LevitovShytov,Kopietz}. Here
we follow the sigma--model calculation of~\cite{KamenevAndreev}.

We are interested in the  single--particle Green function at
coinciding spatial points
\begin{equation}
\mathcal{G}^{ab}(t-t')=-i\big\langle\big\langle\psi_{a}\rt
\bar{\psi}_{b}(\mathbf{r},t')\big\rangle\big\rangle\,,
\end{equation}
where $\langle\langle\ldots\rangle\rangle$ denotes both the quantum
and disorder averaging. One may evaluate it introducing a
corresponding source term into the action which is directly coupled
to the bilinear combination of the fermion operators. Following the
same algebra as in the Section~\ref{sec_NLSM}, performing Keldysh
rotation and disorder averaging, one finds that this source term
enters into the logarithm in~\eqref{NLSM-action}. Differentiating
the latter with respect to the source and putting it to zero, one
obtains for the Green's function
\begin{equation}\label{Part-III-G-def}
\hat{\mathcal{G}}(t-t') = \int\!\! \D[\Phi]\,
\exp\big(i\Tr\{\vec{\Phi}^{T}U^{-1}_{0}\hat{\sigma}_{x}\vec{\Phi}\}\big)
\int\D[\hat Q]
\left[\hat{G}^{-1}+\frac{i}{2\tauel}\hat{Q}+
\hat{\Phi}\right]^{-1}_{tt',\mathbf{r}\mathbf{r}}\, \exp\big(iS[\hat Q,\Phi]\big)\,.
\end{equation}
One evaluates the integral over the $\hat Q$ matrix in the saddle
point approximation, neglecting both the massive and the massless
fluctuations around the stationary point. Then, according
to~\eqref{int-ferm-NLSM-saddle-point-eq}, the pre--exponential
factor is simply $-i\pi\nu\underline{\hat{Q}}_{tt'}$. At the saddle
point $\hat Q$ matrix is given by
Eq.~\eqref{int-ferm-NLSM-Q-gauge-general}. As a result, one obtains
for~\eqref{Part-III-G-def} the following representation
\begin{equation}\label{Part-III-G}
\hat{\mathcal{G}}(t-t')=-i\pi\nu\int\D[\Phi]
\exp\left(i\Tr\{\vec{\Phi}^{T}\hat{U}^{-1}_{RPA}\vec{\Phi}\}\right)
\exp\left(i\hat{\EuScript{K}}\rt\right)\hat{\Lambda}_{t-t'}
\exp\left(-i\hat{\EuScript{K}}(\mathbf{r},t')\right)\,.
\end{equation}
Since $\hat{\EuScript{K}}$ is the linear functional of $\hat{\Phi}$,
given by~\eqref{int-ferm-NLSM-K-q-cl}, the remaining functional
integral is Gaussian. To calculate the latter, one rewrites phase
factors of the gauge field as\footnote{Equation
\eqref{Part-III-Phase-factors} is based on the following property:
consider an arbitrary function which is linear form in Pauli
matrices $f(a+\mathbf{b\hat{\sigma}})$, where $a$ is some arbitrary
number and $\mathbf{b}$ some vector. The observation is that
$f(a+\mathbf{b}\hat{\sigma})=A+\mathbf{B}\hat{\sigma}$, where $A$ is
some new number and $\mathbf{B}$ a new vector. To see this, let us
choose $z$--axis along the direction of the $\mathbf{b}$ vector.
Then the eigenvalues of the operator $a+\mathbf{b}\hat{\sigma}$ are
$a\pm b$, and corresponding eigenvalues of the operator
$f(a+\mathbf{b\hat{\sigma}})$ are $f(a\pm b)$. Thus one concludes
that $A=\frac{1}{2}[f(a+b)+f(a-b)]$ and
$\mathbf{B}=\frac{\mathbf{b}}{2b}[f(a+b)-f(a-b)]$.}
\begin{equation}\label{Part-III-Phase-factors}
e^{\pm i\EuScript{K}^{\alpha}\gamma^{\alpha}}=\frac{1}{2}
\left[\,e^{\pm i(\EuScript{K}^{cl}+\EuScript{K}^{q})}+e^{\pm
i(\EuScript{K}^{cl}-\EuScript{K}^{q})}\right]\hat{\gamma}^{cl}+
\frac{1}{2} \left[\,e^{\pm
i(\EuScript{K}^{cl}+\EuScript{K}^{q})}-e^{\pm
i(\EuScript{K}^{cl}-\EuScript{K}^{q})} \right]\hat{\gamma}^{q}\,.
\end{equation}
Performing Gaussian integration in~\eqref{Part-III-G} with the help
of~\eqref{Part-III-Phase-factors}, the result may be conveniently
expressed in the form
\begin{equation}\label{Part-III-G-1}
\hat{\mathcal{G}}(t)=-i\pi\nu\sum_{\alpha\beta}
\big(\hat{\gamma}^{\alpha}\,\hat{\Lambda}_{t}\,\hat{\gamma}^{\beta}\big)\,
\mathbb{B}^{\alpha\beta}(t)\,,
\end{equation}
where the auxiliary propagator $\mathbb{B}^{\alpha\beta}(t)$ has the
standard bosonic structure [as in, e.g.,~\eqref{int-ferm-V-matrix}]
with
\begin{subequations}\label{Part-III-B-RAK}
\begin{equation}
\hskip-.75cm
\mathbb{B}^{R(A)}(t)=i\exp\big(i[\mathcal{V}^{K}(t)-\mathcal{V}^{K}(0)]/2\big)
\sin\big(\mathcal{V}^{R(A)}(t)/2\big)\,,
\end{equation}
\begin{equation}
\mathbb{B}^{K}(t)=\exp\big(i[\mathcal{V}^{K}(t)-\mathcal{V}^{K}(0)]/2\big)
\cos\big([\mathcal{V}^{R}(t)-\mathcal{V}^{A}(t)]/2\big)\,.
\end{equation}
\end{subequations}
The gauge fields propagator, $\hat{\mathcal{V}}\rt$, defined
by~\eqref{int-ferm-V-matrix} and \eqref{int-ferm-V-RAK},
enters~\eqref{Part-III-B-RAK} at coinciding spatial points
\begin{equation}\label{Part-III-V-t}
\hat{\mathcal{V}}(t)=\int\frac{\d\omega}{2\pi}\exp(-i\omega
t)\sum_{q}\hat{\mathcal{V}}\qo\,.
\end{equation}
Knowledge of the Green's function \eqref{Part-III-G-1} allows to
determine the density of states according to the standard definition
\begin{equation}\label{Part-III-DOS-def}
\nu(\epsilon)=\frac{i}{2\pi}\big[\mathcal{G}^{R}(\epsilon)
-\mathcal{G}^{A}(\epsilon)\big]\,.
\end{equation}
In the thermal equilibrium, the Green's functions obey FDT
(see~\eqref{fermion-FDT}), which together with the relations
$\mathcal{G}^{K}(\epsilon)=\mathcal{G}^{>}(\epsilon)+\mathcal{G}^{<}(\epsilon)$
and
$\mathcal{G}^{>}(\epsilon)=-\exp(\epsilon/T)\mathcal{G}^{<}(\epsilon)$
allows to rewrite~\eqref{Part-III-DOS-def} in the equivalent form
\begin{equation}\label{Part-III-DOS}
\nu(\epsilon)=\frac{i}{2\pi}\ \mathcal{G}^{>}
(\epsilon)[1+\exp(-\epsilon/T)]\,.
\end{equation}
Using \eqref{Part-III-G-1} one relates greater (lesser) Green's
functions $\mathcal{G}^{>(<)}$ to the corresponding components of
the auxiliary propagators $\mathbb{B}^{>(<)}$:
\begin{equation}\label{Part-III-G-B}
\mathcal{G}^{>(<)}(t)=-i\pi\nu\Lambda^{>(<)}_{t}\mathbb{B}^{>(<)}(t)\,.
\end{equation}
The latter are found explicitly to be
\begin{equation}\label{Part-III-B-lesser-greater}
\mathbb{B}^{>(<)}(t)=\frac{1}{2}\exp\left(\int\frac{\d\omega}{2\pi}
\left[\coth\frac{\omega}{2T}(1-\cos\omega t)\pm i\sin\omega t\right]
\mathrm{Im}\sum_{q}\mathcal{V}^{R}\qo\right)\,,
\end{equation}
where we employed~\eqref{Part-III-B-RAK} along with the bosonic FDT
relations
$\mathbb{B}^{R}(t)-\mathbb{B}^{A}(t)=\mathbb{B}^{>}(t)-\mathbb{B}^{<}(t)$,
and $\mathbb{B}^{K}(t)=\mathbb{B}^{>}(t)+\mathbb{B}^{<}(t)$.
Finally, combining~\eqref{Part-III-DOS} and~\eqref{Part-III-G-B}
together, one finds for the density of states
\begin{equation}\label{Part-III-DOS-fin}
\nu(\epsilon)=\frac{\nu}{\tanh(\epsilon/2T)}\int\d t\ F_{t}\
\mathbb{B}^{K}(t)\exp(i\epsilon t)\,.
\end{equation}
Expanding~\eqref{Part-III-B-lesser-greater} to the first order in
the interaction, $\mathcal{V}\qo$, and substituting
into~\eqref{Part-III-DOS-fin}, one recovers  Altshuler and Aronov
result for the zero--bias anomaly~\cite{AltshulerAronov}.

We restrict ourselves to the analysis of the non--perturbative
result, \eqref{Part-III-B-lesser-greater} and
\eqref{Part-III-DOS-fin}, only at zero temperature. Noting that for
$T=0$, $F_{t}=(i\pi t)^{-1}$, one obtains
\begin{eqnarray}\label{Part-III-DOS-T0}
\nu(\epsilon)=\frac{\nu}{\pi}\int\d t\ \frac{\sin|\epsilon|t}{t}\
\exp\left(\int^{\infty}_{0}\frac{\d\omega}{\pi}\
\mathrm{Im}\sum_{q}\mathcal{V}^{R}\qo(1-\cos\omega
t)\right)\nonumber\\
\times\cos\left(\int^{\infty}_{0}\frac{\d\omega}{\pi}\
\mathrm{Im}\sum_{q}\mathcal{V}^{R}\qo\sin\omega t\right)\,.
\end{eqnarray}
In the two--dimensional case~\eqref{int-ferm-V-RAK} with $U_{0}=2\pi
e^{2}/q$ leads to
\begin{equation}\label{Part-III-exponent-integral}
\int^{+\infty}_{0}\frac{\d\omega}{\pi}\sum_{q}\mathrm{Im}
\big[\mathcal{V}^{R}\qo\big]\left(\begin{array}{c}1-\cos\omega t \\
\sin\omega t\end{array}\right)=-\frac{1}{8\pi^{2}g}\left\{
\begin{array}{l}\ln(t/\tauel)\ln(t\tauel\omega^{2}_{0})+2\mathbb{C}\ln(t\omega_{0})
\\ \pi\ln(t\omega_{0})\end{array}\right.\,,
\end{equation}
where $g=\nu D$ is the dimensionless conductance,
$\omega_{0}=D\kappa^{2}$, $\kappa^{2}=2\pi e^{2}\nu$ is the inverse
Thomas--Fermi screening radius and $\mathbb{C}=0.577...$ is the
Euler constant. Since the fluctuations $\hat{\mathcal{W}}$ of the
$\hat Q$ matrix were neglected, while calculating functional
integral in~\eqref{Part-III-G-def}, the obtained
result~\eqref{Part-III-DOS-T0} does not capture corrections, which
are of the order of $\sim g^{-1}\ln(t/\tau_{\mathrm{el}})$ (in
$d=2$), see Section~\ref{app_Part-III-2}. Therefore,
\eqref{Part-III-DOS-T0} can only be trusted for $\epsilon$ not too
small, such that $(8\pi^{2}g)^{-1}\ln(\epsilon\tauel)^{-1}\ll 1$,
however, $\ln^{2}(t/\tauel)$ terms have been accounted correctly by
the preceding procedure. If, in addition,
$g^{-1}\ln(\omega_{0}\tauel)\ll1$, the time integral in
\eqref{Part-III-DOS-T0} may be performed by the stationary point
method, resulting in
\begin{equation}
\nu(\epsilon)=\nu\exp\left\{-\frac{1}{8\pi^{2}g}\ln(|\epsilon|\tauel)^{-1}
\ln(\tauel\omega^{2}_{0}/|\epsilon|)\right\}\,.
\end{equation}
Thus, one achieved a non--perturbative resummation of anomalously
divergent, $\propto\ln^{2}(\epsilon\tauel)$, terms for a
single--particle Green's function. The non--perturbative expression
for the density of states essentially arises from the gauge
non--invariance of the single--particle Green's function. The
calculations above are in essence the Debye--Waller
factor~\cite{Finkel'stein94} owing to the almost pure gauge
fluctuations of electric potential, cf.~\eqref{Part-III-G}.
Gauge--invariant characteristics (such as conductivity, for example)
do not carry phase factors, and therefore are not affected by the
interactions on this level of accuracy (fluctuations of $\hat Q$
matrix should be retained, see next section).

\subsubsection{Altshuler--Aronov correction}\label{app_Part-III-2}

Here we consider yet another example where interactions are
essential, namely electron--electron interactions correction
$\delta\sigma_{\mathrm{AA}}$ to the Drude conductivity $\sigma_{D}$
of the disordered
metal~\cite{AltshulerAronov,AAL,AltshulerAronov-review}. In contrast
to the previous example, where density of states of an interacting
disordered electron liquid was considered
(Section~\ref{app_Part-III-1}), the correction to the conductivity
is not affected by the interactions at the level of trial saddle
point $\Qk=\hat{\Lambda}$ and fluctuations $\hat{\mathcal{W}}$ must
be retained. In what follows, we restrict our consideration to the
lowest non--vanishing order in the expansion of the
action~\eqref{int-ferm-S-012} over $\hat{\mathcal{W}}$,
\eqref{NLSM-S-W} and \eqref{NLSM-d-d}, and identify those terms of
the action which are responsible for interaction correction
$\delta\sigma_{\mathrm{AA}}$.

One starts from the  part of the action $S_{1}[\hat
Q_{\EuScript{K}},\partial_{\mathbf{r}}\EuScript{K}]$ given
by~(\ref{int-ferm-S-012}b). To the linear order in fluctuations
$\hat{\mathcal{W}}$ one finds
\begin{equation}\label{Part-III-S1-W}
iS_{1}[\hat{\mathcal{W}},\partial_{\mathbf{r}}\EuScript{K}]=
-\frac{i\pi\nu}{2}\Tr\left\{\left[D\partial^{2}_{\mathbf{r}}\EuScript{K}^{\alpha}
\big(\hat{\Lambda}\hat{\gamma}^{\alpha}\hat{\Lambda}-\hat{\gamma}^{\alpha}\big)
+\big(\Phi^{\alpha}-\partial_{t}\EuScript{K}^{\alpha}\big)
\big(\hat{\gamma}^{\alpha}\hat{\Lambda}-\hat{\Lambda}\hat{\gamma}^{\alpha}\big)\right]
\hat{W}\right\}\,,
\end{equation}
where
$\hat{W}=\hat{\mathcal{U}}\circ\hat{\mathcal{W}}\circ\hat{\mathcal{U}}^{-1}$,
see~(\ref{NLSM-Q-U-W-parametrization}) and (\ref{NLSM-U-W}). Note
that in  thermal equilibrium
$iS_{1}[\hat{\mathcal{W}},\partial_{\mathbf{r}}\EuScript{K}]\equiv0$.
Indeed, the expression in the square brackets on the right--hand
side of~\eqref{Part-III-S1-W} coincides
with~\eqref{int-ferm-NLSM-Phi-K-1}, which was used to determine the
$\hat{\EuScript{K}}[\Phi]$ functional. In equilibrium it was
possible to solve~\eqref{int-ferm-NLSM-Phi-K-1} by an appropriate
choice of $\hat{\EuScript{K}}[\Phi]$,
see~\eqref{int-ferm-NLSM-Phi-K-matrix}. This was precisely the
motivation behind looking for the saddle point for each realization
of the field $\hat{\Phi}$ to cancel terms linear in
$\hat{\mathcal{W}}$. Since it was not possible to find the exact
saddle point, such terms do appear, however, only in the second
order in $\partial_{\mathbf{r}}\hat{\EuScript{K}}$. These latter
terms originate from the
$S_{2}[\Qk,\partial_{\mathbf{r}}\EuScript{K}]$ part of the action.
Expanding~(\ref{int-ferm-S-012}c) to the linear order in
$\hat{\mathcal{W}}$ one finds
\begin{eqnarray}\label{Part-III-S2-W}
iS_{2}[\hat{\mathcal{W}},\partial_{\mathbf{r}}\EuScript{K}] &=& \frac{\pi\nu D}{2}\,
\Tr\left\{\partial_{\mathbf{r}}\EuScript{K}^{\alpha}(\epsilon_{1}-\epsilon_{2})
\left[\hat{\gamma}^{\alpha}\hat{\Lambda}_{\epsilon_{2}}
\hat{\gamma}^{\beta}\hat{\Lambda}_{\epsilon_{3}}-
\hat{\Lambda}_{\epsilon_{1}}\hat{\gamma}^{\alpha}
\hat{\Lambda}_{\epsilon_{2}}\hat{\gamma}^{\beta}\right]
\hat{W}_{\epsilon_{3}\epsilon_{1}}
\partial_{\mathbf{r}}\EuScript{K}^{\beta}(\epsilon_{2}-\epsilon_{3})\right\}
\nonumber  \\
&=& \pi\nu D\,
\Tr\left\{\partial_{\mathbf{r}}\vec{\EuScript{K}}^{T}(\epsilon_{1}-\epsilon_{2})
\left[M^{d}_{\epsilon_{2}}d_{\epsilon_{3}\epsilon_{1}}+
M^{\bar{d}}_{\epsilon_{1}\epsilon_{2}\epsilon_{3}}
\bar{d}_{\epsilon_{3}\epsilon_{1}}\right]
\partial_{\mathbf{r}}\vec{\EuScript{K}}(\epsilon_{2}-\epsilon_{3})\right\}\,,
\end{eqnarray}
where we used notation
$\vec{\EuScript{K}}^{T}=\big(\EuScript{K}^{cl},\EuScript{K}^{q}\big)$,
and introduced  coupling matrices between diffusons $\{d,\bar{d}\}$
and the gauge fields $\EuScript{K}^{cl(q)}$
\begin{equation}\label{Part-III-M}
M^{d}_{\epsilon_{2}}=\left(\begin{array}{cc}0 & 0
\\ 0 & -2F_{\epsilon_{2}}\end{array}\right)\,,\qquad
M^{\bar{d}}_{\epsilon_{1}\epsilon_{2}\epsilon_{3}}=
\left(\begin{array}{ll}
2F_{\epsilon_{2}}-F_{\epsilon_{1}}-F_{\epsilon_{3}}&
1+F_{\epsilon_{1}}F_{\epsilon_{3}}-2F_{\epsilon_{2}}F_{\epsilon_{3}}\\
-1-F_{\epsilon_{1}}F_{\epsilon_{3}}+2F_{\epsilon_{2}}F_{\epsilon_{1}}\quad&
F_{\epsilon_{1}}+F_{\epsilon_{3}}-2F_{\epsilon_{1}}F_{\epsilon_{2}}F_{\epsilon_{3}}
\end{array}\right)\,.
\end{equation}
Employing now the general expression for the
conductivity~\eqref{Part-II-sigma-def}, we show that
Altshuler--Aronov interaction correction to the conductivity
$\delta\sigma_{AA}$ is obtained from~\eqref{Part-III-S2-W}
\begin{equation}\label{Part-III-sigma-def}
\delta\sigma_{\mathrm{AA}}=-\frac{e^{2}}{2}\lim_{\Omega\to 0}
\frac{1}{\Omega} \left\langle\frac{\delta^{2}}
{\delta\big(\partial_{\mathbf{r}} \EuScript{K}^{cl}(\Omega)\big)
\delta\big(\partial_{\mathbf{r}} \EuScript{K}^{q}(-\Omega)\big)}\
\exp\left(iS_{2}[\hat{\mathcal{W}},\partial_{\mathbf{r}}\EuScript{K}]\right)
\right\rangle_{\mathcal{W},\EuScript{K}}\, ,
\end{equation}
where the averaging goes over the diffusive modes as well as over
the fluctuations of the electric potential. Note also that as
compared to~\eqref{Part-II-sigma-def} here we perform
differentiation over $\partial_{\mathbf{r}}\EuScript{K}$ and not the
vector potential $\mathbf{A}$ itself. The two definitions are the
same since the vector potential and the gauge field enter the
action~\eqref{int-ferm-action-interacting} in the gauge invariant
combination~\eqref{int-ferm-NLSM-C}.

\begin{figure}
\begin{center}\includegraphics[width=10cm]{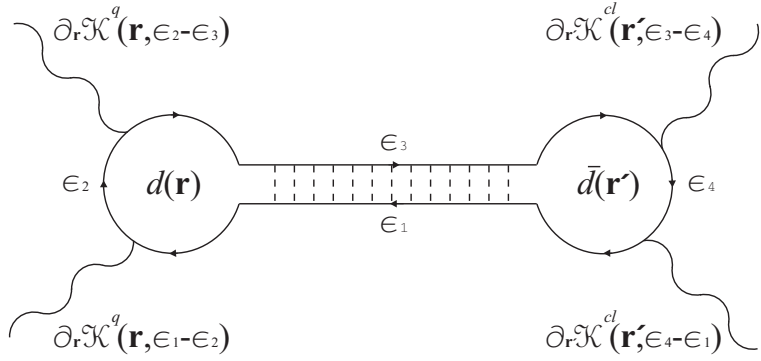}\end{center}
\caption{Diagrammatic representation of an effective four--leg
vertex $\mathbb{V}_{\mathrm{AA}}$, see~\eqref{Part-III-AA-Vertex},
which generates Altshuler--Aronov correction to the
conductivity.\label{Fig-AA-vertex}}
\end{figure}
Having~\eqref{NLSM-S-W} and \eqref{Part-III-S2-W} we deal with a
Gaussian theory of the diffuson modes $d$ and $\bar d$ fluctuations,
which allows for a straightforward averaging
in~\eqref{Part-III-sigma-def}. Integrating over the diffuson modes,
one finds
\begin{equation}
\left\langle\exp\big(iS_{2}[\hat{\mathcal{W}},\partial_{\mathbf{r}}\EuScript{K}]\big)
\right\rangle_{\mathcal{W}}=
\exp\big(i\mathbb{V}_{\mathrm{AA}}[\EuScript{K}]\big)\,.
\end{equation}
This way the $\big(\partial_{\mathbf{r}}\EuScript{K}\big)^4$
effective four--gauge--field vertex is generated
\begin{eqnarray}\label{Part-III-AA-Vertex}
\mathbb{V}_{\mathrm{AA}}[\EuScript{K}]=4\pi\nu
D^{2}\Tr\left\{\,F_{\epsilon_{2}}(2F_{\epsilon_{4}}-F_{\epsilon_{1}}-F_{\epsilon_{3}})
\partial_{\mathbf{r}}\EuScript{K}^{q}(\mathbf{r},\epsilon_{1}-\epsilon_{2})
\partial_{\mathbf{r}}\EuScript{K}^{q}(\mathbf{r},\epsilon_{2}-\epsilon_{3})
\phantom{\mathcal{D}^{A}}\right. \nonumber\\
\left. \times\,
\mathcal{D}^{R}(\mathbf{r}-\mathbf{r}',\epsilon_{3}-\epsilon_{1})
\partial_{\mathbf{r}'}\EuScript{K}^{cl}(\mathbf{r}',\epsilon_{3}-\epsilon_{4})
\partial_{\mathbf{r}'}\EuScript{K}^{cl}(\mathbf{r}',\epsilon_{4}-\epsilon_{1})\right\}\,.
\end{eqnarray}
Its diagrammatic representation is depicted in
Figure~\ref{Fig-AA-vertex}. This vertex originates from
$\Tr\{\partial_{\mathbf{r}}\EuScript{K}
M^{d}d\partial_{\mathbf{r}}\EuScript{K}\}$ and
$\Tr\{\partial_{\mathbf{r}}\EuScript{K}
M^{\bar{d}}\bar{d}\partial_{\mathbf{r}}\EuScript{K}\}$ parts of the
action~\eqref{Part-III-S2-W} after we pair $d$ and $\bar{d}$ by the
diffuson  propagator
$\langle\bar{d}d\rangle_{\mathcal{W}}\propto\mathcal{D}^{A}$. The
factor $F_{\epsilon_{2}}$ originates from $q-q$ element of the
matrix $M^{d}$, while the  combination
$2F_{\epsilon_{4}}-F_{\epsilon_{1}}-F_{\epsilon_{3}}$ of the
distribution functions in~\eqref{Part-III-AA-Vertex} is the $cl-cl$
element of the matrix $M^{\bar{d}}$. By writing
$\mathbb{V}_{\mathrm{AA}}[\EuScript{K}]$ in the form
of~\eqref{Part-III-AA-Vertex} we kept only contributions with the
lowest possible number of quantum gauge fields
$\partial_{\mathbf{r}}\EuScript{K}^{q}$. However, matrix
$M^{\bar{d}}$ has all four non--zero elements, thus
$\mathbb{V}_{\mathrm{AA}}[\EuScript{K}]$ in principle also contains
contributions with four and three  legs carrying the quantum gauge
fields. The latter are to be employed in calculations of the
corresponding interactions corrections to the shot--noise power,
see~\cite{GutmanGefen} for details.

\begin{figure}
\begin{center}\includegraphics[width=10cm]{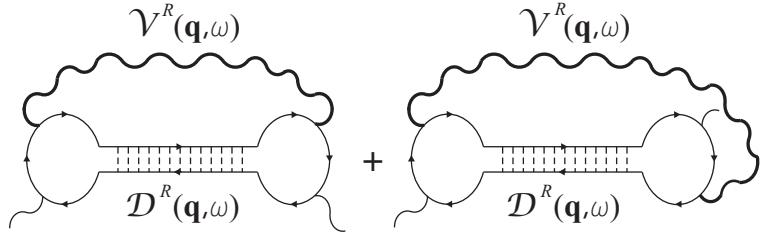}\end{center}
\caption{Diagrams for the interaction correction to the conductivity
$\delta\sigma_{\mathrm{AA}}$. These diagrams are constructed from
the effective vertex $\mathbb{V}_{\mathrm{AA}}[\EuScript{K}]$ by
keeping one classical and one quantum leg to be external, while
connecting the remaining two by the interaction propagator
$\mathcal{V}^{R}\qo$. \label{Fig-AA}}
\end{figure}

Having performed $\hat{\mathcal{W}}$ averaging, one brings now
$\mathbb{V}_{AA}[\EuScript{K}]$ into~\eqref{Part-III-sigma-def} and
integrates out $\EuScript{K}$ field. For the conductivity correction
this gives
\begin{equation}\label{Part-III-sigma-AA}
\delta\sigma_{\mathrm{AA}}= 4\pi e^{2}\nu D^{2}\,
\sum_{q}\iint\frac{\d\epsilon\d\omega}{4\pi^{2}}
\big(F_{\epsilon_{+}}+F_{\epsilon_{-}}\big)
\big(\partial_{\epsilon}F_{\epsilon_{+}}-
\partial_{\epsilon}F_{\epsilon_{-}}\big)
\mathcal{D}^{R}(\mathbf{q},\omega)
\left\langle\partial_{\mathbf{r}}\EuScript{K}^{cl}(\mathbf{q},\omega)
\partial_{\mathbf{r}}\EuScript{K}^{q}(-\mathbf{q},-\omega)
\right\rangle_{\EuScript{K}}\,,
\end{equation}
where new integration variables
$\epsilon=(\epsilon_{3}+\epsilon_{1})/2$ and
$\omega=\epsilon_{3}-\epsilon_{1}$ were introduced. The
$\EuScript{K}$ averaging produces two diagrams, Fig.~\ref{Fig-AA},
for $\delta\sigma_{\mathrm{AA}}$, which follows naturally from the
effective vertex shown in  Figure~\ref{Fig-AA-vertex}, after one
pairs two external legs by the interaction propagator. In the
universal limit of strong interactions $U^{-1}_{0}\to0$ the
propagator $\mathcal{V}^{R}\qo$ takes the simple form. As a result,
\begin{equation}\label{Part-III-V}
\left\langle\partial_{\mathbf{r}}\EuScript{K}^{cl}\qo
\partial_{\mathbf{r}}\EuScript{K}^{q}\qom\right\rangle_{\EuScript{K}}=
\frac{iq^2}{2}\, \mathcal{V}^{R}\qo =
-\frac{i}{2\nu D}\, \frac{1}{Dq^{2} - i\omega}\,,
\end{equation}
which follows from~\eqref{int-ferm-K-corr-fun} and
\eqref{int-ferm-V-RAK}. Inserting~\eqref{Part-III-V} into
Eq.~\eqref{Part-III-sigma-AA} and carrying $\epsilon$ integration
one finds
\begin{equation}
\frac{\delta\sigma_{\mathrm{AA}}}{\sigma_{D}} =
\frac{2i}{\pi\nu}\sum_{q} \int\d\omega\,
\frac{\partial}{\partial\omega}\left[\omega\coth\frac{\omega}{2T}\right]\,
\frac{1}{\big(Dq^{2} - i\omega\big)^{2}}\ .
\end{equation}
In two dimensions this expression leads to the logarithmically
divergent negative correction to the conductivity:
$\delta\sigma_{\mathrm{AA}}= -\frac{e^{2}}{2\pi^{2}}\ln(1/T\tauel)$,
where the elastic scattering rate $\tauel^{-1}$ enters as an upper
cutoff in the integral over the frequency $\omega$. A detailed
review of the effects of the interaction corrections on disordered
conductors can be found in~\cite{AltshulerAronov-review}, see
also~\cite{Zala}.

\subsubsection{Relaxation rate}\label{app_Part-III-3}

Kinetic equation discussed in Section~\ref{sec_int_ferm-5} may be
used to find energy relaxation
rate~\cite{AltshulerAronov-KE,AltshulerAronov-review,Schmid,Blanter}.
Focusing on the "out" term of the collision integral
in~\eqref{int-ferm-col-int-in-out}, one may introduce the out
relaxation rate for an electron of energy $\epsilon$, as
\begin{equation}\label{Part-III-out-rate}
\frac{1}{\tau_{\mathrm{out}}(\epsilon)}=-\sum_{q}\int\d\omega\d\epsilon'\,
\mathbb{K}\qo\,
\mathbf{n}_{F}(\epsilon)[1-\mathbf{n}_{F}(\epsilon-\omega)]
\mathbf{n}_{F}(\epsilon')[1-\mathbf{n}_{F}(\epsilon+\omega)]\,,
\end{equation}
where all electron distributions were substituted by Fermi
functions. This is appropriate if one is interested in small
(linear) deviations of $\mathbf{n}_{\epsilon}$ from its equilibrium
value $\mathbf{n}_{F}(\epsilon)$. Equation \eqref{Part-III-out-rate}
simplifies considerably at zero temperature, $T=0$. Indeed, Fermi
distribution functions limit energy integration to two ranges
$-\omega<\epsilon'<0$ and $0<\omega<\epsilon$, where the
product of all occupation numbers is just unity. In the universal limit
of strong interactions, $U^{-1}_{0}\to 0$,  the
kernel acquires a form, see Eq.~\eqref{int-ferm-col-int-kernel}
\begin{equation}
\mathbb{K}\qo=-\frac{4}{\pi\nu}\frac{1}{(Dq^2)^{2}+\omega^2}\,.
\end{equation}
Inserting $\mathbb{K}\qo$ into~\eqref{Part-III-out-rate}, one finds
for the out relaxation rate the following expression
\begin{equation}
\frac{1}{\tau_{\mathrm{out}}(\epsilon)}=\frac{4}{\pi\nu}\sum_{q}\int^{|\epsilon|}_{0}\d\omega
\int^{0}_{-\omega}\d\epsilon'\frac{1}{(Dq^2)^{2}+\omega^2}=\frac{|\epsilon|}{4\pi
g}\,,
\end{equation}
where $g=\nu D$ and momentum integral was performed for the
two--dimensional case. For an arbitrary dimensionality $d$, out rate
scales with energy as
$\tau^{-1}_{\mathrm{out}}(\epsilon)\propto(1/\nu_{d})(\epsilon/D)^{d/2}$,
see~\cite{AltshulerAronov-review} for further details.

\subsubsection{Third order drag effect}\label{app_Part-III-4}

Discussing Coulomb drag in Section~\ref{app_Part-I-3} it was
emphasized that the effect appears already in the second order in
inter--circuit interactions and the particle--hole asymmetry is
crucial. In the linear response at small temperatures the drag
conductance appears to be quadratic in temperature,
see~(\ref{Part-I-ID-linear-1}). Here we discuss the third order in
the inter--layer interaction contribution to the drag conductance.
Although, being subleading in the interaction strength, it does not
rely on the electron--hole asymmetry (in bulk systems the latter is
due to the curvature of dispersion relation near the Fermi energy
and thus very small). We show that such a third order drag is
temperature independent and thus may be a dominant effect at small
enough  temperatures \cite{Drag-Third-order}. Technically the
third--order contributions originate from the four--leg vertices
(see Figure~\ref{Fig-AA-vertex} and
corresponding~\eqref{Part-III-AA-Vertex}), which describe  induced
non--linear interactions of electromagnetic fields through
excitations of electron--hole pairs in each of the layers.

Following~\cite{Drag-Third-order} we consider two--dimensional
electron gas bilayer and apply  NLSM to calculate the drag
conductivity. From the general expression~\eqref{Part-II-sigma-def}
with the help of~\eqref{Part-III-AA-Vertex} one defines drag
conductivity as
\begin{equation}\label{Part-III-sigma-drag}
\sigma_{\mathrm{drag}}=-\frac{e^2}{2}\lim_{\Omega\to0}\frac{1}{\Omega}
\left\langle\frac{\delta\mathbb{V}_{\mathrm{AA}}[\EuScript{K}]}
{\delta\big(\partial_{\mathbf{r}}\EuScript{K}^{cl}_{1}(\Omega)\big)}
\frac{\delta\mathbb{V}_{\mathrm{AA}}[\EuScript{K}]}
{\delta\big(\partial_{\mathbf{r}}\EuScript{K}^{q}_{2}(-\Omega)\big)}
\right\rangle_{\EuScript{K}}\,,
\end{equation}
where indices $1,2$ refer to the drive and dragged layers,
respectively, following notations of Section~\ref{app_Part-I-3}. The
averaging over the fluctuating gauge field $\EuScript{K}$ is
performed with the help of the correlation function
\begin{equation}\label{Part-III-V-drag}
\mathcal{V}_{ab}^R\qo=2i\big\langle
\EuScript{K}^{cl}_{a}\qo\EuScript{K}^{q}_{b}\qom\big\rangle_{\EuScript{K}}=
\frac{q^2 {U}^{R}_{ab}\qo}{\big(D_{a}q^2-i\omega\big)\big(D_{b}q^{2}-i\omega\big)}\,,
\end{equation}
where $a,b=(1,2)$ and $U^R_{ab}\qo$ is  $2\times2$ matrix of
retarded screened intra-- and inter--layer interactions calculated
within RPA. It is a solution of the following matrix Dyson equation,
$\hat{U}^R=\hat{U}_{0}+\hat{U}_{0}\hat{\Pi}^R\hat{U}^R$, where
\begin{equation}\label{Part-III-RPA}
\hat{U}_{0}=\frac{2\pi e^{2}}{q}\!\left(\begin{array}{cc}
1 & e^{-qd}
\\ e^{-qd} & 1
\end{array}\right)\,,\qquad\quad
\hat{\Pi}^R=\left(\begin{array}{cc} \frac{\nu_{1}D_{1}q^{2}}
{D_{1}q^{2}-i\omega} & 0 \\ 0 & \frac{\nu_{2}D_{2}q^{2}}
{D_{2}q^{2}-i\omega}\end{array}\right)\,.
\end{equation}
Off--diagonal components of $\hat{U}_{0}$ matrix represent bare
Coulomb interaction between the layers, where $d$ is the
inter--layer spacing. Note also that the polarization operator
matrix $\hat{\Pi}^R\qo$ is diagonal, reflecting the absence of
tunneling between the layers.

\begin{figure}
\begin{center}\includegraphics[width=10cm]{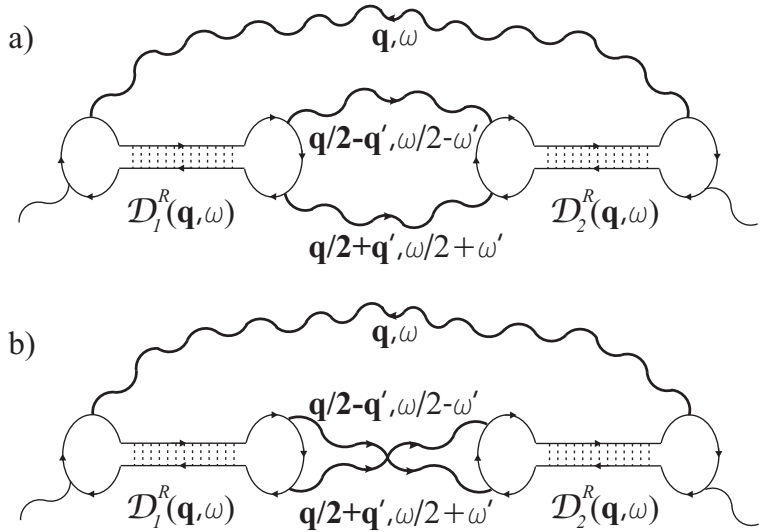}\end{center}
\caption{Two diagrams for the drag conductivity
$\sigma_{\mathrm{drag}}$ in the third order in the inter--layer
interactions, $\mathcal{V}^{R}_{12}\qo$, denoted by  wavy lines. The
intra--layer diffusion propagators
$\mathcal{D}^{R}_{a}\qo=(D_{a}q^2-i\omega)^{-1}$ are denoted by
ladders. \label{Fig-Drag-3rd}}
\end{figure}

We are now on the position to evaluate the third--order drag
conductivity. Inserting~\eqref{Part-III-AA-Vertex}
into~\eqref{Part-III-sigma-drag} and performing averaging with the
help of~\eqref{Part-III-RPA}, one finds the following expression for
drag conductivity
\begin{eqnarray}\label{Part-III-drag}
&&\hskip-1cm
\sigma_{\mathrm{drag}}=32e^{2}T\nu_{1}\nu_{2}D^{2}_{1}D^{2}_{2}\int^{\infty}_{0}
\frac{\mathrm{d}\omega\mathrm{d}\omega'}{4\pi^{2}}\,\,
\mathcal{H}_{1}(\omega,\omega')\mathcal{H}_{2}(\omega,\omega')\nonumber\\
&&\hskip-1cm
\times \sum_{q,q'}\mathrm{Im}\left[\mathcal{D}_{1}^R\qo\mathcal{D}_{2}^R\qo
\mathcal{V}_{12}^R\qo \mathcal{V}_{12}^R\left({\mathbf{q}\over
2}-\mathbf{q}',{\omega\over 2}-\omega'\right)
\mathcal{V}_{12}^R\left({\mathbf{q}\over 2}+\mathbf{q}',{\omega\over
2}+\omega'\right)\right]\,.
\end{eqnarray}
The two functions $\mathcal{H}_{1}(\omega,\Omega)$ and
$\mathcal{H}_{2}(\omega,\Omega)$ originate from the integration over
the fast electronic energy $\varepsilon$, Figure~\ref{Fig-AA}, in
the active and passive layers, respectively. In the dc limit they
are given by
\begin{subequations}\label{Part-III-Spectral-F}
\begin{equation}
\mathcal{H}_{1}(\omega,\omega')=2-\mathcal{B}(\omega'+\omega/2)
-\mathcal{B}(\omega'-\omega/2)+\mathcal{B}(\omega)\,,
\end{equation}
\begin{equation}
\hskip-.5cm
\mathcal{H}_{2}(\omega,\omega')=T\frac{\partial}{\partial\omega'}
\left[\mathcal{B}(\omega'+\omega/2)-\mathcal{B}(\omega'-\omega/2)\right]\,,
\end{equation}
\begin{equation}
\hskip-4.3cm
\mathcal{B}(\omega)=\frac{\omega}{T}\coth\left(\frac{\omega}{2T}\right)\,.
\end{equation}
\end{subequations}
The corresponding  diagrams are constructed from the two vertices of
Figure~\ref{Fig-AA-vertex}: one for each of the layers, see
Figure~\ref{Fig-Drag-3rd}. It turns out that there are only two ways
to connect them, using the propagators $\mathcal{V}_{ab}\qo$, since
$\langle \EuScript{K}^{q}_{a} \EuScript{K}^{q}_b \rangle=0$.

In the following we assume identical layers and consider the
experimentally most relevant case of the long--ranged coupling,
where $\kappa d\gg 1$. Here $\kappa=2\pi e^2\nu$ is the
Thomas--Fermi inverse screening radius. In this limit the effective
interlayer interaction potential, Eqs.~\eqref{Part-III-V-drag},
acquires a simple form
\begin{equation}\label{Part-III-V12}
\mathcal{V}_{12}^R\qo=\frac{1}{g}\, \frac{1}{\kappa d
Dq^{2}-2i\omega}\,,
\end{equation}
where $g=\nu D$. Next, we substitute $\mathcal{D}^{R}\qo$ along
with~\eqref{Part-III-Spectral-F} and \eqref{Part-III-V12}
into~\eqref{Part-III-drag} and perform the energy and momentum
integrations. Inspection of the integrals shows that both energies
$\omega$ and $\omega'$ are of the order of the temperature
$\omega\sim\omega'\sim T$. On the other hand, the characteristic
value   of the transferred momenta is $q\sim q'\sim \sqrt{T/(D\kappa
d)}\ll \sqrt{T/D}$, cf.~\eqref{Part-III-V12}. Therefore, we may
disregard $Dq^2$ in comparison with $i\omega$ in the expressions for
${\mathcal D}_{a}^R\qo$, approximating the product ${\mathcal
D}_1^R{\mathcal D}_2^R$ in~\eqref{Part-III-drag} by $-\omega^{-2}$.
Such  a scale separation implies that the four--leg vertices are
effectively spatially local, while the three inter--layer
interaction lines are long--ranged.

Rescaling  energies by $T$ and momenta by $\sqrt{T/(D\kappa d)}$,
one may reduce expression \eqref{Part-III-drag} for the drag
conductivity  to $\sigma_{\mathrm{drag}}= R_Q^{-1}g^{-1}(\kappa
d)^{-2}\times $ ({\rm dimensionless integral}). The latter integral
does not contain any parameters, and may be evaluated
numerically~\cite{Drag-Third-order}. In the limit
$\sigma_{\mathrm{drag}}\ll g/R_Q$ the drag resistance
$\rho_{\mathrm{drag}}$ is given by
$\rho_{\mathrm{drag}}=\sigma_{\mathrm{drag}}R_Q^2/g^2$, resulting
finally in $\rho_{\mathrm{drag}}\approx 0.27R_{Q}\,g^{-3}(1/\kappa
d)^2$. This is the temperature--independent drag resistivity, which
may be larger than the second order (in the inter--layer
interactions) contribution. The latter goes to zero at small
temperatures as $T^2$. Further details and discussions can be found
in~\cite{Drag-Third-order}.

\section{Superconducting correlations}\label{sec_SuperCond}
\subsection{Generalization of the $\sigma$--model}\label{sec_SuperCond-1}

So far we have been discussing the {\em unitary} version of  Keldysh
$\sigma$--model, i.e. the one, where the time--reversal symmetry was
supposed to be broken by, e.g., external magnetic field. We now
switch to the {\em orthogonal} symmetry class, with the unbroken
time--reversal invariance. The case in point is superconducting
fluctuations in disordered metals. The Keldysh sigma--model,
generalized for the disordered superconductors was developed by
Feigel'man, Larkin and
Skvortsov~\cite{Feigel'manLarkinSkvortsov,SkvortsovLarkinFeigel'man}.
It is also applicable for treating weak--localization effects in
normal metals.

We proceed to describe disordered superconductors by adding the BCS
term to the Hamiltonian of a metal
$$\hat{H}_{\mathrm{BCS}}=-\frac{\lambda}{\nu}\int\d\mathbf{r}\,\hat{\psi}^{\dag}_{\uparrow}(\mathbf{r})
\hat{\psi}^{\dag}_{\downarrow}(\mathbf{r})
\hat{\psi}_{\downarrow}(\mathbf{r})
\hat{\psi}_{\uparrow}(\mathbf{r})\,,$$ which corresponds to the
short--range attraction in the particle--particle (Cooper) channel
mediated by electron--phonon interactions, where $\lambda$ is
dimensionless coupling constant. In a standard way
$\hat{H}_{\mathrm{BCS}}$ translates into the Keldysh action
$$S_{\mathrm{BCS}}=\frac{\lambda}{\nu}\int_{\mathcal{C}}\d
t\int\d\mathbf{r}\ \bar{\psi}_{\uparrow}\rt
\bar{\psi}_{\downarrow}\rt\psi_{\downarrow}\rt\psi_{\uparrow}\rt\,,$$
where the time integral is calculated along the Keldysh contour.
This four--fermion interaction term may be  decoupled via
Hubbard--Stratonovich transformation, by introducing an auxiliary
functional integral over the complex field $\Delta\rt$:
\begin{eqnarray}\label{SuperCond-BCS-HS}
\exp(iS_{\mathrm{BCS}})=\int\D[\Delta]\exp\left(i\int\d
x\left[-\frac{\nu}{\lambda}|\Delta(x)|^{2}+\Delta(x)
\bar{\psi}_{\uparrow}(x)\bar{\psi}_{\downarrow}(x)+
\Delta^{*}(x)\psi_{\downarrow}(x)\psi_{\uparrow}(x)\right]\right)\,,
\end{eqnarray}
here $x=\rt$ and $\int\d x=\int_{\mathcal{C}}\d t\int\d\mathbf{r}$.
To make further notations compact it is convenient to introduce a
bispinor fermionic vectors
$\Psi=1/\sqrt{2}(\psi_{\uparrow},\psi_{\downarrow},
\bar{\psi}_{\downarrow},-\bar{\psi}_{\uparrow})^{T}$ and
$\Psi^{+}=1/\sqrt{2}(\bar{\psi}_{\uparrow},
\bar{\psi}_{\downarrow},-\psi_{\downarrow},\psi_{\uparrow})$ defined
in the four--dimensional space $\Omega$, which can be viewed as the
direct product $S\otimes T$ of the spin
$(\psi_{\uparrow},\psi_{\downarrow})$ and time--reversal spaces
$(\psi,\bar{\psi})$. In principle, choice of the bispinors is not
unique. One can rearrange components of the bispinors in a different
manner, separating explicitly the
Gor'kov--Nambu~\cite{Gor'kov,Nambu} ($N$)
$(\psi_{\uparrow},\bar{\psi}_{\downarrow})$ and spin spaces. Finally
one may equally think of $\Psi$ as acting in the direct product of
the Nambu and time--reversal subspaces. These three representations
are equivalent $\Omega=S\otimes T\propto N\otimes S\propto N\otimes
T$ and the choice between them is dictated by convenience in
calculations for a particular problem at hand. In most cases we use
$N\otimes S$ choice and omit the $S$ part, since the theory is
diagonal in spin subspace. Vectors $\Psi$ and $\Psi^{+}$ are not
independent and related to each other
$\Psi^{+}=(\check{C}\Psi)^{T}$, by the charge--conjugation matrix
$\check{C}\equiv i\hat{\tau}_{y}\otimes\hat{s}_{x}$, where
$\hat{\tau}_{i}$ and $\hat{s}_{i}$, for $i=0,x,y,z$, are Pauli
matrices acting in the Nambu and spin subspaces, respectively; $\hat
\sigma_i$ matrices, as before, act in the Keldysh sub--space. To
avoid confusions, we shall specify, where appropriate, Keldysh and
Nambu sub--spaces by subscripts $K$ and $N$ correspondingly.

After the Hubbard--Stratonovich
transformation~\eqref{SuperCond-BCS-HS}, along with the standard
treatment of disorder and Coulomb interactions, the action appears
to be quadratic in fermion operators. Performing thus  Gaussian
Grassmann integration, one obtains for the disordered averaged
partition function
\begin{eqnarray}\label{SuperCond-Z}
&&\hskip-.5cm
\mathcal{Z}=\int\D[\Phi,\Delta]\exp\left(\frac{i}{2}\Tr\big\{\check{\Phi}
U^{-1}_{0}\check{\Upsilon}\check{\Phi}\big\}-
\frac{i\nu}{2\lambda}\Tr\big\{\check{\Delta}^{\dag}
\check{\Upsilon}\check{\Delta}\big\}\right)
\int\D[\check Q]\exp\big(iS[\check Q,\Delta,\mathbf{A},\Phi]\big)\,,\nonumber\\
&&\hskip-.5cm
 iS[\check{Q},\Delta,\mathbf{A},\Phi]=-\frac{\pi\nu}{4\tauel}
\Tr\big\{\check{Q}^{2}\big\}+\Tr\ln\left[\check{G}^{-1}+
\frac{i}{2\tauel}\check{Q}+\check{\Phi}+
\mathbf{v}_{F}\check{\Xi}\check{\mathbf{A}}+\check{\Delta}\right]\,,
\end{eqnarray}
which generalizes~\eqref{int-ferm-NLSM-action}. In the last equation
and throughout the rest of this chapter we use the check symbol
$\check{O}$ to denote $4\times4$ matrices acting in the $K\otimes N$
space, while hat symbol $\hat{O}$ for the $2\times2$ matrices acting
in  Nambu and Keldysh  subspaces. Equation \eqref{SuperCond-Z}
contains matrices
$\check{\Upsilon}=\hat{\sigma}_{x}\otimes\hat{\tau}_{0}$,
$\check{\Xi}=\hat{\sigma}_{0}\otimes\hat{\tau}_{z}$,
$\check{G}^{-1}=i\check{\Xi}\partial_{t}+\partial^{2}_{\mathbf{r}}/2m+\mu$,
and matrix fields
\begin{eqnarray}\label{SuperCond-A-Phi-Delta}
\check{\Phi}\rt=[\Phi^{cl}\rt\hat{\sigma}_{0}+\Phi^{q}\rt\hat{\sigma}_{x}]\otimes\hat{\tau}_{0}\,,\qquad
\check{\mathbf{A}}\rt=[\mathbf{A}^{cl}\rt\hat{\sigma}_{0}+\mathbf{A}^{q}\rt\hat{\sigma}_{x}]
\otimes\hat{\tau}_{0}\,,
\nonumber\\
\check{\Delta}\rt=[\Delta^{cl}\rt\hat{\sigma}_{0}+\Delta^{q}\rt\hat{\sigma}_{x}]\otimes\hat{\tau}_{+}-
[\Delta^{*cl}\rt\hat{\sigma}_{0}+\Delta^{*q}\rt\hat{\sigma}_{x}]\otimes\hat{\tau}_{-}\,,
\end{eqnarray}
with $\hat{\tau}_{\pm}=(\hat{\tau}_{x}\pm i\hat{\tau}_{y})/2$;
$\check{Q}$ matrix also has $4\times4$ structure in Keldysh and
Nambu spaces along with the matrix structure in the time domain.

We next perform the gauge transformation in~\eqref{SuperCond-Z} with
the help of $\EuScript{K}^{cl(q)}\rt$ fields, as in\footnote{In the
superconducting case the gauge transformation contains phase factors
$\exp(\pm i\check{\Xi}\check{\EuScript{K}})$, which is different
from~\eqref{int-ferm-NLSM-Q-K-gauge} by an extra matrix
$\check{\Xi}$ in the exponential.}~\eqref{int-ferm-NLSM-Q-K-gauge},
and expand the logarithm under the trace operation in gradients of
$\check{Q}_{\EuScript{K}}$ matrix (similar to the calculation
presented in  Section~\ref{sec_NLSM}). As a result, one obtains  the
action of disordered superconductors in the following form
\begin{subequations}\label{SuperCond-S}
\begin{equation}
S[\check Q,\Delta,\mathbf{A},\Phi]=S_{\Delta}+S_{\Phi}+S_{\sigma}\,,
\end{equation}
\begin{equation}
S_{\Delta}=-\frac{\nu}{2\lambda}
\Tr\big\{\check{\Delta}^{\dag}_{\EuScript{K}}
\check{\Upsilon}\check{\Delta}_{\EuScript{K}}\big\}\,,\qquad
S_{\Phi}=\frac{\nu}{2}\Tr\big\{\check{\Phi}_{\EuScript{K}}
\check{\Upsilon}\check{\Phi}_{\EuScript{K}}\big\}\,,
\end{equation}
\begin{equation}
S_{\sigma}=\frac{i\pi\nu}{4}\,
\Tr\big\{D\,(\hat{\bm{\partial}}_{\mathbf{r}}
\check{Q}_{\EuScript{K}})^{2}-4\check{\Xi}\partial_{t}\check{Q}_{\EuScript{K}}+
4i\check{\Phi}_{\EuScript{K}}\check{Q}_{\EuScript{K}}+4i
\check{\Delta}_{\EuScript{K}}\check{Q}_{\EuScript{K}}\big\}\,.
\end{equation}
\end{subequations}
Here gauged electromagnetic potentials $\check{\Phi}_{\EuScript{K}}$
and $\check{\mathbf{A}}_{\EuScript{K}}$ are related to the bare ones
$\check{\Phi}$ and $\check{\mathbf{A}}$ by~\eqref{int-ferm-NLSM-C},
while the gauged order parameter field is given by
\begin{equation}\label{SuperCond-Delta-K}
\check{\Delta}_{\EuScript{K}}\rt=\exp\Big(-i\check{\Xi}\check
{\EuScript{K}}\rt\Big)\check{\Delta}\rt
\exp\Big(i\check{\EuScript{K}}\rt\check{\Xi}\Big)\,.
\end{equation}
As compared with~\eqref{int-ferm-covariant-derivative} the covariant
spatial derivative in~(\ref{SuperCond-S}c) contains an extra
$\check{\Xi}$ matrix due to Nambu structure, i.e.
\begin{equation}\label{SuperCond-covariant-derivative}
\hat{\bm{\partial}}_{\mathbf{r}}\check{Q}_{\EuScript{K}}=
\partial_{\mathbf{r}}\check{Q}_{\EuScript{K}}
-i[\check{\Xi}\check{\mathbf{A}}_{\EuScript{K}},\check{Q}_{\EuScript{K}}]\,.
\end{equation}

Varying the action~\eqref{SuperCond-S} with respect to
$\check{Q}_{\EuScript{K}}$, under the constraint
$\check{Q}_{\EuScript{K}}^{2}=\check{1}$, yields the saddle point
equation
\begin{equation}\label{SuperCond-Usadel}
\hat{\bm{\partial}}_{\mathbf{r}}\big(D\,\check{Q}_{\EuScript{K}}\circ
\hat{\bm{\partial}}_{\mathbf{r}}\check{Q}_{\EuScript{K}}\big)-
\big\{\check{\Xi}\partial_{t},\check{Q}_{\EuScript{K}}\big\}_+ +
i\big[\check{\Phi}_{\EuScript{K}}+\check{\Delta}_{\EuScript{K}},
\check{Q}_{\EuScript{K}}\big]=0\,,
\end{equation}
which for $\check{\EuScript{K}}=0$ coincides with the dynamic Usadel
equation~\cite{LarkinOvchinnikov}. The classical solution of this equation is to be
sought in the  form
\begin{equation}
\check{Q}_{\EuScript{K}}=\left(\begin{array}{cc}\hat{Q}^{R}_{\EuScript{K}}
& \hat{Q}^{K}_{\EuScript{K}}
\\ 0 & \hat{Q}^{A}_{\EuScript{K}} \end{array}\right)_K\,,
\end{equation}
with retarded, advanced and Keldysh components being matrices in
Nambu subspace.

Varying the action with respect to the quantum component
$\Delta^{*q}\rt$ of the order parameter field, one finds the
self--consistency equation for the classical component of the order parameter
\begin{equation}\label{SuperCond-Self-consistency}
\Delta^{cl}_{\EuScript{K}}\rt=\pi\lambda
\Tr\big\{(\hat{\sigma}_{x}\otimes\hat{\tau}_{-})
\check{Q}_{\EuScript{K}}\big\}\,.
\end{equation}
Finally, varying the action with respect to the quantum components
$\Phi^{q}$ and $\mathbf{A}^{q}$ of the electromagnetic potentials
one obtains set of  Maxwell equations, which together with the
dynamic Usadel equation~\eqref{SuperCond-Usadel} and self--consistency
condition~\eqref{SuperCond-Self-consistency} represent the
closed system of equations governing dynamics of the superconductor.

In the generalized $\sigma$--model action~\eqref{SuperCond-S}, and
subsequent dynamical equations for $\check{Q}_{tt'}(\mathbf{r})$ and
$\check{\Delta}\rt$, all the relevant low--energy excitations have
been kept indiscriminately. The price one pays for this is the
technical complexity of the theory. In many practical cases this
exhaustive description is excessive and the theory may be
significantly simplified. For example, one often considers a
superconductor in the deep superconducting state $T\to0$, with well
defined gap $|\Delta|$, and studies dynamical responses when
perturbing frequency $\omega$ of the external field is small
$\omega\ll|\Delta|$, thus dealing with the quasi--stationary
conditions. For this case quasiclassical kinetic equations of
superconductor can be derived from~\eqref{SuperCond-Usadel}. As an
alternative, one may consider temperature range in the vicinity of
the transition $|T-T_{c}|\ll T_{c}$, where the order parameter is
small $|\Delta|\ll T_c$, and develop an effective theory of the
$\Delta\rt$ dynamics, i.e Ginzburg--Landau theory. Both
approximations follow naturally from the general $\sigma$--model
theory and will be considered in the next sections.

\subsection{Quasiclassical approximation}\label{sec_SuperCond-2}

In the superconducting state, choosing an optimal gauge field
$\vec{\EuScript{K}}(\mathbf{r},\epsilon)$ that is valid in the whole
energy range is a complicated task. However, it had been shown in
the~\cite{NarozhnyAleiner} that in the deep subgap limit
($\epsilon\ll|\Delta|$) the effect of the electric potential on the
quasiclassical Green's function $\check{Q}$ is small in the
parameter $\epsilon/|\Delta|\ll1$ and hence as an approximation one
may set $\vec{\EuScript{K}}(\mathbf{r},\epsilon)=0$. This assumption
will be used below~\footnote{Within this section the subscript
$\EuScript{K}$ is suppressed in the notations of
$\check{Q}_{\EuScript{K}}$ matrix,
$\check{Q}_{\EuScript{K}}\to\check{Q}$, and all other gauged
fields.}.

In a spatially uniform, equilibrium superconductor the saddle point
Usadel equation is solved by the the following $\check{Q}$--matrix
\begin{equation}\label{SuperCond-Q-uniform}
\hat{Q}^{R(A)}(\epsilon)=\pm\frac{1}{\sqrt{(\epsilon\pm
i0)^{2}-|\Delta|^{2}}}\left(\begin{array}{cc}\epsilon &\Delta
\\ -\Delta^{*} & -\epsilon\end{array}\right)_N\,,
\end{equation}
while $\hat Q^K=\tanh\frac{\epsilon}{2T}(\hat Q^R-\hat Q^A)$. We
have suppressed superscript $cl$, writing the order parameter as
$\Delta$ (its quantum component will not appear within this
section). Substituting~\eqref{SuperCond-Q-uniform} into the
self--consistency condition~\eqref{SuperCond-Self-consistency}, one
obtains the standard BCS gap equation
\begin{equation}\label{SuperCond-BCS-gap}
\Delta=\lambda\Delta\int^{\omega_{D}}_{|\Delta|}\frac{\d\epsilon}
{\sqrt{\epsilon^{2}-|\Delta|^{2}}}\tanh\frac{\epsilon}{2T}\,,
\end{equation}
which has a non--zero solution for $|\Delta|$ below a critical
temperature $T_{c}$.

In  presence of  boundaries or proximity to a normal metal one faces
the problem of spatially non--uniform superconductivity. In this
case, both $\Delta$ and $\hat Q^{R(A)}$ acquire a coordinate
dependence and one should look for a solution
of~\eqref{SuperCond-Usadel} and \eqref{SuperCond-Self-consistency}.
In doing so, we will assume that $\check{Q}_{tt'}$ is static, i.e.
independent of  the central time and pass to the Wigner transform
representation. From the retarded block of the $4\times4$ matrix
Usadel equation at $\Phi=0$ and $\mathbf{A}=0$ we obtain
\begin{equation}\label{SuperCond-Usadel-Q-R}
\partial_{\mathbf{r}}\big(D\,\hat{Q}^{R}\partial_{\mathbf{r}}
\hat{Q}^{R}\big)+i\epsilon[\hat{\tau}_{z},\hat{Q}^{R}]+
i[\hat{\Delta},\hat{Q}^{R}]=0\,.
\end{equation}
With the similar equation  for the advanced block of the matrix
Usadel Equation~\eqref{SuperCond-Usadel}. The Keldysh sector
provides another equation, which is
\begin{equation}\label{SuperCond-Usadel-Q-K}
\partial_{\mathbf{r}}\big(D\, \hat{Q}^{R}\partial_{\mathbf{r}}
\hat{Q}^{K}+ D\, \hat{Q}^{K}\partial_{\mathbf{r}}\hat{Q}^{A}\big)
+i\epsilon[\hat{\tau}_{z},\hat{Q}^{K}]+i[\hat{\Delta},\hat{Q}^{K}]=0\,.
\end{equation}
The non--linear constraint $\check{Q}^{2}=\check{1}$  imposes the
following conditions
\begin{equation}\label{SuperCond-Normalizations}
\hat{Q}^{R}\hat{Q}^{R}=\hat{Q}^{A}\hat{Q}^{A}=\hat{1}\,,\quad
\hat{Q}^{R}\hat{Q}^{K}+\hat{Q}^{K}\hat{Q}^{A}=0\,.
\end{equation}
They may be explicitly resolved by the angular
parametrization~\cite{BelzigWilhelm} for the retarded and advanced
blocks of the Green's function matrix:
\begin{subequations}\label{SuperCond-Q-R-A}
\begin{equation}
\hskip-2cm \hat{Q}^{R}(\mathbf{r},\epsilon)=\left(\begin{array}{cc}
\cosh\theta & \sinh\theta \exp(i\chi)\\-\sinh\theta \exp(-i\chi) &
-\cosh\theta
\end{array}\right)_N\,,
\end{equation}
\begin{equation}
\hat{Q}^{A}(\mathbf{r},\epsilon)=-\hat{\tau}_{z}\big[\hat{Q}^{R}\big]^{\dag}\hat{\tau}_{z}=
\left(\begin{array}{cc} -\cosh\bar{\theta} & -\sinh\bar{\theta}
\exp(i\bar{\chi})\\ \sinh\bar{\theta} \exp(-i\bar{\chi}) &
\cosh\bar{\theta}
\end{array}\right)_N\, ,
\end{equation}
\end{subequations}
where $\theta(\mathbf{r},\epsilon)$ and $\chi(\mathbf{r},\epsilon)$
are complex, coordinate-- and energy--dependent scalar functions. As
to the Keldysh component, it can be always chosen as
\begin{equation}\label{SuperCond-Q-K}
\hat{Q}^{K}=\hat{Q}^{R}\circ\hat{F}-\hat{F}\circ\hat{Q}^{A},
\end{equation}
where $\hat{F}$ may be thought of as a generalized matrix
distribution function. Following
Schmidt--Sch\"{o}n~\cite{SchmidtSchon}, and
Larkin--Ovchinnikov~\cite{LarkinOvchinnikov-F} we choose
\begin{equation}\label{SuperCond-F}
\hat{F}(\mathbf{r},\epsilon)=\left(
\begin{array}{cc}
F_{L}(\mathbf{r},\epsilon)+F_{T}(\mathbf{r},\epsilon) & 0 \\ 0 &
F_{L}(\mathbf{r},\epsilon)-F_{T}(\mathbf{r},\epsilon)
\end{array}\right)_{N}=F_{L}(\mathbf{r},\epsilon)\hat{\tau}_{0}+
F_{T}(\mathbf{r},\epsilon)\hat{\tau}_{z}\,,
\end{equation}
where abbreviations $F_{L(T)}$ refer to the \textit{longitudinal}
and \textit{transverse} components of the distribution function with
respect to the order parameter. Physically $F_{T}$ corresponds to the charge
mode of the system and  determines the electric current density, while
$F_{L}$ corresponds to the energy mode, determining the heat (energy) current
(further discussions  may be found in  books of
Tinkham~\cite{Tinkham} and Kopnin~\cite{Kopnin}).

Substituting  $\hat{Q}^{R}$ in the form of~\eqref{SuperCond-Q-R-A}
into~\eqref{SuperCond-Usadel-Q-R}, one finds from the diagonal
elements of the corresponding matrix equation
\begin{equation}\label{SuperCond-Usadel-1}
D\, \partial_{\mathbf{r}}\big(
\sinh^{2}\theta\,\partial_{\mathbf{r}}\chi\big)
=2i|\Delta|\sinh\theta\sin(\varphi-\chi)\,,
\end{equation}
where the order parameter is parameterized as
$\Delta(\mathbf{r})=|\Delta(\mathbf{r})|\exp\{i\varphi(\mathbf{r})\}$.
From the off--diagonal block of the matrix
Equation~\eqref{SuperCond-Usadel-Q-R},
using~\eqref{SuperCond-Usadel-1}, one obtains
\begin{equation}\label{SuperCond-Usadel-2}
D\,\partial^{2}_{\mathbf{r}}\theta+2i\epsilon\sinh\theta-2i|\Delta|\cosh\theta\cos(\varphi-\chi)
=\frac{D}{2}\big(\partial_{\mathbf{r}}\chi\big)^{2}\sinh2\theta\,.
\end{equation}
We proceed with the equation for the Keldysh component of the
Green's function matrix $\hat{Q}^{K}$. Using
decomposition~\eqref{SuperCond-Q-K} and substituting it into
Eq.~\eqref{SuperCond-Usadel-Q-K},  one obtains
\begin{eqnarray}
D\left(\partial^{2}_{\mathbf{r}}\hat{F}+
\hat{Q}^{R}\partial_{\mathbf{r}}\hat{Q}^{R}\partial_{\mathbf{r}}\hat{F}-
\partial_{\mathbf{r}}\hat{F}\hat{Q}^{A}\partial_{\mathbf{r}}\hat{Q}^{A}-
\partial_{\mathbf{r}}\big(\hat{Q}^{R}\partial_{\mathbf{r}}\hat{F}\hat{Q}^{A}\big)\right)
+i\epsilon\left(\hat{Q}^{R}\big[\hat{\tau}_{z},\hat{F}]-
[\hat{\tau}_{z},\hat{F}\big]\hat{Q}^{A}\right)\nonumber \\
+i\left(\hat{Q}^{R}\big[\hat{\Delta},\hat{F}]-
[\hat{\Delta},\hat{F}\big]\hat{Q}^{A}\right)=0\,.
\end{eqnarray}
Now using~\eqref{SuperCond-F} for $\hat{F}$ and: (i) taking Nambu
trace of the above matrix equation; (ii) multiplying the above
equation by $\hat{\tau}_{z}$ and then tracing it; one finds two
coupled kinetic equations for the non--equilibrium distribution
junctions $F_{L(T)}$, which can be written in the form of
conservation laws~\cite{LarkinOvchinnikov-KinEq}
\begin{subequations}\label{SuperCond-Usadel-3}
\begin{equation}
\partial_{\mathbf{r}}\big(\EuScript{D}_{L}\partial_{\mathbf{r}}
F_{L}-D\partial_{\mathbf{r}} F_{T}Y\big)+D\partial_{\mathbf{r}}
F_{T}\mathcal{J}_{S}=\mathcal{I}^{a}_{\mathrm{coll}}\,,
\end{equation}
\begin{equation}
\partial_{\mathbf{r}}(\EuScript{D}_{T}\partial_{\mathbf{r}}
F_{T}+D\partial_{\mathbf{r}} F_{L}Y)+D\partial_{\mathbf{r}}
F_{L}\mathcal{J}_{S}=\mathcal{I}^{b}_{\mathrm{coll}}\,.
\end{equation}
\end{subequations}
Here we have introduced  energy-- and coordinate--dependent
diffusion coefficients
\begin{subequations}\label{SuperCond-kinetics-D}
\begin{equation}
\hskip-1.5cm \EuScript{D}_{L}(\mathbf{r},\epsilon)=\frac{D}{4}
\Tr\left\{\hat{\tau}_{0}-\hat{Q}^{R}\hat{Q}^{A}\right\}_N
=\frac{D}{2}\left[1+|\cosh\theta|^{2}-
|\sinh\theta|^{2}\cosh\big(2\mathrm{Im}[\chi]\big)\right]\,,
\end{equation}
\begin{equation}
\EuScript{D}_{T}(\mathbf{r},\epsilon)=\frac{D}{4}\Tr\left\{\hat{\tau}_{0}-\hat{\tau}_{z}
\hat{Q}^{R}\hat{\tau}_{z}\hat{Q}^{A}\right\}_N = \frac{D}{2}\left[1+|\cosh\theta|^{2}+
|\sinh\theta|^{2}\cosh\big(2\mathrm{Im}[\chi]\big)\right]\,,
\end{equation}
\end{subequations}
density of the supercurrent carrying states
\begin{equation}\label{SuperCond-kinetics-Js}
\mathcal{J}_{S}(\mathbf{r},\epsilon)=\frac{1}{4}\Tr
\left\{\hat{\tau}_{z}\big(\hat{Q}^{R}\partial_{\mathbf{r}}\hat{Q}^{A}-
\hat{Q}^{A}\partial_{\mathbf{r}}\hat{Q}^{R}\big)\right\}_N
=-\mathrm{Im}\left(\sinh^{2}\theta\,\partial_{\mathbf{r}}\chi\right)\,,
\end{equation}
and the spectral function
\begin{equation}\label{SuperCond-kinetics-Gamma}
Y(\mathbf{r},\epsilon)=\frac{1}{4}\Tr\left\{\hat{Q}^{R}\hat{\tau}_{z}\hat{Q}^{A}\right\}_N
=\frac{1}{2}|\sinh\theta|^{2}\sinh\big(2\mathrm{Im}[\chi]\big)\,.
\end{equation}
Finally, the right hand side of~\eqref{SuperCond-Usadel-3} contains
the collision integrals
\begin{subequations}\label{SuperCond-kinetics-Coll}
\begin{equation}
\mathcal{I}^{a}_{\mathrm{coll}}=\frac{F_{T}}{2}\, \Tr\left\{\hat{\tau}_{z}
\big(\hat{Q}^{R}\hat{\Delta}+\hat{\Delta}\hat{Q}^{A}\big)\right\}_N =2F_{T}|\Delta|
\mathrm{Re}\left[\sinh\theta\sin(\varphi-\chi)\right]\,,
\end{equation}
\begin{equation}
\mathcal{I}^{b}_{\mathrm{coll}}=\frac{F_{T}}{2}\, \Tr\left\{
\hat{Q}^{R}\hat{\Delta}+\hat{\Delta}\hat{Q}^{A} \right\}_N
=-2F_{T}|\Delta|\mathrm{Im}
\left[\sinh\theta\cos(\varphi-\chi)\right]\,.
\end{equation}
\end{subequations}
Collision integrals associated with the inelastic electron--electron
and electron--phonon interactions are not discussed here, one may
find corresponding derivations in the book of Kopnin~\cite{Kopnin}.
Equations~\eqref{SuperCond-Usadel-1}, \eqref{SuperCond-Usadel-2} and
\eqref{SuperCond-Usadel-3}, together with the spectral
quantities~\eqref{SuperCond-kinetics-D}--\eqref{SuperCond-kinetics-Coll}
represent a complete set of kinetic equations for disordered
superconductors applicable within quasi--classical approximation.
These equations are supplemented by the self--consistency relation,
see~\eqref{SuperCond-Self-consistency}
\begin{equation}\label{SuperCond-Usadel-4}
\Delta(\mathbf{r})=\frac{\lambda}{2}\int\d\epsilon\,\Big\{[\sinh\theta
\exp(i\chi)+\sinh\bar{\theta}\exp(i\bar{\chi})]F_{L}-\big[\sinh\theta
\exp(i\chi)-\sinh\bar{\theta}\exp(i\bar{\chi})\big]F_{T}\Big\}\,,
\end{equation}
and the boundary conditions for the Green's functions, expressing
the current
continuity~\cite{Nazarov-TunnelingAction,Zaitsev,KuprianovLukichev,Lambert},
\begin{equation}\label{SuperCond-BoundaryConditions}
\sigma_{L}\mathcal{A}_{L}\check{Q}_{L}\partial_{\mathbf{r}}\check{Q}_{L}=
\sigma_{R}\mathcal{A}_{R}\check{Q}_{R}\partial_{\mathbf{r}}\check{Q}_{R}=
\mathrm{g}_{T}[\check{Q}_{L},\check{Q}_{R}]\,,
\end{equation}
where $\sigma$ and $\mathcal{A}$ are the bulk normal--state
conductivity and the cross section of the wire next to the
interface, $L/R$ denote left/right from the interface, respectively,
and $\mathrm{g}_{T}$ is the interface tunneling conductance.

An analytic solution of the system of kinetic equations
\eqref{SuperCond-Usadel-1}--\eqref{SuperCond-Usadel-3} is rarely
possible. In general, one has to rely on numerical methods. To find
solution for a given transport problem, one should proceed as
follows~\cite{BelzigWilhelm}.
\begin{enumerate}
  \item Start with a certain $\Delta(\mathbf{r})$. Usually one takes
  $\Delta=\mathrm{const}$ everywhere in the superconductors and
  $\Delta=0$ in the normal metals.
  \item Solve Usadel equations \eqref{SuperCond-Usadel-1}--\eqref{SuperCond-Usadel-2}
  for the retarded Green function, thus determining spectral angles
  $\theta(\mathbf{r},\epsilon)$ and $\chi(\mathbf{r},\epsilon)$.
  \item Use these solutions to calculate spectral kinetic
  quantities $\EuScript{D}_{L,T}(\mathbf{r},\epsilon)$, $\mathcal{J}_{S}(\mathbf{r},\epsilon)$
  and $Y(\mathbf{r},\epsilon)$.
  \item Solve kinetic equations \eqref{SuperCond-Usadel-3} for $F_{L/T}(\mathbf{r},\epsilon)$.
  \item Calculate  new $\Delta(\mathbf{r})$ from
  Equation \eqref{SuperCond-Usadel-4}, and iterate
  this procedure until the self--consistency is achieved.
\end{enumerate}
Having solved the kinetic equations one may determine physical
quantities of interest. For example, for the electric current one
finds $\mathbf{j}=\mathbf{j}_{n}+\mathbf{j}_{s}$, where
$\mathbf{j}_{n}(\mathbf{r})=\nu\int\d\epsilon\,
\EuScript{D}_{T}(\mathbf{r},\epsilon)
\partial_{\mathbf{r}} F_{T}(\mathbf{r},\epsilon)$ is the normal component and
$\mathbf{j}_{s}(\mathbf{r})=\nu D\int\d\epsilon
F_{L}(\mathbf{r},\epsilon)\mathcal{J}_{S}(\mathbf{r},\epsilon)$
is the supercurrent density.

The quasiclassical kinetic theory of disordered superconductors,
outlined above,  may be applied to study various phenomena. To name
a few: the proximity related problems in the superconductor--normal
metal heterostructures~\cite{BelzigBruder,Gueron,Zhou,Hammer},
non--equilibrium Josephson effect~\cite{Heikkila,Wilhelm}, Hall
effect~\cite{ZhouSpivak}, thermoelectric
phenomena~\cite{StoofNazarov,Virtanen} in superconductors, shot
noise~\cite{Stenberg}, engineering of non--equilibrium distribution
functions~\cite{Crosser} and many other problems may be successfully
tackled with the help
of~\eqref{SuperCond-Usadel-1}--\eqref{SuperCond-Usadel-3}. Several
relatively simple (equilibrium) examples are considered  in
Section~\ref{app_Part-IV} for illustration.

\subsection{Time dependent Ginzburg--Landau theory}\label{sec_SuperCond-3}

Gor'kov~\cite{Gor'kov-GL} had shown that the phenomenological
Ginzburg--Landau (GL) theory~\cite{GinzburgLandau} follows naturally
from the microscopic BCS model in the limit when temperature is
close to the critical one $|T-T_{c}|\ll T_{c}$. Later Gor'kov and
Eliashberg~\cite{Gor'kovEliashberg} extended application of the
Ginzburg--Landau theory to include time dependent dynamical
phenomena. It was revisited in a number of  subsequent
publications~\cite{HoughtonMaki,HuThompson,Kramer,SchonAmbegaokar,Hu,Krempasky,Otterlo}
and books~\cite{Tinkham,Kopnin,Larkin-Varlamov}. Within the
$\sigma$--model terminology the static GL functional may be obtained
by means of supersymmetric~\cite{AltlandSimons} or
replica~\cite{YurkevichLerner} approaches. Here we discuss the
dynamic theory in Keldysh formulation~\cite{LevchenkoKamenev}.

The way dynamical time dependent Ginzburg--Landau (TDGL) theory is
derived from~\eqref{SuperCond-S} allows to formulate it in terms of
the effective action, rather than the equation for the order
parameter only, as it is done in a traditional way. As a result, in
addition to the average quantities one has an access to fluctuation
effects, since TDGL action  contains the stochastic noise term,
which serves to satisfy the fluctuation--dissipation theorem.
Moreover, one may naturally and unmistakably identify an anomalous
Gor'kov--Eliashberg (GE) term~\cite{Gor'kovEliashberg}, which
preserves gauge invariance of the theory, along with the
Aslamazov--Larkin (AL)~\cite{AslamazovLarkin} , Maki--Thompson
(MT)~\cite{Maki}  and density of states (DOS) terms~\cite{Hurault},
which renormalize the conductivity and  single particle density of
states owing to superconductive fluctuations. Although
Aslamazov--Larkin term is correctly captured by most of the
approaches to TDGL equation, Gor'kov--Eliashberg, Maki--Thompson and
DOS are frequently lost in many  works on TDGL.

The strategy of deriving the  effective TDGL theory starting from
the general $\sigma$--model action~\eqref{SuperCond-S} is as
follows. (i) Choose a parametrization of a saddle point $\check Q$
matrix manifold, which resolves the non--linear constraint $\check
Q^2=\check{1}$. (ii) Integrates out Gaussian fluctuations around the
saddle point and (iii) keeps terms up to the second order in all
\textit{quantum} fields (the order parameter $\Delta$ and
electromagnetic potentials $\Phi$ and $\mathbf{A}$) in the resulting
action. (iv) Rely on the assumption that the electronic system is
always in a local thermal equilibrium. This in turn implies that the
external fields are not too large. More precisely, the electric
field $\mathbf{E}$ is such that $e|\mathbf{E}|\xi_0\ll T_c$, while
the magnetic field $\mathbf{H}$ is restricted by the condition
$e|\mathbf{H}|\xi_0\ll1/\xi_0$, where $\xi_{0}=\sqrt{D/T_{c}}$ is
superconductive coherence length. The restrictions on spatial and
temporal scales of the external fields along with the fact that
electrons are in local equilibrium considerably simplify the theory.
In particular, most of the terms in the effective action acquire a
local form in space and time. Nevertheless, the effective theory
does not take a completely local form.

This procedure is relatively straightforward in the case of gapless
superconductivity. The latter occurs either in the presence of
magnetic impurities, or in the fluctuating regime above the critical
temperature $T\gtrsim T_{c}$. In the gapped phases, $T\lesssim
T_{c}$, the situation becomes more complicated. As noted by Gor'kov
and Eliashberg~\cite{Gor'kovEliashberg}, the difficulty stems from
the singularity of the BCS density of states at the gap edge. The
latter leads to a slowly decaying oscillatory response at frequency
$2\Delta/\hbar$ in the time domain. As a result, the expansion in
powers of the small parameter $\Delta/T_{c}\ll 1$ fails. In
principle, it may be augmented by an expansion in
$\Delta/(\hbar\omega)$, in case of high--frequency external fields.
To describe low--frequency responses in the gapped phase, one needs
a time \textit{non--local} version of the TDGL theory. The analysis
is greatly simplified in the presence of a pair--breaking mechanism,
such as magnetic impurities or energy relaxation. Such a mechanism
may eliminate singularity in the density of states, leading to
gapless phase in the presence of finite $\Delta$. Under these
conditions, an expansion in powers of $\Delta\tau_{\phi}/\hbar\ll 1$
and $\omega\tau_{\phi}\ll 1$ is justified and thus a time--local
TDGL equation may be recovered (here $\tau_{\phi}$ is the
pair--breaking time). Within this section only fluctuating regime,
$T\gtrsim T_c$, will be considered. In this case the spectrum is
gapless automatically and there is no need in an explicit
pair--breaking mechanism.

Proceeding along the steps (i)--(iv), outlined above, one recalls
that at $T>T_{c}$ energy gap self--consistency Equation
\eqref{SuperCond-Self-consistency} has only trivial solution with
$\langle\Delta^{cl}\rangle=0$. Thus the trial saddle point of the
action \eqref{SuperCond-S} collapses back to the metallic state
$\check{Q}_{\EuScript{K}}=\check{\Lambda}=\hat{\Lambda}\otimes\hat{\tau}_{z}$,
where $\hat{\Lambda}$ is defined by~\eqref{NLSM-Lambda}. The
Gaussian integration around this $\check{Q}_{\EuScript{K}}$ includes
Cooper modes, which are accounted for in the following
parametrization of $\Qk$--matrix:
\begin{equation}\label{SuperCond-Q-W}
    \check{Q}_{\EuScript{K}}=\check{\mathcal{U}}\circ
e^{-\check{\mathcal{W}}/2}\circ(\hat{\sigma}_{z}\otimes\hat{\tau}_{z})\circ
e^{\hat{\mathcal{W}}/2}\circ\check{\mathcal{U}}^{-1}\,,
\end{equation}
with the following choice of the fluctuation matrix
\begin{equation}\label{SuperCond-W}
\check{\mathcal{W}}_{tt'}(\mathbf{r})= \left(\begin{array}{cc}
c_{tt'}(\mathbf{r})\hat{\tau}_{+}-c^{*}_{tt'}(\mathbf{r})\hat{\tau}_{-}
&
d_{tt'}(\mathbf{r})\hat{\tau}_{0}+d^{z}_{tt'}(\mathbf{r})\hat{\tau}_{z}
\\
\bar{d}_{tt'}(\mathbf{r})\hat{\tau}_{0}+
\bar{d}^{z}_{tt'}(\mathbf{r})\hat{\tau_{z}} &
\bar{c}_{tt'}(\mathbf{r})\hat{\tau}_{+}-
\bar{c}^{*}_{tt'}(\mathbf{r})\hat{\tau}_{-}
\end{array}\right)_K\,,\qquad
\check{\mathcal{U}}=\check{\mathcal{U}}^{-1}=
\left(\begin{array}{cc} 1 & F \\ 0 & -1
\end{array}\right)_K\otimes\hat{\tau}_{0}\,.
\end{equation}
As compared with~\eqref{NLSM-U-W}, $\check{\mathcal{W}}$ contains
twice as many diffusive modes, which are described by four Hermitian
matrices in time subspace: $\{d,\bar{d}\}$ and
$\{d^{z},\bar{d}^{z}\}$. It also contains the Cooper modes described
by two  independent {\em complex} matrix fields $\{c,\bar{c}\}$. Now
substitutes the $\check{\mathcal{W}}$--dependent $\Qk$ matrix
$\check{Q}_{\EuScript{K}}[\check{\mathcal{W}}]$
into~\eqref{SuperCond-S} and expands the action up to the second
order in $\check{\mathcal{W}}$ fluctuations: $S[\check
Q,\Delta,\mathbf{A},\Phi]\Rightarrow
S[\check{\mathcal{W}},\Delta,\mathbf{A},\Phi]$. After this step the
Gaussian integration over $\check{\mathcal{W}}$ is possible (see
details of this procedure in
Appendix~\ref{app_FluctuationExpansion})
\begin{equation}\label{SuperCond-W-integration}
 \int\D[\check{\mathcal{W}}]\exp\Big(iS[\check{\mathcal{W}},\Delta,\mathbf{A},\Phi]\Big)=
 \exp\Big(iS_{\mathrm{eff}}[\Delta,\mathbf{A},\Phi]\Big)\,,
\end{equation}
which leads eventually to the effective TDGL action. It consists of
several contributions:
\begin{eqnarray}\label{SuperCond-S-eff}
S_{\mathrm{eff}}[\Delta,\mathbf{A},\Phi]=S_{\mathrm{N}}[\mathbf{A},\Phi]+
S_{\mathrm{GL}}[\Delta,\mathbf{A},\Phi]+
S_{\mathrm{SC}}[\Delta,\mathbf{A},\Phi]+S_{\mathrm{MT}}[\Delta,\mathbf{A},\Phi]+
S_{\mathrm{DOS}}[\Delta,\mathbf{A},\Phi]\,,
\end{eqnarray}
which we describe in order.

The action $S_{\mathrm{N}}[\mathbf{A},\Phi]$ is the normal metal
part of~\eqref{SuperCond-S}, which is obtained from $S[\check
Q,\Delta,\mathbf{A},\Phi]$ by setting
$\check{Q}_{\EuScript{K}}=\check{\Lambda}$ and $\check{\Delta}=0$.
It reads as~\footnote{Note that in~\eqref{SuperCond-S-N} and
throughout the rest of this section we have restored electron charge
$e$ accompanying source fields $\mathbf{A}\to e\mathbf{A}$ and
$\Phi\to e\Phi$, such that $\mathbf{A}$ and $\Phi$ are now actual
electromagnetic potentials, see Note 8.}
\begin{equation}\label{SuperCond-S-N}
S_{\mathrm{N}}[\mathbf{A},\Phi]=e^{2}\nu
D\,\Tr\left\{\vec{\mathbf{A}}^{T}_{\EuScript{K}}
\left(\begin{array}{cc}0 &
D\,\partial^{2}_{\mathbf{r}}-\overleftarrow{\partial}_{t}
\\ D\,\partial^{2}_{\mathbf{r}}-\overrightarrow{\partial}_{t} & 4iT
\end{array}\right)_K \vec{\mathbf{A}}_{\EuScript{K}}\right\}\,,
\end{equation}
where arrows on top of the time derivative indicate direction of differentiation. Since
our starting point is the normal saddle point \eqref{NLSM-Lambda},
 $\vec{\EuScript{K}}[\Phi]$ functional is  given by
Eq.~\eqref{int-ferm-NLSM-K-functional} and gauged vector potential
$\mathbf{A}_{\EuScript{K}}$ is defined by
Eq.~\eqref{int-ferm-NLSM-C}.

The $S_{\mathrm{GL}}$ is the time dependent Ginzburg--Landau part of
the action
\begin{equation}\label{SuperCond-S-GL}
S_{\mathrm{GL}}[\Delta,\mathbf{A},\Phi]=
2\nu\Tr\left\{\vec{\Delta}^{\dag}_{\EuScript{K}}\rt\hat{L}^{-1}
\vec{\Delta}_{\EuScript{K}}\rt\right\}\, ,
\end{equation}
which governs time and space variations of the order parameter under
the influence of external potentials. The effective propagator
$\hat{L}^{-1}$ has the typical bosonic structure in the Keldysh space
\begin{equation}
\hat{L}^{-1}=\left(\begin{array}{cc} 0 & L^{-1}_{A} \\ L^{-1}_{R} &
L^{-1}_K
\end{array}\right)_K\,,
\end{equation}
with the components given by
\begin{subequations}\label{SuperCond-L-R-A}
\begin{equation}
L^{-1}_{R(A)}=
\frac{\pi}{8T_c}\left[\mp\partial_{t}-\tau^{-1}_{\mathrm{GL}}+D\big(\partial_{\mathbf{r}}-
2ie\mathbf{A}^{cl}_{\EuScript{K}}\big)^{2}-\frac{7\zeta(3)}{\pi^{3}T_{c}}
|\Delta^{cl}_{\EuScript{K}}|^{2}\right]\,,
\end{equation}
\begin{equation}
\hskip-2.8cm L^{-1}_{K}= \coth{\omega\over 2T}\big[L_R^{-1}(\omega)
- L_A^{-1}(\omega)\big] \approx  {i\pi\over 2}\,,
\end{equation}
\end{subequations}
where $\omega\ll T\approx T_c$ and Ginzburg--Landau relaxation time
is defined as $\tau_{\mathrm{GL}}=\pi/8(T-T_{c})$. Note that under
the assumption $T-T_c\ll T_c$,  GL part of the action acquires a
time--local form.

The $S_{\mathrm{SC}}$ part of the action is responsible for the
super--current
\begin{equation}\label{SuperCond-S-SC}
S_{\mathrm{SC}}[\Delta,\mathbf{A},\Phi]=\frac{\pi e\nu
D}{2T}\Tr\left\{\mathbf{A}^{q}_{\EuScript{K}}\mathrm{Im}
\big[\Delta^{*cl}_{\EuScript{K}}
\big(\partial_{\mathbf{r}}-2ie\mathbf{A}^{cl}_{\EuScript{K}}\big)
\Delta^{cl}_{\EuScript{K}}\big]\right\}\,.
\end{equation}
The abbreviation is due to the fact that $S_{\mathrm{SC}}$, being
differentiated with respect to $\mathbf{A}^{q}$, provides standard
expression for the super--current in terms of the order
parameter~\cite{Tinkham}.

The Maki--Thompson part of the action, $S_{\mathrm{MT}}$, is
responsible for renormalization of the diffusion coefficient in the
normal action $S_\mathrm{N}$ owing to the superconductive
fluctuations. It reads as
\begin{equation}\label{SuperCond-S-MT}
S_{\mathrm{MT}}[\Delta,\mathbf{A},\Phi]=
e^{2}\nu\Tr\left\{\vec{\mathbf{A}}^{T}_{\EuScript{K}}\rt\hat{\mathcal{T}}_{\delta
D}(t,t')\vec{\mathbf{A}}_{\EuScript{K}}\rtp\right\}\,,
\end{equation}
where the operator $\hat{\mathcal{T}}_{\delta D}(t,t')$ is given by
\begin{equation}
\hat{\mathcal{T}}_{\delta D}=\left(\begin{array}{cc} 0 &
-\overleftarrow{\partial_{t}}\,
\delta D^{\mathrm{MT}}_{\mathbf{r},t',t} \\
-\delta D^{\mathrm{MT}}_{\mathbf{r},t,t'}\,
\overrightarrow{\partial_{t'}} & 2iT \left( \delta
D^{\mathrm{MT}}_{\mathbf{r},t,t'} + \delta
D^{\mathrm{MT}}_{\mathbf{r},t',t} \right)
\end{array}\right)_K\,.
\end{equation}
The diffusion coefficient correction $\delta
D^{\mathrm{MT}}[\Delta_{\EuScript{K}}]$ is the non--local functional
of the fluctuating order parameter
\begin{equation} \label{SuperCond-D-MT}
\delta D^{\mathrm{MT}}_{\mathbf{r},t,t'}=\frac{\pi
D}{4T}\int\mathrm{d} \mathbf{r}'\mathrm{d}\mathbf{r}''\,
 \mathcal{C}^{\,\mathbf{r},\mathbf{r}'}_{\tau,t,t'}
 {\Delta}^{*cl}_\EuScript{K}\left(\mathbf{r}',\tau\right)
 {\Delta}^{cl}_\EuScript{K}(\mathbf{r}'',\tau)\
\bar{\mathcal{C}}^{\,\mathbf{r}'',\mathbf{r}}_{\tau,t',t}\,,
\end{equation}
where $\tau=(t+ t')/2\,$. The retarded
$\mathcal{C}^{\,\mathbf{r},\mathbf{r}'}_{\tau,t,t'}\sim
\theta(t-t')$  and advanced
$\bar{\mathcal{C}}^{\,\mathbf{r},\mathbf{r}'}_{\tau,t,t'}\sim
\theta(t'-t)$ Cooperon propagators are Green's functions of the
following equations:
\begin{subequations}
\begin{equation}
\left\{\phantom{-}\partial_t
-ie\Phi^{cl}_{\EuScript{K}}(\mathbf{r},\tau_+)+
ie\Phi^{cl}_{\EuScript{K}}(\mathbf{r},\tau_-)-D\left[\partial_{\mathbf{r}}-
ie\mathbf{A}^{cl}_\EuScript{K}(\mathbf{r},\tau_+)-
ie\mathbf{A}^{cl}_\EuScript{K}(\mathbf{r},\tau_-)\right]^2 \right\}
\mathcal{C}^{\mathbf{r},\mathbf{r}'}_{\tau,t,t'} =
\delta_{\mathbf{r}-\mathbf{r}'}\delta_{t-t'}\ ,
\end{equation}
\begin{equation}
\left\{-\partial_t+ie\Phi^{cl}_{\EuScript{K}}(\mathbf{r},\tau_+)-
ie\Phi^{cl}_{\EuScript{K}}(\mathbf{r},\tau_-)-D\left[\partial_{\mathbf{r}}-
ie\mathbf{A}^{cl}_\EuScript{K}(\mathbf{r},\tau_+)-
ie\mathbf{A}^{cl}_\EuScript{K}(\mathbf{r},\tau_-)\right]^2 \right\}
\bar{\mathcal{C}}^{\mathbf{r},\mathbf{r}'}_{\tau,t,t'} =
\delta_{\mathbf{r}-\mathbf{r}'}\delta_{t-t'}\,,
\end{equation}
\end{subequations}
with $\tau_\pm=\tau\pm t/2$. Note that MT
action~\eqref{SuperCond-S-MT}, has exactly the same structure as the
normal action $S_{\mathrm{N}}$. It therefore can be incorporated
into~\eqref{SuperCond-S-N} by adding a renormalization of the normal
diffusion constant $D\delta_{t-t'}\to D\delta_{t-t'}+\delta
D^{\mathrm{MT}}_{\mathbf{r},t,t'}$ that is non--local in time.

\begin{figure}
\begin{center}\includegraphics[width=10cm]{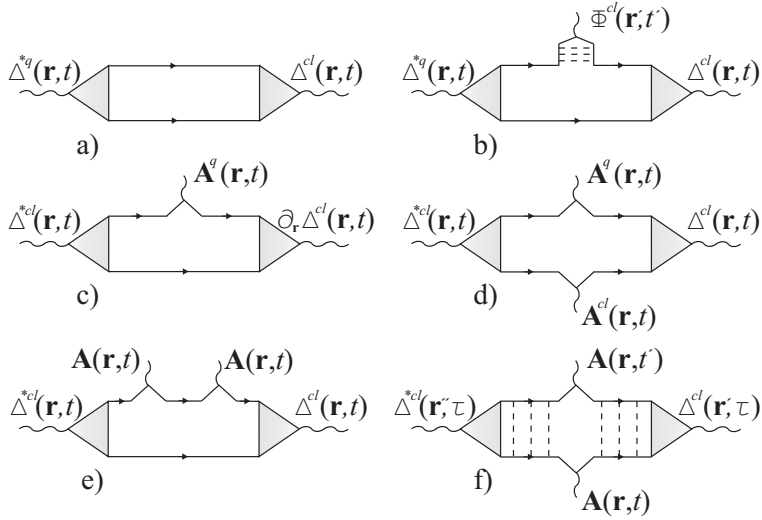}\end{center}
\caption{Diagrammatic representation of the effective action
$S_{\mathrm{eff}}[\Delta,\mathbf{A},\Phi]$. a) Conventional
Ginzbirg--Landau functional $S_{\mathrm{GL}}$,
see~\eqref{SuperCond-S-GL}. b) Anomalous Gor'kov--Eliashberg
coupling between the scalar potential and the order parameter
(see~\eqref{SuperCond-TDGL} and discussions below). c) Paramagnetic
and d) diamagnetic parts of the super--current action
$S_{\mathrm{SC}}$. e) Local DOS term $S_{\mathrm{DOS}}$. f) Nonlocal
MT term $S_{\mathrm{MT}}$. In the case of diagrams e) and f) there
are two possible choices for the vector potentials:
\textit{classical--quantum}, which is a part of the current, and
\textit{quantum--quantum}, which is its FDT counterpart.
\label{Fig-TDGL}}
\end{figure}

Finally, $S_{\mathrm{DOS}}$ has similar structure to
$S_{\mathrm{MT}}$ in~\eqref{SuperCond-S-MT}
\begin{equation}\label{SuperCond-S-DOS}
S_{\mathrm{DOS}}[\Delta,\mathbf{A},\Phi]=e^{2}D
\Tr\left\{\delta\nu^{\mathrm{DOS}}_{\mathbf{r},t}
\left[\vec{\mathbf{A}}^{T}_{\EuScript{K}}\rt
\left(\begin{array}{cc}0 &-\overleftarrow{\partial}_{t} \\
-\overrightarrow{\partial}_{t} & 4iT
\end{array}\right)_{K}\vec{\mathbf{A}}_{\EuScript{K}}\rt\right]\right\}\,,
\end{equation}
with locally renormalized density of states
\begin{equation}\label{SuperCond-DOS-corr}
\delta\nu^{\mathrm{DOS}}_{\mathbf{r},t}=-\nu\frac{7\zeta(3)}{4\pi^{2}T^{2}}
|\Delta^{cl}_{\EuScript{K}}\rt|^{2}\,.
\end{equation}
Each term of the effective action \eqref{SuperCond-S-eff} admits a
transparent diagrammatic representation, shown in
Figure~\ref{Fig-TDGL}.

An equivalent way to display the same information, which is encoded
in the effective action~\eqref{SuperCond-S-eff}, is to use the set
of stochastic time dependent Ginzburg--Landau equations. To derive
those one needs to get rid of terms quadratic in quantum components
of the fields: $\Delta^{q}_{\EuScript{K}}$ in $S_{\mathrm{GL}}$, and
$\mathbf{A}^{q}_{\EuScript{K}}$ in
$S_{\mathrm{N}}+S_{\mathrm{MT}}+S_{\mathrm{DOS}}$. For the first
one, this is achieved with the Hubbard--Stratonovich transformation
\begin{equation}
\exp\left(-{\pi\nu\over 2}
\mathrm{Tr}\big\{|\Delta^{q}_{\EuScript{K}}|^{2}\big\}\right)=\int\D
[\xi_{\Delta}]\exp\left(-\frac{\pi\nu}{8T}\mathrm{Tr}\left\{
\frac{|\xi_{\Delta}|^{2}}{4T}
-i\xi^{*}_{\Delta}\Delta^{q}_{\EuScript{K}}-i\xi_{\Delta}
\Delta^{*q}_{\EuScript{K}}\right\} \right)\,.
\end{equation}
As a result, the effective action $S_{\mathrm{eff}}$
in~\eqref{SuperCond-S-eff} acquires the form linear in quantum
components of the order parameter. Integration over the latter leads
to the functional delta--function, imposing the stochastic equation
of motion. This way the TDGL equation is derived
\begin{equation}\label{SuperCond-KTDGL}
\left[\partial_{t}+\tau^{-1}_{\mathrm{GL}}-D\,
\big[\partial_{\mathbf{r}}-2ie\mathbf{A}^{cl}_{\EuScript{K}}\rt\big]^{2}
+\frac{7\zeta(3)}{\pi^{3}T}|\Delta^{cl}_{\EuScript{K}}\rt|^{2}\right]\Delta^{cl}_{\EuScript{K}}\rt=
\xi_{\Delta}\rt\,.
\end{equation}
The complex Gaussian noise $\xi_{\Delta}\rt$ has white noise
correlation function
\begin{equation}\label{SuperCond-xi-Delta}
\langle\xi_{\Delta}\rt\xi^{*}_{\Delta}(\mathbf{r}',t')\rangle=\frac{16T^{2}}{\pi\nu}\,
\delta(\mathbf{r}-\mathbf{r}')\delta(t-t')\,.
\end{equation}

In a similar way one decouples quadratic in
$\mathbf{A}^{q}_{\EuScript{K}}$ terms in the
action~\eqref{SuperCond-S-eff} by introducing vectorial
Hubbard--Stratonovich field $\xi_{\mathbf{j}}\rt$
\begin{equation}
\exp\left(-4T\Tr\big\{\sigma_{\mathbf{r},t,t'}
[\mathbf{A}^{q}_{\EuScript{K}}]^{2}\big\}\right)=\int\D[\xi_{\mathbf{j}}]\,
\exp\left(-\Tr\left\{\frac{\xi^{2}_{\mathbf{j}}}{4T\sigma_{\mathbf{r},t,t'}}
+2i\mathbf{A}^{q}_{\EuScript{K}}\xi_{\mathbf{j}}\right\}\right)\,,
\end{equation}
where
$\sigma_{\mathbf{r},t,t'}=\sigma_{D}+e^{2}D\delta\nu^{\mathrm{DOS}}_{\mathbf{r},t}+e^{2}\nu\delta
D^{\mathrm{MT}}_{\mathbf{r},t,t'}$ is the complete conductivity
including both DOS and MT renormalizations. The resulting action is
now linear in both $\Phi^{q}_{\EuScript{K}}$ and
$\mathbf{A}^{q}_{\EuScript{K}}$ fields, allowing us to define the
charge $\varrho\rt=(1/2)\delta S_{\mathrm{eff}}/\delta\Phi^{q}\rt$
and current $\mathbf{j}\rt=(1/2)\delta
S_{\mathrm{eff}}/\delta\mathbf{A}^{q}\rt$ densities. It is important
to emphasize that the differentiation here is performed over the
bare electromagnetic potentials $\{\mathbf{A},\Phi\}$, while the
action $S_{\mathrm{eff}}$ in~\eqref{SuperCond-S-eff} is written in
terms of the gauged ones
$\{\mathbf{A}_{\EuScript{K}},\Phi_{\EuScript{K}}\}$. The connection
between the two $\{\Phi,\mathbf{A}\}\rightleftarrows
\{\mathbf{A}_{\EuScript{K}},\Phi_{\EuScript{K}}\}$ is provided by
the functional $\EuScript{K}[\Phi]$, which is implicit
in~\eqref{int-ferm-NLSM-K-functional}. A simple algebra then leads
to a set of the continuity equation
$\partial_{t}\varrho\rt+\mathrm{div}\,\mathbf{j}\rt=0$, and
expression for the current density
\begin{eqnarray}\label{SuperCond-current}
\mathbf{j}\rt&=&\int\d t'\big[D\delta_{t-t'}+\delta
D^{\mathrm{MT}}_{\mathbf{r},t,t'}\big]\big[e^{2}\left(\nu +
\delta\nu^{\mathrm{DOS}}_{\mathbf{r},t'}\right) \mathbf{E}
(\mathbf{r},t')-\partial_{\mathbf{r}}\varrho(\mathbf{r},t')\big]\nonumber\\
&+&\frac{\pi e\nu D}{4T}\,
\mathrm{Im}\Big\{\Delta^{*cl}_{\EuScript{K}}\rt
\big[\partial_{\mathbf{r}}-2ie\mathbf{A}^{cl}_{\EuScript{K}}\rt\big]
\Delta^{cl}_{\EuScript{K}}\rt\Big\}+\xi_{\mathbf{j}}\rt\,,
\end{eqnarray}
where
$\mathbf{E}\rt=\partial_{t}\mathbf{A}_{\EuScript{K}}-
\partial_{\mathbf{r}}\Phi_{\EuScript{K}}$ is electric field.
The current fluctuations are induced by  vector Gaussian white noise
with the correlator
\begin{equation}\label{SuperCond-xi-j}
\big\langle \xi^{\mu}_{\mathbf{j}}(\mathbf{r},t)\,
\xi^{\nu}_{\mathbf{j}}(\mathbf{r'},t') \big\rangle
=\delta_{\mu\nu}\,T e^2
\left(2(\nu+\delta\nu^{\mathrm{DOS}}_{\mathbf{r},t} )D \delta_{t-t'}
+\nu \delta D^{\mathrm{MT}}_{\mathbf{r},t,t'} + \nu \delta
D^{\mathrm{MT}}_{\mathbf{r},t',t}\right)
\delta(\mathbf{r}-\mathbf{r'})\,,
\end{equation}
guaranteeing the validity of FDT. Equations \eqref{SuperCond-KTDGL}
and \eqref{SuperCond-current} together with the continuity relation
must be also supplemented by Maxwell equations for the
electromagnetic potentials.

It is  instructive to rewrite TDGL equation \eqref{SuperCond-KTDGL}
back in the original  gauge. This is achieved by the substitution of
the gauged order parameter
$\Delta^{cl}_{\EuScript{K}}=\Delta^{cl}\exp\big(-2ie\EuScript{K}^{cl}\big)$
into~\eqref{SuperCond-KTDGL}. This way one finds for the bare order
parameter $\Delta^{cl}$ the following equation
\begin{equation}\label{SuperCond-TDGL}
\big[\partial_{t}-2ie\partial_{t}\EuScript{K}^{cl}\rt\big]\Delta^{cl}\rt=
\left[D\,\big[\partial_{\mathbf{r}}-2ie\mathbf{A}^{cl}\rt\big]^{2}-\tau^{-1}_{\mathrm{GL}}-
\frac{7\zeta(3)}{\pi^{3}T}|\Delta^{cl}\rt|^{2}\right]\Delta^{cl}\rt+\xi_{\Delta}\rt\,,
\end{equation}
where we have redefined the order parameter noise as
$\xi_{\Delta}\to\xi_{\Delta}\exp\big(2ie\EuScript{K}^{cl}\big)$,
which, however, does not change its correlation
function~\eqref{SuperCond-xi-Delta}. Unlike TDGL equations
frequently found in the literature, the left hand side
of~\eqref{SuperCond-TDGL} contains Gor'kov--Eliashberg (GE)
anomalous term $\partial_{t}\EuScript{K}^{cl}\rt$ instead of the
scalar potential $\Phi^{cl}\rt$, see Figure~\ref{Fig-TDGL}b. In a
generic case $\EuScript{K}^{cl}\rt$ is a non--local functional of
the scalar and the longitudinal vector potentials, given
by~\eqref{int-ferm-NLSM-Phi-K-matrix}. For the classical
component~\eqref{int-ferm-NLSM-Phi-K-matrix} provides
\begin{equation}\label{SuperCond-K-Phi-functional}
\big(\partial_{t}-D\partial^{2}_{\mathbf{r}}\big)\EuScript{K}^{cl}\rt=\Phi^{cl}\rt-
D\,\mathrm{div}\mathbf{A}^{cl}\rt\,.
\end{equation}
Fields $\partial_t \EuScript{K}^{cl}$ and $\Phi^{cl}$ coincide for
spatially uniform potentials, however in general they are distinct.
The standard motivation behind writing the scalar potential
$\Phi^{cl}\rt$ on the left--hand side of TDGL equation is the gauge
invariance. Note, however, that a local gauge transformation
\begin{eqnarray}\label{SuperCond-gauge}
\Delta^{cl} &\to& \Delta^{cl}\,e^{-2ie\chi}\,,\quad\quad
\Phi^{cl} \to \Phi^{cl} - \partial_t\chi\,,\nonumber \\
 \mathbf{A}^{cl} &\to& \mathbf{A}^{cl} - \partial_{\mathbf{r}}\chi\,, \quad\,\,\,\,\,\,
\EuScript{K}^{cl} \to  \EuScript{K}^{cl} - \chi\,,
 \end{eqnarray}
leaves~\eqref{SuperCond-TDGL} unchanged and therefore this form of
TDGL equation is perfectly gauge invariant. The last expression
in~\eqref{SuperCond-gauge} is an immediate consequence
of~\eqref{SuperCond-K-Phi-functional} and the rules of the gauge
transformation for $\Phi\rt$ and $\mathbf{A}\rt$. In the
$\EuScript{K}$ gauge, specified by $\chi\rt=\EuScript{K}^{cl}\rt$,
the anomalous GE term disappears from  TDGL Equation
\eqref{SuperCond-TDGL}, and one returns back to
Eq.~\eqref{SuperCond-KTDGL}.

\subsection{Applications IV: Non--uniform and fluctuating superconductivity}\label{app_Part-IV}

\subsubsection{Proximity effect}\label{app_Part-IV-1}

Close to the interface with a superconductor a normal metal acquires
partial superconducting properties. At the same time  the
superconductor is weakened by the normal metal. This mutual
influence is called \textit{proximity} effect. The quasiclassical
Usadel and kinetic equations discussed in the
Section~\ref{sec_SuperCond-2} give full account of proximity related
phenomena for superconductor--normal metal structures. One example
of this kind is considered in this section.

Consider a normal diffusive wire of the length $L$ placed between
two bulk superconductors,  forming superconductor--normal
metal--superconductor (SNS) junction. We are interested to study how
the proximity to the superconductor modifies quasiparticle energy
spectrum in the normal wire. It follows from the Usadel equation
\eqref{SuperCond-Usadel-2} that the density of states in the wire
acquires an energy gap $\epsilon_{g}$ and exhibits square--root
non--analytic behavior $\sim \sqrt{\epsilon-\epsilon_{g}}$ above it,
at $\epsilon>\epsilon_{g}$~\cite{Zhou,GolubovKupriyanov}. To see
this explicitly we assume that the wire  cross--section dimension is
much smaller than the superconductive coherence length
$\xi=\sqrt{D/\Delta}$. In this case the wire may be thought of as
being quasi--one--dimensional, such that all the variations occur
along the $x$ coordinate of the wire. If there are no attractive
interactions in the wire, $\lambda=0$, then according to the
self--consistency Equation \eqref{SuperCond-Usadel-4} pair potential
$\Delta(\mathbf{r})=0$ within the wire $-L/2\leqslant x\leqslant
L/2$, and $\Delta(\mathbf{r})=\Delta$ outside this interval. If in
addition there is no phase difference between the two
superconductors, $\partial_{x}\chi=0$, the Usadel equation
\eqref{SuperCond-Usadel-2} simplifies considerably and reads as
\begin{equation}\label{Part-IV-Usadel}
D\,\partial^{2}_{x}\theta(x,\epsilon)+2i\epsilon\sinh\theta(x,\epsilon)=0\,.
\end{equation}
At the interfaces with the superconductors, $x=\pm L/2$, this
equation is supplemented by the  boundary conditions $\theta(\pm
L/2,\epsilon)=\theta_{BCS}(\epsilon)$, where
$\tanh\theta_{BCS}(\epsilon)=\Delta/\epsilon$. It is assumed here
that superconductors are very large and negligibly perturbed by the
wire, such that one can use coordinate--independent
$\theta_{BCS}(\epsilon)$ everywhere inside the superconductors.
Having solved~\eqref{Part-IV-Usadel} one finds density of states as
$\nu(x,\epsilon)=\nu\mathrm{Re}\big[\cosh\theta(x,\epsilon)\big]$.

It is convenient to perform rotation
$\theta(x,\epsilon)=i\pi/2-\vartheta(x,\epsilon)$ such that
Eq.~\eqref{Part-IV-Usadel} becomes real and allows the straightforward
integration
\begin{equation}\label{Part-IV-Usadel-sol}
\sqrt{\frac{\epsilon}{E_{Th}}}=\int^{\vartheta_{0}}_{\vartheta_{BCS}}
\frac{\d\vartheta}{\sqrt{\sinh\vartheta_{0}-\sinh\vartheta}}
\equiv K(\vartheta_{0},\epsilon)\,,
\end{equation}
where $E_{Th}= D/L^2$, $\vartheta_{0}=\vartheta(0,\epsilon)$
and $\sinh\vartheta_{BCS}=\epsilon/\sqrt{\Delta^2-\epsilon^{2}}$.
Equation \eqref{Part-IV-Usadel-sol} defines $\vartheta_{0}$ as a
function of energy $\epsilon$. Knowing $\vartheta_{0}(\epsilon)$ one
determines density of states in the middle of the wire as
$\nu(0,\epsilon)=\nu\mathrm{Im}[\sinh\vartheta_{0}(\epsilon)]$.

In the limit of the long wire, $\xi\ll L$,  modifications of the
density of states occur in the deep sub--gap limit,
$\epsilon\ll\Delta$.  One may thus approximate
$\vartheta_{BCS}\approx0$ and the function on the right--hand side
of~\eqref{Part-IV-Usadel-sol} is essentially energy independent
$K(\vartheta_{0},\epsilon)\approx K(\vartheta_0,0)$. It exhibits the
maximum $K_{\mathrm{max}}=K(\vartheta^{*}_{0})\approx1.75$ at
$\vartheta^{*}_{0}\approx1.5$, whereas the left--hand side
of~\eqref{Part-IV-Usadel-sol} can be larger than $K_{\mathrm{max}}$
for $\epsilon>K^{2}_{\mathrm{max}}E_{Th}=\epsilon_{g}$. Thus, for
all the energies $\epsilon<\epsilon_{g}$, Equation
\eqref{Part-IV-Usadel-sol} has only real solution for
$\vartheta_{0}$ and $\nu(0,\epsilon)\equiv0$, since
$\nu(0,\epsilon)\propto\mathrm{Im}\big[\sinh\vartheta_{0}\big]$. For
$\epsilon>\epsilon_{g}$ function $\vartheta_{0}$ becomes complex and
gives finite density of states. Right above the gap, $0 <
\epsilon-\epsilon_{g}\ll\epsilon_{g}$, one finds with the help of
Eq.~\eqref{Part-IV-Usadel-sol}
\begin{equation}
\nu(\epsilon)=3.7\delta^{-1}\sqrt{\frac{\epsilon}{\epsilon_{g}}-1}\,,
\end{equation}
where $\nu(\epsilon)=\mathcal{A}\int\nu(x,\epsilon)\d x$ is global
density of states, integrated over the volume of the wire
($\mathcal{A}$ is the wire cross--section area, and $\delta=1/(\nu
\mathcal{A}L)$ is its level spacing). Note that since
$\epsilon_{g}\sim E_{Th}\ll\Delta$ the approximation
$\vartheta_{BCS}(\epsilon\sim\epsilon_{g})\approx0$ is well
justified.

In the opposite limit of the short wire, $L\ll\xi$, or equivalently,
$E_{Th}\gg\Delta$, Equation \eqref{Part-IV-Usadel-sol} is still
applicable. However, one must keep the full energy dependence of
$\vartheta_{BCS}(\epsilon)$. One may show that the energy gap is
given by $\epsilon_{g}=\Delta-\Delta^{3}/8E^{2}_{Th}$ and is only
slightly smaller than the bulk gap $\Delta$. This is natural, since
the proximity effect for the short wire is expected to be strong.
Immediately  above the induced gap, the density of states again
exhibits the  square--root non--analyticity. The coefficient in
front of it, however, is large, $\nu(\epsilon)\sim \delta^{-1}
(E_{Th}/\Delta)^{2} \sqrt{\epsilon/\epsilon_{g}-1}$,
(see~\cite{Levchenko-DOS}).

\subsubsection{Josephson current}\label{app_Part-IV-2}

Another example which may be  treated with the help of Usadel
Equations \eqref{SuperCond-Usadel-1} and \eqref{SuperCond-Usadel-2}
is the Josephson effect. Consider the same geometry of  SNS
junction, as in the previous section,
 assuming   a finite phase difference between the
pair potentials on the boundaries  of the junction, i.e.
$\chi(L/2,\epsilon)-\chi(-L/2,\epsilon)=\phi$. Under this condition
Josephson super--current $I_{S}(\phi)$ may flow across the junction.
The aim of this section is to illustrate how Josephson
phase--current relation may be obtained from  the Usadel equations.

For the model of step--function pair potential, $\Delta(x)=\Delta$
for $|x|>L/2$  and $\Delta=0$ for $|x|<L/2$, equations
\eqref{SuperCond-Usadel-1}, \eqref{SuperCond-Usadel-2} acquire the
form
\begin{subequations}\label{Part-IV-Usadel-coupled}
\begin{equation}
\hskip-2.2cm
D\,\partial_{x}\big(\sinh^{2}\theta\partial_{x}\chi\big)=0\,,
\end{equation}
\begin{equation}
D\,\partial^{2}_{x}\theta+2i\epsilon\sinh\theta=
\frac{D}{2}(\partial_{x}\chi)^{2}\sinh2\theta\,.
\end{equation}
\end{subequations}
The latter are supplemented by the boundary conditions $\theta(\pm
L/2,\epsilon)=\theta_{BCS}(\epsilon)$, while boundary condition for
the $\chi$--function is determined by the fixed phase $\phi$ across
the junction mentioned above,
$\chi(L/2,\epsilon)-\chi(-L/2,\epsilon)=\phi$. For the short wire,
$L\ll\xi$, the second term on the left--hand side
of~(\ref{Part-IV-Usadel-coupled}b) is smaller than the gradient term
by $\epsilon/E_{Th}\ll1$ and thus may be neglected. Since
(\ref{Part-IV-Usadel-coupled}a) allows for the first integral
$\sinh^{2}\theta\partial_{x}\chi=\mathcal{J}/L$, one may eliminate
$\partial_{x}\chi$ from Eq.~(\ref{Part-IV-Usadel-coupled}b) and find
$L^2
\partial^{2}_{x}\theta=\mathcal{J}^{2}\cosh\theta/\sinh^{3}\theta$.
This equation may be solved exactly
\begin{equation}\label{Part-IV-Usadel-coupled-sol}
\cosh\theta(z,\epsilon)=\cosh\theta_{0}
\cosh\left(\frac{\mathcal{J}z}{\sinh\theta_{0}}\right)\,,
\end{equation}
where $\theta_{0}=\theta(0,\epsilon)$ and $z=x/L$. Knowing
$\theta(x,\epsilon)$, one inserts it back into the first integral
of~(\ref{Part-IV-Usadel-coupled}a), $\phi= \int^{L/2}_{-L/2}\d
x\partial_x\chi =   \mathcal{J}\int^{1/2}_{-1/2}\d
z/\sinh^{2}\theta(z,\epsilon)$, to find
\begin{equation}
\tan(\phi/2)=\frac{1}{\sinh\theta_{0}}
\tanh\left(\frac{\mathcal{J}}{2\sinh\theta_{0}}\right)\,.
\end{equation}
This last equation along with Eq.~\eqref{Part-IV-Usadel-coupled-sol}
taken at the NS interfaces, $z=\pm1/2$, constitutes the system of
the two algebraic equations for the two unknown quantities:
$\mathcal{J}$ and $\theta_{0}$. Such an algebraic problem may be
easily solved, resulting in
$\mathcal{J}(\epsilon,\phi)=2\sinh\theta_{0}\mathrm{arctanh}\big[\sinh\theta_{0}\tan(\phi/2)\big]$
and
$\sinh\theta_{0}=\sinh\theta_{BCS}/\sqrt{1+\tan^{2}(\phi/2)\cosh^{2}\theta_{BCS}}$,
where $\cosh\theta_{BCS}=\epsilon/\sqrt{\epsilon^{2}-\Delta^{2}}$.
Knowing $\mathcal{J}(\epsilon,\phi)$ one finds Josephson current
with the help of
\begin{equation}\label{Part-IV-Josephson-def}
I_{S}(\phi)=\frac{\mathrm{g}_{D}}{e}\int^{\infty}_{0}\d\epsilon\,
\tanh\left(\frac{\epsilon}{2T}\right)\,\mathrm{Im}\,\mathcal{J}(\epsilon,\phi)\,,
\end{equation}
where $\mathrm{g}_{D}$ is the wire conductance. Using the obtained
solution for $\mathcal{J}(\epsilon,\phi)$ one concludes that
\begin{equation}\label{Part-IV-Im-J}
\mathrm{Im}\, \mathcal{J}(\epsilon,\phi)=\frac{\pi\Delta\cos(\phi/2)}
{\sqrt{\epsilon^{2}-\Delta^{2}\cos^{2}(\phi/2)}}
\end{equation}
for $\Delta\cos(\phi/2)<\epsilon<\Delta$, and
$\mathrm{Im}\,\mathcal{J}(\epsilon,\phi)=0$ otherwise.
Employing~\eqref{Part-IV-Josephson-def} and \eqref{Part-IV-Im-J},
one arrives at the  result derived by Kulik and
Omelyanchuk~\cite{KulikOmelyanchuk} for the zero--temperature
Josephson current of the short diffusive SNS junction
\begin{equation}
I_{S}(\phi)=\frac{\pi\mathrm{g}_{D}\Delta}{e}\cos(\phi/2)\,
\mathrm{arctanh}\big[\sin(\phi/2)\big]\,.
\end{equation}
In the original work~\cite{KulikOmelyanchuk} imaginary time
technique was used to derive $I_{S}(\phi)$. This result was
reproduced later in~\cite{Heikkila,Levchenko-Js} with the help of
real time (energy) Usadel equation.

\subsubsection{Supression of the density of states above $T_c$}\label{app_Part-IV-3}

Superconductor below $T_{c}$ has an energy gap $|\Delta(T)|$ in the
excitation spectrum. Superconductor above and far away from $T_c$
has metallic, constant density of states. One of the  manifestations
of superconducting fluctuations  in the vicinity of the transition,
$0<T-T_c\ll T_{c}$, is the depletion of the density of states near
the Fermi energy. Fluctuations mediated suppression of the density
of states increases with the lowering of temperature and eventually
transforms into the full gap. In this section we calculate the
temperature dependence of this effect employing Keldysh formalism
and compare it to the original works~\cite{Hurault,Abrahams}, where
Matsubara technique and analytic continuation procedure was used.
For comprehensive discussions one may consult the recent book of
Larkin and Varlamov~\cite{Larkin-Varlamov}.
\begin{figure}
\begin{center}\includegraphics[width=10cm]{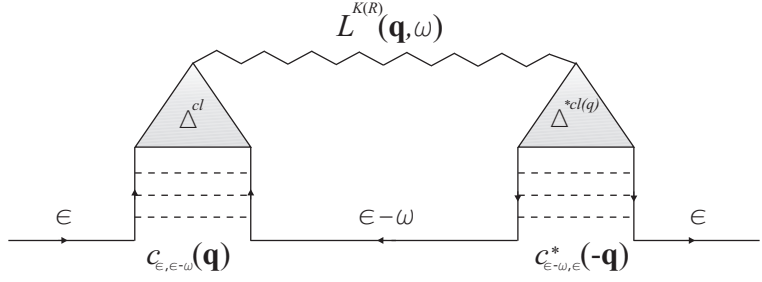}\end{center}
\caption{Diagram for the density of states correction,
\eqref{Part-IV-Dos-corr}, in the vicinity of the critical
temperature $T_{c}$. Two Cooperon fields $c$ and $c^{*}$, shown by
the ladders, are connected to the order parameter $\Delta^{cl(q)}$,
shown as a filled triangle, which are paired by the fluctuations
propagator. \label{Fig-DOS}}
\end{figure}

Our starting point is the expression for the density of states given
in terms of the $\check{Q}$ matrix
$$\nu(\varepsilon)=\frac{\nu}{4}\left\langle\Tr\{\hat{\sigma}_{z}\otimes\hat{\tau}_{z}
\check{Q}_{\varepsilon\varepsilon}\}\right\rangle_{Q}\,,$$ cf.
Section~\ref{app_Part-II-2}. By taking $\check{Q}=\check{\Lambda}$
one finds $\nu(\varepsilon)=\nu$, as it should be for a normal
metal. Expanding $\check{Q}$ to the quadratic order in the Cooperon
fluctuations $\check{\mathcal{W}}$, see~(\ref{SuperCond-Q-W}), one
finds for the density of states correction
\begin{equation}\label{Part-IV-Dos-corr-def}
\delta\nu(\varepsilon)=\frac{\nu}{4}\sum_{q}\int\frac{\d\varepsilon'}{2\pi}
\left\langle\left\langle c_{\varepsilon\varepsilon'}(\mathbf{q})
c^{*}_{\varepsilon'\varepsilon}(-\mathbf{q})+
\bar{c}_{\varepsilon\varepsilon'}(\mathbf{q})
\bar{c}^{*}_{\varepsilon'\varepsilon}(-\mathbf{q})
\right\rangle\right\rangle_{\mathcal{W},\Delta}\,.
\end{equation}
The next step is to perform averaging over fluctuating $c$ and $\bar
c$ fields. For this purpose one uses~\eqref{app_FE-c-extremal},
which relates Cooper modes $c$ and $\bar c$ with the fluctuations of
the order parameter. The latter are governed by the following
correlation functions
\begin{eqnarray}\label{Part-IV-Delta-averages}
&&\left\langle
\Delta^{cl}(\mathbf{q},\omega)\Delta^{*cl}(-\mathbf{q},-\omega)\right\rangle_{\Delta}=
\frac{i}{2\nu}\, L^{K}(\mathbf{q},\omega) \,,\quad \left\langle
\Delta^{cl}(\mathbf{q},\omega)\Delta^{*q}(-\mathbf{q},-\omega)\right\rangle_{\Delta}=
\frac{i}{2\nu}\, L^{R}(\mathbf{q},\omega) \,,\nonumber\\
&&\left\langle
\Delta^{q}(\mathbf{q},\omega)\Delta^{*cl}(-\mathbf{q},-\omega)\right\rangle_{\Delta}=
\frac{i}{2\nu}\, L^{A}(\mathbf{q},\omega) \,,\quad \left\langle
\Delta^{q}(\mathbf{q},\omega)\Delta^{*q}(-\mathbf{q},-\omega)\right\rangle_{\Delta}=0\,,
\end{eqnarray}
which follow from the time--dependent Ginzburg--Landau action
\eqref{SuperCond-S-GL}. As a result one finds for the correlators of
the Cooperon fields
\begin{subequations}\label{Part-IV-c-averages}
\begin{equation}
\left\langle\left\langle
c_{\varepsilon,\varepsilon-\omega}(\mathbf{q})
c^{*}_{\varepsilon-\omega,\varepsilon}(-\mathbf{q})\right\rangle\right\rangle=\frac{2i}{\nu}\,
\frac{L^{K}+F_{\varepsilon-\omega}L^{R}+F_{\varepsilon}L^{A}}
{\big(Dq^{2}-2i\varepsilon+i\omega\big)^{2}}\,,
\end{equation}
\begin{equation}
\left\langle\left\langle
\bar{c}_{\varepsilon,\varepsilon-\omega}(\mathbf{q})
\bar{c}^{*}_{\varepsilon-\omega,\varepsilon}(-\mathbf{q})\right\rangle\right\rangle=\frac{2i}{\nu}\,
\frac{L^{K}-F_{\varepsilon-\omega}L^{A}-F_{\varepsilon}L^{R}}
{\big(Dq^{2}+2i\varepsilon-i\omega\big)^{2}}\,.
\end{equation}
\end{subequations}
Inserting these into~\eqref{Part-IV-Dos-corr-def} and summing up the
two contributions, one obtains
\begin{equation}\label{Part-IV-Dos-corr}
\delta\nu(\varepsilon)=\mathrm{Im}\sum_{q}
\int^{+\infty}_{-\infty}\frac{\d\omega}{2\pi}\,
\frac{L^{K}\qo+F_{\varepsilon-\omega}L^{R}\qo}
{\big(Dq^{2}-2i\varepsilon+i\omega\big)^{2}}\,,
\end{equation}
where terms proportional to $F_{\varepsilon}L^{A(R)}(\omega)$ in the
averages $\langle\langle cc^{*}\rangle\rangle$ and
$\langle\langle\bar{c}\bar{c}^{*}\rangle\rangle$ drop out
from~\eqref{Part-IV-Dos-corr} upon $\omega$ integration, as being
integrals of purely advanced and retarded functions, respectively.
Equation \eqref{Part-IV-Dos-corr} allows convenient diagrammatic
representation shown in Figure~\ref{Fig-DOS}. Using now fluctuation
propagator in the form of~\eqref{app_FE-L-R} and approximating
bosonic distribution function as $B_{\omega}\approx 2T_{c}/\omega$,
since the relevant frequencies $\omega\sim T-T_c\ll T_c$, the
density of states correction \eqref{Part-IV-Dos-corr} reduces to
\begin{equation}\label{Part-IV-Dos-intermediate}
\delta\nu(\varepsilon)=-\frac{16T^{2}_{c}}{\pi^{2}}\mathrm{Re}\sum_{q}
\int^{+\infty}_{-\infty}\frac{\d\omega}
{\big[(Dq^{2}+\tau^{-1}_{\mathrm{GL}})^{2}+\omega^{2}\big]
\big[Dq^{2}-2i\varepsilon+i\omega\big]^{2}}\,,
\end{equation}
where $\tau_{\mathrm{GL}}^{-1}=8(T-T_c)/\pi$.

The further analysis of this expression depends on the effective
dimensionality of the system. We focus on quasi--two--dimensional
case: a metal film with the thickness $b$ which is much smaller then
superconducting coherence length
$b\ll\xi(T)=\sqrt{D\tau_{\mathrm{GL}}}$. Replace the momentum
summation by the integration $\sum_{q}\to \frac{1}{b}\int\frac{\d
q^{2}}{4\pi^2}$, introduce the dimensionless parameters
$x=Dq^{2}/T_{c}$ and $y=\omega/T_{c}$, integrates over $y$ using
residue theorem and finds
\begin{equation}\label{Part-III-DOS-result}
\frac{\delta\nu(\varepsilon)}{\nu}=-\frac{\mathrm{Gi}}{16}
\left(\frac{T_{c}}{T-T_{c}}\right)^{2}\mathcal{Y}(\varepsilon\tau_{\mathrm{GL}}),\qquad
\mathcal{Y}(z)=\mathrm{Re}\int^{+\infty}_{0}\frac{\d
x}{(1+x)(1+2x-2iz)^{2}}\,,
\end{equation}
where $\mathrm{Gi}=\hbar /\nu Db$ is the Ginzburg number. For small
deviations from the Fermi energy,
$\varepsilon\tau_{\mathrm{GL}}\ll1$, the DOS suppression scales as
$\delta\nu(0)\propto - (T/T_{c}-1)^{-2}$, while at larger energies
$\varepsilon\tau_{\mathrm{GL}}\gg1$ DOS approaches its its normal
value as $\delta\nu(\varepsilon)\propto -
(T_{c}/\varepsilon)^{2}\ln(\varepsilon\tau_{\mathrm{GL}})$. Note
also that $\int\d\varepsilon\,\delta\nu(\varepsilon)\equiv0$, which
is expected, since the fluctuations only redistribute  states around
the Fermi energy.

\subsubsection{Fluctuation corrections to the conductivity}\label{app_Part-IV-4}

Superconductive fluctuations above $T_c$ modify not only the density
of states, but also  transport properties. In the case of
conductivity, there are  three types of the corrections, density of
states (DOS) $\delta\sigma_{\mathrm{DOS}}$, Aslamazov--Larkin (AL)
$\delta\sigma_{\mathrm{AL}}$ and Maki--Thompson (MT)
$\delta\sigma_{\mathrm{MT}}$ terms,
see.~\cite{AslamazovLarkin,Maki,Thompson,Hurault}. Although we have
already partially discussed this topic in
Section~\ref{sec_SuperCond-3}, the goal of this section is to show
explicitly how all of them are obtained within Keldysh
$\sigma$--model approach.

According to the definition given by~\eqref{Part-II-sigma-def}, to
find conductivity one needs to know partition function
$\mathcal{Z}[\mathbf{A}^{cl},\mathbf{A}^{q}]$ to the quadratic order
in vector potential. Using~\eqref{SuperCond-S} one
finds~\footnote{Since  Coulomb interactions do not lead to a
singular temperature dependence for kinetic coefficients in the
vicinity of $T_c$, we set $\Phi_{\EuScript{K}}=0$ and suppress
subscript $\EuScript{K}$ throughout this section.}
\begin{equation}\label{Part-IV-Z}
\mathcal{Z}[\mathbf{A}^{cl},\mathbf{A}^{q}]\approx\int\D[\check Q,\Delta]\left[1+\frac{\pi\nu
D}{2}\Tr\big\{\check{\Xi}\check{\mathbf{A}}\check{Q}
\check{\Xi}\check{\mathbf{A}}\check{Q}\big\}-\frac{(\pi\nu
D)^{2}}{8}\left(\Tr\big\{\partial_{\mathbf{r}}\check{Q}
[\check{\Xi}\check{\mathbf{A}},\check{Q}]\big\}\right)^{2}\right]
\exp\big(iS_{\sigma}[\check Q,\Delta]\big)\,,
\end{equation}
where diamagnetic contribution
$\Tr\{\check{\Xi}\check{\mathbf{A}}\check{\Xi}\check{\mathbf{A}}\}$
was omitted. As it was demonstrated in the Section~\ref{Part-II-3},
by taking $\check{Q}=\check{\Lambda}$ and
using~\eqref{Part-II-sigma-def} one finds Drude conductivity
$\sigma_{D}$. To capture superconductive corrections $\delta\sigma$
to normal metal conductivity $\sigma_{D}$ one has to expand
$\check{Q}$--matrix in fluctuations $\check{\mathcal{W}}$ to the
leading (quadratic) order and analyze all  possible contributions.

\begin{figure}
\begin{center}\includegraphics[width=10cm]{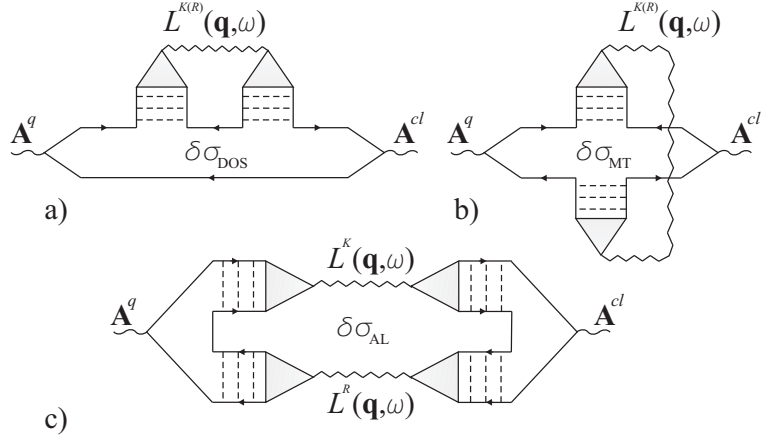}\end{center}
\caption{Diagrams for superconductive fluctuation corrections  to
the conductivity in a vicinity of $T_{c}$: a) density of states
term; b) Maki--Thompson correction; c) Aslamazov--Larkin correction.
\label{Fig-AL-MT-DOS}}
\end{figure}

From the first trace on the right--hand side of~\eqref{Part-IV-Z} by
taking one of the $\check{Q}$ matrices to be $\check{\Lambda}$,
while expanding the other one to $\check{\mathcal{W}}^{2}$ order,
one finds
\begin{equation}\label{Part-IV-Z-DOS}
\mathcal{Z}_{\mathrm{DOS}}[\mathbf{A}^{cl},\mathbf{A}^{q}]=\frac{\pi\nu
D}{2}\left\langle\left\langle\Tr\{\check{\mathbb{A}}_{\varepsilon_{1}\varepsilon_{2}}
(\hat{\sigma}_{z}\otimes\hat{\tau}_{z})
\check{\mathbb{A}}_{\varepsilon_{2}\varepsilon_{3}}
(\hat{\sigma}_{z}\otimes\hat{\tau}_{z})
\check{\mathcal{W}}_{\varepsilon_{3}\varepsilon_{4}}
\check{\mathcal{W}}_{\varepsilon_{4}\varepsilon_{1}}\}
\right\rangle\right\rangle_{\mathcal{W},\Delta}\,,
\end{equation}
where the current vertex matrix is
\begin{equation}
\check{\mathbb{A}}_{\varepsilon\varepsilon'}\equiv
\check{\mathcal{U}}_{\varepsilon}^{-1}
\check{\Xi}\check{\mathbf{A}}_{\varepsilon-\varepsilon'}\check{\mathcal{U}}_{\varepsilon'}
=\left(\begin{array}{cc}
\mathbf{A}^{cl}_{\varepsilon-\varepsilon'}+F_{\varepsilon}
\mathbf{A}^{q}_{\varepsilon-\varepsilon'} &
\mathbf{A}^{q}_{\varepsilon-\varepsilon'}[F_{\varepsilon}F_{\varepsilon'}
- 1]
+\mathbf{A}^{cl}_{\varepsilon-\varepsilon'}[F_{\varepsilon'}-F_{\varepsilon}]\\
-\mathbf{A}^{q}_{\varepsilon-\varepsilon'} &
\mathbf{A}^{cl}_{\varepsilon-\varepsilon'} -
F_{\varepsilon'}\mathbf{A}^{q}_{\varepsilon-\varepsilon'}
\end{array}\right)_K\otimes\hat \tau_{z}\,.
\end{equation}
It will be shown momentarily, that $\mathcal{Z}_{\mathrm{DOS}}$
defines density of states type contribution to the conductivity in
the vicinity of the critical temperature. Indeed, one
substitutes~\eqref{Part-IV-Z-DOS} into~\eqref{Part-II-sigma-def},
carries differentiation over the vector potentials, takes the dc
limit $\Omega\to0$ and evaluates matrix traces. As a result, one
fids
\begin{equation}\label{Part-IV-sigma-DOS-cc}
\delta\sigma_{\mathrm{DOS}}=\frac{\pi e^2\nu D}{2}
\sum_{q}\iint\frac{\d\varepsilon_{2}\d\varepsilon_{4}}{4\pi^2}\,
\partial_{\varepsilon_{2}}F_{\varepsilon_{2}}
\left\langle\left\langle
c_{\varepsilon_{2}\varepsilon_{4}}(\mathbf{q})
c^{*}_{\varepsilon_{4}\varepsilon_{2}}(-\mathbf{q})+
\bar{c}_{\varepsilon_{2}\varepsilon_{4}}(\mathbf{q})
\bar{c}^{*}_{\varepsilon_{4}\varepsilon_{2}}(-\mathbf{q})
\right\rangle\right\rangle_{\mathcal{W},\Delta} \,.
\end{equation}
As the next step, one uses~\eqref{app_FE-c-extremal} and performs
$\Delta$ averaging with the help of correlation functions
Eq.~\eqref{Part-IV-Delta-averages}. Changing integration variables
$\varepsilon_{2}\to\varepsilon$ and
$\varepsilon_{4}\to\varepsilon-\omega$, correction
$\delta\sigma_{\mathrm{DOS}}$ becomes
\begin{equation}\label{Part-IV-sigma-DOS-analytic}
\delta\sigma_{\mathrm{DOS}}=\frac{e^{2}D}{2\pi}\,\mathrm{Im}\sum_{q}\int
^{+\infty}_{-\infty}\d\varepsilon\d\omega\,
\partial_{\varepsilon}F_{\varepsilon}\, \frac{L^{K}\qo+
F_{\varepsilon-\omega}L^{R}\qo}{\big(Dq^{2}-2i\varepsilon+i\omega\big)^{2}}\,
.
\end{equation}
By comparing this expression to~\eqref{Part-IV-Dos-corr} one
concludes that $\delta\sigma_{\mathrm{DOS}}\propto
\int\d\varepsilon\,\partial_{\varepsilon}F_{\varepsilon}
\delta\nu(\varepsilon)$, which establishes connection between
$\delta\sigma_{\mathrm{DOS}}$ and density of states suppression
$\delta\nu(\varepsilon)$, see also Figure~\ref{Fig-AL-MT-DOS}a for
diagrammatic representation. In order to extract the most divergent
part of $\delta\sigma_{\mathrm{DOS}}$, in powers of the deviation
$T-T_{c}$, one only needs to keep
in~\eqref{Part-IV-sigma-DOS-analytic} Keldysh propagator. The
$F_{\varepsilon-\omega}L^{R}$ term gives parametrically smaller
contribution. Using~\eqref{app_FE-L-R} one finds
\begin{equation}
\delta\sigma_{\mathrm{DOS}}=-\frac{16e^{2}DT^{2}_{c}}{\pi^{2}}\,\mathrm{Re}\sum_{q}
\iint^{+\infty}_{-\infty}\d\varepsilon\d\omega\,
\frac{\partial_{\varepsilon}F_{\varepsilon}}
{\big[(Dq^{2}+\tau^{-1}_{\mathrm{GL}})^{2}+\omega^{2}\big]
\big[Dq^{2}-2i\varepsilon+i\omega\big]^{2}}\,.
\end{equation}
After remaining frequency and momentum integrations, for the
quasi--two--dimensional case, one finds
\begin{equation}\label{Part-IV-sigma-DOS}
\frac{\delta\sigma_{\mathrm{DOS}}}{\sigma_{D}}=
-\frac{7\zeta(3)\mathrm{Gi}}{\pi^{4}}
\ln\left(\frac{T_{c}}{T-T_{c}}\right)\,.
\end{equation}
This correction is negative as expected, which stems from the
depletion of the density of states by fluctuations, and has
relatively weak temperature dependence. It is worth emphasizing that
$\delta\sigma_{\mathrm{DOS}}$ can be extracted from the effective
time dependent Ginzburg--Landau theory, which was discussed in the
Section~\ref{sec_SuperCond-3}. Indeed, one can show that
$\delta\sigma_{\mathrm{DOS}}=e^2D\langle\delta\nu^{\mathrm{DOS}}_{\mathbf{r},t}\rangle_{\Delta}$,
where $\delta\nu^{\mathrm{DOS}}_{\mathbf{r},t}$ is taken
from~\eqref{SuperCond-DOS-corr},
reproduces~\eqref{Part-IV-sigma-DOS}.

Let us return back to~\eqref{Part-IV-Z} and look for different
possible contributions. Focusing again on the first trace on the
right--hand side of~\eqref{Part-IV-Z}, one may expand now each of
the $\check{Q}$ matrices to the first order in fluctuations
$\check{\mathcal{W}}$. This way one identifies
\begin{equation}\label{Part-IV-Z-MT}
\mathcal{Z}_{\mathrm{MT}}[\mathbf{A}^{cl},\mathbf{A}^{q}]=\frac{\pi\nu
D}{2}\left\langle\left\langle\Tr\big\{\check{\mathbb{A}}_{\varepsilon_{1}\varepsilon_{2}}
(\hat{\sigma}_{z}\otimes\hat{\tau}_{z})
\check{\mathcal{W}}_{\varepsilon_{2}\varepsilon_{3}}
\check{\mathbb{A}}_{\varepsilon_{3}\varepsilon_{4}}
(\hat{\sigma}_{z}\otimes\hat{\tau}_{z})
\check{\mathcal{W}}_{\varepsilon_{4}\varepsilon_{1}}
\big\}\right\rangle\right\rangle_{\mathcal{W},\Delta}\,,
\end{equation}
which leads to Maki--Thompson  correction to the conductivity. After
differentiation of
$\mathcal{Z}_{\mathrm{MT}}[\mathbf{A}^{cl},\mathbf{A}^{q}]$ over the
vector potential, and evaluation of the traces, in the dc limit, one
finds
\begin{equation}\label{Part-IV-sigma-MT-cc}
\delta\sigma_{\mathrm{MT}}=\frac{\pi e^2\nu D}{2}
\sum_{q}\iint\frac{\d\varepsilon_{2}\d\varepsilon_{4}}{4\pi^2}\,
\partial_{\varepsilon_{2}}F_{\varepsilon_{2}}
\left\langle\left\langle
c_{\varepsilon_{2}\varepsilon_{4}}(\mathbf{q})
\bar{c}^{*}_{\varepsilon_{4}\varepsilon_{2}}(-\mathbf{q})
+c^{*}_{\varepsilon_{2}\varepsilon_{4}}(\mathbf{q})
\bar{c}_{\varepsilon_{4}\varepsilon_{2}}(-\mathbf{q})
\right\rangle\right\rangle_{\mathcal{W},\Delta}.
\end{equation}
As compared to $\delta\sigma_{\mathrm{DOS}}$
in~\eqref{Part-IV-sigma-DOS-cc} $\delta\sigma_{\mathrm{MT}}$
consists of products of mixed retarded $c$ and advanced $\bar{c}$
Cooperons, while $\delta\sigma_{\mathrm{DOS}}$ contains Cooperon
fields of the same causality. Using~\eqref{Part-IV-Delta-averages}
and \eqref{app_FE-c-extremal} one carries averaging
in~\eqref{Part-IV-sigma-MT-cc} over $\Delta$ fluctuations, then
changes integration variables in the same way as
in~\eqref{Part-IV-sigma-DOS-analytic} and arrives at
\begin{equation}\label{Part-IV-sigma-MT-analytic}
\delta\sigma_{\mathrm{MT}}=-\frac{e^{2}D}{\pi}\sum_{q}
\int^{+\infty}_{-\infty}\d\varepsilon\d\omega\,
\partial_{\varepsilon}F_{\varepsilon}\,\frac{\mathrm{Im}[L^{R}\qo]
(B_{\omega}-F_{\varepsilon-\omega})}
{(Dq^{2})^{2}+(2\varepsilon+\omega)^{2}} \,.
\end{equation}
The corresponding diagram is shown in  Figure~\ref{Fig-AL-MT-DOS}b.
With the same accuracy as earlier, approximating $B_{\omega}\approx
2T_{c}/\omega$, neglecting $F_{\varepsilon-\omega}$ and
using~\eqref{app_FE-L-R} for the fluctuations propagator, the latter
expression for $\delta\sigma_{MT}$ reduces to
\begin{equation}
\delta\sigma_{\mathrm{MT}}=\frac{16e^{2}DT^{2}_{c}}{\pi^{2}}\sum_{q}
\iint^{+\infty}_{-\infty}\d\varepsilon\d\omega\,
\frac{\partial_{\varepsilon}F_{\varepsilon}}
{\big[(Dq^{2}+\tau^{-1}_{\mathrm{GL}})^{2}+\omega^{2}\big]
\big[(Dq^{2})^{2}+(2\varepsilon+\omega)^{2}\big]}\,.
\end{equation}
Finally, after the remaining integrations for
quasi--two--dimensional case, one finds
\begin{equation}\label{Part-IV-sigma-MT}
\frac{\delta\sigma_{\mathrm{MT}}}{\sigma_{D}}=\frac{\mathrm{Gi}}{8}\left(\frac{T_{c}}{T-T_{c}}\right)
\left(\frac{1}{1-\tau_{\mathrm{GL}}/\tau_{\phi}}\right)\ln\left(\frac{\tau_{\phi}}
{\tau_{\mathrm{GL}}}\right)\,,
\end{equation}
where infrared divergency in  momentum integral was cut off by a
dephasing rate $Dq^{2}_{\mathrm{min}}=\tau^{-1}_{\phi}$. This
divergency is a well--known feature of the Maki--Thompson diagram.
It can be regularized by some phase--braking mechanism in the Cooper
channel. For example, if magnetic impurities are present in the
system, then the role of $\tau_{\phi}$ is played by the spin--flip
time. In contrast to $\delta\sigma_{\mathrm{DOS}}$ Maki--Thompson
correction \eqref{Part-IV-sigma-MT} is positive and has much
stronger (power law) temperature dependence. Interestingly, that
$\delta\sigma_{\mathrm{MT}}$ follows from the effective
Ginzburg--Landau theory as well. Indeed, defining
$\delta\sigma_{\mathrm{MT}}=e^{2}\nu\langle\delta
D^{\mathrm{MT}}_{\mathbf{r},t,t'}\rangle_{\Delta}$,
employing~\eqref{SuperCond-D-MT} and performing averaging over
$\Delta$, one recovers~\eqref{Part-IV-sigma-MT}.

There is yet another correction to conductivity, called
Aslamazov--Larkin contribution. It is obtained from the second trace
on the right--hand side of~\eqref{Part-IV-Z}. Indeed, expanding each
$\check{Q}$ matrix to the linear order in $\check{\mathcal{W}}$, one
finds
\begin{equation}\label{Part-IV-Z-AL}
\mathcal{Z}_{\mathrm{AL}}[\mathbf{A}^{cl},\mathbf{A}^{q}]=-\frac{(\pi\nu
D)^{2}}{2}\,
\left\langle\left\langle\left(\Tr\big\{\check{\mathbb{A}}_{\varepsilon_{1}\varepsilon_{2}}
(\hat{\sigma}_{z}\otimes\hat{\tau}_{z})
\check{\mathcal{W}}_{\varepsilon_{2}\varepsilon_{3}}
\partial_{\mathbf{r}}\check{\mathcal{W}}_{\varepsilon_{3}\varepsilon_{1}}\big\}
\right)^{2}\right\rangle\right\rangle_{\mathcal{W},\Delta}\,.
\end{equation}
It is convenient to introduce two vertices, which follows
from~\eqref{Part-IV-Z-AL} after differentiation over the vector
potential
\begin{subequations}\label{Part-IV-V-AL}
\begin{equation}
\hskip-5cm \,\,\,\,\,
\mathbb{V}^{cl}_{\mathrm{AL}}[\check{\mathcal{W}}] =
\frac{\delta}{\delta\mathbf{A}^{cl}(\Omega)}\,
\Tr\big\{\check{\mathbb{A}}_{\varepsilon_{1}\varepsilon_{2}}
(\hat{\sigma}_{z}\otimes\hat{\tau}_{z})
\check{\mathcal{W}}_{\varepsilon_{2}\varepsilon_{3}}
\partial_{\mathbf{r}}\check{\mathcal{W}}_{\varepsilon_{3}\varepsilon_{1}}\big\}
\nonumber
\end{equation}
\begin{equation}
 \!\!\!\! = \tr\Big\{ c_{\varepsilon_{2}\varepsilon_{3}}(\mathbf{r})
\partial_{\mathbf{r}}c^{*}_{\varepsilon_{3}\varepsilon_{2}+\Omega}(\mathbf{r})
+c^{*}_{\varepsilon_{2}\varepsilon_{3}}(\mathbf{r})
\partial_{\mathbf{r}}c_{\varepsilon_{3}\varepsilon_{2}+\Omega}(\mathbf{r})
- (c\to\bar{c}) \Big\}\,,
\end{equation}
\begin{equation}
\hskip-4.8cm \mathbb{V}^{q}_{\mathrm{AL}}[\check{\mathcal{W}}] =
\frac{\delta}{\delta\mathbf{A}^{q}(0)}\,
\Tr\big\{\check{\mathbb{A}}_{\varepsilon_{1}\varepsilon_{2}}
(\hat{\sigma}_{z}\otimes\hat{\tau}_{z})
\check{\mathcal{W}}_{\varepsilon_{2}\varepsilon_{3}}
\partial_{\mathbf{r}}\check{\mathcal{W}}_{\varepsilon_{3}\varepsilon_{1}}\big\}
\nonumber
\end{equation}
\begin{equation}
 \!\! = -\tr\Big\{F_{\varepsilon_{2}}
\big(c_{\varepsilon_{2}\varepsilon_{3}}(\mathbf{r})
\partial_{\mathbf{r}}c^{*}_{\varepsilon_{3}\varepsilon_{2}}(\mathbf{r})+
c^{*}_{\varepsilon_{2}\varepsilon_{3}}(\mathbf{r})\partial_{\mathbf{r}}
c_{\varepsilon_{3}\varepsilon_{2}}(\mathbf{r})
+(c\to\bar{c})\big)\Big\}\,.
\end{equation}
\end{subequations}
Notice  that for $\mathbb{V}^{q}_{\mathrm{AL}}$ it is sufficient to
take external frequency to be zero right away, $\Omega=0$, while for
$\mathbb{V}^{cl}_{\mathrm{AL}}$ it is important to keep finite
$\Omega$ and take the dc limit, $\Omega\to 0$, only after
$\check{\mathcal{W}}$ averaging. Performing averaging over
Cooperons, one uses~\eqref{app_FE-c-extremal}. In the case of
$\mathbb{V}^{q}_{\mathrm{AL}}[\check{\mathcal{W}}]$, for the product
of two Cooper fields it is sufficient to retain only contributions
with classical components of the order parameter,
$\mathbb{V}^{q}_{\mathrm{AL}}[\check{\mathcal{W}}]\propto
\Tr\big\{F[c\partial_{\mathbf{r}}c^{*}+c^{*}\partial_{\mathbf{r}}c]\big\}
\propto\Delta^{cl}\partial_{\mathbf{r}}\Delta^{*cl}-
\Delta^{*cl}\partial_{\mathbf{r}}\Delta^{cl}$. In contrast, for the
$\mathbb{V}^{cl}_{\mathrm{AL}}[\check{\mathcal{W}}]$ vertex, it is
crucial to keep at least one quantum component of the order
parameter $\Delta^{q}$, since the corresponding contribution with
two classical components vanishes owing to causality structure.  As
a result, the leading contribution is
$\mathbb{V}^{cl}_{\mathrm{AL}}[\check{\mathcal{W}}]\propto
\Tr\big\{c\partial_{\mathbf{r}}c^{*}+c^{*}\partial_{\mathbf{r}}c\big\}
\propto\Delta^{cl}\partial_{\mathbf{r}}\Delta^{*q}+
\Delta^{q}\partial_{\mathbf{r}}\Delta^{*cl}-c.c.$. The remaining
$\Delta$ averaging of the product
$\big\langle\mathbb{V}^{cl}_{\mathrm{AL}}[\check{\mathcal{W}}]
\mathbb{V}^{q}_{\mathrm{AL}}[\check{\mathcal{W}}]\big\rangle_{\Delta}$
is done with the help of~\eqref{Part-IV-Delta-averages}. Passing to
the momentum representation and collecting all the factors,
Aslamazov--Larkin type correction to conductivity in the dc limit
takes the form
\begin{equation}
\delta\sigma_{\mathrm{AL}}=\frac{\pi^{2}
e^{2}D}{8T^{2}_{c}}\sum_{q}Dq^{2}
\int^{+\infty}_{-\infty}\frac{\d\omega}{2\pi}\,
\frac{\partial}{\partial\omega}\left[\coth\frac{\omega}{2T}\right]
\big[\mathrm{Im}L^{R}\qo\big]^{2}\,.
\end{equation}
The corresponding diagram is shown in  Figure~\ref{Fig-AL-MT-DOS}c.
Since only $Dq^2\sim\omega\sim\tau^{-1}_{GL}\ll T_{c}$ are relevant,
one may approximate
$\partial_{\omega}[\coth\omega/2T]\approx-2T_{c}/\omega^2$ and use
$\mathrm{Im}L^{R}\qo=-(8iT_{c}\omega/\pi)[(Dq^{2}+\tau^{-1}_{\mathrm{GL}})^2+\omega^2]^{-1}$
to obtain
\begin{equation}\label{Part-IV-sigma-AL}
\delta\sigma_{\mathrm{AL}}=\frac{8e^{2}DT_c}{\pi}\sum_{q}\int^{+\infty}_{-\infty}\!\!\d\omega\,
\frac{Dq^{2}}{\big[\big(Dq^{2}+\tau^{-1}_{\mathrm{GL}}\big)^{2}+\omega^{2}\big]^{2}}\,.
\end{equation}
 Performing remaining integrations, one finds for the
 quasi--two--dimensional film
\begin{equation}
\frac{\delta\sigma_{\mathrm{AL}}}{\sigma_{D}}=\frac{\mathrm{Gi}}{16}
\left(\frac{T_{c}}{T-T_{c}}\right)\,.
\end{equation}

At the level of effective time--dependent GL functional,
Aslamazov--Larkin conductivity correction
$\delta\sigma_{\mathrm{AL}}$ appears from the $S_{\mathrm{SC}}$ part
of the action~\eqref{SuperCond-S-SC}. The easiest way to see this is
to use current $$\mathbf{j}_{\mathrm{SC}}=\frac{\pi e\nu
D}{4T_{c}}\mathrm{Im}[\Delta^{*cl}\partial_{\mathbf{r}}\Delta^{cl}]\,,$$
which follows from $S_{\mathrm{SC}}$, along with the
fluctuation--dissipation relation
$\delta\sigma_{\mathrm{AL}}\propto\big\langle
\mathbf{j}_{\mathrm{SC}}\cdot\mathbf{j}_{\mathrm{SC}}\big\rangle_{\Delta}
\propto\sum_{q\omega}Dq^2|L^{R}\qo|^2$. The latter
reproduces~\eqref{Part-IV-sigma-AL}.

The technique which was employed within this section allows to
reproduce all the results for fluctuations--induced conductivity,
known from conventional Matsubara diagrammatic approach. The
simplification here is that no analytical continuation was needed.
Although it is not so complicated for the problem at hand, in some
cases avoiding the  analytical continuation may be an advantage.

\subsubsection{Tunneling  conductance above $T_c$}\label{app_Part-IV-5}

Consider voltage--biased superconductor--normal metal tunnel
junction, where the superconductor is assumed to be at the
temperature just above the transition $T_c$, i.e. in the fluctuating
regime. It is natural to expect that depletion in the density of
states, mediated by fluctuations (see Section~\ref{app_Part-IV-3}),
modifies current--voltage characteristics of the
junction~\cite{VarlamovDorin,DiCastro,Reizer}. This effect can be
studied within $\sigma$--model, using tunneling  part of the action
$S_{T}[\check{Q}_{L},\check{Q}_{R}]$.

One starts from~\eqref{NLSM-action-S-T} and performs gauge
transformation $\check{Q}_{a}\to\exp(-i\check{\Xi}\check{\Phi}_{a})
\check{Q}_{a}\exp(i\check{\Xi}\check{\Phi}_{a})$, for $a=L,R$, where
$\check{\Phi}_{a}(t)=\int^{t}\check{V}_{a}(t)\d
t=[\Phi^{cl}_{a}(t)\hat{\sigma}_{0}+
\Phi^{q}_{a}(t)\hat{\sigma}_{x}]\otimes\hat{\tau}_{0}$, and
$\Phi^{cl}_{L}-\Phi^{cl}_{R}=eVt$, which moves an applied voltage
$V$ from the Keldysh blocks of the $\check{Q}$ matrices, to the
tunneling  part of the action
\begin{equation}\label{Part-IV-S-T}
iS_{T}[\check{Q}_{L},\check{Q}_{R}]=\frac{\mathrm{g}_{T}}{4\mathrm{g}_{Q}}
\Tr\big\{\check{Q}_{L}e^{-i\check{\Xi}\check{\Phi}}
\check{Q}_{R}e^{i\check{\Xi}\check{\Phi}}\big\}\,,
\end{equation}
here $\check{\Phi}=\check{\Phi}_{L}-\check{\Phi}_{R}$, and
$\Phi^{q}(t)$ serves as the generating field. Indeed, since the phase
$\check{\Phi}$ is quantum canonical conjugate to the number of
particles $\check{N}=i\partial/\partial\Phi$ the tunneling current
is obtained by differentiating the partition function
$\mathcal{Z}_{T}[\Phi]=\exp\big(iS_{T}[\check{Q}_{L},\check{Q}_{R}]\big)$
with respect to the quantum component of the phase
\begin{equation}\label{Part-IV-I-T}
I_{T}(t)=ie\left(\frac{\delta\mathcal{Z}_{T}[\Phi]}{\delta\Phi^{q}(t)}
\right)_{\Phi^{q}=0}.
\end{equation}
Applying this definition to~\eqref{Part-IV-S-T}, using
\begin{equation}\label{Part-IV-derivative}
\left.\frac{\delta \exp(\pm
i\check{\Xi}\check{\Phi})}{\delta\Phi^{q}(t')}\right|_{\Phi^{q}=0}=\pm
i\delta(t-t')\,\big(\hat{\sigma}_{x}\otimes\hat{\tau}_{z}\big)\,\exp\big[\pm
ieVt\check{\Xi}\big]\,
\end{equation}
and taking $\check{Q}_{L}=\check{Q}_{R}=\check{\Lambda}$, one finds
 Ohm's law $I_{T}=\mathrm{g}_{T}V$, as it should be, for the tunneling
junction in the normal state. One may account now for the
fluctuation effects by expanding one of the $\check{Q}$ matrices
in~\eqref{Part-IV-S-T} over Cooper modes $\check{\mathcal{W}}$. This
leads to the  correction of the form
\begin{equation}\label{Part-IV-I-T-corr}
\delta I_{T}(V)=-\frac{\pi
\mathrm{g}_{T}}{2e}\sum_{q}\iint\frac{\d\varepsilon\d\varepsilon'}{4\pi^{2}}
\big(F_{\varepsilon+eV}-F_{\varepsilon-eV}\big)
\left\langle\left\langle c_{\varepsilon\varepsilon'}(\mathbf{q})
c^{*}_{\varepsilon'\varepsilon}(-\mathbf{q})+
\bar{c}_{\varepsilon\varepsilon'}(\mathbf{q})
\bar{c}^{*}_{\varepsilon'\varepsilon}(-\mathbf{q})
\right\rangle\right\rangle_{\mathcal{W},\Delta}\,,
\end{equation}
which is physically expected result. Indeed, from the combination of
the Cooper modes in~\eqref{Part-IV-I-T-corr} one recognizes density
of states correction $\delta\nu(\varepsilon)$,
see~\eqref{Part-IV-Dos-corr-def}. The latter is convoluted
in~\eqref{Part-IV-I-T-corr} with the difference of Fermi functions,
leading to the correction to the tunneling current of the form
$\delta
I_T(V)\sim\int\d\varepsilon[F_{\varepsilon+eV}-F_{\varepsilon-eV}]
\delta\nu_{L}(\varepsilon)\nu_{R}$. Using  previous result for
$\delta\nu(\varepsilon)$ from~\eqref{Part-IV-Dos-intermediate},
bringing it into~\eqref{Part-IV-I-T-corr} and transforming to the
dimensionless units $x=Dq^{2}/T$, $y=\omega/T$, $z=\varepsilon/2T$
one finds for the tunneling differential conductance correction
$\delta \mathrm{g}_{T}(V)=\partial\delta I_{T}(V)/\partial V$ the
following expression:
\begin{eqnarray}
\frac{\delta\mathrm{g}_{T}(V)}{\mathrm{g}_{T}}=\frac{4\mathrm{Gi}}{\pi^{3}}
\int^{\infty}_{0}\d x\iint^{+\infty}_{-\infty}\d y\d
z\left[\frac{1}{\cosh^{2}(z+u)}+\frac{1}{\cosh^{2}(z-u)}\right]\\
\times\,\mathrm{Re}\left[\frac{1}{\big(x+iy-4iz\big)^{2}
\big((x+1/T\tau_{\mathrm{GL}})^{2}+y^{2}\big)}\right]\,,
\end{eqnarray}
where $u=eV/2T$ and we assumed quasi--two--dimensional geometry.
Remaining integrations can be done in the closed form, resulting
in~\cite{VarlamovDorin}
\begin{equation}
\frac{\delta\mathrm{g}_{T}(V)}{\mathrm{g}_{T}}=\frac{\mathrm{Gi}}{\pi^{4}}
\ln\left(\frac{T_{c}}{T-T_{c}}\right)
\mathrm{Re}\,\psi^{[2]}\left(\frac{1}{2}-\frac{ieV}{2\pi T}\right),
\end{equation}
where $\psi^{[2]}(x)$ is the second order derivative of the digamma
function $\psi(x)$. Notice, that although having direct relation to
the density of states suppression $\delta\nu(\varepsilon)$, the
differential conductance correction $\delta\mathrm{g}_{T}$ exhibits
much weaker temperature dependence. The sharp suppression in the
density of states $\delta\nu(0)\propto(T-T_{c})^{-2}$ translates
only into the moderate logarithmic in temperature correction
$\delta\mathrm{g}_{T}\propto\ln(T_{c}\tau_{\mathrm{GL}})$. Another
interesting feature is that suppression  of the
$\delta\nu(\varepsilon)$ occurs at the energies
$\varepsilon\sim\tau^{-1}_{\mathrm{GL}}\sim T-T_{c}$, while
corresponding suppression of the differential conductance happens at
voltages $V\sim T_{c}$, and not at $V\sim T-T_c$. Finally one should
mention, that  more singular in $(T-T_c)$ MT and AL corrections
appear  only in the fourth order in the tunneling matrix elements,
while the discussed DOS effect is linear in  $\mathrm{g}_{T}$ (i.e.
it is of the second order in the tunneling matrix elements).

\subsubsection{Current noise in fluctuating regime}\label{app_Part-IV-6}

Apart from the density of states related effects, there are
interesting consequences of superconducting fluctuations on the
current noise of the tunneling
junction~\cite{Larkin-Varlamov,Kulik,Scalapino,Martin-Balatsky,Levchenko-J-Noise}.
Assume now that both sides of the junction are made from identical
superconductors that are kept right above $T_c$. While there is no
average Josephson current in this case, the  noise power turns out
to be sensitive to the Jesephson frequency, $\omega_J=2eV/\hbar$,
and exhibits sharp peak at $\omega=\omega_{J}$. The height and shape
of this peak have a singular temperature dependence near $T_c$,
which makes its experimental detection possible. To show this we
establish an expression for the fluctuating part of the tunneling
current $\delta I_{T}(t)$ in terms of the product of fluctuating
order parameters $\Delta_{L(R)}\rt$ residing on the different sides
of the junction, namely $\delta
I_{T}(t)\propto\int\d\mathbf{r}\,[\Delta_{R}\rt\Delta^{*}_{L}\rt\exp(-i\omega_{J}t)-c.c.]$.
Since $\langle\Delta_{L(R)}\rangle=0$ above $T_{c}$, it is  clear
that $\langle\delta I_{T}(t)\rangle=0$. However, the average square
of the current $\langle\delta I_{T}(t)\delta I_{T}(t')\rangle$ is
not vanishing and its Fourier transform displays a peak at the
Josephson frequency. In what follows we calculate its temperature
dependence.

One starts from the  definition of the current--current
correlation function
\begin{equation}\label{Part-IV-noise-def}
\mathcal{S}_{T}(\omega)=-e^{2}\int^{+\infty}_{-\infty}\d(t-t')
\left(\frac{\delta^{2}\mathcal{Z}_{T}[\Phi]}{\delta
\Phi^{q}(t)\delta
\Phi^{q}(t')}\right)_{\Phi^{q}=0}e^{-i\omega(t-t')}\,.
\end{equation}
In the normal state $\check{Q}_{L}=\check{Q}_{R}=\check{\Lambda}$
and the noise power of the tunneling  junction, as it follows
from~\eqref{Part-IV-noise-def}, is given by the Schottky formula
$\mathcal{S}_{T}(\omega)=2\mathrm{g}_{T}T\sum_{\pm}v_{\pm}\coth
v_{\pm}$, where $v_{\pm}=(eV\pm\omega)/2T$. To account for the
superconductive fluctuations on  both sides of the junction one has
to expand each of the $\check{Q}$ matrices in~\eqref{Part-IV-S-T} to
the leading (linear) order in Cooper modes. This gives for the
fluctuation correction to the current
\begin{equation}\label{Part-IV-I-J}
\delta I_{T}(t)=\frac{i\pi\mathrm{g}_{T}}{4e}\frac{\delta}{\delta
\Phi^{q}(t)}\Tr\left\{e^{i\check{\Xi}\check{\Phi}}
\check{\mathcal{U}} (\hat{\sigma}_{z}\otimes\hat{\tau}_{z})
\check{\mathcal{W}}_{L}\check{\mathcal{U}}^{-1}
e^{-i\check{\Xi}\check{\Phi}}
\check{\mathcal{U}} (\hat{\sigma}_{z}\otimes\hat{\tau}_{z})
\check{\mathcal{W}}_{R}\check{\mathcal{U}}^{-1}\right\}.
\end{equation}
To proceed further, one simplifies~\eqref{Part-IV-I-J}, exploring
separation of  time scales between electronic and order parameter
degrees of freedom. Indeed, one should notice that, as follows
from~\eqref{app_FE-L-R}, the relevant energies and momenta for the
order--parameter variations are
$Dq^{2}\sim\omega\sim\tau^{-1}_{\mathrm{GL}}$, while the relevant
fermionic energies entering the Cooperons are
$\epsilon\sim\epsilon'\sim T\gg \tau^{-1}_{\mathrm{GL}}$.  As a
result, non--local relations between Cooper modes
\eqref{SuperCond-W} and the order parameter, see
Eqs.~\eqref{app_FE-c-extremal}, may be approximated as
\begin{eqnarray}\label{Part-IV-c-Delta-approx}
&& \check{\mathcal{W}}^{a}_{tt'}(\mathbf{r})\approx-i\,
\hat{\Theta}_{tt'}\otimes\hat{\Delta}^{a}_{tt'}(\mathbf{r}),\quad\quad
\hat{\Theta}_{tt'}= \left(\begin{array}{cc} \theta(t-t') & 0 \\
0 & -\theta(t'-t),
\end{array}\right)_K,\nonumber \\
&&
\hat{\Delta}^{a}_{tt'}(\mathbf{r})=\Delta^{cl}_{a}\left(\mathbf{r},\frac{t+t'}{2}\right)
\hat{\tau}_{+}+
\Delta^{*cl}_{a}\left(\mathbf{r},\frac{t+t'}{2}\right)\hat{\tau}_{-}\,,\quad
a=L,R\,,
\end{eqnarray}
where $\theta(t)$ is the step function.
Physically~\eqref{Part-IV-c-Delta-approx} implies that Cooperon is
short--ranged, having characteristic length scale
$\xi_{0}=\sqrt{D/T_{c}}$, as compared to the long--ranged
fluctuations of the order parameter, which propagates to the
distances of the order of
$\xi_{\mathrm{GL}}=\sqrt{D\tau_{\mathrm{GL}}}\gg\xi_{0}$. Thus,
relations \eqref{app_FE-c-extremal} are effectively local, which
considerably simplifies the further analysis. Equations
\eqref{Part-IV-c-Delta-approx} allow us to trace Keldysh subspace
in~\eqref{Part-IV-I-J} explicitly to arrive at
\begin{equation}\label{Part-IV-I-J-1}
\delta I_{T}(t)=-\frac{\pi
\mathrm{g}_{T}}{e}\Tr\left\{\theta(t_{2}-t_{1})F_{t_{1}-t}\theta(t-t_{2})
\hat{\Delta}^{L}_{tt_{2}}\hat{\tau}_{z}
\hat{\Delta}^{R}_{t_{2}t_{1}}e^{ieV(t+t_{2})\hat{\tau}_{z}}\right\}_N\,,
\end{equation}
where we have used~\eqref{Part-IV-derivative} and wrote trace in the
real space--time representation (note that $\mathrm{Tr}\{\ldots\}$
here does not imply time $t$ integration). Changing integration
variables $t_{1}=t-\mu$ and $t_{3}=t-\eta$, and rescaling $\eta,\mu$
in the units of temperature $T\eta\to\eta, T\mu\to\mu$, one finds
for Eq.~\eqref{Part-IV-I-J-1} an equivalent representation,
\begin{equation}\label{Part-IV-I-J-2}
\delta I_{T}(t)=-\frac{i\pi
\mathrm{g}_{T}}{eT}\iint^{+\infty}_{-\infty}d\eta d\mu\,\,
\frac{\theta(\eta)\theta(\mu-\eta)}{\sinh(\pi\mu)}\,
\Tr \left\{\hat{\Delta}^{L}_{t,t-\frac{\eta}{T}}\hat{\tau}_{z}
\hat{\Delta}^{R}_{t-\frac{\eta}{T},t-\frac{\mu}{T}}
e^{ieV\big(2t-\frac{\eta}{T}\big)\hat{\tau}_{z}}\right\}_N\,,
\end{equation}
where we used equilibrium fermionic distribution function in the
time domain $F_{t}=-iT/\sinh(\pi Tt)$. The most significant
contribution to the above integrals comes from
$\eta\sim\mu\lesssim1$. At this range ratios $\{\eta,\mu\}/T$ change
on the scale of inverse temperature, while as we already discussed,
order--parameter variations are set by
$t\sim\tau_{\mathrm{GL}}\gg1/T$. Thus, performing $\eta$ and $\mu$
integrations one may neglect $\{\eta,\mu\}/T$ dependence of the
order parameters and the exponent. As a result one finds
\begin{equation}\label{Part-IV-I-J-3}
\delta I_T(t)=\frac{i\pi
\mathrm{g}_{T}}{4eT}\int\frac{\mathrm{d}^{2}\mathbf{r}}{\mathcal{A}}\left[
\Delta^{cl}_{R}(\mathbf{r},t)\Delta^{*cl}_{L}(\mathbf{r},t)
e^{-i\omega_{J}t}-c.c.\right]\,,
\end{equation}
where the spatial integration runs over the junction area
$\mathcal{A}$ and $\omega_J=2eV/\hbar$. Finally one is ready to
calculate corresponding contribution to the current noise. One
substitutes two currents in the form of~\eqref{Part-IV-I-J-3}
into~\eqref{Part-IV-noise-def} and pairs fluctuating order
parameters using the correlation function, which follows
from~\eqref{Part-IV-Delta-averages},
$$\langle\Delta^{cl}_{a}(\mathbf{r},t)\Delta^{*cl}_{b}(\mathbf{r}',t')\rangle_{\Delta}=
\frac{i}{2\nu}\delta_{ab}L^{K}(\mathbf{r}-\mathbf{r}',t-t')\,.$$ As
a result, superconducting  fluctuation correction to the noise power
is given by
\begin{equation}\label{Part-IV-noise-correction}
\delta\mathcal{S}_{T}(\omega)=-\frac{1}{4\nu^{2}}\left(\frac{\pi
\mathrm{g}_{T}}{4eT_{c}}\right)^{2}\sum_{\pm}
\int\frac{\mathrm{d}^{2}\mathbf{r}}{\mathcal{A}}
\int^{+\infty}_{-\infty}\mathrm{d}t\,
\big[L^{K}(\mathbf{r},t)\big]^{2}\, \exp(-i\omega_{\pm}t)\,,
\end{equation}
where $\omega_{\pm}=\omega\pm\omega_{J}$. Performing the remaining
integrations one finds first Keldysh component of the fluctuation
propagator in the mixed momentum/time representation
$L^{K}(\mathbf{q},t)=\int L^{K}(\mathbf{q},\omega)e^{-i\omega
t}\d\omega/2\pi$, which is
\begin{equation}
L^{K}(\mathbf{q},t)=-\frac{2iT^{2}_{c}}{T-T_{c}}
\frac{e^{-\varkappa_{q}|t|/\tau_{\mathrm{GL}}}}{\varkappa_{q}},\quad\quad
\varkappa_{q}=(\xi_{\mathrm{GL}}q)^{2}+1\,.
\end{equation}
Then insert $L^{K}(\mathbf{r},t)=\int
L^{K}(\mathbf{q},t)e^{i\mathbf{qr}}\d q^{2}/4\pi$
into~\eqref{Part-IV-noise-correction}, introduces dimensionless time
$\tau=t/\tau_{\mathrm{GL}}$, and changes from $q$ to $\varkappa_q$
integration $\d q^{2}=\d\varkappa_q/\xi^{2}_{\mathrm{GL}}$, which
gives altogether~\cite{Levchenko-J-Noise}
\begin{equation}
\delta\mathcal{S}_{T}(\omega)=\sum_{\pm}\frac{\pi
\mathrm{Gi}^{2}}{64
T_c}\left(\frac{\mathrm{g}_{T}T_c}{e}\right)^{2}\frac{\xi^{2}_{0}}{\mathcal{A}}
\left(\frac{T_{c}}{T-T_{c}}\right)^{2}\mathcal{N}(\omega_{\pm}\tau_{GL})\,,
\end{equation}
where the spectral function is given by
\begin{equation}
\mathcal{N}(z)=\int^{+\infty}_{-\infty}d\tau\int^{+\infty}_{1}
\frac{\d\varkappa}{\varkappa^{2}}\, \exp(-2\varkappa|\tau|-iz
\tau)=\frac{4}{z^{2}}\ln\sqrt{1+z^{2}/4}\,.
\end{equation}
The noise power correction $\delta\mathcal{S}_{T}(\omega)$ is peaked
at the Josephson frequency $\omega=\pm\omega_{J}$ and has strong
temperature dependence, which makes its experimental detection
possible in a vicinity of the superconducting transition.

\section{Concluding remarks and acknowledgments}\label{sec_Summary}

We have attempted to review various  ingredients and elements of the
Keldysh technique in applications to systems of interacting  bosons
and fermions. The emphasis has been on the functional integral
representation of microscopic  models and some modern developments,
such as non--linear $\sigma$--model. Our motivation was not to
review some specific  area of physics, where Keldysh technique may
be applied successfully, but rather to focus on exposing the method
itself. The goal was to give a broad perspective of the technique
and its applications. To accomplish this goal  we have (rather
subjectively) chosen examples  from mesoscopic physics of normal
metals as well superconducting kinetics. We hope that this review
may serve as a learning source for interested graduate students and
a reference point for experts.

We benefited greatly from collaboration and discussions with many of
our colleagues: I.~Aleiner, A.~Altland, A.~Andreev, M.~Feigel'man, Y.~Gefen,
L.~Glazman, A.~I.~Larkin, I.~Lerner, M.~Skvortsov and many others who shaped
this paper. While writing this review, communications with
D.~Bagrets, I.~Burmistrov, G.~Catelani, A.~Chudnovskiy, M.~Khodas, and A.~Varlamov were
especially helpful. This work was supported by NSF Grants  DMR-0405212 and
DMR-0804266.  A.~L. was supported by the
 Doctoral Dissertation Fellowship from the University of Minnesota.

\newpage
\appendix
\section{Gaussian integrals for bosons and fermions}\label{app_Gaussian}

For any complex $N\times N$ matrix $A_{ij}$, where $i,j=1,\ldots N$,
such that all its eigenvalues, $\lambda_i$, have a positive real
part, $\mathrm{Re} \lambda_i>0$, the following statement holds
\begin{equation}\label{app-Gauss-int-bos}
Z[J]=\iint_{-\infty}^{+\infty}\prod\limits_{j=1}^N
\frac{\d(\mathrm{Re} z_j)  \d(\mathrm{Im} z_j)}{\pi}\,
\exp\left(-\sum_{ij}^{N} \bar z_i A_{ij} z_j+ \sum_{j}^{N}\left[\bar
z_j J_j + \bar J_j z_j\right]\right) =\frac{\exp\left(\sum_{ij}^{N}
\bar J_i (A^{-1})_{ij} J_j\right)}{\mathrm{Det}(A)}\, ,
\end{equation}
where $J_j$ is an arbitrary complex vector. To prove it, one may
start from a Hermitian matrix, that is diagonalized by a unitary
transformation: $A=U^\dagger\Lambda U$, where
$\Lambda=\mbox{diag}\{\lambda_j\}$. The identity is then easily
proven by a  change of  variables (with unit Jacobian) to
$w_i=U_{ij}z_j$. Finally, one notices that the right--hand side
of~\eqref{app-Gauss-int-bos} is an analytic function of both
$\mathrm{Re} A_{ij}$ and $\mathrm{Im} A_{ij}$. Therefore, one may
continue them analytically to the complex plane to reach an
arbitrary complex matrix $A_{ij}$. The identity
\eqref{app-Gauss-int-bos} is thus valid as long as the integral is
well defined, that is all the eigenvalues of $A_{ij}$ have a
positive real part.

The Wick theorem deals with the average value of a string  $
z_{a_1}\ldots z_{a_k}\bar z_{b_1}\ldots \bar z_{b_k}$ weighted with
the factor $\exp\big(-\sum_{ij}\bar z_i A_{ij} z_j\big)$. The
theorem states  that this average is given by the sum of all
possible products of pair--wise averages. For example,
\begin{eqnarray}\label{app-Gauss-Wick}
&&\left\langle z_{a} \bar z_{b}\right\rangle \equiv \left.
\frac{1}{Z[0]}\frac{\delta^{2}Z[J]}{\delta \bar J_a\delta
J_b}\right|_{J=0}= \big(A^{-1}\big)_{ab}\, ,\\
&&\left\langle z_{a}z_{b} \bar z_{c}\bar z_{d} \right\rangle
\equiv\left.\frac{1}{Z[0]}\frac{\delta^4Z[J]}{\delta \bar
J_{a}\delta\bar{J}_{b} \delta J_{c} \delta J_{d} }\right|_{J=0}=
A^{-1}_{ac}A^{-1}_{bd}+A^{-1}_{ad}A^{-1}_{bc} ,\nonumber
\end{eqnarray}
and so on.

The Gaussian identity for integration over real variables  has the form
\begin{equation}\label{app-Gauss-Real}
Z[J]=\int_{-\infty}^{+\infty}\prod\limits_{j=1}^N \frac{d
x_j}{\sqrt{\pi}}\,\exp\left(-\sum\limits_{ij}^N  x_i A_{ij} x_j +
2\sum\limits_{j}^N   x_j J_j\right)=\frac{ \exp\left(\sum_{ij}^{N}
J_i (A^{-1})_{ij} J_j\right)}{\sqrt{\mathrm{Det} (A)}}\, ,
\end{equation}
where $A$ is a {\em symmetric} complex matrix with all its
eigenvalues having a positive real part. The proof is similar to the
proof in the case of complex variables: one starts from a real
symmetric matrix, that may be diagonalized by an orthogonal
transformation. The identity \eqref{app-Gauss-Real} is then easily
proved by the change of variables. Finally, one may analytically
continue the right--hand side (as long as the integral is well
defined) from a real symmetric matrix $A_{ij}$, to a {\em complex
symmetric} matrix.

For an integration over two sets of {\em independent} Grassmann
variables, $\bar \xi_j$ and $\xi_j$, where $j=1,2,\ldots,N$, the
Gaussian identity is valid for {\em any invertible} complex matrix
$A$
\begin{equation}\label{app-Gauss-int-ferm}
Z[\bar\chi,\chi]=\iint\prod\limits_{j=1}^N \d\bar\xi_j \d\xi_j
\exp\left(-\sum\limits_{ij}^N \bar \xi_i A_{ij} \xi_j +
\sum\limits_{j}^N \left[\bar \xi_j \chi_j + \bar \chi_j
\xi_j\right]\right)= \mathrm{Det}(A) \exp\left(\sum\limits_{ij}^N
\bar \chi_i (A^{-1})_{ij} \chi_j\right)\, .\quad
\end{equation}
Here $\bar \chi_j$ and $\chi_j$ are two additional mutually
independent (and independent from $\bar \xi_j$ and $\xi_j$) sets of
Grassmann numbers. The proof may be obtained by, e.g., brute force
expansion of the exponential factors, while noticing that only terms
that are linear in {\em all} $2N$ variables $\bar \xi_j$ and $\xi_j$
are non--zero. The Wick theorem is formulated in the same manner  as
for the bosonic case, with the exception that every combination is
multiplied by the parity of the corresponding permutation. For
example, the first term on the right--hand side of the second
expression of~\eqref{app-Gauss-Wick} comes with the minus sign.

\section{Single particle quantum mechanics}
\label{app_SPQM}

The simplest many--body system of a single bosonic state (considered
in Section~\ref{sec_bosons}) is, of course, equivalent to a
single--particle harmonic oscillator. To make this connection
explicit,  consider the Keldysh contour action~\eqref{boson-Z-1}
with the correlator~\eqref{boson-G-continious} written in terms of
the complex field $\phi(t)$. The latter may be parameterized by its
real and imaginary parts as
\begin{equation}\label{app-SPQM-qp}
\phi(t) = {1\over\sqrt{\,2\omega_0}}\,\big({p(t)} -i\, {\omega_0}\,
q(t)  \big)\, , \qquad \bar \phi(t) =
{1\over\sqrt{\,2\omega_0}}\,\big( {p(t)}+i\, {\omega_0}\, q(t)
\big)\, .
\end{equation}
In terms of the real fields $p(t)$ and $q(t)$ the
action~\eqref{boson-Z-1} takes the form
\begin{equation}\label{app-SPQM-S}
S[p,q]=\int_{{\cal C}}\d t\left[p\,\dot q - {1\over
2}\left(p^2+\omega_0^2 q^2\right) \right]\, ,
\end{equation}
where the full time derivatives of $p^2$, $q^2$ and $p\,q$ were
omitted, since they contribute only to the boundary terms, not
written  explicitly in the continuum  notation (they have to be kept
for the proper regularization). Equation \eqref{app-SPQM-S} is
nothing but the action of the quantum harmonic oscillator in the
Hamiltonian form. One may perform the Gaussian integration over the
$p(t)$ field to obtain
\begin{equation}\label{app-SPQM-S-1}
S[q]={1\over 2}\int_{{\cal C}}\d t\left[ \dot q^2 -
\omega_0^2 \, q^2 \right]\, .
\end{equation}
This is the Feynman Lagrangian action of the harmonic oscillator,
written on the Keldysh contour. It may be generalized for  an
arbitrary single--particle potential $U(q)$
\begin{equation}\label{app-SPQM-S-2}
S[q(t)]= \int_{{\cal C}}\d t\left[ {1\over 2}\,\big(\dot q(t)\big)^2
- U\big(q(t)\big) \right]\, .
\end{equation}
One may split the $q(t)$ field into two components, $q_+(t)$ and
$q_-(t)$, residing on the forward and backward branches of the
contour, and then perform the Keldysh rotation: $q_{\pm}=q^{cl}\pm
q^{q}$. In terms of these fields the action takes the form
\begin{equation}\label{app-SPQM-S-3}
S[q^{cl},q^q]=\int_{-\infty}^{+\infty}\d t\left[-2\,q^{q}{\d^{\,2}
q^{cl}\over \d t^2} - U\big(q^{cl}+q^{q}\big) + U\big(q^{cl}-
q^{q}\big) \right],
\end{equation}
where  integration by parts was performed in the term $\dot
q^{q}\dot q^{cl}$. This is the Keldysh form of the Feynman path
integral. The omitted boundary terms provide a convergence factor of
the form $\sim i0(q^q)^2$.

If the fluctuations of the quantum component $q^{q}(t)$ are regarded
as small, one may expand  the potential to the first order and find
for the action
\begin{equation}\label{app-SPQM-S-4}
S[q^{cl},q^{q}]=\int_{-\infty}^{+\infty}\d
t\left[-2\,q^q\left({\d^{\,2} q^{cl}\over \d t^2} +{\partial
U\big(q^{cl}\big)\over \partial q^{cl} }\right) + i0 (q^{q})^{2}
+O\big[(q^q)^3\big]\right].
\end{equation}
In this limit the integration over the quantum component, $q^{q}$,
may be explicitly performed, leading to a functional
$\delta$--function of the expression in the round brackets. This
$\delta$--function enforces the classical Newtonian dynamics of
$q^{cl}$
\begin{equation}\label{app-SPQM-Newton}
{\d^{\,2} q^{cl}\over \d t^2} =- {\partial U\big(q^{cl}\big)\over
\partial q^{cl} }\, .
\end{equation}
For this  reason  the symmetric (over forward and backward branches)
part of the Keldysh field is called  the classical component. The
quantum--mechanical information is contained in the higher--order
terms in $q^{q}$, omitted in~\eqref{app-SPQM-S-4}. Note that for the
harmonic oscillator potential the terms denoted as $O[(q^{q})^3]$
are absent identically. The quantum (semiclassical) information
resides, thus, in the convergence term, $i0(q^{q})^2$, as well as in
the retarded regularization of the $\d^{\,2}/(\d t^2)$ operator
in~\eqref{app-SPQM-S-4}.

One may generalize the single--particle quantum mechanics onto a
chain (or lattice) of harmonically coupled particles by assigning an
index ${\bf r}$ to particle coordinates, $q_{\bf r}(t)$, and adding
the spring potential energy, ${v_s^2\over 2}(q_{{\bf r}+{\bf
1}}(t)-q_{\bf r}(t))^2$. Changing to spatially continuum  notation,
$\phi(\mathbf{r},t)\equiv q_{\mathbf{r}}(t)$, one finds for the
Keldysh action of the real (e.g. phonon) field
\begin{equation}\label{app-SPQM-S-Phonons}
S[\phi]= \int\d\mathbf{r}\int_{{\cal C}}\d t\left[ {1\over
2}\,\dot \phi^{\,2} - {v_s^2\over 2}\,(\partial_\mathbf{r}
\phi )^2 - U\big(\phi\big) \right]\, ,
\end{equation}
where the constant $v_s$ has the meaning of the sound velocity.
Finally, splitting the field into $(\phi_+,\phi_-)$ components and
performing the Keldysh transformation, $\phi_\pm=\phi^{cl}\pm
\phi^q$, and integrating by parts, one obtains
\begin{equation}\label{app-SPQM-S-Phonons-1}
 S[\phi^{cl},\phi^{q}]= \int\d\mathbf{r}
\int_{-\infty}^{+\infty}\d t\left[ 2\phi^{q}\big( -\partial_t^2 +
v_s^2\,\partial_\mathbf{r}^2 \big) \phi^{cl}
-U(\phi^{cl}+\phi^{q})+U(\phi^{cl}-\phi^{q}) \right] .
\end{equation}
According to the general structure of the Keldysh theory the
differential operator  $\big(-\partial_t^2 +
v_s^2\,\partial_\mathbf{r}^2\big)$, should be understood as the
retarded one. This means it is a lower--triangular matrix in the
time domain. Actually, one may symmetrize the action by performing
the integration by parts, and write it as
$\phi^{q}\big(-\partial_t^2 + v_s^2\,\partial_\mathbf{r}^2\big)^R
\phi^{cl}+ \phi^{cl}\big( -\partial_t^2 +
v_s^2\,\partial_\mathbf{r}^2\big)^A \phi^{q}$, with the advanced
regularization in the second term.

\section{Gradient expansion of the $\sigma$--model}
\label{app_GradientExpansion}

This Appendix serves as the complementary material for
Section~\ref{sec_NLSM-2}. Its purpose is to provide  technical
details hidden behind the transition from~\eqref{NLSM-action-SAV}
to~\eqref{NLSM-action-general}. For the gradient expansion of the
logarithm in~\eqref{NLSM-action-SAV} one uses $\hat Q$ matrix in the
form of~\eqref{NLSM-Q-Rotated} and finds in analogy
with~\eqref{NLSM-action-trlog-R}:
\begin{equation}\label{app-GE-S-trlog}
iS[\hat Q,\mathbf{A},V]=\Tr\ln\left[\hat{1}+i\hat{\mathcal{G}}
\hat{\mathcal{R}}\partial_{t}\hat{\mathcal{R}}^{-1}+i\hat{\mathcal{G}}\hat{\mathcal{R}}
\mathbf{v}_{F}\partial_{\mathbf{r}}\hat{\mathcal{R}}^{-1}+
\hat{\mathcal{G}}\hat{\mathcal{R}}\hat{V}\hat{\mathcal{R}}^{-1}+
\hat{\mathcal{G}}\hat{\mathcal{R}}\mathbf{v}_{F}
\hat{\mathbf{A}}\hat{\mathcal{R}}^{-1}\right]\,.
\end{equation}
Expanding this expression to the linear order in $\hat{\mathcal{G}}
\hat{\mathcal{R}}\partial_{t}\hat{\mathcal{R}}^{-1}$ and quadratic
in $\hat{\mathcal{G}}\hat{\mathcal{R}}
\mathbf{v}_{F}\partial_{\mathbf{r}}\hat{\mathcal{R}}^{-1}$,  one
reproduces~\eqref{NLSM-action-trlog-Approx} for $S[\hat Q]$, which
leads eventually to~\eqref{NLSM-action-noninteracting}. To the
linear order in $\hat{V}$ and $\hat{\mathbf{A}}$ one finds
from~\eqref{app-GE-S-trlog}
\begin{equation}\label{app-GE-S1-trlog}
iS_{1}[\hat Q,\mathbf{A},V]=\Tr\big\{\hat{\mathcal{G}}
\hat{\mathcal{R}}\hat{V}\hat{\mathcal{R}}^{-1}\big\}-
i\Tr\big\{\hat{\mathcal{G}}(\hat{\mathcal{R}}\mathbf{v}_{F}
\partial_{\mathbf{r}}\hat{\mathcal{R}}^{-1})\hat{\mathcal{G}}
(\hat{\mathcal{R}}\mathbf{v}_{F}\hat{\mathbf{A}}\hat{\mathcal{R}}^{-1})\big\}\,.
\end{equation}
In  view of
$\sum_{p}\hat{\mathcal{G}}(\mathbf{p},\epsilon)=-i\pi\nu\hat{\Lambda}_{\epsilon}$,
which follows from the saddle point
Equation~\eqref{NLSM-saddle-point-eq}, for the first term on the
right--hand side of~\eqref{app-GE-S1-trlog} one finds, using cyclic
property of trace $\Tr\big\{\hat{\mathcal{G}}
\hat{\mathcal{R}}\hat{V}\hat{\mathcal{R}}^{-1}\!\big\}\!\!=\!
-i\pi\nu\Tr\big\{\hat{\mathcal{R}}^{-1}\hat{\Lambda}\hat{\mathcal{R}}\hat{V}\big\}\!\!=
-i\pi\nu\Tr\big\{\hat{V}\hat{Q}\big\}$. As to the second term on the
right--hand side of~\eqref{app-GE-S1-trlog}, retaining
retarded--advanced products of the Green functions
$\sum_{p}\mathcal{G}^{R}(\mathbf{p},\epsilon)\mathbf{v}_{F}
\mathcal{G}^{A}(\mathbf{p},\epsilon)\mathbf{v}_{F}=2\pi\nu D$, one
finds $$\Tr\big\{\hat{\mathcal{G}}(\hat{\mathcal{R}}\mathbf{v}_{F}
\partial_{\mathbf{r}}\hat{\mathcal{R}}^{-1})\hat{\mathcal{G}}
(\hat{\mathcal{R}}\mathbf{v}_{F}\hat{\mathbf{A}}\hat{\mathcal{R}}^{-1})\big\}=-
\pi\nu
D\Tr\big\{(\hat{\mathcal{R}}^{-1}\partial_{\mathbf{r}}\hat{\mathcal{R}}+
\hat{\mathcal{R}}^{-1}\hat{\Lambda}\hat{\mathcal{R}}\partial_{\mathbf{r}}
\hat{\mathcal{R}}^{-1}\hat{\Lambda}\hat{\mathcal{R}})\hat{\mathbf{A}}\big\}=
-\pi\nu
D\Tr\big\{\hat{\mathbf{A}}\hat{Q}\partial_{\mathbf{r}}\hat{Q}\big\}\,,$$
where
$\hat{\mathcal{R}}\circ\partial_{\mathbf{r}}\hat{\mathcal{R}}^{-1}=-
\partial_{\mathbf{r}}\hat{\mathcal{R}}\circ\hat{\mathcal{R}}^{-1}$ was used.
All together it gives for~\eqref{app-GE-S1-trlog}
\begin{equation}
iS_{1}[\hat Q,\mathbf{A},V]=-i\pi\nu\Tr\big\{\hat{V}\hat{Q}\big\}+i\pi\nu
D\Tr\big\{\hat{\mathbf{A}}\hat{Q}\partial_{\mathbf{r}}\hat{Q}\big\}\,.
\end{equation}

To the second order in $\hat{V}$ and $\hat{\mathbf{A}}$ one finds
\begin{equation}\label{app-GE-S2-trlog}
iS_{2}[\hat Q,\mathbf{A},V]=-\frac{1}{2}
\Tr\big\{\hat{\mathcal{G}}\hat{V}\hat{\mathcal{G}}\hat{V}\big\}-\frac{1}{2}
\Tr\big\{\hat{\mathcal{G}}(\hat{\mathcal{R}}\mathbf{v}_{F}
\hat{\mathbf{A}}\hat{\mathcal{R}}^{-1})
\hat{\mathcal{G}}(\hat{\mathcal{R}}\mathbf{v}_{F}
\hat{\mathbf{A}}\hat{\mathcal{R}}^{-1})\big\}\,.
\end{equation}
Note that in the term $\sim\hat{V}^2$ we took
$\hat{\mathcal{R}}=\hat{\mathcal{R}}^{-1}=\hat{1}$. This is because
$\sim\hat{V}^{2}$ contribution represents essentially static
compressibility of the electron gas which is determined by the
entire energy band, while $\hat{\mathcal{R}}$ and
$\hat{\mathcal{R}}^{-1}$ matrices are different from unit matrix
only in the narrow energy strip around the Fermi energy. Thus, for
the first term on the right--hand side of~\eqref{app-GE-S2-trlog}
one can write
$\Tr\big\{\hat{\mathcal{G}}\hat{V}\hat{\mathcal{G}}\hat{V}\big\}
=\Tr\big\{V^{\alpha}\hat{\Upsilon}^{\alpha\beta}V^{\beta}\big\}$,
where
\begin{equation}
\hat{\Upsilon}^{\alpha\beta}=-\frac{1}{2}\sum_{p}\int\frac{\d\epsilon}{2\pi}
\Tr\big\{\hat{\mathcal{G}}(\mathbf{p},\epsilon_{+})\hat{\gamma}^{\alpha}
\hat{\mathcal{G}}(\mathbf{p},\epsilon_{-})\hat{\gamma}^{\beta}\big\}\,,
\qquad \epsilon_{\pm}=\epsilon\pm\omega/2\,,
\end{equation}
and trace spans only over the Keldysh matrix structure.
Using~\eqref{NLSM-G-impurity-dressed} for the matrix Green function,
and retaining only retarded--retarded and advanced--advanced
products one finds
\begin{equation}
\hat{\Upsilon}^{\alpha\beta}=-\frac{1}{8}\sum_{p}\int\frac{\d\epsilon}{2\pi}
\Tr\left\{\big(\mathcal{G}^{R}\big)^{2}\big[\hat{1}+\hat{\Lambda}_{\epsilon_{+}}\big]
\hat{\gamma}^{\alpha}\big[\hat{1}+\hat{\Lambda}_{\epsilon_{-}}\big]\hat{\gamma}^{\beta}
+\big(\mathcal{G}^{A}\big)^{2}\big[\hat{1}-\hat{\Lambda}_{\epsilon_{+}}\big]
\hat{\gamma}^{\alpha}\big[\hat{1}-\hat{\Lambda}_{\epsilon_{-}}\big]\hat{\gamma}^{\beta}\right\}
=i\nu\hat{\sigma}^{\alpha\beta}_{x}\,.
\end{equation}
This result is derived noticing that
$\big[\mathcal{G}^{R(A)}(\mathbf{p},\epsilon)\big]^{2}=
-\partial_{\epsilon}\mathcal{G}^{R(A)}(\mathbf{p},\epsilon)$, and
integrating by parts
\begin{equation}
\int\d\epsilon\,
F_{\epsilon}\sum_{p}\left[\big[\mathcal{G}^{R}(\mathbf{p},\epsilon)\big]^{2}-
\big[\mathcal{G}^{A}(\mathbf{p},\epsilon)\big]^{2}\right]=
\int\d\epsilon\, \frac{\partial
F_{\epsilon}}{\partial\epsilon}\sum_{p}
\left[\mathcal{G}^{R}(\mathbf{p},\epsilon)-\mathcal{G}^{A}(\mathbf{p},\epsilon)\right]=
-4i\pi\nu\,,
\end{equation}
using $\sum_{p}\big( \mathcal{G}^{R}(\mathbf{p},\epsilon) -
\mathcal{G}^{A}(\mathbf{p},\epsilon)\big) = - 2\pi i\nu$ and
assuming that $F_{\epsilon\to\pm\infty}\to\pm1$. An additional
contribution to $\hat{\Upsilon}^{\alpha\beta}$, originating from the
retarded--advanced products of Green's functions, although
non--zero, contains an extra small factor $\omega\tauel\ll 1$, and
thus neglected.

For the second term on the right hand side
of~\eqref{app-GE-S2-trlog} one finds
$\Tr\big\{\hat{\mathcal{G}}(\hat{\mathcal{R}}\mathbf{v}_{F}
\hat{\mathbf{A}}\hat{\mathcal{R}}^{-1})
\hat{\mathcal{G}}(\hat{\mathcal{R}}\mathbf{v}_{F}
\hat{\mathbf{A}}\hat{\mathcal{R}}^{-1})\big\}= \pi\nu
D\Tr\big\{[\hat{1}+\hat{\Lambda}]
\hat{\mathcal{R}}\hat{\mathbf{A}}\hat{\mathcal{R}}^{-1}
[\hat{1}-\hat{\Lambda}]\hat{\mathcal{R}}\hat{\mathbf{A}}\hat{\mathcal{R}}^{-1}\big\}=\pi\nu
D\Tr\big\{\hat{\mathbf{A}}^2-\hat{\mathbf{A}}\hat{Q}\hat{\mathbf{A}}\hat{Q}\big\}$,
which finally gives for the $S_{2}[\hat Q,\mathbf{A},V]$ part of the
action
\begin{equation}
iS_{2}[\hat Q,\mathbf{A},V]=-\frac{\nu}{2}
\Tr\big\{\hat{V}\hat{\sigma}_{x}\hat{V}\big\}+
\frac{\pi\nu
D}{2}\Tr\big\{\hat{\mathbf{A}}\hat{Q}
\hat{\mathbf{A}}\hat{Q}-\hat{\mathbf{A}}^{2}\big\}\,.
\end{equation}
Combining now~\eqref{NLSM-action-noninteracting} together with
$S_{1}[\hat Q,\mathbf{A},V]$, and $S_{2}[\hat Q,\mathbf{A},V]$, and
taking into account that
$\Tr\big\{(\partial_{\mathbf{r}}\hat{Q})^{2}-
4i\hat{\mathbf{A}}\hat{Q}\partial_{\mathbf{r}}\hat{Q}-
2(\hat{\mathbf{A}}\hat{Q}
\hat{\mathbf{A}}\hat{Q}-\hat{\mathbf{A}}^{2})\big\}
=\Tr\big\{(\hat{\bm{\partial}}_{\mathbf{r}}\hat{Q})^{2}\big\}$,
where covariant derivative is defined according
to~\eqref{NLSM-covariant-deriv}, one finds  the  full action in the
form of~\eqref{NLSM-action-general}.

\section{Expansion over superconducting fluctuations}
\label{app_FluctuationExpansion}

In this section we provide details of the Gaussian integration over
the Cooper modes performed in~\eqref{SuperCond-W-integration}.
Throughout this section we suppress subscript--$\EuScript{K}$ in
$\check{Q}_{\EuScript{K}}$ and $\Delta_{\EuScript{K}}$ for brevity.
As a first step one expands~\eqref{SuperCond-S} in fluctuations
$\check{\mathcal{W}}$ around the metallic saddle point
$\check{Q}=\check{\Lambda}$: $S[\check Q,\Delta]\Rightarrow
S[\check{\mathcal{W}},\Delta]$. To this end, we take
$\check{\mathcal{W}}$ from~\eqref{SuperCond-W} and substitute it
into~(\ref{SuperCond-S}c). For the spatial gradient part of the
action $S_{\sigma}$ one finds in quadratic order
$\Tr\big\{\big(\partial_{\mathbf{r}}\check{Q}\big)^{2}\big\}=
\Tr\big\{\check{\mathcal{W}}_{\varepsilon\varepsilon'}
\partial^{2}_{\mathbf{r}}\check{\mathcal{W}}_{\varepsilon'\varepsilon}\big\}$. Tracing
the latter over Keldysh$\otimes$Nambu space gives
\begin{equation}\label{app_FE-S-expandsion-1}
D\,\Tr\big\{\big(\partial_{\mathbf{r}}\check{Q}\big)^{2}\big\}=
2\sum_{q}\iint\frac{\d\varepsilon\d\varepsilon'}{4\pi^{2}}\,Dq^{2}
[c^{*}_{\varepsilon\varepsilon'}(\mathbf{q})
c_{\varepsilon'\varepsilon}(-\mathbf{q})+
\bar{c}^{*}_{\varepsilon\varepsilon'}(\mathbf{q})
\bar{c}_{\varepsilon'\varepsilon}(-\mathbf{q})]\,,
\end{equation}
where we kept only Cooper modes $c$ and $\bar c$, while omitting the
diffuson modes $d$ and $\bar d$, since expansion for the latter was
already given in~\eqref{NLSM-S-W}.  For the time derivative term in
the action $S_{\sigma}$ one finds
$\Tr\{\check{\Xi}\partial_{t}\check{Q}\}=-\frac{i}{2}\Tr\{\varepsilon
(\hat{\sigma}_{z}\otimes\hat{\tau}_{0})\check{\mathcal{W}}_{\varepsilon\varepsilon'}
\check{\mathcal{W}}_{\varepsilon'\varepsilon}\}$, where we took
$\partial_{t}\to-i\varepsilon$ in the energy space. The latter,
after  evaluation of the trace reduces to
\begin{equation}\label{app_FE-S-expandsion-2}
\Tr\{\check{\Xi}\partial_{t}\check{Q}\}=
\frac{i}{2}\sum_{q}\iint\frac{\d\varepsilon\d\varepsilon'}{4\pi^2}\,
(\varepsilon+\varepsilon')[c^{*}_{\varepsilon\varepsilon'}(\mathbf{q})
c_{\varepsilon'\varepsilon}(-\mathbf{q})-
\bar{c}^{*}_{\varepsilon\varepsilon'}(\mathbf{q})
\bar{c}_{\varepsilon'\varepsilon}(-\mathbf{q})]\,.
\end{equation}
To the leading order in $\check{\mathcal{W}}$ the coupling term
between Cooper modes and the order parameter, $\Delta$, reads as
$\Tr\big\{\check{\Delta}\check{Q}\big\}=
\Tr\big\{\check{\mathcal{U}}_{\varepsilon}
\check{\Delta}_{\varepsilon-\varepsilon'}\check{\mathcal{U}}^{-1}_{\varepsilon'}
\big(\hat{\sigma}_{z}\otimes\hat{\tau}_{z}\big)\check{\mathcal{W}}_{\varepsilon'\varepsilon}\big\}
+O(\Delta \mathcal{W}^{2})$, where $\check{\mathcal{U}}$ is given by
Eq.~\eqref{SuperCond-W}. Evaluating traces,  one finds
\begin{equation}\label{app_FE-S-expandsion-3}
\Tr\big\{\check{\Delta}\check{Q}\big\}=
\sum_{q}\iint\frac{\d\varepsilon\d\varepsilon'}{4\pi^2}\, \big[
\mathbf{\Delta}^{c}_{\varepsilon\varepsilon'}(\mathbf{q})
c^{*}_{\varepsilon'\varepsilon}(-\mathbf{q})-
\mathbf{\Delta}^{\bar{c}}_{\varepsilon\varepsilon'}(\mathbf{q})
\bar{c}^{*}_{\varepsilon'\varepsilon}(-\mathbf{q})-c.c.\big]\,,
\end{equation}
where the following form factors were introduced
\begin{equation}
\mathbf{\Delta}^{c}_{\varepsilon\varepsilon'}(\mathbf{q})=
\Delta^{cl}(\mathbf{q},\varepsilon-\varepsilon')+F_{\varepsilon}
\Delta^{q}(\mathbf{q},\varepsilon-\varepsilon'),\qquad
\mathbf{\Delta}^{\bar{c}}_{\varepsilon\varepsilon'}(\mathbf{q})=
\Delta^{cl}(\mathbf{q},\varepsilon-\varepsilon')-F_{\varepsilon'}
\Delta^{q}(\mathbf{q},\varepsilon-\varepsilon')\,.
\end{equation}
It is important to emphasize, that the diffusion modes
$\{d,\bar{d}\}$ couple to $\Delta$ only starting from the quadratic
order  in $\check{\mathcal{W}}$. These terms  produce non--local and
non--linear interaction vertices between the order parameter
components and will not be considered here,
see~\cite{LevchenkoKamenev} for more details. Combining
now~\eqref{app_FE-S-expandsion-1}--\eqref{app_FE-S-expandsion-3},
one finds for the quadratic part of the action $S_{\sigma}[
\check{\mathcal{W}},\Delta]=S^{c}_{\sigma}[\check{\mathcal{W}},\Delta]+
S^{\bar{c}}_{\sigma}[\check{\mathcal{W}},\Delta]$, where
\begin{subequations}\label{app_FE-S-expansion-Gaussian}
\begin{equation}
iS^{c}_{\sigma}[\check{\mathcal{W}},\Delta]=-\frac{\pi\nu}{2}
\tr\Big\{c^{*}_{\varepsilon\varepsilon'}(\mathbf{q})
[Dq^{2}-i(\varepsilon+\varepsilon')]c_{\varepsilon'\varepsilon}(-\mathbf{q})
+2i\mathbf{\Delta}^{\!c}_{\varepsilon\varepsilon'}(\mathbf{q})
c^{*}_{\varepsilon'\varepsilon}(-\mathbf{q})-
2i\mathbf{\Delta}^{*c}_{\varepsilon\varepsilon'}(\mathbf{q})
c_{\varepsilon'\varepsilon}(-\mathbf{q})\Big\}\,,
\end{equation}
\begin{equation}
iS^{\bar{c}}_{\sigma}[\check{\mathcal{W}},\Delta]=-\frac{\pi\nu}{2}
\tr \Big\{\bar{c}^{*}_{\varepsilon\varepsilon'}(\mathbf{q})
[Dq^{2}+i(\varepsilon+\varepsilon')]\bar{c}_{\varepsilon'\varepsilon}(-\mathbf{q})
-2i\mathbf{\Delta}^{\!\bar{c}}_{\varepsilon\varepsilon'}(\mathbf{q})
\bar{c}^{*}_{\varepsilon'\varepsilon}(-\mathbf{q})+
2i\mathbf{\Delta}^{*\bar{c}}_{\varepsilon\varepsilon'}(\mathbf{q})
\bar{c}_{\varepsilon'\varepsilon}(-\mathbf{q})\Big\}\,,
\end{equation}
\end{subequations}
and traces stand for energy and momentum integrations $\tr =
\sum_{q}\iint\frac{\d\varepsilon\d\varepsilon'}{4\pi^2}$. At this
stage, one is prepared to perform Gaussian integration over the
Cooper modes $c$ and $\bar c$. Quadratic forms
in~\eqref{app_FE-S-expansion-Gaussian} are extremized by
\begin{equation}\label{app_FE-c-extremal}
c_{\varepsilon\varepsilon'}(\mathbf{q})=
\frac{-2i\mathbf{\Delta}^{c}_{\varepsilon\varepsilon'}(\mathbf{q})}
{Dq^{2}-i(\varepsilon+\varepsilon')},\qquad
\bar{c}_{\varepsilon\varepsilon'}(\mathbf{q})=
\frac{2i\mathbf{\Delta}^{\bar{c}}_{\varepsilon\varepsilon'}(\mathbf{q})}
{Dq^{2}+i(\varepsilon+\varepsilon')}\,.
\end{equation}
Similar equations for the conjugated fields,  are obtained
from~\eqref{app_FE-c-extremal} by replacing $\Delta\to\Delta^{*}$
and flipping an overall sign. The Gaussian integral
$\int\D[\check{\mathcal{W}}]\exp(iS_{\sigma}[\check{\mathcal{W}},\Delta])=
\exp(iS_{\sigma}[\Delta])$, where $S_{\sigma}[\Delta]$ is calculated
on the extremum~\eqref{app_FE-c-extremal}:
\begin{equation}\label{app_FE-S-Delta}
iS_{\sigma}[\Delta]=4\pi\nu\sum_{q}\iint\frac{\d\epsilon\d\omega}{4\pi^2}\,
\frac{\big[\Delta^{cl}_{+}+ F_{\epsilon_{+}}\Delta^{q}_{+}\big]
[\Delta^{*cl}_{-}+
F_{\epsilon_{-}}\Delta^{*q}_{-}]}{Dq^{2}-2i\epsilon}\,,
\end{equation}
where $\Delta^{cl(q)}_{\pm}=\Delta^{cl(q)}(\pm\mathbf{q},\pm\omega)$
and $\epsilon_{\pm}=\epsilon\pm\omega/2$. We have also introduced
new integration variables $\omega=\varepsilon-\varepsilon'$,
$\epsilon=(\varepsilon+\varepsilon')/2$ and employed the fact that
$F_\epsilon$ is an odd function to change  variables as $\epsilon\to
-\epsilon$ in the contribution coming from $\bar c$   fields. The
contribution to $iS_{\sigma}[\Delta]$ with the two classical
components of the order parameter
$\sim\Delta^{cl}_{+}\Delta^{*cl}_{-}$ vanishes identically after the
$\epsilon$--integration as being an integral of the purely retarded
function. This is nothing, but manifestation of the normalization
condition for the Keldysh--type action (see
Section~\ref{sec_bosons-3} for discussions). Adding to
$iS_{\sigma}[\Delta]$ zero in the form
$-4\pi\nu\,\tr\big\{\Delta^{q}_{+}\Delta^{*q}_{-}/[Dq^{2}-2i\epsilon]\big\}$,
which vanishes after $\epsilon$ integration by causality, and
combining~\eqref{app_FE-S-Delta} with $S_{\Delta}$ from
Eq.~(\ref{SuperCond-S}b), one finds for
$S_{\mathrm{GL}}[\Delta]=S_{\sigma}[\Delta]+S_{\Delta}[\Delta]$ the
following result
\begin{equation}\label{app_FE-S-GL}
S_{\mathrm{GL}}[\Delta]=2\nu\sum_{q}\int\frac{\d\omega}{2\pi}
\left[\Delta^{*q}_{-}L^{-1}_{R}\Delta^{cl}_{+}+
\Delta^{*cl}_{-}L^{-1}_{A}\Delta^{q}_{+}
+\Delta^{*q}_{-}B_{\omega}[L^{-1}_{R}-L^{-1}_{A}]\Delta^{q}_{+}\right]\,,
\end{equation}
where superconducting fluctuations propagator is given by the
integral
\begin{equation}\label{app_FE-L-integral}
L^{-1}_{R(A)}(\mathbf{q},\omega)=-\frac{1}{\lambda}-i\int\d\epsilon
\,\, \frac{F_{\epsilon\mp\omega/2}}{Dq^{2}-2i\epsilon}\,.
\end{equation}
This expression for $L(\mathbf{q},\omega)$ can be reduced to the
more familiar form. Indeed, adding and subtracting right--hand side
of~\eqref{app_FE-L-integral} taken at zero frequency and momentum
one writes
\begin{equation}\label{app_FE-L-intermediate}
L^{-1}_{R}(\mathbf{q},\omega)=-\frac{1}{\lambda}+\int^{+\omega_{D}}_{-\omega_{D}}
\mathrm{d}\varepsilon\,\frac{F_{\varepsilon}}{2\varepsilon}-
i\int^{+\infty}_{-\infty}\mathrm{d}\varepsilon\left[
\frac{F_{\varepsilon}}{Dq^{2}-i\omega-2i\varepsilon}+
\frac{F_{\varepsilon}}{2\varepsilon}\right]\,,
\end{equation}
where the second term on the right--hand side is the logarithmically
divergent integral which is to be cut in the standard way by the
Debye frequency $\omega_{D}$. Introducing dimensionless variable
$x=\varepsilon/2T$, and performing the integration in the  last term
on the right--hand side of~\eqref{app_FE-L-intermediate} by parts
with the help of the identity $\int^{\infty}_{0}\mathrm{d}x\ln(x)
\mathrm{sech}^{2}(x)=-\ln\frac{4\gamma}{\pi}$, where
$\gamma=e^{\mathbb{C}}$ with $\mathbb{C}=0.577$ is the Euler
constant, and using the definition of the superconductive transition
temperature $T_{c}=(2\gamma\omega_{D}/\pi)\exp(-1/\lambda\nu)$, one
finds for~\eqref{app_FE-L-intermediate}
\begin{equation}
L^{-1}_{R}(\mathbf{q},\omega)=\ln\frac{T_{c}}{T}-\frac{i}{2}
\int^{+\infty}_{-\infty}\mathrm{d}x
\left[\frac{\tanh(x)}{\frac{Dq^{2}
-i\omega}{4T}-ix}+\frac{\tanh(x)}{ix}\right].
\end{equation}
With the help of the expansion
\begin{equation}\label{tanh-series}
\tanh(x)=\sum^{\infty}_{n=0}\frac{2x}{x^{2}+x^{2}_{n}},\quad
x_{n}=\pi(n+1/2),
\end{equation}
one may perform the $x$--integration explicitly interchanging the
order of summation and integration
\begin{eqnarray}
\int^{+\infty}_{-\infty}\frac{\mathrm{d}x}{x^{2}+x^{2}_{n}}=
\frac{\pi}{x_{n}},\quad\quad
\int^{+\infty}_{-\infty}\frac{x\mathrm{d}x}{\left[x^{2}+x^{2}_{n}\right]
\left[\frac{Dq^{2}-i\omega}{4T}-ix\right]}=\frac{i\pi}
{\frac{Dq^{2}-i\omega}{4T}+x_{n}}\,.
\end{eqnarray}
Recalling now the definition of the digamma function
\begin{equation}
\psi(x)=-\mathbb{C}-\sum^{\infty}_{n=0}\left[\frac{1}{n+x}-
\frac{1}{n+1}\right]\,,
\end{equation}
one transforms~\eqref{app_FE-L-integral} to the final result
\begin{equation}\label{app_FE-L-R}
L^{-1}_{R}(\mathbf{q},\omega)=\ln\frac{T_{c}}{T}-\psi\left(
\frac{Dq^{2}-i\omega}{4\pi T}+\frac{1}{2}\right)+
\psi\left(\frac{1}{2}\right)\approx-\frac{\pi}{8T}
\Big(Dq^2+\tau^{-1}_{\mathrm{GL}}-i\omega\Big)\,,
\end{equation}
where $\tau_{\mathrm{GL}}^{-1}=8(T-T_c)/\pi$. Since according to the
last expression $Dq^2\sim \omega\sim \tau_{\mathrm{GL}}^{-1}\ll T$,
the expansion of the digamma function is justified.

As a result, the time dependent Ginzburg--Landau part of the
effective action~\eqref{SuperCond-S-eff} is obtained
(compare~\eqref{app_FE-S-GL} and (\ref{app_FE-L-R})
with~\eqref{SuperCond-S-GL}). The non--linear contribution $\sim
|\Delta|^2$ in~\eqref{SuperCond-L-R-A} can be restored once
$\sim\Tr\{\check{\mathcal{W}}^{3}\check{\Delta}\}$ is kept in the
expansion of $\Tr\{\check{Q}\check{\Delta}\}$ part of the action.
Furthermore, for $Dq^{2}\to-D\partial^{2}_{\mathbf{r}}$
in~\eqref{app_FE-L-R}, one actually has
$D\big(\partial_{\mathbf{r}}-2ie\mathbf{A}^{cl}_{\EuScript{K}}\big)^{2}$,
once the vector potential is kept explicitly in the action.

Let us comment now on the origin of the other terms in the effective
action~\eqref{SuperCond-S-eff}. The supercurrent part of the action
$S_{SC}$ emerges from the
$\Tr\big\{\partial_{\mathbf{r}}\check{Q}_{\EuScript{K}}
[\check{\Xi}\check{\mathbf{A}}_{\EuScript{K}},
\check{Q}_{\EuScript{K}}]\big\}$ upon second order expansion over
the Cooper modes, namely
\begin{equation}\label{app_FE-S-SC}
S_{\mathrm{SC}}[\Delta,\mathbf{A},\Phi]=\frac{i\pi\nu}{4}\Tr\big\{c^{*}_{tt'}(\mathbf{r})
\EuScript{N}^{\mathrm{SC}}_{tt'}c_{t't}(\mathbf{r})+\bar{c}^{*}_{tt'}(\mathbf{r})
\EuScript{N}^{\mathrm{SC}}_{tt'}\bar{c}_{t't}(\mathbf{r})\big\}\,,
\end{equation}
where
\begin{equation}\label{app_FE-N-SC}
\EuScript{N}^{\mathrm{SC}}_{tt'}=-\delta_{t-t'}\frac{2eD}{T}\left[
\frac{1}{2}\mathrm{div}\mathbf{A}^{q}_{\EuScript{K}}\rt+\mathbf{A}^{q}_{\EuScript{K}}\rt
\big[\partial_{\mathbf{r}}-2ie\mathbf{A}^{cl}_{\EuScript{K}}\rt\big]\right]\,.
\end{equation}
Deriving $\EuScript{N}^{\mathrm{SC}}_{tt'}$ one uses an
approximation for the equilibrium Fermi function
\begin{equation}\label{app_FE-F-approx}
F_t= -\frac{iT}{\sinh(\pi T t)} \stackrel{t\gg 1/T}{\longrightarrow}
{i\,\over 2T}\,\, \delta\,'(t)\, ,
\end{equation}
which is applicable for slowly varying external fields. Performing
integration over the Cooper modes one
substitutes~\eqref{app_FE-c-extremal} into~\eqref{app_FE-S-SC}.
Noticing that in the real--space
representation~\eqref{app_FE-c-extremal} reads as
\begin{subequations}\label{app_FE-c-extremal-realspace}
\begin{equation}
\hskip-.5cm
c_{tt'}(\mathbf{r})=-i\theta(t-t')\Delta^{cl}_{\EuScript{K}}\left(\mathbf{r},\frac{t+t'}{2}\right)
+\chi(t-t')\Delta^{q}_{\EuScript{K}}\left(\mathbf{r},\frac{t+t'}{2}\right)\,,
\end{equation}
\begin{equation}
\hskip-.5cm
\bar{c}_{tt'}(\mathbf{r})=i\theta(t-t')\Delta^{cl}_{\EuScript{K}}\left(\mathbf{r},\frac{t+t'}{2}\right)
-\chi(t-t')\Delta^{q}_{\EuScript{K}}\left(\mathbf{r},\frac{t+t'}{2}\right)\,,
\end{equation}
\begin{equation}
\chi(t)=\int^{+\infty}_{-\infty}\frac{\d\epsilon}{2\pi}\tanh\left(\frac{\epsilon}{2T}\right)
\frac{e^{-i\epsilon
t}}{\epsilon+i0}=\frac{2}{\pi}\mathrm{arctanh}\big(\exp(-\pi
T|t|)\big)\,,
\end{equation}
\end{subequations}
and keeping contributions only with the classical components of
fluctuating order parameter, since $\EuScript{N}^{\mathrm{SC}}$ is
already linear in quantum field $\mathbf{A}^{q}_{\EuScript{K}}$, one
can perform $t'$ integration in~\eqref{app_FE-S-SC} explicitly and
recover $S_{\mathrm{SC}}$ in the form given
by~\eqref{SuperCond-S-SC}.

The Maki--Thompson part of the effective action $S_{\mathrm{MT}}$
emerges from
$\Tr\big\{([\check{\Xi}\check{\mathbf{A}}_{\EuScript{K}},\check{Q}_{\EuScript{K}}])^{2}\big\}$
when each $\check{Q}_{\EuScript{K}}$ matrix is expanded to the first
order in fluctuations $\check{\mathcal{W}}$:
\begin{equation}\label{app_FE-S-MT}
S_{\mathrm{MT}}[\Delta,\mathbf{A},\Phi]=-\frac{\pi\nu}{4}\Tr\big\{c^{*}_{tt'}(\mathbf{r})
\EuScript{N}^{\mathrm{MT}}_{tt'}\bar{c}_{t't}(\mathbf{r})+
\bar{c}^{*}_{tt'}(\mathbf{r})\EuScript{N}^{\mathrm{MT}}_{tt'}c_{t't}(\mathbf{r})\big\}\,,
\end{equation}
where
\begin{equation}\label{app_FE-N-MT}
\EuScript{N}^{\mathrm{MT}}_{tt'}=-2e^{2}D\left[\mathbf{A}^{q}_{\EuScript{K}}\rt+\frac{i}{2T}\partial_{t}
\mathbf{A}^{cl}_{\EuScript{K}}\rt\right]\mathbf{A}^{q}_{\EuScript{K}}(\mathbf{r},t')\,,
\end{equation}
and we again used~\eqref{app_FE-F-approx}. With the help
of~\eqref{app_FE-c-extremal-realspace} one should perform now
integration over Cooper modes in~\eqref{app_FE-S-MT}. Observe,
however, that in contrast to~\eqref{app_FE-S-SC}, where we had
product of either two retarded or two advanced Cooperon fields,
which restricted integration over one of the time variables, in the
case of MT contribution \eqref{app_FE-S-MT}, we end up with the
product between one retarded and one advanced Cooperon and the time
integration running over the entire range $t>t'$. Precisely, this
difference between~\eqref{app_FE-S-SC} and \eqref{app_FE-S-MT} makes
contribution $S_{\mathrm{SC}}$ to be local, while $S_{\mathrm{MT}}$
non--local. Finally, in each of the Cooperon fields
$c,\bar{c}$,~\eqref{app_FE-S-MT}, one keeps only contribution with
the classical component of the order parameter and recovers
$S_{\mathrm{MT}}$ in the form given by~\eqref{SuperCond-S-MT}.

The remaining  density of states part of the effective action
$S_{\mathrm{DOS}}$ emerges, similarly to $S_{\mathrm{MT}}$, from
$\Tr\big\{([\check{\Xi}\check{\mathbf{A}}_{\EuScript{K}},\check{Q}_{\EuScript{K}}])^{2}\big\}$.
This time one of the $\check{Q}_{\EuScript{K}}$ matrices is kept at
the saddle point $\check{\Lambda}$, while another is expanded to the
second order in $\check{\mathcal{W}}$:
\begin{equation}\label{app_FE-S-DOS}
S_{\mathrm{DOS}}[\Delta,\mathbf{A},\Phi]=\frac{i\pi\nu}{4}\Tr\big\{c^{*}_{tt'}(\mathbf{r})
\EuScript{N}^{\mathrm{DOS}}_{tt't''}c_{t't''}(\mathbf{r})+\bar{c}^{*}_{tt'}(\mathbf{r})
\EuScript{N}^{\mathrm{DOS}}_{tt't''}\bar{c}_{t't''}(\mathbf{r})\big\}\,,
\end{equation}
where
\begin{equation}\label{app_FE-N-DOS}
\EuScript{N}^{\mathrm{DOS}}_{tt't''}=2e^{2}D\left[\mathbf{A}^{q}_{\EuScript{K}}\rt
\big[\mathbf{A}^{cl}_{\EuScript{K}}\rt-\mathbf{A}^{cl}_{\EuScript{K}}(\mathbf{r},t'')\big]F_{t-t''}
+\int\d t'''\mathbf{A}^{q}_{\EuScript{K}}\rt
F_{t-t''}\mathbf{A}^{q}_{\EuScript{K}}(\mathbf{r},t''')F_{t'''-t''}\right]\,.
\end{equation}
It is important to emphasize here, that as compared
to~\eqref{app_FE-N-SC} and~\eqref{app_FE-N-MT}, when deriving
$\EuScript{N}^{\mathrm{DOS}}$ it is \textit{not} sufficient to take
the approximate form of the distribution
function~\eqref{app_FE-F-approx}, but rather one should keep full
$F_{t}$. In what follows, we deal with the part of the action
\eqref{app_FE-S-DOS} having one classical and one quantum components
of the vector potential. The other one, having two quantum fields
can be restored via FDT. To this end, we substitute  Cooperon
generators in the form \eqref{app_FE-c-extremal-realspace} into the
action \eqref{app_FE-S-DOS}. We keep only classical components of
$\Delta_{\EuScript{K}}$ (the quantum one produce insignificant
contributions) and account for an additional factor of 2 due to
identical contributions from $c$ and $\bar{c}$ Cooperons. Changing
time integration variables $t-t''=\tau$ and $t+t''=2\eta$, one finds
\begin{eqnarray}
\hskip-.5cm
S_{\mathrm{DOS}}[\Delta,\mathbf{A},\Phi]\!\!\!\!&=&\!\!\!\!i\pi
e^{2}\nu D\ \mathrm{Tr} \left[
\mathbf{A}^{q}_{\EuScript{K}}(\mathbf{r},\eta+\tau/2)
[\mathbf{A}^{cl}_{\EuScript{K}}(\mathbf{r},\eta+\tau/2)-
\mathbf{A}^{cl}_{\EuScript{K}}(\mathbf{r},\eta-\tau/2)]
F_{\tau}\right.\nonumber
\\ &\times&\!\!\!\! \left.\theta(\eta+\tau/2-t')\theta(t'-\eta+\tau/2)
\Delta^{*cl}_{\EuScript{K}}
\left(\mathbf{r},\frac{\eta+\tau/2-t'}{2}\right)
\Delta^{cl}_{\EuScript{K}}
\left(\mathbf{r},\frac{\eta-\tau/2-t'}{2}\right)\right] \,.
\end{eqnarray}
Note that owing to the step functions, integration over $t'$ is
restricted to be in the range $\eta+\tau/2>t'>\eta-\tau/2$. Since
$F_{\tau}$ is a rapidly decreasing function of its argument, the
main contribution to the $\tau$ integral comes from the range
$\tau\sim 1/T\ll\eta$. Keeping this in mind, one makes use of the
following approximations:
$\mathbf{A}^{q}_{\EuScript{K}}(\mathbf{r},\eta+\tau/2)
[\mathbf{A}^{cl}_{\EuScript{K}}(\mathbf{r},\eta+\tau/2)-
\mathbf{A}^{cl}_{\EuScript{K}}(\mathbf{r},\eta-\tau/2)]\approx\tau
\mathbf{A}^{q}_{\EuScript{K}}(\mathbf{r},\eta)\partial_{\eta}
\mathbf{A}^{cl}_{\EuScript{K}}(\mathbf{r},\eta)$ and
$\Delta^{*cl}_{\EuScript{K}}
\left(\mathbf{r},\frac{\eta+\tau/2-t'}{2}\right)
\Delta^{cl}_{\EuScript{K}}
\left(\mathbf{r},\frac{\eta-\tau/2-t'}{2}\right)
\approx|\Delta^{cl}_{\EuScript{K}}(\mathbf{r},\eta)|^{2}$, which
allows to integrate over $t'$ explicitly
$\int\mathrm{d}t'\theta(\eta+\tau/2-t')\theta(t'-\eta+\tau/2)=\tau\theta(\tau)$.
Using fermionic distribution function \eqref{app_FE-F-approx} and
collecting all factors, we find
\begin{equation}
S_{\mathrm{DOS}}[\Delta,\mathbf{A},\Phi]=\pi e^{2}\nu DT\,
\mathrm{Tr}\left[ \mathbf{A}^{q}_{\EuScript{K}}\rt
\partial_{t}\mathbf{A}^{cl}_{\EuScript{K}}\rt
|\Delta^{cl}_{\EuScript{K}}\rt|^{2}\right] \int^{\infty}_{0}
\frac{\tau^{2}\mathrm{d}\tau}{\sinh(\pi T\tau)}
\end{equation}
where we set $\eta\rightarrow t$. Performing remaining integration
over $\tau$ and restoring
$S_{\mathrm{DOS}}\sim\mathbf{A}^{q}_{\EuScript{K}}\mathbf{A}^{q}_{\EuScript{K}}$
via FDT, we arrive at $S_{\mathrm{DOS}}$ in the form given
by~\eqref{SuperCond-S-DOS}. Additional details of the derivation of
the effective action~\eqref{SuperCond-S-eff} can be found
in~\cite{LevchenkoKamenev}.


\end{document}